\def\spacingNumerator{5}
\def\spacingDenominator{4}

\def\ifundefined#1{\expandafter\ifx\csname#1\endcsname\relax}
\ifundefined{ftmagnification}  \def\ftmagnification{1200} \fi
\ifundefined{spacingNumerator}  \def\spacingNumerator{5} \fi
\ifundefined{spacingDenominator}  \def\spacingDenominator{4} \fi


\magnification\ftmagnification
\tolerance=10000
\hsize=17truecm\vsize=23truecm

\parindent=40pt
\mathsurround=0pt
     \multiply\baselineskip by \spacingNumerator
     \divide \baselineskip by \spacingDenominator 

%
%
\def\today{\ifcase\month\or January\or February\or March\or April\or
     May\or June\or July\or August\or September\or October\or November\or
     December\fi\space\number\day, \number\year}
%
%
\def\dst{\displaystyle}
\def\sst{\scriptstyle}
\def\tst{\textstyle}
\def\ssst{\scriptscriptstyle}
%
%
\def\frac#1#2{\dst {#1\over#2}}     
\def\sfrac#1#2{{\tst{#1\over#2}}}   

\def\deqalign#1{\vcenter{\openup1\jot \mathsurround=0pt \ialign{
                \strut\hfil$\displaystyle{##}$&&$\displaystyle{{}##}$\hfil
                \crcr
                #1\crcr}}}         

\def\meqalign#1{\vcenter{\openup1\jot \mathsurround=0pt \ialign{
                &\strut\hfil$\displaystyle{##}$&$\displaystyle{{}##}$\hfil&
                \quad$##$\crcr
                #1\crcr}}}         

%
%
\def\al{\alpha}
\def\be{\beta}
\def\ga{\gamma}
\def\de{\delta}

\def\ze{\zeta}
\def\et{\eta}

\def\ka{\kappa}

\def\si{\sigma}

\def\ch{\chi}

\def\Th{\Theta}
\def\La{\Lambda}
\def\Si{\Sigma}

\def\Om{\Omega}   
%
%
\def\pmb#1{\setbox0=\hbox{#1}       
     \kern-.025em\copy0\kern-\wd0
     \kern.05em\copy0\kern-\wd0
     \kern-.025em\box0}             
\def\0{{\bf 0}}

\def\k{{\bf k}}

\def\t{{\bf t}}

\def\x{{\bf x}}

\def\cB{{\cal B}}

\def\cF{{\cal F}}
\def\cG{{\cal G}}

%
%
\font\tenfrak                 = eufm10
\font\sevenfrak               = eufm7
\font\fivefrak                = eufb5
\newfam\frakfam
     \textfont\frakfam=\tenfrak
     \scriptfont\frakfam=\sevenfrak   
     \scriptscriptfont\frakfam=\fivefrak
\def\frak{\fam\frakfam\tenfrak}
\font \tensans                = cmss10
\font \fivesans               = cmss10 at 5pt
\font \sevensans              = cmss10 at 7pt
\newfam\sansfam
     \textfont\sansfam=\tensans
     \scriptfont\sansfam=\sevensans
     \scriptscriptfont\sansfam=\fivesans
\def\sans{\fam\sansfam\tensans}
%
%
\def\bbbr{{\rm I\!R}}  
\def\bbbn{{\rm I\!N}}

\def\bbbc{{\mathchoice {\setbox0=\hbox{$\displaystyle\rm C$}\hbox{\hbox 
to0pt{\kern0.4\wd0\vrule height0.9\ht0\hss}\box0}}
{\setbox0=\hbox{$\textstyle\rm C$}\hbox{\hbox
to0pt{\kern0.4\wd0\vrule height0.9\ht0\hss}\box0}}
{\setbox0=\hbox{$\scriptstyle\rm C$}\hbox{\hbox
to0pt{\kern0.4\wd0\vrule height0.9\ht0\hss}\box0}}
{\setbox0=\hbox{$\scriptscriptstyle\rm C$}\hbox{\hbox
to0pt{\kern0.4\wd0\vrule height0.9\ht0\hss}\box0}}}}
\def\bbbq{{\mathchoice {\setbox0=\hbox{$\displaystyle\rm               
Q$}\hbox{\raise
0.15\ht0\hbox to0pt{\kern0.4\wd0\vrule height0.8\ht0\hss}\box0}}
{\setbox0=\hbox{$\textstyle\rm Q$}\hbox{\raise
0.15\ht0\hbox to0pt{\kern0.4\wd0\vrule height0.8\ht0\hss}\box0}}
{\setbox0=\hbox{$\scriptstyle\rm Q$}\hbox{\raise
0.15\ht0\hbox to0pt{\kern0.4\wd0\vrule height0.7\ht0\hss}\box0}}
{\setbox0=\hbox{$\scriptscriptstyle\rm Q$}\hbox{\raise
0.15\ht0\hbox to0pt{\kern0.4\wd0\vrule height0.7\ht0\hss}\box0}}}}
\def\bbbz{{\mathchoice {\hbox{$\sans\textstyle Z\kern-0.4em Z$}}       
{\hbox{$\sans\textstyle Z\kern-0.4em Z$}}
{\hbox{$\sans\scriptstyle Z\kern-0.3em Z$}}
{\hbox{$\sans\scriptscriptstyle Z\kern-0.2em Z$}}}}
%
%
\def\const{{\rm const}\,}
\def\sgn{{\rm sgn}}
\def\half{\sfrac{1}{2}}

\def\optbar#1{\vbox{\ialign{##\crcr\hfil${\scriptscriptstyle(}\mkern -1mu
         \vrule height 1.2pt width 3pt depth -.8pt
         {\scriptscriptstyle)}$\hfil\crcr
          \noalign{\kern-1pt\nointerlineskip}$\hfil\displaystyle{#1}\hfil$\crcr}}}
\def\<{\left<}
\def\>{\right>}

\def\smprod{\mathop{\textstyle\prod}}
\def\smsum{\mathop{\textstyle\sum}}
\def\set#1#2{\big\{ \ #1\ \big|\ #2\ \big\}}
\def\eval#1{\big|\lower4pt\hbox{$\displaystyle\sst #1$}}
%
%
\font \tafontt                = cmbx10 scaled\magstep2
\font \tbfontt                = cmbx10 scaled\magstep1
\def\titlea#1{\centerline{\tafontt #1 }\vskip.5truein}
\def\titleb#1{\removelastskip\vskip.3truein%
\noindent{\tbfontt #1 }\vskip.25truein}

%
%
\def\newenvironment#1#2#3#4{\long\def#1##1##2{%
\removelastskip\penalty-100\vskip\baselineskip%
\noindent{#3#2\if!##1!.\else\unskip\ \ignorespaces
##1\unskip\fi\ }{#4\ignorespaces##2\vskip\baselineskip}}}
\newenvironment\lemma{Lemma}{\bf}{\it}
\newenvironment\proposition{Proposition}{\bf}{\it}
\newenvironment\theorem{Theorem}{\bf}{\it}
\newenvironment\corollary{Corollary}{\bf}{\it}
\newenvironment\example{Example}{\bf}{\rm}
\newenvironment\problem{Problem}{\bf}{\rm}
\newenvironment\definition{Definition}{\bf}{\rm}
\newenvironment\remark{Remark}{\bf}{\rm}
\newenvironment\hypothesis{Hypothesis}{\bf}{\it}
\newenvironment\convention{Convention}{\bf}{\it}

\def\Item{\vskip.1in\noindent}

%
%
\long\def\proof#1{\removelastskip\penalty-100\vskip\baselineskip\noindent{\bf
            Proof\if!#1!\else\ \ignorespaces#1\fi:\ }\ \ \ignorespaces}
\long\def\prf{\removelastskip\penalty-100\vskip\baselineskip\noindent{\bf
            Proof:\ }\ \ \ignorespaces}
\def\endproof{\hfill\vrule height .6em width .6em depth 0pt\goodbreak\vskip.25in }

\ifundefined{warnForwardRef}  \def\warnForwardRef{n} \fi
\newcount\chapno
\newcount\sectno
\newcount\equano
\newcount\theono
\newcount\probno

\def\IgNoRe#1{}

\chapno=0
\sectno=0
\equano=0
\theono=0
\probno=0
\def\eqhead{}
\def\frefwarning{\if\warnForwardRef y\immediate\write16{   Forward reference on line \the\inputlineno}\fi}
\def\qqqrefwarning{\immediate\write16{   ??? reference on line \the\inputlineno}}

\def\chap#1{\equano=0\sectno=0\theono=0\probno=0\global\advance\chapno by 1%
\def\eqhead{\ifcase\chapno\or I\or II\or III\or IV\or V\or VI\or VII\or
VIII\or IX\or X\or XI\or XII\or XIII\or XIV\or XV\or XVI\or XVII\or XVIII\or
XIX\or XX\or XXI\or XXII\or XXIII\or XXIV\or XXV\or XXVI\or XXVII\or XXVIII\or XXIX\or XXX\or XXXI\or XXXII\or XXXIII\or XXXIV\or XXXV\or XXXVI\or XXXVII\or XXXVIII\or XXXIX\fi.}%
\titlea{\eqhead \hglue 5pt #1}%
}

\def\sect#1{\global\advance\sectno by 1%
\titleb{\eqhead\number\sectno  \hglue 5pt #1}%
}%

\def\appendix#1#2{\equano=0\sectno=0\theono=0\probno=0\def\eqhead{#1.}
\titlea{Appendix #1: #2}%
}

\def\:#1{\def\temp{\expandafter\IgNoRe\string#1}%
\expandafter\ifx\csname\temp\endcsname\relax%
\expandafter\gdef#1{\qqqrefwarning ???}\fi#1}

\def\Eqn{{\hbox{\global\advance\equano by 1}}%
\eqno ({\rm \eqhead\number\equano})}%

\def\Eqno{{\hbox{\global\advance\equano by 1}}%
 ({\rm \eqhead\number\equano})}%

\def\EQN#1{\Eqn\edef\Zwi{\eqhead\number\equano}%
\global\let #1=\Zwi
}

\def\EQNO#1{\Eqno\edef\Zwi{\eqhead\number\equano}%
\global\let #1=\Zwi
}

\def\STM#1{{\global\advance \theono by 1}%
\eqhead\number\theono
\edef\Zwi{\eqhead\number\theono }
\global\let#1=\Zwi
}

\def\PRB#1{{\global\advance \probno by 1}%
\eqhead\number\probno
\edef\Zwi{\eqhead\number\probno }
\global\let#1=\Zwi
}

\def\PG#1{\def\Zwi{\number\pageno }
\global\let#1=\Zwi
}

\def\Stm{{\global\advance \theono by 1}%
\eqhead\number\theono
}

\def\Prb{{\global\advance \probno by 1}%
\eqhead\number\probno
}

\def\EDEF#1#2{
\def\tEmP{#1}\expandafter\gdef\tEmP{#2}
}



\def\suffix{ps}
\newcount\system
\global\system=3   

\def\ifundefined#1{\expandafter\ifx\csname#1\endcsname\relax}
\ifundefined{figdir}\def\figdir{}\fi
%
%
\newcount\firstline
\newdimen\pswidth  \newdimen\xleft
\newdimen\psheight \newdimen\ytop \newdimen\ybot
\newcount\justx \newcount\justy
\global\justx=0 \global\justy=0
\newdimen\vpos \newtoks\labeL 
\newread\labeLfile \newdimen\xcoord \newdimen\ycoord
\newif\ifdoit 
\newbox\labox
\newdimen\xdvikwid 
\newdimen\xdvikht
\newdimen\pspoints
\newdimen\rwi
\pspoints=1bp
\newcount\temp
\def\readdim#1{\global\read\labeLfile to \temp
\global #1=\temp pt}
%
%
%
%
\def\figcrop#1{\par
\openin\labeLfile=\figdir#1.lbl                                              
\global\read\labeLfile to\firstline\message{#1}               
\global\read\labeLfile to\temp
\readdim{\ybot}
\readdim{\xleft}
\readdim{\ytop}
\global\read\labeLfile to\justx
\global\read\labeLfile to\justy
\global\read\labeLfile to\labeL
\readdim{\pswidth}
\global\advance\pswidth by -\xleft
\readdim{\psheight}
\global\advance\ybot by -\psheight
\global\advance\psheight by -\ytop
\global\read\labeLfile to\justx
\global\read\labeLfile to\justy
\global\read\labeLfile to\labeL
\vbox to\psheight{\vfill
\ifnum\system=1
\fi 
\ifnum\system=2
\fi 
\ifnum\system=3
                                                 \fi         
\ifnum\system=4
\fi         
\ifnum\system=1
\hbox to \pswidth{\kern-\xleft\special{postscriptfile \figdir#1.\suffix }\hfil}\fi
\ifnum\system=2
\hbox to \pswidth{\kern-\xleft\special{ps: plotfile \figdir#1.\suffix }\hfil}\fi
\ifnum\system=3
\hbox to \pswidth{\kern-\xleft\includegraphics{\figdir#1.\suffix}\hfil}\fi
\ifnum\system=4
\hbox to \pswidth{\kern-\xleft\includegraphics{\figdir#1.\suffix}\hfil}\fi
\ifnum\system=5
\hbox to \pswidth{\kern-\xleft\includegraphics{\figdir#1.\suffix}\hfil}\fi 
\ifnum\system=6
   \xdvikwid=\pswidth
   \xdvikht=\psheight
   {\global\divide\xdvikwid by \pspoints}
   {\global\divide\xdvikht by \pspoints}
   \rwi=\xdvikwid
    {\global\multiply\rwi by 10}
\hbox to \pswidth{\kern-\xleft\includegraphics{\figdir#1.\suffix\space}\hfil}\fi                   
\vskip -\baselineskip
\vskip -\ybot 
\vskip-\psheight %
\hbox to\pswidth  {\hss}%
\parindent=0pt\offinterlineskip                                       
\vpos=0 pt%
\loop\readdim{\xcoord}                                 
\ifdim \xcoord < -999pt \doitfalse\else\doittrue\fi                        
\ifdoit \advance \xcoord by -\xleft
\readdim{\ycoord}
\advance \ycoord by -\ytop                              
\global\read\labeLfile to\justx                                       
\global\read\labeLfile to\justy                                       
\global\read\labeLfile to\labeL
\global\setbox\labox=\hbox{\labeL\hskip-0.3em}%
\advance\vpos by-\ycoord                                              
\vskip-\vpos \vpos=\ycoord                                         
\hbox to\pswidth{\hskip\xcoord %
\hbox to 0pt{\ifnum\justx>0\hss\fi%
\vbox to0pt{%
\ifnum\justy<2\vss\fi%
\copy\labox\kern0pt%
\ifnum\justy>0\vss\fi}%
\ifnum\justx<2\hss\fi}%
\hss}%
\repeat%
\advance\vpos by-\psheight%
\vskip-\vpos %
}\closein\labeLfile}
%
%
%
\def\figplace#1#2#3{
\openin\labeLfile=\figdir#1.lbl
\ifeof \labeLfile
       \immediate\write16{***Can't find \figdir#1.lbl; Skipping it.***}
\else  \closein\labeLfile
       \null\hskip#2\raise #3 \hbox{\figcrop{#1}}
\fi
}
%
%
%
%
\def\figput#1{
\openin\labeLfile=\figdir#1.lbl
\ifeof \labeLfile
       \immediate\write16{***Can't find \figdir#1.lbl; Skipping it.***}
\else  \closein\labeLfile
       \hbox{\figcrop{#1}}
\fi
}


    \def\squiggle{\raise2pt\hbox{${\scriptstyle\sim}$}}
    \def\stoday{\number\day\space\ifcase\month\or Jan\or Feb\or 
                      Mar\or Apr\or May\or Jun\or Jul\or Aug\or Sep\or 
                      Oct\or Nov\or Dec\fi, \number\year}

    \def\veps{{\varepsilon}}
    
    \def\abcst{{\sst const}}
    \def\cst#1#2{{\rm const}^{#1}_{#2}\,}
    \def\Cont#1#2#3{\mathop{{\rm\ \, {\cal C}on}_{#3}}\limits_{#1\rightarrow#2}}

    \def\abcst{{\sst const}}
    \def\cb{{\frak c}}
    \def\ib{{\rm b}}
    \def\IB{{\rm\sst B}}
    \def\cl{;}
    
    \def\fcirc{\circ}
    
    \def\ord{{\rm Ord}\,}
    
    \def\dunion{\cup\kern-0.7em\cdot\kern0.45em}

    \def\cC{{\cal C}}
    \def\cD{{\cal D}}
    \def\rD{{\rm D}}
    \def\cP{{\cal P}}
    
    \def\cV{{\cal V}}    
    \def\cW{{\cal W}}
    \def\n{{\bf n}}
    
    \def\fe{{\frak e}}
    \def\fl{{\frak l}}
    \def\fn{{\frak n}}
    
    \def\fN{{\frak N}}
    \def\fX{{\frak X}}

    \def\bde{{\mathchoice{\pmb{$\de$}}{\pmb{$\de$}}
                              {\pmb{$\sst\de$}}{\pmb{$\ssst\de$}}}}

    \def\rw{\mathclose{:}}
    \def\lw{\mathopen{:}}
    \def\lW{\mathopen{{\tst{\hbox{.}\atop\raise 2.5pt\hbox{.}}}}}
    \def\rW{\mathclose{{\tst{{.}\atop\raise 2.5pt\hbox{.}}}}}
    \def\lww{\mathopen{{\tst{\raise 1pt\hbox{.}\atop\raise 1pt\hbox{.}}}}}
    \def\rww{\mathclose{{\tst{\raise 1pt\hbox{.}\atop\raise 1pt\hbox{.}}}}}

    \def\sv{\pmb{$\sst\vert$}}
    \def\v{\pmb{$\vert$}}
    \def\V{\pmb{$\big\vert$}}
    \def\VV{\pmb{$\Big\vert$}}

    \def\tn{|\kern-1pt|\kern-1pt|}
    \def\TN{\big|\kern-1.5pt\big|\kern-1.5pt\big|}
    \def\TTN{\Big|\kern-2pt\Big|\kern-2pt\Big|}

    \def\cnorm{\kern8pt\check{\kern-8pt\|}}
    \def\Cnorm{\kern8pt\check{\kern-8pt\big\|}}
    \def\CNorm{\kern8pt\check{\kern-8pt\Big\|}}

    \def\tnorm{\kern8pt\tilde{\kern-8pt\|}}
    \def\Tnorm{\kern8pt\tilde{\kern-8pt\big\|}}
    \def\TNorm{\kern8pt\tilde{\kern-8pt\Big\|}}

    \def\tv{\kern8pt\tilde{\kern-8pt\pmb{$\vert$}}}
    \def\tV{\kern8pt\tilde{\kern-8pt\pmb{$\big\vert$}}}
    \def\tVV{\kern8pt\tilde{\kern-8pt\pmb{$\Big\vert$}}}

    \def\il{\jbar}
    \def\jbar{{\mathchoice
                   {{\smash{\lower1ex\hbox{$\mathchar'26$}}\mkern-9mu j}}
                   {{\smash{\lower1ex\hbox{$\mathchar'26$}}\mkern-9mu j}}
                   {{\smash{\lower1.2ex\hbox{$\mathchar'26$}}\mkern-10.2mu j}}
                   {{\smash{\lower1.2ex\hbox{$\mathchar'26$}}\mkern-10.2mu j}}}}

     \def\fcirc{\circ}

\def\Eqnb{{\hbox{\global\advance\equano by 1}}%
\eqno ({\rm \eqhead\number\equano}}%

\def\EQNB#1{\Eqnb\edef\Zwi{\eqhead\number\equano}%
\global\let #1=\Zwi
}

   \font\sixrm=cmr6   \font\eightrm=cmr8  
   \font\sixi=cmmi6   \font\eighti=cmmi8  
  \font\sixsy=cmsy6  \font\eightsy=cmsy8 
  \font\sixbf=cmbx6  \font\eightbf=cmbx8 
                     \font\eightit=cmti8 
                     \font\eightsl=cmsl8 
                     \font\eighttt=cmtt8 

\font\eightfrak=eufm7 at 8pt

\def\eightpoint{\def\rm{\fam0\eightrm}
 \textfont0=\eightrm \scriptfont0=\sixrm \scriptscriptfont0=\fiverm
 \textfont1=\eighti \scriptfont1=\sixi \scriptscriptfont1=\fivei
 \textfont2=\eightsy \scriptfont2=\sixsy \scriptscriptfont2=\fivesy
 \textfont3=\tenex \scriptfont3=\tenex \scriptscriptfont3=\tenex
 \textfont\itfam=\eightit \def\it{\fam\itfam\eightit}%
 \textfont\slfam=\eightsl \def\sl{\fam\slfam\eightsl}%
 \textfont\ttfam=\eighttt \def\tt{\fam\ttfam\eighttt}%
 \textfont\frakfam=\eightfrak \def\frak{\fam\frakfam\tenfrak}%
 \textfont\bffam=\eightbf \scriptfont\bffam=\sixbf
 \scriptscriptfont\bffam=\fivebf \def\bf{\fam\bffam\eightbf}%
 \normalbaselineskip=9pt
 \setbox\strutbox=\hbox{\vrule height7pt depth2pt width0pt}%
 \let\sc=\sixrm \let\big=\eightbig \normalbaselines\rm}
\catcode`@=11
\def\footnote#1{\edef\@sf{\spacefactor\the\spacefactor}#1\@sf
     \insert\footins\bgroup\eightpoint
     \interlinepenalty100 \let\par=\endgraf
     \leftskip=0pt \rightskip=0pt
     \splittopskip=10pt plus 1pt minus 1pt \floatingpenalty=20000
     \smallskip\item{#1}\bgroup\strut\aftergroup\@foot\let\next}
\skip\footins=12pt plus 2pt minus 4pt
\dimen\footins=30pc
\catcode`@=12


  \IgNoRe{PG}
  \IgNoRe{STM Assertion }
  \IgNoRe{PG}
  \IgNoRe{PG}
  \IgNoRe{STM Assertion }
  \IgNoRe{PG}
  \IgNoRe{STM Assertion }
  \IgNoRe{STM Assertion }
  \IgNoRe{EQN}
 \def\defContract{\frefwarning II.5} \IgNoRe{STM Assertion }
  \IgNoRe{STM Assertion }
  \IgNoRe{PG}
  \IgNoRe{STM Assertion }
  \IgNoRe{STM Assertion }
  \IgNoRe{EQN}
  \IgNoRe{STM Assertion }
  \IgNoRe{STM Assertion }
  \IgNoRe{STM Assertion }
  \IgNoRe{STM Assertion }
  \IgNoRe{STM Assertion }
  \IgNoRe{STM Assertion }
  \IgNoRe{PG}
  \IgNoRe{STM Assertion }
  \IgNoRe{STM Assertion }
  \IgNoRe{STM Assertion }
  \IgNoRe{STM Assertion }
  \IgNoRe{STM Assertion }
  \IgNoRe{STM Assertion }
  \IgNoRe{STM Assertion }
  \IgNoRe{STM Assertion }
 \def\deffunctnorm{\frefwarning II.23} \IgNoRe{STM Assertion }
  \IgNoRe{STM Assertion }
 \def\defcontractintbound{\frefwarning II.25} \IgNoRe{STM Assertion }
  \IgNoRe{STM Assertion }
  \IgNoRe{PG}
  \IgNoRe{EQN}
  \IgNoRe{STM Assertion }
 \def\theorII{\frefwarning II.28} \IgNoRe{STM Assertion }
  \IgNoRe{STM Assertion }
  \IgNoRe{PG}
  \IgNoRe{STM Assertion }
  \IgNoRe{STM Assertion }
 \def\corwicknorm{\frefwarning II.32} \IgNoRe{STM Assertion }
  \IgNoRe{STM Assertion }
  \IgNoRe{EQN}
  \IgNoRe{EQN}
 \def\remrenschw{\frefwarning III.1} \IgNoRe{STM Assertion }
  \IgNoRe{PG}
  \IgNoRe{PG}
  \IgNoRe{STM Assertion }
  \IgNoRe{EQN}
 \def\remfunctor{\frefwarning III.3} \IgNoRe{STM Assertion }
  \IgNoRe{STM Assertion }
  \IgNoRe{STM Assertion }
  \IgNoRe{EQN}
  \IgNoRe{STM Assertion }
  \IgNoRe{STM Assertion }
  \IgNoRe{STM Assertion }
  \IgNoRe{PG}
  \IgNoRe{STM Assertion }
  \IgNoRe{STM Assertion }
  \IgNoRe{PG}
  \IgNoRe{STM Assertion }
  \IgNoRe{STM Assertion }
  \IgNoRe{STM Assertion }
  \IgNoRe{PG}
 \def\remjointanalyticitywick{\frefwarning IV.3} \IgNoRe{STM Assertion }
 \def\theoremIVb{\frefwarning IV.4} \IgNoRe{STM Assertion }
  \IgNoRe{STM Assertion }
  \IgNoRe{STM Assertion }
  \IgNoRe{STM Assertion }
  \IgNoRe{STM Assertion }
  \IgNoRe{STM Assertion }
 \def\propBII{\frefwarning A.2} \IgNoRe{STM Assertion }
  \IgNoRe{PG}
  \IgNoRe{STM Assertion }
  \IgNoRe{STM Assertion }
  \IgNoRe{STM Assertion }
  \IgNoRe{STM Assertion }
  \IgNoRe{STM Assertion }
  \IgNoRe{STM Assertion }
  \IgNoRe{STM Assertion }
  \IgNoRe{STM Assertion }
  \IgNoRe{PG}
  \IgNoRe{STM Assertion }
  \IgNoRe{STM Assertion }
  \IgNoRe{PG}
  \IgNoRe{PG}
 \def\defimprnorm{\frefwarning VI.1} \IgNoRe{STM Assertion }
 \def\lemGrassimprnorm{\frefwarning VI.2} \IgNoRe{STM Assertion }
  \IgNoRe{PG}
  \IgNoRe{PG}
  \IgNoRe{STM Assertion }
  \IgNoRe{STM Assertion }
  \IgNoRe{EQN}
  \IgNoRe{STM Assertion }
  \IgNoRe{PG}
 \def\theoremVa{\frefwarning VI.6} \IgNoRe{STM Assertion }
  \IgNoRe{STM Assertion }
  \IgNoRe{STM Assertion }
  \IgNoRe{PG}
  \IgNoRe{STM Assertion }
  \IgNoRe{STM Assertion }
 \def\deftildeimprconf{\frefwarning VI.11} \IgNoRe{STM Assertion }
  \IgNoRe{EQN}
 \def\lemtildeimprconfig{\frefwarning VI.12} \IgNoRe{STM Assertion }
  \IgNoRe{PG}
  \IgNoRe{EQN}
 \def\defimprconf{\frefwarning VI.13} \IgNoRe{STM Assertion }
  \IgNoRe{STM Assertion }
 \def\lemimprconfig{\frefwarning VI.15} \IgNoRe{STM Assertion }
  \IgNoRe{STM Assertion }
  \IgNoRe{EQN}
  \IgNoRe{EQN}
  \IgNoRe{PG}
  \IgNoRe{PG}
  \IgNoRe{STM Assertion }
  \IgNoRe{STM Assertion }
  \IgNoRe{STM Assertion }
  \IgNoRe{STM Assertion }
  \IgNoRe{STM Assertion }
  \IgNoRe{STM Assertion }
  \IgNoRe{STM Assertion }
  \IgNoRe{STM Assertion }
  \IgNoRe{PG}
  \IgNoRe{STM Assertion }
  \IgNoRe{STM Assertion }
  \IgNoRe{STM Assertion }
  \IgNoRe{STM Assertion }
  \IgNoRe{STM Assertion }
  \IgNoRe{STM Assertion }
  \IgNoRe{STM Assertion }
  \IgNoRe{STM Assertion }
  \IgNoRe{STM Assertion }
  \IgNoRe{STM Assertion }
  \IgNoRe{PG}
  \IgNoRe{STM Assertion }
  \IgNoRe{STM Assertion }
  \IgNoRe{STM Assertion }
  \IgNoRe{PG}
  \IgNoRe{STM Assertion }
  \IgNoRe{STM Assertion }
  \IgNoRe{STM Assertion }
  \IgNoRe{STM Assertion }
  \IgNoRe{PG}
  \IgNoRe{STM Assertion }
  \IgNoRe{STM Assertion }
  \IgNoRe{PG}
  \IgNoRe{STM Assertion }
  \IgNoRe{PG}
  \IgNoRe{STM Assertion }
  \IgNoRe{PG}
  \IgNoRe{STM Assertion }
  \IgNoRe{STM Assertion }
  \IgNoRe{STM Assertion }
  \IgNoRe{STM Assertion }
  \IgNoRe{PG}
  \IgNoRe{PG}
  \IgNoRe{STM Assertion }
  \IgNoRe{STM Assertion }
  \IgNoRe{EQN}
  \IgNoRe{EQN}
  \IgNoRe{STM Assertion }
  \IgNoRe{STM Assertion }
  \IgNoRe{STM Assertion }
  \IgNoRe{STM Assertion }
  \IgNoRe{PG}
  \IgNoRe{STM Assertion }
  \IgNoRe{STM Assertion }
  \IgNoRe{STM Assertion }
  \IgNoRe{STM Assertion }
  \IgNoRe{STM Assertion }
  \IgNoRe{STM Assertion }
  \IgNoRe{STM Assertion }
  \IgNoRe{PG}
  \IgNoRe{STM Assertion }
  \IgNoRe{EQN}
  \IgNoRe{EQN}
  \IgNoRe{PG}
  \IgNoRe{STM Assertion }
  \IgNoRe{EQN}
  \IgNoRe{STM Assertion }
  \IgNoRe{STM Assertion }
  \IgNoRe{STM Assertion }
  \IgNoRe{PG}
  \IgNoRe{STM Assertion }
  \IgNoRe{EQN}
 \def\propAIV{\frefwarning C.4} \IgNoRe{STM Assertion }
  \IgNoRe{PG}
  \IgNoRe{PG}


  \IgNoRe{STM Assertion }
  \IgNoRe{PG}
  \IgNoRe{EQN}
  \IgNoRe{EQN}
  \IgNoRe{EQN}
  \IgNoRe{EQN}
  \IgNoRe{EQN}
  \IgNoRe{STM Assertion }
  \IgNoRe{STM Assertion }
  \IgNoRe{STM Assertion }
  \IgNoRe{STM Assertion }
  \IgNoRe{STM Assertion }
  \IgNoRe{STM Assertion }
  \IgNoRe{STM Assertion }
  \IgNoRe{STM Assertion }
  \IgNoRe{STM Assertion }
  \IgNoRe{STM Assertion }
  \IgNoRe{STM Assertion }
  \IgNoRe{PG}
  \IgNoRe{PG}
  \IgNoRe{PG}
  \IgNoRe{PG}
  \IgNoRe{EQN}
  \IgNoRe{EQN}
  \IgNoRe{EQN}
  \IgNoRe{EQN}
  \IgNoRe{PG}
  \IgNoRe{PG}
  \IgNoRe{PG}
  \IgNoRe{EQN}
  \IgNoRe{PG}
  \IgNoRe{PG}
  \IgNoRe{EQN}
  \IgNoRe{EQN}
  \IgNoRe{EQN}
  \IgNoRe{EQN}
  \IgNoRe{EQN}
  \IgNoRe{EQN}
  \IgNoRe{EQN}
  \IgNoRe{EQN}
  \IgNoRe{EQN}
  \IgNoRe{EQN}
  \IgNoRe{PG}
  \IgNoRe{PG}
  \IgNoRe{EQN}
  \IgNoRe{EQN}
  \IgNoRe{EQN}
  \IgNoRe{STM Assertion }
  \IgNoRe{PG}
  \IgNoRe{EQN}
  \IgNoRe{EQN}
  \IgNoRe{EQN}
  \IgNoRe{EQN}
  \IgNoRe{STM Assertion }
  \IgNoRe{STM Assertion }
  \IgNoRe{STM Assertion }
  \IgNoRe{STM Assertion }
  \IgNoRe{STM Assertion }
  \IgNoRe{STM Assertion }
  \IgNoRe{STM Assertion }
  \IgNoRe{STM Assertion }
  \IgNoRe{STM Assertion }
  \IgNoRe{EQN}
  \IgNoRe{EQN}
  \IgNoRe{EQN}
  \IgNoRe{EQN}
  \IgNoRe{STM Assertion }
  \IgNoRe{EQN}
  \IgNoRe{STM Assertion }
  \IgNoRe{EQN}
  \IgNoRe{STM Assertion }
  \IgNoRe{PG}
  \IgNoRe{EQN}
  \IgNoRe{STM Assertion }
  \IgNoRe{STM Assertion }
  \IgNoRe{EQN}
  \IgNoRe{PG}
  \IgNoRe{PG}
  \IgNoRe{STM Assertion }
  \IgNoRe{STM Assertion }
  \IgNoRe{PG}
  \IgNoRe{STM Assertion }
  \IgNoRe{STM Assertion }
  \IgNoRe{STM Assertion }
  \IgNoRe{STM Assertion }
  \IgNoRe{STM Assertion }
  \IgNoRe{STM Assertion }
  \IgNoRe{STM Assertion }
  \IgNoRe{STM Assertion }
  \IgNoRe{PG}
  \IgNoRe{STM Assertion }
  \IgNoRe{STM Assertion }
  \IgNoRe{STM Assertion }
  \IgNoRe{STM Assertion }
  \IgNoRe{EQN}
  \IgNoRe{STM Assertion }
  \IgNoRe{STM Assertion }
 \def\remNPnvsnj{\frefwarning VI.8} \IgNoRe{STM Assertion }
  \IgNoRe{EQN}
  \IgNoRe{STM Assertion }
  \IgNoRe{STM Assertion }
  \IgNoRe{STM Assertion }
  \IgNoRe{STM Assertion }
  \IgNoRe{PG}
  \IgNoRe{STM Assertion }
  \IgNoRe{STM Assertion }
  \IgNoRe{STM Assertion }
  \IgNoRe{STM Assertion }
  \IgNoRe{STM Assertion }
  \IgNoRe{STM Assertion }
  \IgNoRe{STM Assertion }
  \IgNoRe{STM Assertion }
  \IgNoRe{STM Assertion }
  \IgNoRe{EQN}
  \IgNoRe{EQN}
  \IgNoRe{EQN}
  \IgNoRe{STM Assertion }
  \IgNoRe{PG}
  \IgNoRe{STM Assertion }
  \IgNoRe{STM Assertion }
  \IgNoRe{STM Assertion }
  \IgNoRe{EQN}
  \IgNoRe{EQN}
  \IgNoRe{EQN}
  \IgNoRe{STM Assertion }
  \IgNoRe{STM Assertion }
  \IgNoRe{EQN}
  \IgNoRe{EQN}
  \IgNoRe{EQN}
  \IgNoRe{STM Assertion }
  \IgNoRe{PG}
  \IgNoRe{PG}
  \IgNoRe{STM Assertion }
  \IgNoRe{STM Assertion }
  \IgNoRe{PG}
  \IgNoRe{STM Assertion }
  \IgNoRe{STM Assertion }
  \IgNoRe{EQN}
  \IgNoRe{EQN}
  \IgNoRe{EQN}
  \IgNoRe{EQN}
  \IgNoRe{EQN}
  \IgNoRe{EQN}
  \IgNoRe{EQN}
  \IgNoRe{EQN}
  \IgNoRe{EQN}
  \IgNoRe{EQN}
  \IgNoRe{EQN}
  \IgNoRe{EQN}
  \IgNoRe{EQN}
  \IgNoRe{EQN}
  \IgNoRe{STM Assertion }
  \IgNoRe{EQN}
  \IgNoRe{PG}
  \IgNoRe{EQN}
  \IgNoRe{STM Assertion }
  \IgNoRe{EQN}
  \IgNoRe{STM Assertion }
  \IgNoRe{STM Assertion }
  \IgNoRe{EQN}
  \IgNoRe{EQN}
  \IgNoRe{EQN}
  \IgNoRe{STM Assertion }
  \IgNoRe{EQN}
  \IgNoRe{EQN}
  \IgNoRe{EQN}
  \IgNoRe{EQN}
  \IgNoRe{EQN}
  \IgNoRe{EQN}
  \IgNoRe{EQN}
  \IgNoRe{EQN}
  \IgNoRe{EQN}
  \IgNoRe{EQN}
  \IgNoRe{EQN}
  \IgNoRe{EQN}
  \IgNoRe{EQN}
  \IgNoRe{EQN}
  \IgNoRe{EQN}
  \IgNoRe{EQN}
  \IgNoRe{EQN}
  \IgNoRe{EQN}
  \IgNoRe{EQN}
  \IgNoRe{EQN}
  \IgNoRe{EQN}
  \IgNoRe{STM Assertion }
  \IgNoRe{PG}
  \IgNoRe{PG}
  \IgNoRe{STM Assertion }
  \IgNoRe{PG}
  \IgNoRe{EQN}
  \IgNoRe{EQN}
  \IgNoRe{EQN}
  \IgNoRe{EQN}
  \IgNoRe{STM Assertion }
  \IgNoRe{EQN}
  \IgNoRe{PG}
  \IgNoRe{EQN}
  \IgNoRe{EQN}
  \IgNoRe{EQN}
  \IgNoRe{EQN}
  \IgNoRe{EQN}
  \IgNoRe{EQN}
  \IgNoRe{EQN}
  \IgNoRe{EQN}
  \IgNoRe{EQN}
  \IgNoRe{EQN}
  \IgNoRe{EQN}
  \IgNoRe{EQN}
  \IgNoRe{STM Assertion }
  \IgNoRe{STM Assertion }
  \IgNoRe{EQN}
  \IgNoRe{EQN}
  \IgNoRe{PG}
  \IgNoRe{PG}
  \IgNoRe{STM Assertion }
  \IgNoRe{EQN}
  \IgNoRe{STM Assertion }
  \IgNoRe{PG}
  \IgNoRe{EQN}
  \IgNoRe{EQN}
  \IgNoRe{EQN}
  \IgNoRe{STM Assertion }
  \IgNoRe{STM Assertion }
  \IgNoRe{EQN}
  \IgNoRe{EQN}
  \IgNoRe{EQN}
  \IgNoRe{EQN}
  \IgNoRe{EQN}
  \IgNoRe{STM Assertion }
  \IgNoRe{EQN}
  \IgNoRe{EQN}
  \IgNoRe{EQN}
  \IgNoRe{EQN}
  \IgNoRe{STM Assertion }
  \IgNoRe{STM Assertion }
  \IgNoRe{EQN}
  \IgNoRe{STM Assertion }
  \IgNoRe{STM Assertion }
  \IgNoRe{STM Assertion }
  \IgNoRe{STM Assertion }
  \IgNoRe{PG}
  \IgNoRe{STM Assertion }
  \IgNoRe{STM Assertion }
  \IgNoRe{STM Assertion }
  \IgNoRe{STM Assertion }
  \IgNoRe{STM Assertion }
  \IgNoRe{STM Assertion }
  \IgNoRe{STM Assertion }
  \IgNoRe{STM Assertion }
  \IgNoRe{STM Assertion }
  \IgNoRe{STM Assertion }
  \IgNoRe{STM Assertion }
  \IgNoRe{STM Assertion }
  \IgNoRe{STM Assertion }
  \IgNoRe{STM Assertion }
  \IgNoRe{STM Assertion }
  \IgNoRe{PG}
  \IgNoRe{STM Assertion }
  \IgNoRe{STM Assertion }
  \IgNoRe{STM Assertion }
  \IgNoRe{STM Assertion }
  \IgNoRe{STM Assertion }
  \IgNoRe{STM Assertion }
  \IgNoRe{STM Assertion }
  \IgNoRe{STM Assertion }
  \IgNoRe{STM Assertion }
  \IgNoRe{STM Assertion }
  \IgNoRe{STM Assertion }
  \IgNoRe{STM Assertion }
  \IgNoRe{EQN}
  \IgNoRe{STM Assertion }
  \IgNoRe{STM Assertion }
  \IgNoRe{STM Assertion }
  \IgNoRe{STM Assertion }
  \IgNoRe{STM Assertion }
  \IgNoRe{STM Assertion }
  \IgNoRe{EQN}
  \IgNoRe{STM Assertion }
  \IgNoRe{PG}
  \IgNoRe{PG}
  \IgNoRe{STM Assertion }
  \IgNoRe{STM Assertion }
  \IgNoRe{STM Assertion }
  \IgNoRe{EQN}
  \IgNoRe{STM Assertion }
  \IgNoRe{PG}
  \IgNoRe{EQN}
  \IgNoRe{STM Assertion }
  \IgNoRe{STM Assertion }
  \IgNoRe{EQN}
  \IgNoRe{EQN}
  \IgNoRe{EQN}
  \IgNoRe{EQN}
  \IgNoRe{EQN}
  \IgNoRe{EQN}
  \IgNoRe{EQN}
  \IgNoRe{EQN}
  \IgNoRe{EQN}
  \IgNoRe{EQN}
  \IgNoRe{EQN}
  \IgNoRe{EQN}
  \IgNoRe{EQN}
  \IgNoRe{EQN}
  \IgNoRe{EQN}
  \IgNoRe{EQN}
  \IgNoRe{EQN}
  \IgNoRe{EQN}
  \IgNoRe{EQN}
  \IgNoRe{EQN}
  \IgNoRe{STM Assertion }
  \IgNoRe{STM Assertion }
  \IgNoRe{PG}
  \IgNoRe{EQN}
  \IgNoRe{EQN}
  \IgNoRe{EQN}
  \IgNoRe{EQN}
  \IgNoRe{EQN}
  \IgNoRe{EQN}
  \IgNoRe{EQN}
  \IgNoRe{EQN}
  \IgNoRe{EQN}
  \IgNoRe{EQN}
  \IgNoRe{EQN}
  \IgNoRe{EQN}
  \IgNoRe{EQN}
  \IgNoRe{EQN}
  \IgNoRe{EQN}
  \IgNoRe{EQN}
  \IgNoRe{EQN}
  \IgNoRe{EQN}
  \IgNoRe{EQN}
  \IgNoRe{EQN}
  \IgNoRe{EQN}
  \IgNoRe{STM Assertion }
  \IgNoRe{PG}
  \IgNoRe{STM Assertion }
  \IgNoRe{STM Assertion }
  \IgNoRe{EQN}
  \IgNoRe{EQN}
  \IgNoRe{PG}
  \IgNoRe{EQN}
  \IgNoRe{EQN}
  \IgNoRe{EQN}
  \IgNoRe{EQN}
  \IgNoRe{EQN}
  \IgNoRe{EQN}
  \IgNoRe{EQN}
  \IgNoRe{EQN}
  \IgNoRe{STM Assertion }
  \IgNoRe{STM Assertion }
  \IgNoRe{STM Assertion }
  \IgNoRe{STM Assertion }
  \IgNoRe{STM Assertion }
  \IgNoRe{PG}
  \IgNoRe{PG}


\newcount\CHAPNO
\newcount\APPNO
\CHAPNO=0
\APPNO=1
\def\advCHAPNO{\advance\CHAPNO by 1}
\def\advAPPNO{\advance\APPNO by 1}

\def\caproman#1{\ifcase#1\or I\or II\or III\or IV\or V\or VI\or VII\or
VIII\or IX\or X\or XI\or XII\or XIII\or XIV\or XV\or XVI\or XVII\or XVIII\or
XIX\or XX\or XXI\or XXII\or XXIII\or XXIV\or XXV\or XXVI\or XXVII\or XXVIII\or XXIX\or XXX\or XXXI\or XXXII\or XXXIII\or XXXIV\or XXXV\or XXXVI\or XXXVII\or XXXVIII\or XXXIX\fi}%

\def\capletter#1{\ifcase#1\or A\or B\or C\or D\or E\or F\or G\or
H\or I\or J\or K\or L\or M\or N\or O\or P\or Q\or R\or
S\or T\or U\or V\or W\or X\or Y\or Z\fi}%

\newcount\cHintroI \cHintroI=\CHAPNO \advCHAPNO 
\newcount\cHintroOverview  \cHintroOverview=\CHAPNO \advCHAPNO 
                              \edef\CHintroOverview{\caproman\CHAPNO}  
\newcount\cHrenmap \cHrenmap=\CHAPNO \advCHAPNO 

 \advAPPNO

\newcount\cHintroII \cHintroII=\CHAPNO \advCHAPNO 
                              \edef\CHintroII{\caproman\CHAPNO}
\newcount\cHfirstscale \cHfirstscale=\CHAPNO \advCHAPNO
                              
\newcount\cHnewsectors \cHnewsectors=\CHAPNO \advCHAPNO
                              
\newcount\cHphladders \cHphladders=\CHAPNO \advCHAPNO
                              \edef\CHphladders{\caproman\CHAPNO}
\newcount\cHfinitescale \cHfinitescale=\CHAPNO \advCHAPNO
                              
\newcount\cHstep \cHstep=\CHAPNO \advCHAPNO
                              
\newcount\cHrecurs \cHrecurs=\CHAPNO \advCHAPNO
                              
 \advAPPNO

\newcount\cHintroIII \cHintroIII=\CHAPNO \advCHAPNO
                              
\newcount\cHtildefinitescale \cHtildefinitescale=\CHAPNO \advCHAPNO
                              
\newcount\cHtildenewsectors \cHtildenewsectors=\CHAPNO \advCHAPNO
                              
\newcount\cHtildephladders \cHtildephladders=\CHAPNO \advCHAPNO
                              
\newcount\cHtildestep  \cHtildestep=\CHAPNO \advCHAPNO

 \advAPPNO
 \advAPPNO


  \IgNoRe{PG}
 \def\eqnOSintroI{\frefwarning I.1} \IgNoRe{EQN}
  \IgNoRe{STM Assertion }
  \IgNoRe{PG}
  \IgNoRe{STM Assertion }
 \def\defOSdecayop{\frefwarning II.3} \IgNoRe{STM Assertion }
  \IgNoRe{STM Assertion }
  \IgNoRe{STM Assertion }
 \def\exOSSymmNorm{\frefwarning II.6} \IgNoRe{STM Assertion }
 \def\lemOSelloneinfty{\frefwarning II.7} \IgNoRe{STM Assertion }
  \IgNoRe{EQN}
 \def\corOSelloneinfty{\frefwarning II.8} \IgNoRe{STM Assertion }
 \def\defOSFmn{\frefwarning II.9} \IgNoRe{STM Assertion }
  \IgNoRe{STM Assertion }
 \def\defOScontnorm{\frefwarning III.1} \IgNoRe{STM Assertion }
  \IgNoRe{STM Assertion }
  \IgNoRe{STM Assertion }
 \def\exOSelloneinftycontr{\frefwarning III.4} \IgNoRe{STM Assertion }
 \def\defOSintbnd{\frefwarning III.5} \IgNoRe{STM Assertion }
  \IgNoRe{PG}
  \IgNoRe{STM Assertion }
  \IgNoRe{STM Assertion }
  \IgNoRe{STM Assertion }
 \def\defOSgrnorm{\frefwarning III.9} \IgNoRe{STM Assertion }
  \IgNoRe{STM Assertion }
  \IgNoRe{STM Assertion }
 \def\defIntBndsS{\frefwarning IV.1} \IgNoRe{STM Assertion }
  \IgNoRe{STM Assertion }
  \IgNoRe{EQN}
  \IgNoRe{PG}
  \IgNoRe{PG}
 \def\propIntBndsII{\frefwarning IV.3} \IgNoRe{STM Assertion }
  \IgNoRe{STM Assertion }
  \IgNoRe{STM Assertion }
 \def\defOSderivmom{\frefwarning IV.6} \IgNoRe{STM Assertion }
  \IgNoRe{STM Assertion }
  \IgNoRe{STM Assertion }
  \IgNoRe{PG}
  \IgNoRe{EQN}
  \IgNoRe{EQN}
  \IgNoRe{EQN}
  \IgNoRe{EQN}
  \IgNoRe{EQN}
  \IgNoRe{EQN}
  \IgNoRe{STM Assertion }
 \def\defOScbzero{\frefwarning IV.10} \IgNoRe{STM Assertion }
 \def\propOSrealfirstpropbound{\frefwarning IV.11} \IgNoRe{STM Assertion }
 \def\lemOSscalednorm{\frefwarning V.1} \IgNoRe{STM Assertion }
  \IgNoRe{PG}
 \def\thmOSinsulators{\frefwarning V.2} \IgNoRe{STM Assertion }
  \IgNoRe{EQN}
  \IgNoRe{STM Assertion }
  \IgNoRe{STM Assertion }
  \IgNoRe{STM Assertion }
  \IgNoRe{STM Assertion }
  \IgNoRe{PG}
 \def\lemOSappMonoidIV{\frefwarning A.4} \IgNoRe{STM Assertion }
 \def\corOSappMonoidIV{\frefwarning A.5} \IgNoRe{STM Assertion }
  \IgNoRe{STM Assertion }
 \def\lemOSappMonoidV{\frefwarning A.7} \IgNoRe{STM Assertion }
  \IgNoRe{PG}
 \def\eqnOSjdef{\frefwarning VI.1} \IgNoRe{EQN}
  \IgNoRe{EQN}
  \IgNoRe{PG}
 \def\defOSrengrpmap{\frefwarning VII.1} \IgNoRe{STM Assertion }
  \IgNoRe{STM Assertion }
 \def\lemOStworengrpmaps{\frefwarning VII.3} \IgNoRe{STM Assertion }
  \IgNoRe{EQN}
  \IgNoRe{PG}
 \def\defOSextimpr{\frefwarning VII.4} \IgNoRe{STM Assertion }
  \IgNoRe{STM Assertion }
  \IgNoRe{EQN}
 \def\propOSextimpr{\frefwarning VII.6} \IgNoRe{STM Assertion }
  \IgNoRe{STM Assertion }
  \IgNoRe{STM Assertion }
 \def\defOSscales{\frefwarning VIII.1} \IgNoRe{STM Assertion }
  \IgNoRe{PG}
  \IgNoRe{STM Assertion }
  \IgNoRe{STM Assertion }
 \def\defOSextendedshell{\frefwarning VIII.4} \IgNoRe{STM Assertion }
  \IgNoRe{STM Assertion }
 \def\thmOSfirststep{\frefwarning VIII.6} \IgNoRe{STM Assertion }
  \IgNoRe{STM Assertion }
 \def\remOSneedsectors{\frefwarning VIII.8} \IgNoRe{STM Assertion }
 \def\defOSfourtrans{\frefwarning IX.1} \IgNoRe{STM Assertion }
  \IgNoRe{PG}
  \IgNoRe{STM Assertion }
 \def\defOSftcov{\frefwarning IX.3} \IgNoRe{STM Assertion }
 \def\defOSfourtransII{\frefwarning IX.4} \IgNoRe{STM Assertion }
  \IgNoRe{STM Assertion }
 \def\lemOSprepintup{\frefwarning IX.6} \IgNoRe{STM Assertion }
  \IgNoRe{EQN}
 \def\defOSamptransinv{\frefwarning X.1} \IgNoRe{STM Assertion }
 \def\defOSdiffdecay{\frefwarning X.2} \IgNoRe{STM Assertion }
  \IgNoRe{PG}
 \def\remOSdiffdecay{\frefwarning X.3} \IgNoRe{STM Assertion }
 \def\defOSdiffdecaynorm{\frefwarning X.4} \IgNoRe{STM Assertion }
  \IgNoRe{STM Assertion }
  \IgNoRe{STM Assertion }
 \def\remOSelloneinftyamp{\frefwarning X.7} \IgNoRe{STM Assertion }
 \def\defOScheckcF{\frefwarning X.8} \IgNoRe{STM Assertion }
  \IgNoRe{STM Assertion }
  \IgNoRe{STM Assertion }
 \def\lemOSTZsourceterm{\frefwarning X.11} \IgNoRe{STM Assertion }
  \IgNoRe{EQN}
 \def\thmOSTfirststep{\frefwarning X.12} \IgNoRe{STM Assertion }
  \IgNoRe{STM Assertion }
  \IgNoRe{PG}
  \IgNoRe{STM Assertion }
  \IgNoRe{STM Assertion }
  \IgNoRe{STM Assertion }
  \IgNoRe{STM Assertion }
  \IgNoRe{STM Assertion }
  \IgNoRe{STM Assertion }
  \IgNoRe{STM Assertion }
  \IgNoRe{PG}
  \IgNoRe{STM Assertion }
  \IgNoRe{PG}
 \def\pgOSXI{\frefwarning 1} \IgNoRe{PG}
 \def\defOSsectors{\frefwarning XII.1} \IgNoRe{STM Assertion }
 \def\defOScbj{\frefwarning XII.2} \IgNoRe{STM Assertion }
 \def\pgOSXII{\frefwarning 2} \IgNoRe{PG}
 \def\lemOSsectpartunit{\frefwarning XII.3} \IgNoRe{STM Assertion }
 \def\defOSsectrepr{\frefwarning XII.4} \IgNoRe{STM Assertion }
 \def\exOSsectrepr{\frefwarning XII.5} \IgNoRe{STM Assertion }
 \def\defOStens{\frefwarning XII.6} \IgNoRe{STM Assertion }
 \def\remOSideal{\frefwarning XII.7} \IgNoRe{STM Assertion }
 \def\propOSfunctorialitySect{\frefwarning XII.8} \IgNoRe{STM Assertion }
 \def\defOSsectnorm{\frefwarning XII.9} \IgNoRe{STM Assertion }
 \def\exOScommsectnorms{\frefwarning XII.10} \IgNoRe{STM Assertion }
 \def\remOSdiffnorm{\frefwarning XII.11} \IgNoRe{STM Assertion }
 \def\lemOSNormMom{\frefwarning XII.12} \IgNoRe{STM Assertion }
 \def\remOScommsectnorms{\frefwarning XII.13} \IgNoRe{STM Assertion }
 \def\lemOSelloneinftysectors{\frefwarning XII.14} \IgNoRe{STM Assertion }
 \def\defOSsectcontnorm{\frefwarning XII.15} \IgNoRe{STM Assertion }
 \def\propOScontrintboundsectors{\frefwarning XII.16} \IgNoRe{STM Assertion }
 \def\remOShavesectors{\frefwarning XII.17} \IgNoRe{STM Assertion }
 \def\propOSoverlapploops{\frefwarning XII.18} \IgNoRe{STM Assertion }
 \def\eqnOSsupDbound{\frefwarning XII.1} \IgNoRe{EQN}
 \def\lemOSsectextimpr{\frefwarning XII.19} \IgNoRe{STM Assertion }
 \def\propOSGenDecay{\frefwarning XIII.1} \IgNoRe{STM Assertion }
 \def\pgOSXIII{\frefwarning 17} \IgNoRe{PG}
 \def\lemOSsectorderiv{\frefwarning XIII.2} \IgNoRe{STM Assertion }
 \def\eqnOSsecpropboundI{\frefwarning XIII.1} \IgNoRe{EQN}
 \def\eqnOSpartunit{\frefwarning XIII.2} \IgNoRe{EQN}
 \def\lemOSmorepartunity{\frefwarning XIII.3} \IgNoRe{STM Assertion }
 \def\eqnOSsecpropboundII{\frefwarning XIII.3} \IgNoRe{EQN}
 \def\eqnOSprodcontrbound{\frefwarning XIII.4} \IgNoRe{EQN}
 \def\lemOSplainpropest{\frefwarning XIII.4} \IgNoRe{STM Assertion }
 \def\eqnOSsecpropboundIII{\frefwarning XIII.5} \IgNoRe{EQN}
 \def\propOSrealpropbound{\frefwarning XIII.5} \IgNoRe{STM Assertion }
 \def\eqnOSexpandc{\frefwarning XIII.6} \IgNoRe{EQN}
 \def\lemOSdiffpropbound{\frefwarning XIII.6} \IgNoRe{STM Assertion }
 \def\lemOSumu{\frefwarning XIII.7} \IgNoRe{STM Assertion }
 \def\remOSumu{\frefwarning XIII.8} \IgNoRe{STM Assertion }
 \def\defOSbubbleprop{\frefwarning XIV.1} \IgNoRe{STM Assertion }
 \def\pgOSXIV{\frefwarning 28} \IgNoRe{PG}
 \def\defOSantisymmFour{\frefwarning XIV.2} \IgNoRe{STM Assertion }
 \def\defOSsectbubbleprop{\frefwarning XIV.3} \IgNoRe{STM Assertion }
 \def\lemOSladderfunctoriality{\frefwarning XIV.4} \IgNoRe{STM Assertion }
 \def\lemOSsectorladderfunctoriality{\frefwarning XIV.5} \IgNoRe{STM Assertion }
 \def\eqnOSladderfunctC{\frefwarning XIV.1} \IgNoRe{EQN}
 \def\eqnOSrhomn{\frefwarning XV.1} \IgNoRe{EQN}
 \def\defOSscalednorms{\frefwarning XV.1} \IgNoRe{STM Assertion }
 \def\remOSscalednorms{\frefwarning XV.2} \IgNoRe{STM Assertion }
 \def\pgOSXV{\frefwarning 32} \IgNoRe{PG}
 \def\thOSrengroupestimate{\frefwarning XV.3} \IgNoRe{STM Assertion }
 \def\remOSrengroupestimate{\frefwarning XV.4} \IgNoRe{STM Assertion }
 \def\lemOSconcreteintconst{\frefwarning XV.5} \IgNoRe{STM Assertion }
 \def\eqnOScbprimebnd{\frefwarning XV.2} \IgNoRe{EQN}
 \def\lemOSconcretesourcetermintconst{\frefwarning XV.6} \IgNoRe{STM Assertion }
 \def\eqnOSrengroupestimate{\frefwarning XV.3} \IgNoRe{EQN}
 \def\thOSrengroupdiffestimate{\frefwarning XV.7} \IgNoRe{STM Assertion }
 \def\lemOSconcretediffintconst{\frefwarning XV.8} \IgNoRe{STM Assertion }
 \def\eqnOSsectextimpr{\frefwarning XV.4} \IgNoRe{EQN}
 \def\lemOSconcretesourcetermderivconst{\frefwarning XV.9} \IgNoRe{STM Assertion }
 \def\eqnOSimprderivI{\frefwarning XV.5} \IgNoRe{EQN}
 \def\eqnOSimprderivIa{\frefwarning XV.6} \IgNoRe{EQN}
 \def\eqnOSimprderivIb{\frefwarning XV.7} \IgNoRe{EQN}
 \def\eqnOSimprderivIc{\frefwarning XV.8} \IgNoRe{EQN}
 \def\propOSresidualrengroupest{\frefwarning XV.10} \IgNoRe{STM Assertion }
 \def\remOSchoiceofrep{\frefwarning XV.11} \IgNoRe{STM Assertion }
 \def\defOSdisjointfield{\frefwarning XVI.1} \IgNoRe{STM Assertion }
 \def\defOSdisjointOrd{\frefwarning XVI.2} \IgNoRe{STM Assertion }
 \def\pgOSXVI{\frefwarning 46} \IgNoRe{PG}
 \def\remOSbigdisjointunion{\frefwarning XVI.3} \IgNoRe{STM Assertion }
 \def\defOSsectdiffdecaynorm{\frefwarning XVI.4} \IgNoRe{STM Assertion }
 \def\remOSsecdiffdecaynorm{\frefwarning XVI.5} \IgNoRe{STM Assertion }
 \def\lemOSelloneinftyampsectors{\frefwarning XVI.6} \IgNoRe{STM Assertion }
 \def\defOSsectcheckcF{\frefwarning XVI.7} \IgNoRe{STM Assertion }
 \def\propOSmomcontrintboundsectors{\frefwarning XVI.8} \IgNoRe{STM Assertion }
 \def\defOSsectbubblepropII{\frefwarning XVI.9} \IgNoRe{STM Assertion }
 \def\remOSsectbubblepropII{\frefwarning XVI.10} \IgNoRe{STM Assertion }
 \def\lemOSladderfunctorialityII{\frefwarning XVI.11} \IgNoRe{STM Assertion }
 \def\lemOSsectorladderfunctorialityII{\frefwarning XVI.12} \IgNoRe{STM Assertion }
 \def\eqnOSdisjointladder{\frefwarning XVI.1} \IgNoRe{EQN}
 \def\eqnOStilderhomn{\frefwarning XVII.1} \IgNoRe{EQN}
 \def\defOSmomscalednorms{\frefwarning XVII.1} \IgNoRe{STM Assertion }
 \def\pgOSXVII{\frefwarning 56} \IgNoRe{PG}
 \def\remOSmomscalednorms{\frefwarning XVII.2} \IgNoRe{STM Assertion }
 \def\thOSmomrengroupestimate{\frefwarning XVII.3} \IgNoRe{STM Assertion }
 \def\eqnOSmomimprnorom{\frefwarning XVII.2} \IgNoRe{EQN}
 \def\lemOSmomconcreteintconst{\frefwarning XVII.4} \IgNoRe{STM Assertion }
 \def\lemOStildesourceterm{\frefwarning XVII.5} \IgNoRe{STM Assertion }
 \def\eqnNPderivampprelim{\frefwarning XVII.3} \IgNoRe{EQN}
 \def\eqnNPderivamp{\frefwarning XVII.4} \IgNoRe{EQN}
 \def\eqnOSmomsectgeneral{\frefwarning XVII.5} \IgNoRe{EQN}
 \def\eqnOSmomsecttwopoint{\frefwarning XVII.6} \IgNoRe{EQN}
 \def\eqnOSmomsectfourpoint{\frefwarning XVII.7} \IgNoRe{EQN}
 \def\thOSmomrengroupdiffestimate{\frefwarning XVII.6} \IgNoRe{STM Assertion }
 \def\lemOSmomconcretediffintconst{\frefwarning XVII.7} \IgNoRe{STM Assertion }
 \def\lemOStildesourcetermderiv{\frefwarning XVII.8} \IgNoRe{STM Assertion }
 \def\eqnNPderivampB{\frefwarning XVII.8} \IgNoRe{EQN}
 \def\eqnNPderivampC{\frefwarning XVII.9} \IgNoRe{EQN}
 \def\eqnOSmomimprderivI{\frefwarning XVII.10} \IgNoRe{EQN}
 \def\eqnOSmomimprderivIa{\frefwarning XVII.11} \IgNoRe{EQN}
 \def\eqnOSmomimprderivIb{\frefwarning XVII.12} \IgNoRe{EQN}
 \def\eqnOSmomimprderivIc{\frefwarning XVII.13} \IgNoRe{EQN}
 \def\remOSmomchoiceofrep{\frefwarning XVII.9} \IgNoRe{STM Assertion }
 \def\defOSchannelnorm{\frefwarning D.1} \IgNoRe{STM Assertion }
 \def\lemchannelnorm{\frefwarning D.2} \IgNoRe{STM Assertion }
 \def\pgOSD{\frefwarning 69} \IgNoRe{PG}
 \def\corchannelnorm{\frefwarning D.3} \IgNoRe{STM Assertion }
 \def\lemOSNaivetensor{\frefwarning D.4} \IgNoRe{STM Assertion }
 \def\eqOSNaivetensor{\frefwarning D.1} \IgNoRe{EQN}
 \def\eqOSNaivetensorII{\frefwarning D.2} \IgNoRe{EQN}
 \def\lemOSNaiveBubble{\frefwarning D.5} \IgNoRe{STM Assertion }
 \def\eqnOSNaiveII{\frefwarning D.3} \IgNoRe{EQN}
 \def\eqnOSNaiveIII{\frefwarning D.4} \IgNoRe{EQN}
 \def\corOSchoppedladder{\frefwarning D.6} \IgNoRe{STM Assertion }
 \def\eqnOSchopladder{\frefwarning D.5} \IgNoRe{EQN}
 \def\eqnOSladderdiff{\frefwarning D.6} \IgNoRe{EQN}
 \def\propOSNaiveLadder{\frefwarning D.7} \IgNoRe{STM Assertion }
 \def\remOSnaiveladderest{\frefwarning D.8} \IgNoRe{STM Assertion }
 \def\pgOSIIIref{\frefwarning 78} \IgNoRe{PG}
  \IgNoRe{STM Assertion }
  \IgNoRe{STM Assertion }
  \IgNoRe{STM Assertion }
  \IgNoRe{PG}
  \IgNoRe{STM Assertion }
  \IgNoRe{STM Assertion }
  \IgNoRe{STM Assertion }
  \IgNoRe{STM Assertion }
  \IgNoRe{PG}
  \IgNoRe{STM Assertion }
  \IgNoRe{STM Assertion }
  \IgNoRe{STM Assertion }
  \IgNoRe{STM Assertion }
  \IgNoRe{STM Assertion }
  \IgNoRe{STM Assertion }
  \IgNoRe{STM Assertion }
  \IgNoRe{STM Assertion }
  \IgNoRe{STM Assertion }
  \IgNoRe{STM Assertion }
  \IgNoRe{STM Assertion }
  \IgNoRe{STM Assertion }
  \IgNoRe{STM Assertion }
  \IgNoRe{STM Assertion }
  \IgNoRe{PG}
  \IgNoRe{STM Assertion }
  \IgNoRe{PG}
  \IgNoRe{STM Assertion }
  \IgNoRe{STM Assertion }
  \IgNoRe{PG}
  \IgNoRe{STM Assertion }
  \IgNoRe{STM Assertion }
  \IgNoRe{EQN}
  \IgNoRe{PG}
  \IgNoRe{EQN}
  \IgNoRe{STM Assertion }
  \IgNoRe{STM Assertion }
  \IgNoRe{PG}
  \IgNoRe{EQN}
  \IgNoRe{EQN}
  \IgNoRe{EQN}
  \IgNoRe{EQN}
  \IgNoRe{EQN}
  \IgNoRe{STM Assertion }
  \IgNoRe{STM Assertion }
  \IgNoRe{EQN}
  \IgNoRe{STM Assertion }
  \IgNoRe{PG}
  \IgNoRe{PG}
  \IgNoRe{PG}
  \IgNoRe{STM Assertion }
  \IgNoRe{EQN}
  \IgNoRe{STM Assertion }
  \IgNoRe{PG}
  \IgNoRe{STM Assertion }
  \IgNoRe{EQN}
  \IgNoRe{STM Assertion }
  \IgNoRe{STM Assertion }
  \IgNoRe{PG}
  \IgNoRe{EQN}
  \IgNoRe{EQN}
  \IgNoRe{EQN}
  \IgNoRe{STM Assertion }
  \IgNoRe{STM Assertion }
  \IgNoRe{STM Assertion }
  \IgNoRe{EQN}
  \IgNoRe{STM Assertion }
  \IgNoRe{STM Assertion }
 \def\theoremOSLadA{\frefwarning XXII.8} \IgNoRe{STM Assertion }
  \IgNoRe{STM Assertion }
  \IgNoRe{STM Assertion }
  \IgNoRe{STM Assertion }
  \IgNoRe{PG}
  \IgNoRe{STM Assertion }
  \IgNoRe{STM Assertion }
  \IgNoRe{STM Assertion }
  \IgNoRe{STM Assertion }
  \IgNoRe{STM Assertion }
  \IgNoRe{STM Assertion }
  \IgNoRe{STM Assertion }
  \IgNoRe{PG}
  \IgNoRe{PG}
  \IgNoRe{PG}
 \def\pgOSIIInot{\frefwarning 79} \IgNoRe{PG}
  \IgNoRe{PG}


\newcount\CHAPNO
\newcount\APPNO
\CHAPNO=0
\APPNO=1
\def\advCHAPNO{\advance\CHAPNO by 1}
\def\advAPPNO{\advance\APPNO by 1}

\def\caproman#1{\ifcase#1\or I\or II\or III\or IV\or V\or VI\or VII\or
VIII\or IX\or X\or XI\or XII\or XIII\or XIV\or XV\or XVI\or XVII\or XVIII\or
XIX\or XX\or XXI\or XXII\or XXIII\or XXIV\or XXV\or XXVI\or XXVII\or XXVIII\or XXIX\or XXX\or XXXI\or XXXII\or XXXIII\or XXXIV\or XXXV\or XXXVI\or XXXVII\or XXXVIII\or XXXIX\fi}%

\def\capletter#1{\ifcase#1\or A\or B\or C\or D\or E\or F\or G\or
H\or I\or J\or K\or L\or M\or N\or O\or P\or Q\or R\or
S\or T\or U\or V\or W\or X\or Y\or Z\fi}%

\newcount\cHintroI \cHintroI=\CHAPNO \advCHAPNO 
                              
\newcount\cHnorms  \cHnorms=\CHAPNO \advCHAPNO 
                              \edef\CHnorms{\caproman\CHAPNO}
\newcount\cHproprengrp \cHproprengrp=\CHAPNO \advCHAPNO 
                              
\newcount\cHcovbounds  \cHcovbounds=\CHAPNO \advCHAPNO 
                              
\newcount\cHinsulator \cHinsulator=\CHAPNO \advCHAPNO

 \advAPPNO

\newcount\cHintroII \cHintroII=\CHAPNO \advCHAPNO 
                              \edef\CHintroII{\caproman\CHAPNO}
\newcount\cHamputate \cHamputate=\CHAPNO \advCHAPNO
                              
\newcount\cHscales \cHscales=\CHAPNO \advCHAPNO
                              \edef\CHscales{\caproman\CHAPNO}
\newcount\cHfourier \cHfourier=\CHAPNO \advCHAPNO
                              \edef\CHfourier{\caproman\CHAPNO}
\newcount\cHmomentum \cHmomentum=\CHAPNO \advCHAPNO
                              \edef\CHmomentum{\caproman\CHAPNO}

 \advAPPNO
 \advAPPNO

\newcount\cHintroIII \cHintroIII=\CHAPNO \advCHAPNO
                              
\newcount\cHsectors \cHsectors=\CHAPNO \advCHAPNO
                              \edef\CHsectors{\caproman\CHAPNO}
\newcount\cHsecpropbounds \cHsecpropbounds=\CHAPNO \advCHAPNO
                              \edef\CHsecpropbounds{\caproman\CHAPNO}
\newcount\cHladdersNotn  \cHladdersNotn=\CHAPNO \advCHAPNO
                              \edef\CHladdersNotn{\caproman\CHAPNO}
\newcount\cHestren  \cHestren=\CHAPNO \advCHAPNO
                              \edef\CHestren{\caproman\CHAPNO}
\newcount\cHsecmomnorm \cHsecmomnorm=\CHAPNO \advCHAPNO
                              \edef\CHsecmomnorm{\caproman\CHAPNO}
\newcount\cHmomestren \cHmomestren=\CHAPNO \advCHAPNO
                              \edef\CHmomestren{\caproman\CHAPNO}

\edef\APappNaiveladder{\capletter\APPNO} \advAPPNO

\newcount\cHintroIV  \cHintroIV=\CHAPNO \advCHAPNO
                              
\newcount\cHcomparison   \cHcomparison=\CHAPNO \advCHAPNO
                              
\newcount\cHsumsmom  \cHsumsmom=\CHAPNO \advCHAPNO
                              
\newcount\cHsectorsmom   \cHsectorsmom=\CHAPNO \advCHAPNO
                              
\newcount\cHppladsect    \cHppladsect=\CHAPNO \advCHAPNO
                              \edef\CHppladsect{\caproman\CHAPNO}

 \advAPPNO

\chapno=\cHintroIII


{\nopagenumbers
\multiply\baselineskip by \spacingDenominator\divide \baselineskip by\spacingNumerator

\null\vskip3truecm

%
%
\centerline{\tafontt Single Scale Analysis of Many Fermion Systems}

\vskip0.1in
\centerline{\tbfontt Part 3: Sectorized Norms}

\vskip0.75in
\centerline{Joel Feldman{\parindent=.15in\footnote{$^{*}$}{Research supported 
in part by the
 Natural Sciences and Engineering Research Council of Canada and the Forschungsinstitut f\"ur Mathematik, ETH Z\"urich}}}
\centerline{Department of Mathematics}
\centerline{University of British Columbia}
\centerline{Vancouver, B.C. }
\centerline{CANADA\ \   V6T 1Z2}
\centerline{feldman@math.ubc.ca}
\centerline{http:/\hskip-3pt/www.math.ubc.ca/\squiggle
feldman/}
\vskip0.3in
\centerline{Horst Kn\"orrer, Eugene Trubowitz}
\centerline{Mathematik}
\centerline{ETH-Zentrum}
\centerline{CH-8092 Z\"urich}
\centerline{SWITZERLAND}
\centerline{knoerrer@math.ethz.ch, trub@math.ethz.ch}
\centerline{http:/\hskip-3pt/www.math.ethz.ch/\squiggle
knoerrer/}

\vskip0.75in
\noindent
%
{\bf Abstract.\ \ \ } 
The generic renormalization group map associated to a weakly coupled 
system of fermions at temperature zero is treated by supplementing the 
methods of Part 1. The interplay between position and momentum space is 
captured by `sectors'. It is shown that the difference between the complete 
four legged vertex and its  `ladder' part is irrelevant for the sequence 
of renormalization group maps. 

\vfill
\eject


\titleb{Table of Contents}
\halign{\hfill#\ &\hfill#\ &#\hfill&\ p\ \hfil#&\ p\ \hfil#\cr
\noalign{\vskip0.05in}
\S XI&\omit Introduction to Part 3                         \span&\:\pgOSXI\cr
\noalign{\vskip0.05in}
\S XII&\omit Sectors and Sectorized Norms                \span&\:\pgOSXII\cr
\noalign{\vskip0.05in}
\S XIII&\omit Bounds for Sectorized Propagators        \span&\:\pgOSXIII\cr
\noalign{\vskip0.05in}
\S XIV&\omit Ladders                                   \span&\:\pgOSXIV\cr
\noalign{\vskip0.05in}
\S XV&\omit Norm Estimates on the Renormalization Group Map\span&\:\pgOSXV\cr
\noalign{\vskip0.05in}
\S XVI&\omit Sectorized Momentum Space Norms            \span&\:\pgOSXVI\cr
\noalign{\vskip0.05in}
\S XVII&\omit The Renormalization Group Map and Norms in Momentum Space  
                                                          \span&\:\pgOSXVII\cr
\noalign{\vskip0.05in}
{\bf Appendices}\span\cr
\noalign{\vskip0.05in}
\S D&\omit Naive Ladder Estimates                    \span&\:\pgOSD\cr
\noalign{\vskip0.05in}
 &\omit References                                    \span&\:\pgOSIIIref \cr
\noalign{\vskip0.05in}
 &\omit Notation                                      \span&\:\pgOSIIInot \cr
}
\vfill\eject
\multiply\baselineskip by \spacingNumerator\divide \baselineskip by\spacingDenominator}
\pageno=1


\chap{Introduction to Part 3}\PG\pgOSXI

 We use ``sectors'' to construct 
norms that allow nonperturbative control of renormalization group maps 
for two dimensional many  fermion systems.
Thus from Section \CHsectors\ on, we  assume that the dimension $d$ of 
our system is two. Notation tables are provided at the end of the paper.

We assume that the dispersion relation $e(\k)$  is 
$r+d+1$ times differentiable, with $r\ge 2$, and that its 
gradient does not vanish on the Fermi surface 
$$
F = \set{(k_0,\k) \in \bbbr\times\bbbr^d}{k_0=0,\,e(\k)=0}
$$ 
It follows from these hypotheses that the gradient of
the dispersion relation $e(\k)$ does not vanish in a neighbourhood of $F$
and that there is an $r+d+1$ times differentiable projection $\pi_F$ to $F$ 
 in a neighbourhood of the Fermi surface. We
assume that the scale parameter $M$ of Section \CHscales\ has been chosen so big 
that the ``second doubly extended neighbourhood'' 
$\set{k \in \bbbr \times \bbbr^2}{\bar\nu^{(\ge 2)}(k)\ne 0}$ 
is contained in the two above mentioned neighbourhoods.

\vfill\eject

\chap{ Sectors and Sectorized Norms}\PG\pgOSXII

From now on we consider only $d=2$, so that the Fermi ``surface'' is a curve
in $\bbbr\times\bbbr^2$.

\definition{\STM\defOSsectors (Sectors and sectorizations)}{ 
\Item i)
Let $I$ be an interval on the Fermi surface $F$ and $j\ge 2$. Then
$$
s=\set{k\ {\rm in\ the\ } j^{\rm th}\ {\rm neighbourhood}}{\pi_F(k)\in I}
$$
is called a sector of length $|I|$ at scale $j$. 
Recall that $\pi_F(k)$ is the projection of $k$ on the Fermi surface. 
Two different sectors $s$ and $s'$ are called neighbours if  
$s'\cap s\ne \emptyset$.
\Item ii)
If $s$ is a sector at scale $j$, its extension is 
$$
\tilde s=\set{k\ {\rm in\ the\ } j^{\rm th}\ {\rm extended\ neighbourhood}}
{\pi_F(k)\in s}
$$
\Item iii)
A sectorization of length $\fl$ at scale $j$ is a set $\Si$
of sectors of length $\fl$ at scale $j$ that obeys 
\item{-} the set $\Si$ of sectors covers the Fermi surface
\item{-} each sector in $\Si$ has precisely two neighbours in $\Si$, 
one to its left and one to its right
\item{-} if $s,\ s'\in\Si$ are neighbours then 
$\sfrac{1}{16}\fl\le |s\cap s'\cap F|\le\sfrac{1}{8}\fl$

\noindent Observe that there are at most 
$2\,{\rm length}(F)/\fl$ sectors in $\Si$.

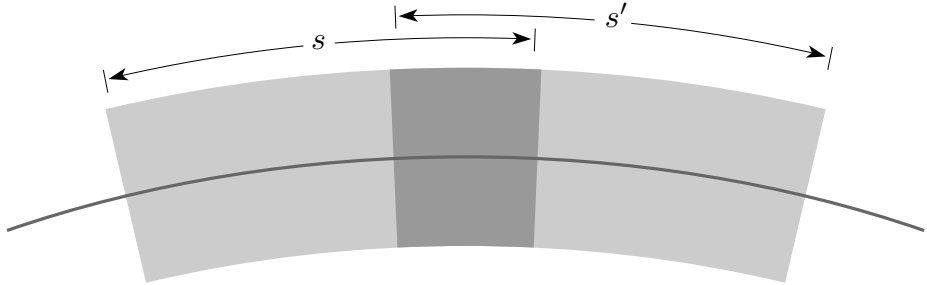
\centerline{\figput{sector2}}
}

We will need partitions of unity for the sectors, as well as functions that
envelope the sectors -- i.e. that are identically one on a sector and are
supported near the sector. Their $L_1$--$L_\infty$ norm will be typical for a 
function with the specified support. To measure it we generalize
Definition \defOScbzero.

\definition{\STM\defOScbj}{ The element $\cb_j$ of $\fN_{d+1}$ is defined as
$$
\cb_j=\sum_{|\bde|\le r\atop |\de_0|\le r_0}  M^{j|\de|}\,t^\de
+\sum_{|\bde|> r\atop {\rm or\ }|\de_0|> r_0}\infty\, t^\de
\in\fN_{d+1}
$$ 
}

\lemma{\STM\lemOSsectpartunit}{
Let $\Si$ be a sectorization of length 
$\sfrac{1}{M^{j-3/2}}\le\fl\le\sfrac{1}{M^{(j-1)/2}}$ at scale $j\ge 2$.
Then there exist $\chi_s(k),\, \tilde \chi_s(k),\ s\in \Si$ 
that take values in $[0,1]$ such that
\Item{i)} $\chi_s$ is supported in the extended sector $\tilde s$ and
$$
\smsum_{s\in \Si} \chi_s(k) =1 \qquad {\rm for\ } k\ {\rm in\ the\ } j^{\rm th}\
{\rm neighbourhood}
$$
\Item{ii)}
$\tilde \chi_s$ is identically one on the extended sector $\tilde s$, is
supported on the $j^{\rm th}$ doubly extended neighbourhood and 
$\tilde \chi_s(k) \cdot \tilde \chi_{s'}(k) =0$ if $s\cap s' = \emptyset$. 
Furthermore, $\int d^3k\,\sfrac{\tilde \chi_s(k)}{|\imath k_0 - e(\k)|} 
\le \const\,\sfrac{\fl}{M^j}$. 
\Item{iii)}
$$
\|\hat\chi_s\|_{1,\infty},\ 
\|\hat{\tilde\chi}_s\|_{1,\infty}\ 
\le\  \abcst\,\cb_{j-1}
\le\  \abcst\,\cb_j
$$
with a constant $\abcst$ that does not depend $M$, $j$, $\Si$ or $s$.
}

The proof of this Lemma is postponed to \S\CHsecpropbounds.

\definition{\STM\defOSsectrepr (Sectorized representatives)}{
Let $\Si$ be a  sectorization at scale $j$, and let $m,n \ge 0$. 
\Item{i)}
The antisymmetrization of a function $\varphi$ on 
$\cB^m\times \big(\cB \times \Si \big)^n$ is
$$
{\rm Ant} \varphi ({\sst \eta_1,\cdots,\eta_m;\,(\xi_1,s_1),\cdots,(\xi_n,s_n)})
=\sfrac{1}{m!\,n!} \smsum_{\pi \in S_m \atop \pi'\in S_n} 
\varphi({\sst
\eta_{\pi(1)},\cdots,\eta_{\pi(m)};
\,(\xi_{\pi'(1)},s_{\pi'(1)}),\cdots,(\xi_{\pi'(n)},s_{\pi'(n)})})
$$
\Item{ ii)} 
Denote by  $\cF_m(n;\Si)$ the space
of all translation invariant, complex valued functions 
$$
\varphi({\sst\eta_1,\cdots,\eta_m;\,(\xi_1,s_1),\cdots,(\xi_n,s_n)} )
$$
on $\cB^m \times  \big( \cB \times\Si \big)^n$ that are antisymmetric in their 
external ($=\eta$) variables and whose Fourier transform
$\check\varphi({\sst\check\eta_1,\cdots,\check\eta_m;
\,(\check\xi_1,s_1),\cdots,(\check\xi_n,s_n)} )$
vanishes unless 
$ k_i\in \tilde s_i$ for all $1\le j\le n$. 
Here, $\check\xi_i=(k_i,\si_i,a_i)$.
\Item{iii)}
Let $f\in \cF_m(n)$ be translation invariant.
A $\Si$--sectorized representative for $f$ is a function
$\varphi\in\cF_m(n;\Si)$ obeying
$$
\check f({\sst\check\eta_1,\cdots,\check\eta_m;\,\check\xi_1,\cdots,\check\xi_n }) 
=\smsum_{s_i\in\Si \atop i=1,\cdots,n} \check \varphi
({\sst\check\eta_1,\cdots,\check\eta_m;
\,(\check\xi_1,s_1),\cdots,(\check\xi_n,s_n) } )
$$
for all $\check\xi_i=(k_i,\si_i,a_i)$ with $k_i$ in the $j^{\rm th}$ neighbourhood. 
\Item iv) Let $u({\sst(\xi,s)},{\sst(\xi',s')})$ be a translation invariant, spin independent, particle number conserving function on
$(\cB\times\Si)^2$. We define $\check u(k)$ by
$$
\de_{\si,\si'}\check u(k)=
\sum_{s,s'\in\Si}\check u({\sst(k,\si,1,s)},{\sst(k,\si',0,s')})
$$
}

\example{\STM\exOSsectrepr}{ Set
$$
\varphi({\sst\eta_1,\cdots,\eta_m;\,(\xi_1,s_1),\cdots,(\xi_n,s_n) } )
= \int\smprod_{i=1}^n\, ({\sst d\xi'_i\ \hat\chi_{s_i}(\xi_i,\xi'_i)})
\ f({\sst\eta_1,\cdots,\eta_m;\,\xi'_1,\cdots,\xi'_n} )
$$
where $\chi_s$ is the partition of unity of Lemma \lemOSsectpartunit\ and
$\hat\chi_s$ was defined in Definition \defOSfourtransII. Then
$\varphi$ is a
$\Si$--sectorized representative for $f$. 
}

Recall that we want
to control the renormalization group map $\Om_C$ on $\bigwedge_A V$, where
$A$ is the Grassmann algebra generated by the fields $\phi({\sst \eta})$ and $V$
is the vector space generated by the fields $\psi({\sst \xi})$. We shall do this
by controlling norms of sectorized representatives of the coefficient functions.
In preparation, we consider a renormalization group map that is adjusted to the
sectorization.

\definition{\STM\defOStens}{
\Item{i)}
$V_\Si$ is the vector space generated by $\psi({\sst \xi,s})$, 
$\xi \in\cB,\,s\in\Si$. If $\varphi\in \cF_m(n;\Si)$ we define the element 
${\rm Tens}(\varphi)$ of $A_m\otimes V_\Si^{\otimes n}$ by
$$
{\rm Tens}(\varphi) =\hskip-6pt \sum_{s_j\in \Si \atop j=1,\cdots,n}
\hskip-6pt \int\hskip-2pt \smprod_{i=1}^m {\sst d\eta_i}\hskip-1pt \smprod_{j=1}^n {\sst d\xi_j}\ 
\varphi({\sst\eta_1,\cdots,\eta_m;\, (\xi_1,s_1),\cdots,(\xi_n,s_n) })\,
 \phi({\sst \eta_1})\cdots\phi({\sst\eta_m})\ 
\psi({\sst\xi_1,s_1}) \otimes \cdots \otimes \psi({\sst \xi_n,s_n})
$$
and the element $Gr(\varphi)$ of $A_m\otimes \bigwedge^n V_\Si$ as
$$
Gr(\varphi) =\hskip-2pt \sum_{s_j\in \Si \atop j=1,\cdots,n}\hskip-2pt
\int \smprod_{i=1}^m d\eta_i\, \smprod_{j=1}^n d\xi_j\ 
\varphi({\sst\eta_1,\cdots,\eta_m;\, (\xi_1,s_1),\cdots,(\xi_n,s_n) })\,
 \phi({\sst \eta_1})\cdots\phi({\sst\eta_m})\ 
\psi({\sst\xi_1,s_1}) \cdots  \psi({\sst \xi_n,s_n})
$$
Elements of $A\otimes \bigwedge V_\Si$ are called sectorized Grassmann functions.
\Item{ii)}
Let 
$$
\cW =\smsum_{m,n \ge0} \int \smprod_{i=1}^m d\eta_i\, \smprod_{j=1}^n d\xi_j\ 
W_{m,n}({\sst\eta_1,\cdots,\eta_m;\, \xi_1,\cdots,\xi_n })\,
 \phi({\sst \eta_1})\cdots\phi({\sst\eta_m})\ 
\psi({\sst\xi_1}) \cdots  \psi({\sst \xi_n})
$$
be a Grassmann function with $W_{m,n} \in \cF_m(n)$ antisymmetric in its internal
($=\psi$) variables. A sectorized representative for $\cW$ is a sectorized
Grassmann function of the form 
$$
w =\smsum_{m,n \ge0} Gr(w_{m,n})
$$
where, for each $m,n$, $\ w_{m,n}$ is a sectorized representative for $W_{m,n}$
that is also antisymmetric in the variables $(\xi_1,s_1),\cdots,(\xi_n,s_n)$.
}

\remark{\STM\remOSideal}{
Let $I_O$ be the ideal in $\bigwedge_A V$ consisting of all
$$
\cW =\smsum_{n >0} \int \smprod_{j=1}^n d\xi_j\ 
W_{n}({\sst\xi_1,\cdots,\xi_n })\,
\psi({\sst\xi_1}) \cdots  \psi({\sst \xi_n})
$$
obeying
$$
\check W_{n}({\sst(k_1,\si_1,a_1),\cdots,(k_n,\si_n,a_n) })=0\qquad
\hbox{for all } k_1,\cdots,k_n\hbox{ in $j^{\rm th}$ neighbourhood}
$$
and let $V_\Si^{\rm eff}$ be the linear subspace of $V_\Si$ consisting of all
$$
\cV= \sum_{s\in \Si}\ \int  d\xi\ \varphi({\sst(\xi,s)})\,\psi({\sst\xi,s})
$$
obeying
$$
\check \varphi({\sst(k,\si,a,s) })=0\qquad\hbox{unless } k\in \tilde s
$$
 Furthermore let 
$\pi:V_\Si\rightarrow V$ be the linear map that sends $\psi(\xi,s) \in V_\Si$ to
$\psi(\xi)\in V$. It induces an algebra homomorphism from $\bigwedge_A V_\Si$
to $\bigwedge_A V$, which we again denote by $\pi$.
Then the sectorized Grassmann function
$w$ is a sectorized representative for the Grassmann function $\cW$ if and only if
$w \in \bigwedge_A V_\Si^{\rm eff}$ and $\pi(w) -\cW \in I_O$.
}

\proposition{\STM\propOSfunctorialitySect (Functoriality)}{
Let $C(\xi,\xi')$ be a skew symmetric function on
$\cB\times\cB$. Assume that there is an antisymmetric function 
$c\big((\xi,s),(\xi',s')\big) \in \cF_0(2;\Si)$
such that
$$
C(\xi,\xi') = \smsum_{s,s'\in\Si} c\big((\xi,s),(\xi',s')\big)
$$
and
$$
\check c\big((k,\si,a,s),(k',\si',a',s') \big) =0
$$
unless\footnote{$^{(1)}$}{The hypothesis $c\left((\xi,s),(\xi',s')\right) \in \cF_0(2;\Si)$ implies that $\check c\left((k,\,\cdot\,,s),(k',\,\cdot\,,s') \right)$
vanishes unless $k\in\tilde s$, $k'\in\tilde s'$. Here we further 
require that $\check c\left((k,\,\cdot\,,s),(k',\,\cdot\,,s') \right)$
vanish unless $k$ and $k'$ are in the $j^{\rm th}$ neighbourhood.}
 $k\in s,\ k'\in s'$. Define a covariance on $V_\Si$ by
$$
C_\Si\big(\psi({\sst\xi,s}),\psi({\sst\xi',s'}) \big) = 
\sum_{t\cap s \ne \emptyset \atop t'\cap s' \ne \emptyset}
c\big((\xi,t),(\xi',t')\big)
$$
\Item i)
If $\varphi\in\cF_0(n;\Si)$ then
$$\eqalign{
&  \smsum_{s_1,\cdots,s_{n} \in \Si} \int \hskip -.1cm
\int {\sst d\xi_1\cdots d\xi_{n}}\, 
\varphi({\sst(\xi_1,s_1),\cdots,(\xi_n,s_n)}) 
\,\psi({\sst \xi_1,s_1})\cdots\psi({\sst \xi_{n},s_n})\,d\mu_{C_\Si}(\psi)\cr
&\hskip1cm= \int \hskip -.1cm
\int {\sst d\xi_1\cdots d\xi_{n}}\, \Big[ \smsum_{s_1,\cdots,s_{n} \in \Si}
\varphi({\sst(\xi_1,s_1),\cdots,(\xi_n,s_n)}) \Big]
\,\psi({\sst \xi_1})\cdots\psi({\sst \xi_{n}})\,d\mu_C(\psi)\cr
}$$
\Item{ii)}
Let $\cW(\phi;\psi)$ be an even Grassmann function and $w$ a 
sectorized representative for $\cW$. Then
 $\Om_{C_\Si}(w)$ is a sectorized representative for $\Om_C(\cW)$.

\noindent For any $\ze(\xi)$, set, with some abuse of notation,
$$
\big(C\ze\big)(\xi,s)
=\int d\xi'\ C(\xi,\xi')\ \ze(\xi')
$$
Then $\half\phi JCJ\phi+\Om_{C_\Si}(w)(\phi,\psi+ C J\phi)$  is a
 sectorized representative for $\tilde\Om_C(\cW)$.

\Item{iii)}
Let $\cW(\phi;\psi)$ be a Grassmann function and $w$ a sectorized 
representative for $\cW$. Then
 $\lw w\rw_{C_\Si}$ is a sectorized representative for 
$\lw \cW\rw_C$.

 }

\prf i) First consider $n=2$. Then 
$$\eqalignno{
&  \smsum_{s_1,s_{2} \in \Si} \int \hskip -.1cm
\int {\sst d\xi_1 d\xi_{2}}\, 
\varphi({\sst(\xi_1,s_1),(\xi_2,s_2)}) 
\,\psi({\sst \xi_1,s_1})\psi({\sst \xi_{2},s_2})\,d\mu_{C_\Si}(\psi)\cr
&\hskip1cm = \smsum_{s_1,s_{2} \in \Si} 
\int {\sst d\xi_1 d\xi_{2}}\, 
\varphi({\sst(\xi_1,s_1),(\xi_2,s_2)}) 
\,C_\Si\big(\psi({\sst \xi_1,s_1}),\psi({\sst \xi_{2},s_2})\big)\cr
&\hskip1cm = \smsum_{s_1,s_{2},t_1,t_2 \in \Si\atop
{t_1\cap s_1\ne\emptyset\atop t_2\cap s_2\ne\emptyset }} 
\int {\sst d\xi_1 d\xi_{2}}\, 
\varphi({\sst(\xi_1,s_1),(\xi_2,s_2)}) 
\,c\big({\sst (\xi_1,t_1)},{\sst (\xi_{2},t_2)}\big)\cr
&\hskip1cm = \smsum_{s_1,s_{2},t_1,t_2 \in \Si} 
\int {\sst d\xi_1 d\xi_{2}}\, 
\varphi({\sst(\xi_1,s_1),(\xi_2,s_2)}) 
\,c\big({\sst (\xi_1,t_1)},{\sst (\xi_{2},t_2)}\big)\cr
&\hskip1cm = 
\int {\sst d\xi_1 d\xi_{2}}\, 
\Big[ \smsum_{s_1,s_{2} \in \Si}
\varphi({\sst(\xi_1,s_1),(\xi_2,s_2)}) \Big] 
\,C\big({\sst \xi_1},{\sst \xi_{2}}\big)\cr
&\hskip1cm= \int \hskip -.1cm
\int {\sst d\xi_1 d\xi_{2}}\, \Big[ \smsum_{s_1,s_{2} \in \Si}
\varphi({\sst(\xi_1,s_1),(\xi_2,s_2)}) \Big]
\,\psi({\sst \xi_1})\psi({\sst \xi_{2}})\,d\mu_C(\psi)\cr
}$$
In the third equality, we used conservation of momentum to imply that
$$
\int {\sst d\xi_1 d\xi_{2}}\, 
\varphi({\sst(\xi_1,s_1),(\xi_2,s_2)}) 
\,c\big({\sst (\xi_1,t_1)},{\sst (\xi_{2},t_2)}\big)=0
$$
unless $\tilde s_1\cap t_1\ne\emptyset$ and $\tilde s_2\cap t_2\ne\emptyset$
and hence unless $s_1\cap t_1\ne\emptyset$ and $s_2\cap t_2\ne\emptyset$.

The claim for general $n$ is now proven by induction on $n$ using
integration by parts (see, for example, \S II.2 of [FKTr1]).

\Item ii)
Set $\cW' =\cW-\pi(w) \in I_O$. By assumption, 
$\check C\big((k,\si,a),(k',\si',a') \big) =0$ unless $k$ and $k'$ both 
lie in the $j^{\rm th}$ neighbourhood. Therefore 
$\int f(\psi+\ze)\ d\mu_C(\ze)\in I_O$ for all $f(\ze)\in I_O$.
Consequently
$$
\int e^{\pi(w)(\phi,\psi+\ze)}d\mu_C(\ze)
-\int e^{\cW(\phi,\psi+\ze)}d\mu_C(\ze)
=\int  e^{\pi(w)(\phi,\psi+\ze)}\Big[1-e^{\cW'(\phi,\psi+\ze)}\Big]d\mu_C(\ze)
\in I_O
$$
 since $1-e^{\cW'(\phi,\psi)}\in I_O$ and $I_O$ is an ideal. In particular
$$ 
Z\big(\pi(w)\big)=\int e^{\pi(w)(0,\ze)}d\mu_C(\ze)=\int e^{\cW(0,\ze)}d\mu_C(\ze)=Z(\cW)
$$
so that
$$
\sfrac{1}{Z(\pi(w))}\int e^{\pi(w)(\phi,\psi+\ze)}d\mu_C(\ze)
-\sfrac{1}{Z(\cW)}\int e^{\cW(\phi,\psi+\ze)}d\mu_C(\ze)
\in I_O
$$
Expanding the power series for $\log(1+x)$, one sees that
$$\eqalign{
\Om_C\big(\pi(w)\big)-\Om_C(\cW)
\in I_O
}$$
As 
$C_\Si(v,v') =C\big(\pi(v),\pi(v')\big)$, for all $v,v'\in V_\Si^{\rm eff}$,
functoriality of the renormalization group map (Remarks \remfunctor\ and 
\remrenschw.i of [FKTr1])
implies that 
$\Om_C\big(\pi(w)\big)=\pi \big(\Om_{C_\Si}(w)\big)$. So 
$\Om_C(\cW)-\pi \big(\Om_{C_\Si}(w)\big)\in I_O$. Also, by construction,
$\Om_{C_\Si}(w) \in \bigwedge_A V_\Si^{\rm eff}$. Hence, by Remark \remOSideal,
$\,\Om_{C_\Si}(w)$ is a sectorized representative for $\Om_C(\cW)$.

 If $v(\phi,\psi)$ is a sectorized representative for
$V(\phi,\psi)$, then $v(\phi,\psi+C J\phi)$ is a sectorized representative 
for $V(\phi,\psi+CJ\phi)$. Therefore the second claim follows from the first
and Lemma \lemOStworengrpmaps. 

\Item iii) is an immediate consequence of part (i) and Proposition \propBII.i
of [FKTr1].

\endproof

\definition{\STM\defOSsectnorm(Norms for sectorized functions)}{
Let $\Si$ be a  sectorization at scale $j\ge 2$ and let $m,\,n \ge 0$
and $p>0$ be integers.

\Item{i)}
For a function $\varphi$ on $\cB^m\times (\cB\times\Si)^n$ we define the seminorm
$\v \varphi\v_{p,\Si}$ to be zero 
if $m\ge 1$, $p\ge 2$ or if $m=0,\ p>n$.

\noindent
In the case $m \ge 1$, $p=1$ we set
$$
\v \varphi\v_{p,\Si} 
= \smsum_{s_i \in \Si}\, 
\|\varphi({\sst\eta_1,\cdots,\eta_m;\,(\xi_1,s_1),\cdots,(\xi_n,s_n)})
\|_{1,\infty} 
$$
 
\noindent
In the case $m =0$, $p\le n$ we set
$$
\v \varphi\v_{p,\Si}\ 
=\max_{1\le i_1<\cdots<i_p\le n}\ \ 
 \max_{s_{i_1},\cdots,s_{i_p} \in \Si} \ \ 
 \sum_{s_i \in \Si \ {\rm for} \atop i \ne i_1,\cdots,i_p} 
 \|\varphi({\sst (\xi_1,s_1),\cdots,(\xi_n,s_n)} ) \|_{1,\infty}
$$

\noindent
In both cases, the $\|\cdot\|_{1,\infty}$ norm (defined in Example
\exOSSymmNorm) applies to all the position space variables. Furthermore,
maxima  acting on a formal power series $\sum_\de a_\de t^\de$ are
to be applied separately to each coefficient $a_\de$.

\Item{ii)}
Let $f\in A_m\otimes\big(V_\Si^{\rm eff}\big)^{\otimes n}$ be a sectorized Grassmann function. 
Then there is a unique 
$\varphi\in \cF_m(n;\Si)$ such that $f={\rm Tens}(\varphi)$.
By definition
$$
\v f\v_{p,\Si} = \cases{
\v \varphi\v_{p,\Si}  & if  $\varphi$ is
translation invariant, conserves particle numbers\cr
&\hskip.5in and is spin independent\cr
\noalign{\vskip.05in}
\infty & otherwise \cr }
$$
}

\example{\STM\exOScommsectnorms}{ Let $f\in\cF_m(n)$ with $m\ge 1$. Then $f$ has a
sectorized representative $\varphi$ fulfilling
$$
\v \varphi\v_{1,\Si} \le \big( \sfrac{ \abcst}{\fl} \big)^n
\,\cb_j^n\,\|f\|_{1,\infty}
$$
}
\prf
Select a sectorized representative $\varphi$ for $f$ as in Example
\exOSsectrepr. For each choice of sectors $s_1,\cdots,s_n$ of sectors in $\Si$
$$\eqalign{
\|\varphi({\sst\eta_1,\cdots,\eta_m;\,(\xi_1,s_1),\cdots,(\xi_n,s_n)}) \|_{1,\infty}
&= \big\| \int\smprod_{i=1}^n\, ({\sst d\xi'_i\ \hat\chi_{s_i}(\xi_i,\xi'_i)})
\ f({\sst\eta_1,\cdots,\eta_m;\,\xi_1,\cdots,\xi_n} ) \big\|_{1,\infty} \cr
& \le \| f\|_{1,\infty}\,\smprod_{i=1}^n \|\hat\chi_{s_i}\|_{1,\infty} \cr
& \le \abcst^n \,\cb_{j-1}^n\,\| f\|_{1,\infty}
}$$
by Lemma \lemOSelloneinfty\ ($n$ times) and Lemma \lemOSsectpartunit.iii. As there are $\sfrac{\abcst}{\fl}$ sectors, the claim follows from the Definition.

\endproof

\remark{\STM\remOSdiffnorm}{
Let $\cD$ be a decay operator and $\varphi$ a function on $(\cB\times\Si)^n$. Then 
$$
\v \cD\varphi\v_{p,\Si}\le\sfrac{\partial^{|\de(\cD)|}}{\partial t^{\de(\cD)}}
\,\v \varphi\v_{p,\Si}
$$

}

\lemma{\STM\lemOSNormMom} {
Let $\varphi\in\cF_0(n;\Si)$ be a sectorized representative for
a translation invariant $f\in\cF_0(n)$. Then, for $p\le n$,
$$
\sum_{\de\in\bbbn_0\times\bbbn_0^2}\sfrac{1}{\de !}\bigg[
\max\limits_{\rD\ {\rm differential\ operator} \atop 
                {\rm with\ } \de(\rD) =\de} 
\sup_{\check\eta_1,\cdots,\check\eta_n\atop 
           {\check\eta_1+\cdots+\check\eta_n=0\atop
           k_1,\cdots,k_n{\rm\ in\ }j^{\rm th}{\rm\ neighbourhhood}}}
\big|\rD\check f(\check\eta_1,\cdots,\check\eta_n)\big|\bigg]\ t^\de
\le 2^p\, \v \varphi\v_{p,\Si}
$$
Here $\check \eta_i=(k_i,\si_i,b_i)$ and the differential operators $\rD$
are defined in Definition \defOSdiffdecay.iii. 
}
\prf
Fix a differential operator 
$\rD = \rD^{\de^{(1)}}_{u_1;v_1}\, \cdots  \rD^{\de^{(q)}}_{u_q;v_q}$ 
and let $\cD = \cD^{\de^{(1)}}_{u_1,v_1} \cdots  \cD^{\de^{(q)}}_{u_q,v_q}$
be the corresponding decay operator as in Definition \defOSdecayop.
Fix any $\check \eta_i=(k_i,\si_i,b_i),\ 1\le i\le n$ 
with $k_1,\cdots,k_n$ in the $j^{\rm th}$ neighbourhood and 
$\sum\limits_{i=1}^{n}(-1)^{b_i} k_i=0$. 
Then
$$\eqalign{
\check f\big(\check \eta_1,\cdots,\check \eta_n\big)
&=\smsum_{s_i\owns k_i\atop 1\le i\le n}
\check \varphi\big((k_1,\si_1,b_1,s_1),\cdots,(k_n,\si_n,b_n,s_n)\big)\cr
}$$
so that
$$\eqalign{
|\rD\check f\big(\check \eta_1,\cdots,\check \eta_n\big)|
&\le\smsum_{s_i\owns k_i\atop 1\le i\le p}
\smsum_{s_i\in\Si\atop p+1\le i\le n}\ 
\int \smprod_{\ell\ne n}d\xi_\ell\ 
\big|\cD\varphi\big((\xi_1,s_1),\cdots,(\xi_{n-1},s_{n-1}),(0,\si_n,b_n,s_n)\big)
\big|\cr
&\le 2^p\max_{s_1,\cdots,s_p}\smsum_{s_i\in\Si\atop p+1\le i\le n}
\TN \cD\varphi\big((\ \cdot\ ,s_1),\cdots,(\ \cdot\ ,s_{n})\big)\TN_{1,\infty}\cr
}$$
since each $k_i$ can be contained in at most two sectors.

\endproof

\remark{\STM\remOScommsectnorms}{
We will use the norms of Definition \defOSsectnorm\ in a multi scale analysis
to prove existence of Green's functions in the position space $L_\infty$--norm.
This is the reason why we take the suprema over the external variables 
$\eta_1,\cdots,\eta_m$, and why we do not ``sectorize'' these variables.

In Section \CHsecmomnorm, we will introduce another set of norms, designed 
to study the smoothness properties of the amputated two and four point functions in momentum space.
} 
\vskip 1cm

\lemma{\STM\lemOSelloneinftysectors}{
Let $\Si$ be a  sectorization and let
$\varphi$ be a function on $\cB^{m}\times \big(\cB\times\Si\big)^{n}$,  $\varphi'$  be a function on $\cB^{m'}\times \big(\cB\times\Si\big)^{n'}$ and 
$1\le i \le n,\  1\le i'\le n'$. Define the function 
$\ga$ on $\cB^{m+m'}\times \big(\cB\times\Si\big)^{n+n'-2}$ by
$$\eqalign{
&\ga({\sst \eta_1,\cdots,\eta_{m+m'};
(\xi_1,s_1),\cdots,(\xi_{i-1},s_{i-1}),\,(\xi_{i+1},s_{i+1}),\cdots,
(\xi_{n+i'-1},s_{n+i'-1}),\,
(\xi_{n+i'+1},s_{n+i'+1}),\cdots,(\xi_{n+n'},s_{n+n'})}) \cr
&\hskip .1cm  = \smsum_{s,s'\in\Si\atop s\cap s'\ne\emptyset}
\int_\cB\!\!{\sst d\zeta}\,
\varphi( {\sst \eta_1,\cdots,\eta_m;\, 
(\xi_1,s_1),\cdots,(\xi_{i-1},s_{i-1}),\,(\zeta,s),\,(\xi_{i+1},s_{i+1}),\cdots,(\xi_n,s_n) })\cr
& \hskip 1.4cm \cdot\ \varphi'( {\sst \eta_{m+1},\cdots,\eta_{m+m'};
\,(\xi_{n+1},s_{n+1}),\cdots,(\xi_{n+i'-1},s_{n+i'-1}),\,(\zeta,s'),\,
(\xi_{n+i'+1},s_{n+i'+1}),\cdots,(\xi_{n+n'},s_{n+n'})})
\cr
}$$
If $m=0$ or $m'=0$, 
$$
\v\ga\,\v_{p,\Si} \le 3 \ 
\max_{p_1+p_2 =p+1 \atop p_1,p_2 \ {\rm odd}}
\v\varphi\v_{p_1,\Si}\ \v\varphi'\v_{p_2,\Si}
$$
for all odd natural numbers $p$.
}

\prf
The variable indices for $\ga$ lie in the set $I\cup I'$, where
$$\eqalign{
I & = \{1,\cdots,i-1,\,i+1,\cdots,n\}\cr
I' & = \{n+1,\cdots,n+i'-1,\,n+i'+1,\cdots,n+n'\}\cr 
}$$ 
Fix $u_1,\cdots,u_q \in I,\ u_{q+1},\cdots u_p \in I'$ and fix sectors 
$s_{u_1},\cdots,s_{u_p} \in \Si$. 

First assume that $q$ is odd. Then $p-q$ is even.  By Lemma \lemOSelloneinfty, 
for each choice of sectors 
$s_\nu,\ \nu \in I\cup I' \setminus \{u_1,\cdots,u_p\}$ one has
$$\eqalign{
\|& \ga({\sst \eta_1,\cdots,\eta_{m+m'}\,;\,
(\xi_1,s_1),\,\cdots\,,(\xi_{i-1},s_{i-1}),\,(\xi_{i+1},s_{i+1}),\,\cdots\,,
(\xi_{n+i'+1},s_{n+i'+1}),\,\cdots\,,(\xi_{n+n'},s_{n+n'})})  \|_{1,\infty}  \cr
& \hskip .2cm \le\smsum_{s,s'\in\Si\atop s\cap s'\ne\emptyset}
\|\varphi( {\sst \eta_1,\cdots,\eta_m;\, 
(\xi_1,s_1),\,\cdots\,,(\xi_{i-1},s_{i-1}),\,(\zeta,s),\,(\xi_{i+1},s_{i+1}),\,
\cdots\,,(\xi_n,s_n) }) \|_{1,\infty} \cr
 & \hskip 2.2cm \cdot\ 
\|\varphi'( {\sst \eta_{m+1},\cdots,\eta_{m+m'};
\,(\xi_{n+1},s_{n+1}),\,\cdots\,,(\xi_{n+i'-1},s_{n+i'-1}),\,(\zeta',s'),\,
\cdots\,,(\xi_{n+n'},s_{n+n'})}) \|_{1,\infty} \cr
}$$
Observe that for every $s\in\Si$ there are at most three sectors $ s'$ 
such that $s'\cap s \ne \emptyset$. Consequently
$$\eqalign{
&\sum_{s_\nu \in \Si \atop \nu \in I\cup I' \setminus \{u_1,\cdots,u_p\}}
 \hskip -.6cm
\| \ga({\sst \eta_1,\cdots,\eta_{m+m'};
(\xi_1,s_1),\cdots,(\xi_{i-1},s_{i-1}),\,(\xi_{i+1},s_{i+1}),\cdots,\,
(\xi_{n+i'+1},s_{n+i'+1}),\cdots,(\xi_{n+n'},s_{n+n'})})  \|_{1,\infty} \cr
& \hskip .5cm \le 3 \hskip -.6cm
\sum_{s_\nu  \in  \Si
 \atop   \nu\in I\setminus \{u_1,\cdots,u_q\}}  \hskip -.4cm
\smsum_{s\in\Si}
\|\varphi( {\sst \eta_1,\cdots,\eta_m;\, 
(\xi_1,s_1),\cdots,(\xi_{i-1},s_{i-1}),\,(\zeta,s),\,(\xi_{i+1},s_{i+1}),\cdots,(\xi_n,s_n) }) \|_{1,\infty} \cr
& \hskip 1.2cm \cdot\ \max_{s'\in \Si}  \hskip -.8cm
\sum_{ s'_\mu \in  \Si
 \atop \mu\in I'\setminus \{u_{q+1},\cdots,u_p\} }  \hskip -.6cm
\|\varphi'( {\sst \eta_{m+1},\cdots,\eta_{m+m'};
\,(\xi_{n+1},s_{n+1})\cdots,(\xi_{n+i'-1},s_{n+i'-1}),\,(\zeta',s'),\,
\cdots,(\xi_{n+n'},s_{n+n'})}) \|_{1,\infty} \cr
& \hskip .5cm \le 3\ \v\varphi\v_{q,\Si}\ \  \v\varphi'\v_{p-q+1,\Si}
}$$
The case that $q$ is even follows as in the case discussed above 
by interchanging the roles of $\varphi$ and $\varphi'$.
\endproof

We define ``contraction'' for sectorized functions as the obvious generalization of Definition \defOScontnorm.
\definition{\STM\defOSsectcontnorm }{ 
Let $c\big((\xi,s),(\xi',s')\big)$ be any skew symmetric function on $\big(\cB\times\Si\big)^2$. Let 
$m,n\ge 0$ and $1\le i < j\le n$. For $\varphi\in \cF_m(n;\Si)$ the contraction 
$\Cont{i}{j}{c} \varphi \in \cF_m(n-2;\Si) $ is defined as
$$\eqalign{
&\Cont{i}{j}{c} \varphi\, ({\sst \eta_1,\cdots,\eta_m;
(\xi_1,s_1),\cdots,(\xi_{i-1},s_{i-1})\,,\,
(\xi_{i+1},s_{i+1}),\cdots,(\xi_{j-1},s_{j-1})\,,\,
(\xi_{j+1},s_{j+1}),\cdots,(\xi_n,s_n)})
\cr
& \hskip1cm= (-1)^{j-i+1} 
\sum_{s_i,s_j,t_i,t_j\in\Si\atop 
            {t_i\cap s_i\ne\emptyset\atop t_j\cap s_j\ne\emptyset}}
\int {\sst d\xi_i\,d\xi_j}\  c({\sst (\xi_i,t_i),(\xi_j,t_j)})\,
\varphi({\sst \eta_1,\cdots,\eta_m;
(\xi_1,s_1),\cdots,(\xi_n,s_n)}) 
}$$  
}

\proposition{\STM\propOScontrintboundsectors}{
Let $\Si$ be a  sectorization of length $\fl$ 
at scale $j\ge 2$ and let $c\big((\xi,s),(\xi',s')\big) \in \cF_0(2;\Si)$ be an antisymmetric function. 
\Item{i)}
Let $p$ be an odd integer, $m,m'\ge 0$, $n,n'\ge 1$
and $\varphi\in\cF_m(n;\Si),\ \varphi'\in\cF_{m'}(n',\Si)$.
If $m=0$ or $m'=0$
$$
\V \Cont{1}{n+1}{c}\, {\rm Ant}_{\rm ext}(\varphi\otimes \varphi')\V_{p,\Si}\
\le 9\,\v c \v_{1,\Si}
\max_{p_1+p_2 =p+1 \atop p_1,p_2 \ {\rm odd}}
\v \varphi\v_{p_1,\Si}\ \v \varphi'\v_{p_2,\Si}
$$ 
If $m\ne 0$ and $m'\ne0$ 
$$
\V \Cont{1}{n+1}{c}\, {\rm Ant}_{\rm ext}(\varphi\otimes \varphi')\V_{1,\Si}\
\le 9\big( \sup_{\xi,\xi',s,s'} \big| c\big((\xi,s),(\xi',s')\big) \big| \big) \
 \v \varphi\v_{1,\Si}\ \v \varphi'\v_{1,\Si}
$$

\Item{ii)} 
Assume that there is a function $C(k)$ 
that is supported in the  $j^{\rm th}$ neighbourhood, such that $c\big((\cdot,s),(\cdot,s') \big)$ is the Fourier transform of
$\chi_s(k)\,C(k)\,\chi_{s'}(k)$ in the sense of Definition \defOSftcov\ 
and that $|C(k)| \le \sfrac{\veps}{|\imath k_0 -e(\k)|}$ for
some $\veps \ge 0$.

\noindent
Let  $\varphi\in \cF_m(n;\Si)$, $n'\le n$ and set, as in 
Definition \defOSintbnd\ of integral bound,
$$\eqalign{
\varphi'&
({\sst \eta_1,\cdots,\eta_m;\,(\xi_{n'+1},s_{n'+1}),\cdots,(\xi_n,s_n)})\cr
& = 
\smsum_{s_i\in\Si \atop i=1,\cdots,n'}
\int \hskip -3pt \int {\sst d\xi_1\cdots d\xi_{n'}}\ 
\varphi({\sst\eta_1,\cdots,\eta_m;\,(\xi_1,s_1),\cdots,(\xi_{n'},s_{n'}),
\cdots,(\xi_n,s_n)})\ 
\psi({\sst\xi_1,s_1})\cdots\psi({\sst \xi_{n'},s_{n'}})\,d\mu_{C_\Si}(\psi)
}$$
where
$$
C_\Si\big(\psi({\sst\xi,s}),\psi({\sst\xi',s'}) \big) = 
\sum_{t\cap s \ne \emptyset \atop t'\cap s' \ne \emptyset}
c\big((\xi,t),(\xi',t')\big)
$$
Then for all $p$
$$
\v \varphi'\v_{p,\Si}\ 
\le\ \big(\veps\,\IB_1\, \sfrac{\fl}{M^j}\big)^{n'/2}\,
\v \varphi\v_{p,\Si}
$$
with a constant $\IB_1$ that is independent of $j$ and $\Si$.
}

\prf
\Item{i)} 
Set 
$$\eqalign{
&\ga({\sst \eta_1,\cdots,\eta_{m+m'};
(\xi_2,s_2),\cdots,(\xi_n,s_n),\,(\xi_{n+2},s_{n+2})\cdots,
(\xi_{n+n'},s_{n+n'})}) \cr
&\hskip 1cm  = \smsum_{s,s',t,t'\in\Si\atop 
            {t\cap s\ne\emptyset\atop t'\cap s'\ne\emptyset}}
\ \int {\sst d\zeta d\eta}\,
\varphi( {\sst \eta_1,\cdots,\eta_m;\, 
(\zeta,s),(\xi_2,s_2),\cdots,(\xi_n,s_n) }) \,
c ({\sst(\zeta,t),(\eta,t')})\cr
\noalign{\vskip-.2in}
& \hskip 6cm \cdot\ \varphi'( {\sst \eta_{m+1},\cdots,\eta_{m+m'};
(\eta,s')\,(\xi_{n+2},s_{n+2})\cdots,(\xi_{n+n'},s_{n+n'})}) \cr
}$$
Then $(-1)^{n+1}\rm Ant_{ext} \,\ga = \Cont{1}{n+1}{c}\, {\rm Ant}_{\rm
ext}(\varphi\otimes \varphi')$. If $m\ne 0$ and $m'\ne 0$ then
$$\eqalign{
\V\Cont{1}{n+1}{c}\, {\rm Ant}_{\rm ext}(\varphi\otimes \varphi')\V_{1,\Si} 
& =\v  \rm Ant_{ext} \,\ga\,\v_{1,\Si}  \cr
&\le 9 \big( \sup_{\xi,\xi',t,t'} | c\big((\xi,t),(\xi',t')\big) | \big) \ 
 \v \varphi\v_{1,\Si}\ \v \varphi'\v_{1,\Si}  \cr
}$$
If $m=0$ or $m'=0$, by iterated application of Lemma \lemOSelloneinftysectors
$$\eqalign{
\v \ga\v_{p,\Si} 
& \le 3\hskip -5pt\max_{p_1+p_2=p+1 \atop p_1,p_2 \ {\rm odd}} 
\VV \smsum_{s,t \in \Si\atop s\cap t\ne\emptyset} \int {\sst d\zeta }\, 
\varphi( {\sst \eta_1,\cdots,\eta_m;\, 
(\zeta,s),(\xi_2,s_2),\cdots,(\xi_n,s_n) }) \,
c ({\sst(\zeta,t),(\eta,t')}) \VV_{p_1,\Si} \ \v\varphi'\v_{p_2,\Si} \cr
& \le 9\,\max_{p_1+p_2=p+1 \atop p_1,p_2 \ {\rm odd}}\  
\v \varphi \v_{p_1,\Si}  \ \ 
\v c\v_{1,\Si} \ \ \v\varphi'\v_{p_2,\Si} \cr
}$$

\Item{ii)}
Define the covariance $C(\xi,\xi')$ to be the Fourier transform of $C(k)$.
By part (i) of Proposition \propOSfunctorialitySect,
$$\eqalign{
&\varphi'({\sst \eta_1,\cdots,\eta_m;\,(\xi_{n'+1},s_{n'+1}),\cdots,(\xi_n,s_n)})\cr
&  \hskip .4cm =  \int \hskip -.1cm
\int {\sst d\xi_1\cdots d\xi_{n'}}\, \Big[ \smsum_{s_1,\cdots,s_{n'} \in \Si}
\varphi({\sst\eta_1,\cdots,\eta_m;\,(\xi_1,s_1),\cdots,\,
(\xi_{n'},s_{n'}),\cdots,(\xi_n,s_n)}) \Big]
\,\psi(\xi_1)\cdots\psi(\xi_{n'})\,d\mu_C(\psi)
}$$
Since 
$\varphi \in \cF_m(n;\Si)$
$$\eqalign{
&\varphi({\sst\eta_1,\cdots,\eta_m;\,(\xi_1,s_1),\cdots,\,
(\xi_{n'},s_{n'}),\,(\xi_{n'+1},s_{n'+1})\cdots,(\xi_n,s_n)}) \cr
& \hskip 2cm= \int {\sst d\xi'_1\cdots d\xi'_{n'}}\,
\varphi({\sst\eta_1,\cdots,\eta_m;\,(\xi'_1,s_1),\cdots,\,
(\xi'_{n'},s_{n'}),\,(\xi_{n'+1},s_{n'+1})\cdots,(\xi_n,s_n)}) \,
\smprod_{i=1}^{n'} \hat{\tilde \chi}_{s_i}(\xi'_i,\xi_i)
}$$
Consequently
$$\eqalign{
&\varphi'({\sst \eta_1,\cdots,\eta_m;\,(\xi_{n'+1},s_{n'+1}),\cdots,(\xi_n,s_n)})\cr
&  \hskip .2cm = \hskip -.5cm\smsum_{s_1,\cdots,s_{n'} \in \Si}\int 
\int {\sst d\xi'_1\cdots d\xi'_{n'}}\,
\varphi({\sst\eta_1,\cdots,\eta_m;\,(\xi'_1,s_1),\cdots,\,
(\xi'_{n'},s_{n'}),\cdots,(\xi_n,s_n)}) \
\psi_{s_1}(\xi'_1)\cdots \psi_{s_{n'}}(\xi'_{n'})\,d\mu_C(\psi)
}$$
where
$$
\psi_s(\xi') = \int  d\xi\,\hat{\tilde \chi}_s(\xi',\xi)\,\psi(\xi)
$$
Therefore, by part (ii) of Proposition \propIntBndsII
$$\eqalign{
&\big| \varphi'({\sst \eta_1,\cdots,\eta_m;\,(\xi_{n'+1},s_{n'+1}),\cdots,(\xi_n,s_n)}) \big|\cr
&  \hskip .6cm \le \hskip -.5cm\smsum_{s_1,\cdots,s_{n'} \in \Si}
\int {\sst d\xi_1\cdots d\xi_{n'}}\,
| \varphi({\sst\eta_1,\cdots,\eta_m;\,(\xi_1,s_1),\cdots,(\xi_n,s_n)})| 
\ \Big| \int 
\psi_{s_1}(\xi_1)\cdots \psi_{s_{n'}}(\xi_{n'})\,d\mu_C(\psi) \Big| \cr
&  \hskip .6cm \le  G^{n'/2}\smsum_{s_1,\cdots,s_{n'} \in \Si}
\int {\sst d\xi_1\cdots d\xi_{n'}}\,
| \varphi({\sst\eta_1,\cdots,\eta_m;\,(\xi_1,s_1),\cdots,(\xi_n,s_n)})|  \cr
}$$
with
$$
G\ =
\ \int \sfrac{d^{d+1}k}{(2\pi)^{d+1}}\,\tilde\chi_s(k)^2\,|C(k)|
\ \le\ \veps \int \sfrac{d^{d+1}k}{(2\pi)^{d+1}}\,\sfrac{\tilde\chi_s(k)^2}{|\imath k_0 -e(\k)|}
\ \le \ \const\ \veps \,\sfrac{\fl}{M^j}
$$
\endproof

\remark{\STM\remOShavesectors}{
If $C$ fulfills the hypothesis of part (i) of the Proposition, then
$$
\cb = 9\,\max \big\{\ \v c \v_{1,\Si},\  
\sup_{\xi,\xi',s,s'} \big| c\big((\xi,s),(\xi',s')\big)\big|\  \big\}
$$ 
is a contraction bound for the system $\v\cdot\v_{1,\Si}$ of seminorms.
We shall show in Proposition \:\propOSrealpropbound, that for 
$C^{(j)}(k) = \sfrac{\nu^{(j)}(k)}{\imath k_0 - e(\k)}$
and $\sfrac{1}{M^{j-3/2}}\le\fl\le\sfrac{1}{M^{(j-1)/2}}$, 
the constant coefficient  $\cb_\0$ of $\cb$ is bounded by $\const M^j$. 
On the other hand, part (ii) of Proposition
\propOScontrintboundsectors\ shows that
$\ib=\const\sqrt{\sfrac{\fl}{M^j}}$ is an integral bound for $C_\Si^{(j)}$ 
with respect to this system of seminorms. 
Thus, if $\cW(\psi)$ is an even Grassmann function with sectorized representative
$$
w(\psi) = \sum_{n=0 }^\infty\  \smsum_{s_1,\cdots,s_{2n} \in \Si}
\int {\sst d\xi_1\cdots d\xi_{2n}}\, w_{2n}({\sst (\xi_1,s_1),\cdots,(\xi_{2n},s_{2n})})\,
\psi({\sst \xi_1,s_1}) \cdots \psi({\sst \xi_{2n}, s_{2n}})
$$
the quantity $N(\cW;\cb,\ib,\al)$ of Definition \deffunctnorm\ of [FKTr1]
(with $V$ replaced by $V_\Si$) has
$$
N(w;\cb,\ib,\al)_\0 \le \const \Big\{ \al^2\,M^j\ \v w_2\v_{1,\Si}\ +\ 
\al^4\,\fl\  \v w_4\v_{1,\Si}\ +\ 
\sfrac{M^{2j}}{\fl} \smsum_{n\ge 3} \big(\const\sfrac{\al^2\,\fl}{M^j}\big)^n
\v w_{2n}\v_{1,\Si} \Big\}
$$
In contrast to the situation of Remark \remOSneedsectors, this norm is
of order one if  $\v w_4\v_{1,\Si}$ is of order $\sfrac{1}{\fl}$, which 
is approximately the number of sectors. As $d=2$, this is a realistic 
estimate for the  original interaction $\cV$, with all momenta restricted 
to the $j^{\rm th}$ shell. Observe, however, that for $d\ge 3$ 
this estimate is not expected to hold.  See [FKTf1, \S \CHintroOverview, 
subsection 8].

\noindent
For more precise control of $W_4$ one also uses the norm $\v w_4\v_{3,\Si}$.

}

To prepare for the application of Theorem \theoremVa\ in [FKTr2] about overlapping loops,
we state

\proposition{\STM\propOSoverlapploops}{
Let $\Si$ be a  sectorization of length $\fl$ at scale $j\ge 2$ and let
$c\big((\xi,s),(\xi',s')\big)\in \cF_0(2;\Si)$ be an antisymmetric function. 
Let $D(k),\,D'(k)$ be functions obeying 
$|D(k)|,\,|D'(k)| \le \sfrac{2}{|\imath k_0 -e(\k)|}$ and let 
$d\big((\cdot,s),(\cdot,s') \big)$ resp. 
$d'\big((\cdot,s),(\cdot,s') \big)$ be the Fourier transform of
$\chi_s(k)\,D(k)\,\chi_{s'}(k)$ resp. $\chi_s(k)\,D'(k)\,\chi_{s'}(k)$
in the sense of Definition \defOSftcov. 

\noindent
Let $1\le i_1,i_2,i_3 \le n$, $1\le i'_1,i'_2,i'_3 \le n'$ with
$i_1\ne i_2 \ne i_3 \ne i_1$, $i'_1\ne i'_2 \ne i'_3 \ne i'_1$, 
and let $p$ be an odd natural number. 
Then for $\varphi\in\cF_0(n;\Si)$, $\varphi'\in \cF_0(n',\Si)$
$$
\V
\Cont{i_1}{n+i_1'}{c}\,\Cont{i_2}{n+i_2'}{d}\,\Cont{i_3}{n+i_3'}{d'}\, 
(\varphi\otimes \varphi')\V_{p,\Si}\
\le\ \big(\IB_2\,\sfrac{\fl}{M^{j}} \big)^2\v c \v_{1,\Si}
\max_{p_1+p_2 =p+3 \atop p_1,p_2 \ {\rm odd}}
\v \varphi\v_{p_1,\Si}\ \v \varphi'\v_{p_2,\Si}
$$
with a constant $\IB_2$ that is independent of $j$ and $\Si$.

}

\prf
By the symmetry of the norms we may assume that 
$i_1=i'_1=1,\ i_2=i'_2=2,\ i_3=i'_3=3$. Set
$$\eqalign{
&\ga({\sst(\xi_4,s_4),\cdots,(\xi_n,s_n),\,(\xi_{n+4},s_{n+4})\cdots,
(\xi_{n+n'},s_{n+n'})}) \cr
&\hskip .2cm  = \hskip -.5cm 
\smsum_{s_i,t_i,t_i',s'_i\in\Si \atop i=1,2,3}\ 
\int \smprod_{i=1}^3 {\sst( d\zeta_i d\eta_i)}  \,
\varphi( {\sst 
(\zeta_1,s_1),(\zeta_2,s_2),(\zeta_3,s_3),(\xi_4,s_4),\cdots,(\xi_n,s_n) }) \,
c({\sst (\zeta_1,t_1),(\eta_1,t'_1)})\cr
& \hskip 2cm \cdot\ 
d({\sst (\zeta_2,t_2),(\eta_2,t'_2)})\,
d'({\sst (\zeta_3,t_3),(\eta_3,t'_3)})\ 
\varphi'( {\sst (\eta_1,s_1'),(\eta_2,s_2'),(\eta_3,s_3'),
(\xi_{n+4},s_{n+4})\cdots,(\xi_{n+n'},s_{n+n'})}) \cr
}$$
Then 
$$
(-1)^{3(n+1)}\rm Ant_{ext} \,\ga = 
\Cont{i_1}{n+i_1'}{c}\,\Cont{i_2}{n+i_2'}{d}\,\Cont{i_3}{n+i_3'}{d'}\, 
(\varphi\otimes \varphi')
$$
Observe that $d({\sst (\zeta,t),(\eta,t')})=0$ if $t\cap t'=\emptyset$ and
that
$$\eqalign{
\sup_{\zeta,\eta;\,t,t'}\
\big| d({\sst (\zeta,t),(\eta,t'})\big|   
& \le \int\sfrac{d^{d+1}k}{(2\pi)^{d+1}}\ \chi_t(k)\,|D(k)| \,\chi_{t'}(k)\ \cr 
& \le \int\sfrac{d^{d+1}k}{(2\pi)^{d+1}}\, 
\sfrac{ 2\,\chi_t(k)\ \chi_{t'}(k)}{|\imath k_o-e(\k)|} \cr
& \le \const \sfrac{\fl}{M^j}
}\EQN\eqnOSsupDbound$$
The same properties hold for $d'$.

Set
$$
\varphi''( {\sst (\ze_1,t_1),(\xi_2,s_2),\cdots,(\xi_{n'},s_{n'})})
= \smsum_{t_1',s_1'\in\Si} \int {\sst d\et_1}\,
c({\sst (\ze_1,t_1),(\et_1,t_1')})\,
\varphi'( {\sst (\et_1,s_1'),(\xi_2,s_2),\cdots,(\xi_{n'},s_{n'})})
$$
By Lemma \lemOSelloneinftysectors
$$
\v\varphi''\,\v_{p,\Si} 
\le 3 \,\v c\v_{1,\Si}\,\v\varphi'\,\v_{p,\Si} 
$$
Furthermore
$\check\varphi''( {\sst (k_1,\si_1,a_1,t_1),(\check\xi_2,s_2),\cdots,(\check\xi_{n'},s_{n'})})=0$
unless $k_1\in \tilde t_1$. Set
$$\eqalign{
\tilde \gamma&({\sst(\xi_4,s_4),\cdots,(\xi_n,s_n),\,(\xi_{n+4},s_{n+4})\cdots,
(\xi_{n+n'},s_{n+n'})};{\sst s_1,t_1,\,s_2,t_2,t_2',s_2',\,s_3,t_3,t_3',s_3'})
\cr
&\hskip 1cm = \int {\sst d\ze_1\, d\zeta_2 d\eta_2\, d\zeta_3 d\eta_3 }  \,
\varphi( {\sst 
(\zeta_1,s_1),(\zeta_2,s_2),(\zeta_3,s_3),(\xi_4,s_4),\cdots,(\xi_n,s_n) })\,
d({\sst (\zeta_2,t_2),(\eta_2,t'_2)})\cr
& \hskip 3.8cm 
\,\cdot\ 
d'({\sst (\zeta_3,t_3),(\eta_3,t'_3)})\ 
\varphi''( {\sst (\ze_1,t_1),(\eta_2,s_2'),(\eta_3,s_3'),
(\xi_{n+4},s_{n+4})\cdots,(\xi_{n+n'},s_{n+n'})}) \cr
}$$
Then
$$\eqalign{
&\ga({\sst(\xi_4,s_4),\cdots,(\xi_n,s_n),\,(\xi_{n+4},s_{n+4})\cdots,
(\xi_{n+n'},s_{n+n'})}) \cr
&\hskip .2cm  = 
\smsum_{s_1,t_1\in \Si} \ \ 
\smsum_{s_i,t_i,t_i',s'_i\in\Si \atop i=2,3}\ 
\tilde \gamma ({\sst(\xi_4,s_4),\cdots,(\xi_n,s_n),\,(\xi_{n+4},s_{n+4})\cdots,
(\xi_{n+n'},s_{n+n'})};{\sst s_1,t_1,\,s_2,t_2,t_2',s_2',\,s_3,t_3,t_3',s_3'})
\cr
}$$
and $\tilde \ga( \cdot\ ;
{\sst s_1,t_1,\,s_2,t_2,t_2',s_2',\,s_3,t_3,t_3',s_3'})=0\ $ unless 
$s_1\cap t_1 \ne \emptyset$ and $s_i\cap t_i\cap t_i'\cap s_i' \ne \emptyset$
for $i=2,3$.  
By Corollary \corOSelloneinfty, for all choices of sectors,
$$\eqalign{
\| \tilde \gamma &
({\sst(\,\cdot\,,s_4),\cdots,(\,\cdot\,,s_n),\,\,\cdot\,,s_{n+4})\cdots,
(\,\cdot\,,s_{n+n'})};{\sst s_1,t_1,\,s_2,t_2,t_2',s_2',\,s_3,t_3,t_3',s_3'})
\|_{1,\infty} \cr
& \hskip 1cm \le \sup |d|\,\sup|d'| \ 
\|\varphi( {\sst(\,\cdot\,,s_1),(\,\cdot\,,s_2),(\,\cdot\,,s_3),(\,\cdot\,,s_4),
\cdots,(\,\cdot\,,s_n) })  \|_{1,\infty} \cr
& \hskip6.6cm \cdot\ 
\|\varphi''( {\sst (\,\cdot\,,t_1),(\,\cdot\,,s_2'),(\,\cdot\,,s_3'),
(\,\cdot\,,s_{n+4})\cdots,(\,\cdot\,,s_{n+n'})}) \|_{1,\infty}
}$$

The variable indices for $\ga$ lie in the set $I\cup I'$, where
$$
I  = \{4,\cdots,n\}\qquad,\qquad
I' = \{n+4,\cdots,n+n'\}
$$ 
Fix $u_1,\cdots,u_q \in I,\ u_{q+1},\cdots u_p \in I'$ and fix sectors 
$s_{u_1},\cdots,s_{u_p} \in \Si$.
First assume that $q$ is odd so that $p-q$ is even. Then, by (\eqnOSsupDbound) and the estimate on $\tilde \ga$ 
above
$$\eqalign{
& \smsum_{s_i\in\Si \atop i\in I\cup I'\setminus\{u_1,\cdots,u_p\}} 
\|\ga({\sst(\,\cdot\,,s_4),\cdots,(\,\cdot\,,s_n),\,(\,\cdot\,,s_{n+4})\cdots,
(\,\cdot\,,s_{n+n'})}) \|_{1,\infty}\cr
&\hskip 1cm  \le  \const \,\sfrac{\fl^2}{M^{2j}}
\smsum_{s_i\in\Si \atop{ i\in \{1,\cdots,n\}\atop i\ne u_1,\cdots,u_q}}
\ \hskip -.3cm
\|\varphi( {\sst(\,\cdot\,,s_1),(\,\cdot\,,s_2),(\,\cdot\,,s_3),(\,\cdot\,,s_4),
\cdots,(\,\cdot\,,s_n) })  \|_{1,\infty}\cr
& \hskip 2.5cm \cdot\ 
\max_{t_1,s'_2,s'_3 \in \Si}\ \hskip -.4cm
\smsum_{s_i\in\Si \atop i\in I'\setminus \{u_{q+1},\cdots,u_p\}}
\|\varphi''( {\sst (\,\cdot\,,t_1),(\,\cdot\,,s_2'),(\,\cdot\,,s_3'),
(\,\cdot\,,s_{n+4})\cdots,(\,\cdot\,,s_{n+n'})}) \|_{1,\infty} \cr
&\hskip 1cm  \le \const \sfrac{\fl^2}{M^{2j}}\
\v \varphi\v_{q,\Si}\ \ \v \varphi''\v_{p-q+3,\Si} \cr
&\hskip 1cm  \le \const \sfrac{\fl^2}{M^{2j}}\ \ \v c \v_{1,\Si}\ \ 
\v \varphi\v_{q,\Si}\ \ \v \varphi'\v_{p-q+3,\Si} \cr
}$$
The case that $q$ is even is similar.
\endproof

To treat source terms, we state, motivated by Definition \defOSextimpr,

\lemma{\STM\lemOSsectextimpr (External improving)}{
Let $C(k)$ be  a function obeying 
$|C(k)|\le \sfrac{2}{|\imath k_0 -e(\k)|}$ and  
$c\big((\cdot,s),(\cdot,s') \big)$ be the Fourier transform of
$\chi_s(k)\,C(k)\,\chi_{s'}(k)$ in the sense of Definition \defOSftcov. 
Let $\varphi\in\cF_m(n;\Si)$, $1\le i\le n$ and set
$$\eqalign{
&\varphi'({\sst\et_1,\cdots,\et_{m+1}\,;
\,(\xi_1,s_1),\cdots,(\xi_{n-1},s_{n-1})})\cr
&\hskip1cm= {\rm Ant_{ext}}\smsum_{s,t,t'\in\Si}\int\!\! d{\sst \ze}d{\sst \ze'}\ 
\varphi({\sst\et_1,\cdots,\et_m\,;\,(\xi_1,s_1),\cdots,
(\xi_{i-1},s_{i-1}),(\ze',t),(\xi_{i},s_i),\cdots,(\xi_{n-1},s_{n-1})})\cr
\noalign{\vskip-.1in}
&\hskip10cm \cdot\ c({\sst(\ze',t',),(\ze,s)})\,J({\sst \ze,\et_{m+1}})\cr
}$$
Then
$$
\v\varphi'\v_{1,\Si}
\le \const \v \varphi\v_{1,\Si}\cases{\sfrac{1}{\fl}\v c\v_{1,\Si}& if $m=0$ \cr
                                \noalign{\vskip.05in}
                               \sfrac{\fl}{M^j}& if $m\ne 0$ \cr}
$$
with a constant that is independent of $j$ and $\Si$.
 }
\prf First consider the case $m=0$.
Define the function $\varphi''$ on $\big(\cB\times\Si\big)
\times\big(\cB\times\Si\big)^{n-1}$ by
$$\eqalign{
&\varphi''({\sst(\et,s)\,;
\,(\xi_1,s_1)\cdots,(\xi_{n-1},s_{n-1})})\cr
&\hskip1cm= \smsum_{t,t'\in\Si}\int\!\! d{\sst \ze}d{\sst \ze'}\ 
\varphi({\sst(\xi_1,s_1)\cdots,
(\xi_{i-1},s_{i-1}),(\ze',t),(\xi_{i},s_i),\cdots,(\xi_{n-1},s_{n-1})})\,
c({\sst(\ze',t'),(\ze,s)})\,J({\sst \ze,\et}) \cr
}$$
Then
$$
\varphi'({\sst\et\,;
\,(\xi_1,s_1)\cdots,(\xi_{n-1},s_{n-1})})
=\sum_{s\in\Si}
\varphi''({\sst(\et,s)\,;\,(\xi_1,s_1)\cdots,(\xi_{n-1},s_{n-1})})
$$
Hence, by Lemma \lemOSelloneinftysectors,
$$
\v\varphi'\v_{1,\Si}
\le|\Si|\,\v\varphi''\v_{1,\Si}
\le\sfrac{\abcst}{\fl}\v \varphi\v_{1,\Si}\ \v cJ\v_{1,\Si}
\le\sfrac{\abcst}{\fl}\v c\v_{1,\Si}\  \v \varphi\v_{1,\Si}
$$
Now suppose that $m\ne 0$. Then
$$\eqalign{
&\V \varphi'({\sst\et_1,\cdots,\et_{m+1}\,;
\,(\xi_1,s_1)\cdots,(\xi_{n-1},s_{n-1})})\V_{1,\Si}\cr
&\hskip.3cm\le \smsum_{s_1,\cdots,s_{n-1}\atop s,t,t'}\sup_{\et_1,\cdots,\et_{m+1}}
\int\!\! d{\sst \ze d\ze' d\xi_1,\cdots d\xi_{n-1}}\ \big|
\varphi({\sst\et_1,\cdots,\et_m\,;\,(\xi_1,s_1),\cdots,
(\ze',t),\cdots,(\xi_{n-1},s_{n-1})})\cr
\noalign{\vskip-.2in}
&\hskip10cm \cdot\ c({\sst(\ze',t'),(\ze,s)})\,J({\sst \ze,\et_{m+1}})\big|\cr
&\hskip.3cm\le \smsum_{s_1,\cdots,s_{n-1},t}
\smsum_{s,t'\atop s\cap t\cap t'\ne \emptyset}
\sup_{\et_{m+1},\ze'}
\big|c({\sst(\ze',t'),(\et_{m+1},s)})\big|\ 
\big\|
\varphi({\sst\et_1,\cdots,\et_m\,;\,(\xi_1,s_1),\cdots,
(\ze',t),\cdots,(\xi_{n-1},s_{n-1})})\big\|_{1,\infty}\cr
&\hskip.3cm\le 9\sup|c|\smsum_{s_1,\cdots,s_{n-1},t}
\ \big\|
\varphi({\sst\et_1,\cdots,\et_m\,;\,(\xi_1,s_1),\cdots,
(\ze',t),\cdots,(\xi_{n-1},s_{n-1})})\big\|_{1,\infty}\cr
&\hskip.3cm\le \const\sfrac{\fl}{M^j}\ \v\varphi\v_{1,\Si}\cr
}$$
by (\eqnOSsupDbound).
\endproof

\vfill\eject

\chap{Bounds for Sectorized Propagators}\PG\pgOSXIII

 In this Section we prove the existence of the partitions of unity, $\{\chi_s(k)\}$, and enveloping functions satisfying Lemma \lemOSsectpartunit. 
We derive bounds on $\ \v c\v_{1,\Si}\ $, for various sectorized 
covariances $c$ whose Fourier transforms are related to  
$\chi_s(k)\sfrac{\nu^{(j)}(k)}{\imath k_0 -e(\k)}\chi_{s'}(k)$, that, together with Proposition \propOScontrintboundsectors, give good contraction bounds.

The reason it is not easy to get good 
$L_1$--$L_\infty$--bounds on the propagators in position space is that
integration by parts in Cartesian coordinates is not well suited to 
the curvature of the Fermi surface and the shells around it. This is 
why we introduce sectorization.
If the sectors are not too long (more precisely, at most of order $\sfrac{1}{M^{j/2}}$), the
curvature of the sector has little effect. The first step in
deriving $L_1$--$L_\infty$--bounds using sectorization is

\proposition{\STM\propOSGenDecay}{ Let $j\ge 2$.
Let $I$ be an interval of the Fermi curve with length $\fl$ and let 
$f(k)$ be a function that is supported on $\set{k\in\bbbr^3}{|ik_0-e(\k)|\le 
\sfrac{2}{M^j},\ \pi_F(k)\in I}$. Set, as in Lemma \lemOSprepintup
$$
f'(x) =\int e^{ \imath<k,x>_-} f(k) \sfrac{d^3k}{(2\pi)^3}  
$$
Fix any point $\k'_c\in I$, let  $\hat\t$ and $\hat \n$ 
be unit tangent and normal vectors to the Fermi curve at $\k'_c$ and let
$\x_{\|}$ be the component of $\x$ parallel to
$\hat\t$ and $\x_{\perp}$ the component parallel to $\hat \n$.
There is a constant,
$\abcst$, depending on $e(\k)$, 
but independent of $M$, $f$,  $j$ and $x$  such that
\Item i) For all multiindices $\ga\in\bbbn_0\times\bbbn_0^2$
$$\eqalign{
&\Big|\big(\sfrac{x_0}{M^j}\big)^{\ga_0}
\big(\sfrac{\x_{\perp} }{M^j}\big)^{\ga_1}
\big(\fl\,\x_{\|}\big)^{\ga_2}  f'(x)\Big| 
\le \abcst\ \sfrac{\fl}{M^{2j}}\ \sup_{k}
\ \sfrac{\fl^{\ga_2}}{M^{j(\ga_0+\ga_1)}}
\Big|\partial_{k_0}^{\ga_0} \big(\hat \n\cdot\nabla_\k\big)^{\ga_1}
\big(\hat \t\cdot\nabla_\k\big)^{\ga_2}f(k)\Big| 
}$$
\Item ii)
$$
\sup_{x}\big| f'(x) \big|
 \le \abcst\ \sfrac{\fl}{M^{2j}}\sup_{k}|f(k)|   
$$
\Item iii) If $\fl\ge\sfrac{1}{M^j}$, then, for all multiindices $\de$
$$
\int dx\,\big|x^{\de} f'(x) \big| 
\le \abcst\ 2^{\de_1+\de_2} M^{|\de| j}
\max_{\ga_0\le\de_0+2\atop\ga_1+\ga_2\le \de_1+\de_2+3}
\sup_{k}
\sfrac{\fl^{\ga_2}}{M^{j(\ga_0+\ga_1)}}
\Big|\partial_{k_0}^{\ga_0} \big(\hat \n\cdot\nabla_\k\big)^{\ga_1}
\big(\hat \t\cdot\nabla_\k\big)^{\ga_2}f(k)\Big| 
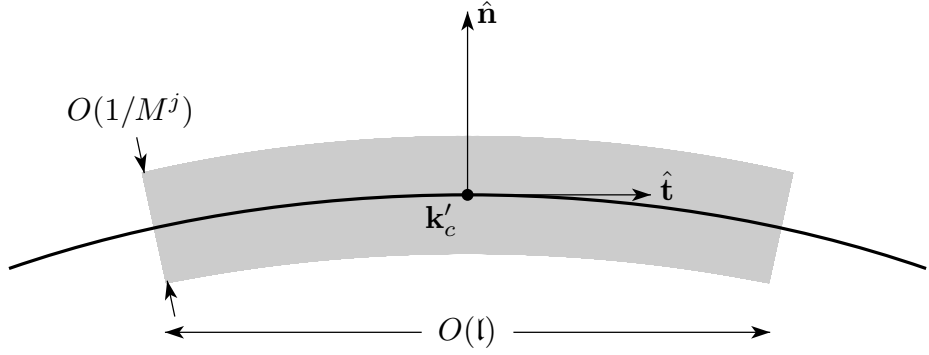
$$ 
}
\goodbreak
\centerline{\figplace{normal3}{0in}{0in}}
\prf i) By integration by parts
$$
\Big|\big(\sfrac{x_0}{M^j}\big)^{\ga_0}
\big(\sfrac{\x_{\perp} }{M^j}\big)^{\ga_1}
\big(\fl\,\x_{\|}\big)^{\ga_2}  f'(x)\Big|
 =\Big|\int \sfrac{d^{3}k}{(2\pi)^{3}} e^{\imath <k,x>_-}
\big(\sfrac{1}{M^j}\partial_{k_0}\big)^{\ga_0} 
\big(\sfrac{1}{M^j}\hat \n\cdot\nabla_\k\big)^{\ga_1}
\big(\fl\hat \t\cdot\nabla_\k\big)^{\ga_2}f(k)\Big|
$$
Use $S$ to denote the support of $f(k)$. 
Observe that $S$ has volume at most $\abcst\, M^{-2j}\fl$, since $k_0$
is supported in an interval of length $\abcst\, M^{-j}$, the distance of $\k$
from $F$ is bounded by $\abcst\, M^{-j}$
and the $\pi_F(\k)$  runs over an interval of length $\fl$. Hence
$$\eqalign{
\Big|\big(\sfrac{x_0}{M^j}\big)^{\ga_0}
\big(\sfrac{\x_{\perp} }{M^j}\big)^{\ga_1}
\big(\fl\,\x_{\|}\big)^{\ga_2}  f'(x)\Big|
& \le {\rm vol}(S) \ \sup_k
\Big|\big(\sfrac{1}{M^j}\partial_{k_0}\big)^{\ga_0} 
\big(\sfrac{1}{M^j}\hat \n\cdot\nabla_\k\big)^{\ga_1}
\big(\fl\hat \t\cdot\nabla_\k\big)^{\ga_2}f(k)\Big|\cr
& \le \abcst\, \sfrac{\fl}{M^{2j}} \ \sup_k
\Big|\big(\sfrac{1}{M^j}\partial_{k_0}\big)^{\ga_0} 
\big(\sfrac{1}{M^j}\hat \n\cdot\nabla_\k\big)^{\ga_1}
\big(\fl\hat \t\cdot\nabla_\k\big)^{\ga_2}f(k)\Big|
}$$
\Item ii) This is simply  a restatement of part (i) with $\ga=\0$.
\Item iii) Set
$$
\rho(x)\ =\  \big[1+M^{-j}|x_{0}|]^2\,
\big[1+M^{-j}|\x_{\perp} | +\fl |\x_{\|}|\big]^3
$$ 
By part (i) 
$$\eqalign{
&\rho(x)\big|x^{\de} f'(x)\big|\cr
&\hskip1cm \le \abcst\,2^{\de_1+\de_2} M^{j|\de|}\sfrac{\fl}{M^{2j}} \ \max_{\ga_0\le\de_0+2\atop\ga_1+\ga_2\le \de_1+\de_2+3}\sup_k
\Big|\big(\sfrac{1}{M^j}\partial_{k_0}\big)^{\ga_0} 
\big(\sfrac{1}{M^j}\hat \n\cdot\nabla_\k\big)^{\ga_1}
\big(\fl\hat \t\cdot\nabla_\k\big)^{\ga_2}f(k)\Big|
}$$
since 
$$ 
\big|x^\de\big|\le M^{|\de|j}\big[M^{-j}|x_{0} |\big]^{\de_0} 
\big[M^{-j}|\x_{\perp}| +\fl |\x_{\|}|\big]^{\de_1+\de_2}
$$
The desired bound now follows from
$$
\int dx\,\sfrac{1}{\rho(x)} \le \abcst\,M^{2j}/\fl
$$
To see this, just make the change of variables
$x_0=M^jz_0,\ \x_\perp=M^jz_1,\ \x_\|=z_2/\fl$.

\endproof

We parametrize the Fermi curve $F$ by arc length, using a real variable $\k'$ for the parametrization. To simplify notation, set $\k'(k) =\pi_F(k)$, the projection on the Fermi surface.

\lemma{\STM\lemOSsectorderiv}{  Let $j>0$. 
Let $I$ be an interval of the Fermi curve
with length $\fl\in\big[\sfrac{1}{M^j},\sfrac{1}{M^{j/2}}\big]$.  Let $\chi(k)=R\big(k_0,e(\k)\big)\Theta\big(\k'(k)\big)$ 
with $R(x,y)$ vanishing unless $|y|\le\sqrt{2}\fl$ and $\Th$ supported
in $I$. Fix any point $\k'_c\in I$ and let  $\hat\t$ and $\hat \n$ 
be unit tangent and normal vectors to the Fermi curve at $\k'_c$.
There is a constant, $\abcst$, depending on $r_0,\ r$ and $e(\k)$, 
but independent of $M$, $\chi$ and  $j$  such that, for all 
$\ga\in\bbbn_0\times\bbbn_0^2$ with $\ga_0\le r_0+2$, $\ga_1+\ga_2\le r+3$,
$$\eqalign{
&\sup_{k}
\ \sfrac{\fl^{\ga_2}}{M^{j(\ga_0+\ga_1)}}
\Big|\partial_{k_0}^{\ga_0} \big(\hat \n\cdot\nabla_\k\big)^{\ga_1}
\big(\hat \t\cdot\nabla_\k\big)^{\ga_2}\chi(k)\Big|\cr 
&\hskip1in\le \abcst\,\max_{m+n\le \ga_1+\ga_2}\ 
\sup_{\k'}\fl^m\big|\partial_{\k'}^m\Th(\k')\big|\ 
\sup_{x,y}\sfrac{1}{M^{j(\ga_0+n)}}
\Big|\sfrac{\partial^{\ga_0+n} R}{\partial x^{\ga_0}\partial^n y }(x,y)\Big|
}$$
}
\prf
Since $\fl\ge\sfrac{1}{M^j}$ and all derivatives of $\k'(k)$ to order
$\ga_1+\ga_2$ are bounded,
$$
\sup_{k}\ 
\Big|
\big(\sfrac{1}{M^j}\hat \n\cdot\nabla_\k\big)^{\ga_1-\be_1}
\big(\fl\hat \t\cdot\nabla_\k\big)^{\ga_2-\be_2}\Th\big(\k'(k)\big)\Big|
\le\abcst\max_{m\le \ga_1+\ga_2\atop \phantom{m\le }-\be_1-\be_2}\sup_{\k'}
\fl^m\big|\partial_{\k'}^m\Th(\k')\big|
$$
for all $\be_1\le\ga_1$ and $\be_2\le\ga_2$.
So, by the product rule, it suffices to prove
$$
\sup_{k\in S}\ 
\Big|\big(\sfrac{1}{M^j}\partial_{k_0}\big)^{\ga_0} 
\big(\sfrac{1}{M^j}\hat \n\cdot\nabla_\k\big)^{\be_1}
\big(\fl\hat \t\cdot\nabla_\k\big)^{\be_2}
R\big(k_0,e(\k)\big)\Big|
\le\abcst\hskip-4pt\max_{n\le \be_1+\be_2}
\sup_{x,y}\sfrac{1}{M^{j(\ga_0+n)}}
\Big|\sfrac{\partial^{\ga_0+n} R}{\partial x^{\ga_0}\partial^n y }(x,y)\Big|
$$
where $S$ is the support of $\chi(k)$.
Set $\pi=\{1,\ \cdots,\ \be_1+\be_2\}$,
$$
d_i=\cases{
      \sfrac{1}{M^j}\hat \n\cdot\nabla_\k & if $1\le i\le \be_1$\cr
\noalign{\vskip.05in}
      \fl\hat \t\cdot\nabla_\k & if $\be_1+1\le i\le \be_1+\be_2$\cr   }
$$
and, for each $\pi'\subset \pi$, $d^{\pi'}=\smprod_{i\in \pi'}d_i$.
By the product and chain rules
$$
\big(\sfrac{1}{M^j}\partial_{k_0}\big)^{\ga_0} d^\pi R\left(k_0,e(\k)\right)
=\sum_{n=1}^{\be_1+\be_2}\sum_{(\pi_1,\cdots,\pi_n)\in\cP_n}
\sfrac{1}{M^{j(\ga_0+n)}}
\sfrac{\partial^{\ga_0+n} R}{\partial x^{\ga_0}\partial^n y }
\left(k_0,e(\k)\right)
\smprod_{i=1}^nM^{j}d^{\pi_i}e(\k)
$$
where $\cP_n$ is the set of all partitions of $\pi$ into
$n$ nonempty subsets $\pi_1,\cdots,\pi_n$ with, for all $i<i'$, the smallest
element of $\pi_i$ smaller than the smallest element of $\pi_{i'}$. 
So to prove the Lemma, it suffices to prove that 
$$
\max_{1\le\be_1+\be_2\le\ga_1+\ga_2}\ \sup_{k\in S}\ 
\Big|M^{j}
\big(\sfrac{1}{M^j}\hat \n\cdot\nabla_\k\big)^{\be_1}
\big(\fl\hat \t\cdot\nabla_\k\big)^{\be_2}
e(\k)\Big|\le\abcst\,
\EQN\eqnOSsecpropboundI$$
If $\be_1\ge 1$ or $\be_2\ge 2$,
this follows from $\sfrac{M^j\fl^{\be_2}}{M^{\be_1 j}}\le 1$. (Recall that
$\fl\le\sfrac{1}{M^{j/2}}$.) The only remaining possibility
is $\be_1=0,\ \be_2=1$.

If $\hat\t\cdot \nabla_\k e(\k)$ is evaluated at $\k=\k'_c$, it vanishes,
since $\nabla_\k e(\k'_c)$ is parallel to $\hat\n$.
The second derivative of $e$ is bounded so that,  
$$
M^j\fl\sup_{k\in S}\big|\hat\t\cdot\nabla_\k e(\k)\big|
\le\abcst\, M^j\fl\sup_{k\in S}|\k-\k'_c|\le\abcst\, M^j\fl^2\le\abcst\,
$$
since $\fl\le\sfrac{1}{M^{j/2}}$. 
\endproof

For the rest of this section, we fix a sectorization $\Si$ of scale 
$j\ge 2$ and length $\sfrac{1}{M^{j-3/2}}\le\fl\le\sfrac{1}{M^{(j-1)/2}}$.
We choose a smooth partition of unity $\Th_s(\k'),\,s\in\Si$ of the Fermi curve $F$ subordinate to the sets $s\cap F$, such that
$\big|\partial_{\k'}^m\,\Th_s(\k')\big| \le  \abcst \sfrac{1}{\fl^m}$
for $m=0,1,\cdots,r+3$. Furthermore choose enveloping functions
$\tilde\Th_s(\k'),\,s\in\Si$ that are identically one on $s\cap F$ and obey
$\big|\partial_{\k'}^m\,\tilde\Th_s(\k')\big| \le  \abcst \sfrac{1}{\fl^m}$
for $m=0,1,\cdots,r+3$ and $\,\tilde\Th_s\,\tilde\Th_{s'}=0\,$ if $\,s\cap s' = \emptyset$. Set
$$\deqalign{
\chi_s(k) &= \tilde\nu^{(\ge j)}(k)\,\Th_s(\k'(k)) 
&= \varphi(M^{2j-2}(k_0^2+e(\k)^2))\,\Th_s(\k'(k))\cr
\tilde \chi_s(k) &= \bar\nu^{(\ge j)}(k)\,\tilde \Th_s(\k'(k)) 
&= \varphi(M^{2j-3}(k_0^2+e(\k)^2))\,\Th_s(\k'(k))\cr
}\EQN\eqnOSpartunit$$
where $\tilde\nu^{(\ge j)},\ \bar\nu^{(\ge j)}$ are the functions of 
Definition \defOSextendedshell.

\lemma{\STM\lemOSmorepartunity}{
Let $s\in\Si$. Set for $x=(x_0,\x)\in\bbbr\times\bbbr^2$
$$\meqalign{
\chi_s'(x)& =\int e^{ \imath<k,x>_-}\chi_s(k) \sfrac{d^3k}{(2\pi)^3} 
\qquad,\qquad
\tilde \chi_s'(x)& =\int e^{ \imath<k,x>_-}\tilde \chi_s(k) \sfrac{d^3k}{(2\pi)^3}
  \cr
\chi_s^o(\x) 
& = \int e^{ \imath \k\cdot\x}\chi_s\big((0,\k)\big) \sfrac{d^2\k}{(2\pi)^2}
\qquad\qquad 
}$$
Then
$$\meqalign{
\| \chi_s'\|_{L^1} & \le \abcst\,\cb_{j-1},&&
\| \sfrac{\partial\ }{\partial x_0} \chi_s'\|_{L^1} 
& \le \abcst\,\sfrac{1}{M^{j-1}}\,\cb_{j-1} ,&&
\|x_0 \sfrac{\partial\ }{\partial x_0} \chi_s'\|_{L^1}
& \le \abcst\,\cb_{j-1} 
\cr
\| \tilde \chi_s'\|_{L^1} & \le \abcst\,\cb_{j-{3\over 2}} ,&&
 \| \sfrac{\partial\ }{\partial x_0}\tilde \chi_s'\|_{L^1} 
& \le \abcst\,\sfrac{1}{M^{j-3/2}}\,\cb_{j-{3\over 2}} ,&&
 \| x_0 \sfrac{\partial\ }{\partial x_0}\tilde \chi_s'\|_{L^1} 
& \le \abcst \,\cb_{j-{3\over 2}}
\cr
}$$ 
and
$$
\| \chi_s^o\|_{L^1} \le \abcst\,\cb_{j-1} 
$$
Here, for a function $f(x)$ on $\bbbr\times\bbbr^2$, 
$$
\|f\|_{L^1} =\smsum_{\de \in \bbbn_0\times\bbbn_0^d}\sfrac{1}{\de!} 
\Big[ \int |x^\de f(x)|\, dx \Big]\,t^\de\ \in\ \fN_{d+1}
$$
is the norm defined before Lemma \lemOSprepintup, and for a function $g(\x)$ on
$\bbbr^2$ we set
$$
\|g\|_{L^1} =
\smsum_{\de \in \bbbn_0\times\bbbn_0^d \atop \de_0=0} \sfrac{1}{\de!}
\Big[ \int |\x^\de f(\x)|\, d\x \Big]\,t^\de\ 
$$

}

\prf
Fix a point $\k'_c\in s\cap F$ and let $\hat\t$ 
and $\hat \n$ be unit tangent and normal vectors to $F$ at $\k'_c$.
By Lemma \lemOSsectorderiv, with $j$ replaced by $j-1$ and
$R(x,y)=\varphi\big(M^{2j-2}(x^2+y^2)\big)$, 
$$
\max_{\ga_0\le\de_0+2\atop\ga_1+\ga_2\le \de_1+\de_2+3} 
\sup_{k}
\sfrac{\fl^{\ga_2}}{M^{(j-1)(\ga_0+\ga_1)}}
\Big|\partial_{k_0}^{\ga_0} \big(\hat \n\cdot\nabla_\k\big)^{\ga_1}
\big(\hat \t\cdot\nabla_\k\big)^{\ga_2}\chi_s(k)\Big| 
\le \abcst
\EQN\eqnOSsecpropboundII
$$
for every multiindex
$\delta\in\bbbn_0\times\bbbn_0^2$ with $\de_0\le r_0$ and 
$\de_1+\de_2\le r$. Here, we have used that $M^{j-1}|x|, M^{j-1}|y|\le\sqrt{2}$
on the support of $R(x,y)$.
Therefore, by Proposition \propOSGenDecay.iii,
$$
\sfrac{1}{M^{(j-1)|\de|}}\int dx\,\big|x^\de \ch'(x) \big| \le \abcst
$$
By Definition of $\|\,\cdot\,\|_{L^1}$, this implies that 
$\,\| \chi_s'\|_{L^1} \le \abcst\,\cb_{j-1}\,$.

\noindent
By Lemma \lemOSsectorderiv\ and the product rule
$$
\max_{\ga_0\le\de_0+2\atop\ga_1+\ga_2\le \de_1+\de_2+3} 
\sup_{k}
\sfrac{\fl^{\ga_2}}{M^{(j-1)(\ga_0+\ga_1)}}
\Big|\partial_{k_0}^{\ga_0} \big(\hat \n\cdot\nabla_\k\big)^{\ga_1}
\big(\hat \t\cdot\nabla_\k\big)^{\ga_2}
\big( k_0 \chi_s(k) \big)\Big| 
\le \abcst\,\sfrac{1}{M^{j-1}}
$$
and as above it follows that $\,\| \sfrac{\partial\ }{\partial x_0} \chi_s'\|_{L^1} \le  \abcst\,\sfrac{1}{M^{j-1}}\,\cb_{j-1}$.

\noindent
Again, by Lemma \lemOSsectorderiv
$$
\max_{\ga_0\le\de_0+2\atop\ga_1+\ga_2\le \de_1+\de_2+3} 
\sup_{k}
\sfrac{\fl^{\ga_2}}{M^{(j-1)(\ga_0+\ga_1)}}
\Big|\partial_{k_0}^{\ga_0} \big(\hat \n\cdot\nabla_\k\big)^{\ga_1}
\big(\hat \t\cdot\nabla_\k\big)^{\ga_2}
\big( \sfrac{\partial\ }{\partial k_0}\,k_0\chi_s(k) \big)\Big| 
\le \abcst
$$
and the proof that 
$\|x_0 \sfrac{\partial\ }{\partial x_0} \chi_s'\|_{L^1} \le  \abcst\,\cb_{j-1}$ is as before.

\noindent
The bounds on $\tilde\chi_s'$ are obtained in the same way, with $j-1$
replaced by $j-\sfrac{3}{2}$.

The estimate on $ \| \chi_s^o\|_{L^1}$ follows from the fact that
$$
 \chi_s^o(\x) = \int dx_0\,\chi_s'(x_0,\x)
$$
\endproof

\proof{ of Lemma \lemOSsectpartunit} 
Parts (i) and (ii) of the Lemma are trivial. To prove part (iii)
observe that 
$$
\hat\chi_s(\xi,\xi') = \de_{\si,\si'}\de_{a,a'} \,\chi_s'({\sst (-1)^a}(x-x'))
$$
for $\xi =(x,\si,a),\ \xi' =(x',\si',a')  \in \cB$. Therefore, by Lemma 
\lemOSmorepartunity
$$
\|\hat\chi_s\|_{1,\infty}\ \le\ \abcst\,\|\chi'_s\|_{L^1}\ \le\  \abcst\,\cb_{j-1}
$$
The estimate for $\| \hat{\tilde\chi}_s\|_{1,\infty}$ is obtained in the same way.
\endproof

From now on, we fix for each sectorization $\Si$, a partition of unity 
$\chi_s,\ s\in\Si$ and a system of functions $\tilde \chi_s,\ s\in\Si$ 
that fulfill the conclusions of Lemma  \lemOSsectpartunit\ and Lemma
\lemOSmorepartunity.

Recall from Definition \defOScbj\ that
$$
\cb_j=\sum_{|\bde|\le r\atop |\de_0|\le r_0}  M^{j|\de|}\,t^\de
+\sum_{|\bde|> r\atop {\rm or\ }|\de_0|> r_0}\infty\, t^\de
\in\fN_{d+1}
$$ 
Observe that by Corollary \corOSappMonoidIV, there is a constant $\abcst$ that is independent of $M$ such that for $2\le i\le j$
$$
\cb_i \,\cb_j \le \abcst\, \cb_j
\EQN\eqnOSprodcontrbound
$$

\lemma{\STM\lemOSplainpropest}{
Set
$$
C^{(j)}(k) = \sfrac{\nu^{(j)}(k)}{\imath k_0 -e(\k)} \qquad,\qquad
\tilde C^{(j)}(k) = \sfrac{\tilde\nu^{(j)}(k)}{\imath k_0 -e(\k)}
$$
and, for $s,s'\in\Si$, let $c^{(j)}({\sst(\xi,s),(\xi',s')})$ resp. 
$\tilde c^{(j)}({\sst(\xi,s),(\xi',s')})$ be the Fourier transforms 
(as in Definition \defOSftcov) of $\chi_s(k)\,C^{(j)}(k)\,\chi_{s'}(k)$ resp. 
$\chi_s(k)\,\tilde C^{(j)}(k)\,\chi_{s'}(k)$.
Then
\Item{i)} 
$\,c^{(j)}({\sst(\cdot,s),(\cdot,s')}) 
= \tilde c^{(j)}({\sst(\cdot,s),(\cdot,s')}) =0\,$ if $s\cap s' =\emptyset\,$.
\Item{ii)}
$$
\V c^{(j)} \V_{1,\Si}\,,\ \V \tilde c^{(j)} \V_{1,\Si}
\le \const M^j\cb_j
$$
\Item{iii)} Let $c_0^{(j)}({\sst(\xi,s),(\xi',s')})$  the Fourier transform 
(as in Definition \defOSftcov) of $\chi_s(k)\,k_0\,C^{(j)}(k)\,\chi_{s'}(k)$
or $\chi_s(k)\,e(\k)\,C^{(j)}(k)\,\chi_{s'}(k)\ $ 
and let
$\tilde c_0^{(j)}({\sst(\xi,s),(\xi',s')})$ be the Fourier transform 
 of either $\chi_s(k)\,k_0\,\tilde C^{(j)}(k)\,\chi_{s'}(k)$ or
$\ \chi_s(k)\,e(\k)\,\tilde C^{(j)}(k)\,\chi_{s'}(k)$.
Then
$$
\V c_0^{(j)} \V_{1,\Si}\,,\ \V \tilde c_0^{(j)} \V_{1,\Si}
\le \const \cb_j
$$
\Item iv)
$$
\sum_{\de\in\bbbn_0\times\bbbn_0^2}\sfrac{1}{\de!}
\max_{s,s'\in\Si}\sup_{\xi,\xi'\in\cB}\big|\cD_{1,2}^\de 
c^{(j)}({\sst(\xi,s),(\xi',s')})\big|t^\de
\le\const \sfrac{\fl}{M^j}\cb_j
$$
}

\prf
Part (i) is obvious. To prove part (ii) fix sectors $s,s'\in \Si$, with $s\cap s' \ne \emptyset$ and a point $\k'_c\in s\cap F$.
Let $\hat\t$ and $\hat \n$ 
be unit tangent and normal vectors to $F$ at $\k'_c$. First we claim that for all $k$ in the intersection of $s$ with the $j^{\rm th}$ shell
$$
\max_{\be_0\le r_0+2\atop \be_1+\be_2\le r+3}\
\Big|\big(\sfrac{1}{M^j}\partial_{k_0}\big)^{\be_0} 
\big(\sfrac{1}{M^j}\hat \n\cdot\nabla_\k\big)^{\be_1}
\big(\fl\,\hat \t\cdot\nabla_\k\big)^{\be_2}
\sfrac{1}{ik_0-e(\k)}\Big|\le\const M^j
\EQN\eqnOSsecpropboundIII
$$
To see this, set $\pi=\{1,\ \cdots,\ |\be|\}$,
$$
d_i=\cases{\sfrac{1}{M^j}\partial_{k_0} & if $1\le i\le \be_0$\cr
\noalign{\vskip.05in}
      \sfrac{1}{M^j}\hat \n\cdot\nabla_\k & if $\be_0+1\le i\le \be_0+\be_1$\cr
\noalign{\vskip.05in}
      \fl\hat \t\cdot\nabla_\k & if $\be_0+\be_1+1\le i\le |\be|$\cr   }
$$
and, for each $\pi'\subset \pi$, $d^{\pi'}=\smprod_{i\in \pi'}d_i$.
By the product and chain rules
$$
d^\pi\sfrac{1}{ik_0-e(\k)}
=M^j\sum_{n=1}^{|\be|}(-1)^nn!\sum_{(\pi_1,\cdots,\pi_n)\in\cP_n}
\left(\sfrac{1/M^j}{ik_0-e(\k)}\right)^{n+1}
\smprod_{i=1}^nM^jd^{\pi_i}(ik_0-e(\k))
$$
In the sector $s$, $|ik_0-e(\k)|\ge\const \sfrac{1}{M^j}$ so that 
$\left(\sfrac{1/M^j}{ik_0-e(\k)}\right)^{n+1}$ is bounded uniformly 
in~$j$. That $M^jd^{\pi_i}(ik_0-e(\k))$  is bounded uniformly in
$j$ follows immediately from (\eqnOSsecpropboundI) and the fact that 
$|k_0|\le\sfrac{\const}{M^j}$ on the $j^{\rm th}$ shell. This proves 
(\eqnOSsecpropboundIII).
 
As in (\eqnOSsecpropboundII), for all $k$ in the intersection of $s$ 
with the $j^{\rm th}$ shell,
$$
\max_{\be_0\le r_0+2\atop \be_1+\be_2\le r+3}\
\Big|\big(\sfrac{1}{M^j}\partial_{k_0}\big)^{\be_0} 
\big(\sfrac{1}{M^j}\hat \n\cdot\nabla_\k\big)^{\be_1}
\big(\fl\,\hat \t\cdot\nabla_\k\big)^{\be_2}
\nu^{(j)}(k)\Big|\le\const 
$$
By Leibniz's rule it follows from this inequality and the inequalities 
(\eqnOSsecpropboundII), (\eqnOSsecpropboundIII) that
$$
\max_{\ga_0\le r_0+2\atop \ga_1+\ga_2\le r+3} 
\sup_{k}
\sfrac{\fl^{\ga_2}}{M^{j(\ga_0+\ga_1)}}
\Big|\partial_{k_0}^{\ga_0} \big(\hat \n\cdot\nabla_\k\big)^{\ga_1}
\big(\hat \t\cdot\nabla_\k\big)^{\ga_2}\chi_s(k)\,C^{(j)}(k)\,\chi_{s'}(k)\Big| 
\le \const M^j
$$
Hence, by Proposition \propOSGenDecay
$$
\v c^{(j)} \v_{1,\Si}\le \const M^j\cb_j
$$
The proof for $\v\tilde c^{(j)} \v_{1,\Si} $ is analogous.

The proof of part (iii) is the same as the proof of part (ii) with
(\eqnOSsecpropboundIII) replaced by
$$\eqalign{
\max_{\be_0\le r_0+2\atop \be_1+\be_2\le r+3}\
\Big|\big(\sfrac{1}{M^j}\partial_{k_0}\big)^{\be_0} 
\big(\sfrac{1}{M^j}\hat \n\cdot\nabla_\k\big)^{\be_1}
\big(\fl\,\hat \t\cdot\nabla_\k\big)^{\be_2}
\sfrac{k_0}{ik_0-e(\k)}\Big|&\le\const \cr
\max_{\be_0\le r_0+2\atop \be_1+\be_2\le r+3}\
\Big|\big(\sfrac{1}{M^j}\partial_{k_0}\big)^{\be_0} 
\big(\sfrac{1}{M^j}\hat \n\cdot\nabla_\k\big)^{\be_1}
\big(\fl\,\hat \t\cdot\nabla_\k\big)^{\be_2}
\sfrac{e(\k)}{ik_0-e(\k)}\Big|&\le\const \cr
}$$

To prove part (iv), observe that, by (\eqnOSsecpropboundIII) and the 
fact that $\chi_s(k)C^{(j)}(k)\chi_{s'}(k)$ is supported in a region of volume
$\const \sfrac{\fl}{M^{2j}}$,
$$
\max_{s,s'\in\Si}\sup_{\xi,\xi'\in\cB}\big|\cD_{1,2}^\de 
c^{(j)}({\sst(\xi,s),(\xi',s')})\big|
\le\const \sfrac{\fl}{M^{2j}}M^{j(1+|\de|)}
$$
for all $\de_0\le r_0$ and $|\bde|\le r$.

\endproof

\proposition{\STM\propOSrealpropbound}{
There are constants $\tau_1,\ \const$ that depend on $e(\k)$ and $M$, but not on $j$ or $\Si$ with the following property:

{\parindent=.25in
\item{}
Let $u({\sst(\xi,s)},{\sst(\xi',s')})\in\cF_0(2;\Si)$ be antisymmetric,
 spin independent and particle number conserving and obey
$\ \v  u \v_{1,\Si} \le \sfrac{\tau_1}{M^j}+\sum_{\de\ne \0}\infty\, t^\de\ $. 
Let
$$
C(k) = \sfrac{\nu^{(j)}(k)}{\imath k_0 -e(\k) -\check u(k)}
$$
$c({\sst(\xi,s),(\xi',s')})$ be the Fourier transform (as in Definition \defOSftcov) of $\chi_s(k)\,C(k)\,\chi_{s'}(k)$
and $c_0({\sst(\xi,s),(\xi',s')})$ be the Fourier transform
 of $\chi_s(k)\,k_0C(k)\,\chi_{s'}(k)$ or $\chi_s(k)\,e(\k)C(k)\,\chi_{s'}(k)$. Then

i)
 $\,c({\sst(\cdot,s),(\cdot,s')}) =0\,$ if $s\cap s' =\emptyset\,$

ii)
$\v c \v_{1,\Si} \le \const \sfrac{M^j\cb_j}{1-M^j\sv  u \sv_{1,\Si}}$

iii)
$
\sum\limits_{\de\in\bbbn_0\times\bbbn_0^2}\sfrac{1}{\de!}
\sup\limits_{\xi,\xi',s,s'}\big|\cD_{1,2}^\de 
c({\sst(\xi,s),(\xi',s')})\big|t^\de
\le\const \sfrac{\fl}{M^j}\sfrac{\cb_j}{1-M^j\sv  u \sv_{1,\Si}}
$

iv)
$\v c_0 \v_{1,\Si} \le \const \sfrac{\cb_j}{1-M^j\sv  u \sv_{1,\Si}}$

}
}

\prf 
Again part (i) is trivial.
To prove part (ii), observe that
$$\eqalign{
C(k) \ &=\  \frac{C^{(j)}(k)}{1 - \sfrac{\check u(k)}{\imath k_0-e(\k)}}
\ =\ \frac{C^{(j)}(k)}{1 - \sfrac{\check u(k)\, \tilde\nu^{(j)}(k)}{\imath k_0-e(\k)}}
\ =\ \frac{C^{(j)}(k)}{1 - \check u(k)\,\tilde C^{(j)}(k)} \cr
&=\  C^{(j)}(k)\,\smsum_{n=0}^\infty \big( \check u(k)\,\tilde C^{(j)}(k) \big)^n 
}$$
Introducing the local notation
$$\eqalign{
 C^{(j)}(k;s,s')&= \chi_s(k)\, C^{(j)}(k)\,\chi_{s'}(k)\cr
\tilde C^{(j)}(k;s,s')&= \chi_s(k)\,\tilde C^{(j)}(k)\,\chi_{s'}(k)\cr
}$$
we have
$$
\chi_s(k)\, C(k)\,\chi_{s'}(k) = C^{(j)}(k;s,s')
 +\smsum_{n=1}^\infty \smsum_{s''\in \Si}\
   \smsum_{t_i,t'_i\in\Si 
     \atop{ {\rm for\ } i=1,\cdots,n \atop {\rm with\ }t_n'=s'}}\hskip-11pt
  C^{(j)}(k;s,s'') \,\smprod_{i=1}^n \check u(k)\,
   \tilde C^{(j)}(k;t_i,t_i')
$$
Define the operator $*$--product of the sectorized functions
$A({\sst(\xi,s)},{\sst(\xi',s')})$ and $B({\sst(\xi',s')},{\sst(\xi'',s'')})$
by
$$
(A*B)({\sst(\xi,s)},{\sst(\xi'',s'')}) 
= \sum_{s',t'\in\Si\atop s'\cap t'\ne \emptyset}\int d\xi'\ 
A({\sst(\xi,s)},{\sst(\xi',s')})\,B({\sst(\xi',t')},{\sst(\xi'',s'')})
$$
Then, by part (i) of Lemma \lemOSplainpropest
$$
c = c^{(j)}\smsum_{n=0}^\infty \big(*u*\tilde c^{(j)}\big)^n
\EQN\eqnOSexpandc$$
so that by iterated application of Lemma \lemOSelloneinftysectors\ 
and part (ii) of Lemma \lemOSplainpropest
$$\eqalign{
\v c \v_{1,\Si} 
\ &\le\  \v c^{(j)} \v_{1,\Si} \
\smsum_{n=0}^\infty \big( 9\,\v  u \v_{1,\Si} \,
   \, \v \tilde c^{(j)} \v_{1,\Si} \big)^n \cr
& \le \ \const M^j\cb_j \ \smsum_{n=0}^\infty 
   \big( \cst{'}{} M^j\cb_j\,\v  u \v_{1,\Si} \big)^n \cr
& =\  \const \sfrac{M^j\cb_j}
{1-\cst{'}{}M^j\cb_j\sv  u \sv_{1,\Si}}
}$$
If $\tau_1<\min\{\sfrac{1}{2\cst{'}{}},1\}$, then,
by Corollary \corOSappMonoidIV.i, with $X=M^j\sv  u \sv_{1,\Si}\,$, $\La=M^j$
and $\mu=\cst{'}{}$,
$$
\v c \v_{1,\Si}  \le\  \const \sfrac{M^j\cb_j}{1-M^j\sv  u \sv_{1,\Si}}
$$

\Item iii) The bound
$$
\sum_{\de}\sfrac{1}{\de!}
\sup_{\xi,\xi',s,s'}\big|\cD_{1,2}^\de 
(A*B)({\sst(\xi,s),(\xi',s')})\big|t^\de
\le 3\Big\{\sum_{\de}\sfrac{1}{\de!}
\sup_{\xi,\xi',s,s'}\big|\cD_{1,2}^\de 
A({\sst(\xi,s),(\xi',s')})\big|t^\de\Big\}\V B\V_{1,\Si}
$$
is proven in much the same way as Lemma \lemOSelloneinfty, but uses
$$
\sup_{\xi,\xi'}\Big|\int d\ze\ A(\xi,\ze)B(\ze,\xi')\Big|
\le\sup_{\xi,\ze}\big|A(\xi,\ze)\big|
\ \sup_{\ze}\int d\xi'\ \big|B(\ze,\xi')\big|
$$
in place of 
$$
\sup_{\xi}\int d\xi'\ \Big|\int d\ze\ A(\xi,\ze)B(\ze,\xi')\Big|
\le\sup_{\xi}\int d\ze\ \big| A(\xi,\ze)\big|
\sup_{\ze}\int d\xi' \big|B(\ze,\xi')\big|
$$
Repeatedly applying this bound 
to (\eqnOSexpandc) and using Lemma \lemOSplainpropest.iv yields
$$\eqalign{
&\sum_{\de\in\bbbn_0\times\bbbn_0^2}\sfrac{1}{\de!}
\sup_{\xi,\xi',s,s'}\big|\cD_{1,2}^\de 
c({\sst(\xi,s),(\xi',s')})\big|t^\de \cr
&\hskip.5in\le\  \Big\{
\sum_{\de\in\bbbn_0\times\bbbn_0^2}\sfrac{1}{\de!}
\sup_{\xi,\xi',s,s'}\big|\cD_{1,2}^\de 
c^{(j)}({\sst(\xi,s),(\xi',s')})\big|t^\de\Big\} \
\smsum_{n=0}^\infty \big( 9\,\v  u \v_{1,\Si} \,
   \, \v \tilde c^{(j)} \v_{1,\Si} \big)^n \cr
&\hskip.5in\le \ \const \sfrac{\fl}{M^j}\cb_j \ \smsum_{n=0}^\infty 
   \big( \cst{'}{} M^j\cb_j\,\v  u \v_{1,\Si} \big)^n \cr
&\hskip.5in =\  \const \sfrac{\fl}{M^j}\sfrac{\cb_j}
{1-\cst{'}{}M^j\cb_j\sv  u \sv_{1,\Si}}\cr
&\hskip.5in \le\  \const \sfrac{\fl}{M^j}\sfrac{\cb_j}
{1-M^j\sv  u \sv_{1,\Si}}
}$$

\Item iv) Repeat the proof of (ii) with (\eqnOSexpandc) replaced by
$$
c_0 = c^{(j)}_0\smsum_{n=0}^\infty \big(*u*\tilde c^{(j)}\big)^n
$$
and using Lemma \lemOSplainpropest.iii. 

\endproof

\lemma{\STM\lemOSdiffpropbound}{
There are constants $\tau_2,\ \const$ that depend on $e(\k)$ and $M$, but not on $j$ or $\Si$ with the following property:

{\parindent=.25in
\item{}
Let, for $\ka$ in a neighbourhood of zero, $u_\ka\in\cF_0(2;\Si)$ be 
antisymmetric, spin independent and particle number conserving and obey
$\ \v  u_0\v_{1,\Si} 
\le \sfrac{\tau_2}{M^j}+\sum_{\de\ne \0}\infty t^\de\ $. 
Let
$$
C_\ka(k) = \sfrac{\nu^{(j)}(k)}{\imath k_0 -e(\k) -\check u_\ka(k)}
$$
and $c_\ka({\sst(\xi,s),(\xi',s')})$ be the Fourier transform of $\chi_s(k)\,C_\ka(k)\,\chi_{s'}(k)$.
Let
$c_{\ka,0}({\sst(\xi,s),(\xi',s')})$ be the Fourier transform
 of $\chi_s(k)\,k_0C_\ka(k)\,\chi_{s'}(k)$ or 
$\chi_s(k)\,e(\k)C_\ka(k)\,\chi_{s'}(k)$. Then

\vskip.1in
i)
$\V \sfrac{d\hfill}{d\ka}c_\ka\big|_{\ka=0} \V_{1,\Si} 
\le \const M^j\cb_j\ \frac{M^j\v  {d\hfill\over d\ka}u_\ka|_{\ka=0} \v_{1,\Si}}
{1-M^j\sv  u_0 \sv_{1,\Si}}$

\vskip.1in
ii)
$\sup_{\xi,\xi',s,s'} \big| \sfrac{d\hfill}{d\ka} c_\ka({\sst(\xi,s),(\xi',s')})
\big|_{\ka=0} \big|\le\const\fl\ \v  {d\hfill\over d\ka}u_\ka|_{\ka=0} \v_{1,\Si}$

\vskip.1in
iii)
$\V \sfrac{d\hfill}{d\ka}c_{\ka,0}\big|_{\ka=0} \V_{1,\Si} 
\le \const \cb_j\ \frac{M^j\v  {d\hfill\over d\ka}u_\ka|_{\ka=0} \v_{1,\Si}}
{1-M^j\sv  u_0 \sv_{1,\Si}}$

}
}

\prf 
The proof is similar to that of Proposition \propOSrealpropbound, using
$$\eqalign{
\frac{d\hfill}{d\ka}C_\ka(k) 
&= \frac{d\hfill}{d\ka}\ \frac{C^{(j)}(k)}{1-\check u_\ka(k)\,\tilde C^{(j)}(k)}
 \cr
&=\ C^{(j)}(k)\frac{1}{1 - \check u_\ka(k)\,\tilde C^{(j)}(k)}
\Big[\sfrac{d\hfill}{d\ka}\check u_\ka(k)\,\tilde C^{(j)}(k)\Big]
\frac{1}{1 - \check u_\ka(k)\,\tilde C^{(j)}(k)} \cr
&=\  C^{(j)}(k)\,\smsum_{m,n=0}^\infty 
\big( \check u_\ka(k)\,\tilde C^{(j)}(k) \big)^m 
\Big[\sfrac{d\hfill}{d\ka}\check u_\ka(k)\,\tilde C^{(j)}(k)\Big]
 \big( \check u_\ka(k)\,\tilde C^{(j)}(k) \big)^n 
}$$
and Lemma \corOSappMonoidIV, which implies that
$\ 
\Big(\sfrac{\cb_j}{1-M^j\sv  u_0 \sv_{1,\Si}}\Big)^2
\le\abcst\ \sfrac{\cb_j}{1-M^j\sv  u_0 \sv_{1,\Si}}
\ $.
\endproof

\lemma{\STM\lemOSumu}{
Let $u({\sst(\xi,s)},{\sst(\xi',s')})$ be a translation invariant
function on $(\cB\times\Si)^2$ with the property that
$\check u\big((k,\si,a,s),(k',\si',a',s')\big)$ vanishes unless 
$\pi_F(k)\in\pi_F(s)$ and $\pi_F(k')\in\pi_F(s')$.  
Let $\mu(t)$ be a $C_0^\infty$ function on $\bbbr$ and set, for each $\La>0$ 
$$\eqalign{
\mu_\La(k)&=\mu\big(\La^2[k_0^2+e(\k)^2]\big)\cr
(u*\hat\mu_\La)({\sst(\xi,s)},{\sst(\xi',s')})
&=\int_\cB d\ze\ u({\sst(\xi,s)},{\sst(\ze,s')})\hat\mu_\La(\ze,\xi')\cr
(\hat\mu_\La*u)({\sst(\xi,s)},{\sst(\xi',s')})
&=\int_\cB d\ze\ u({\sst(\ze,s)},{\sst(\xi',s')})\hat\mu_\La(\ze,\xi)\cr
}$$
where $\hat\mu_\La$ was defined in Definition \defOSfourtransII. Denote
$\ 
j(\La)=\min\set{i\in\bbbn}{M^i\ge\La}
\ $. Then, there is a constant $\abcst$, depending on $\mu$, but not on $M$, $j$ or $\La$, such that
$$
\V u*\hat\mu_\La\V_{1,\Si}\ ,\ 
\V \hat\mu_\La*u\V_{1,\Si}
\le\abcst\,\cb_{j(\La)}\ \v u\v_{1,\Si}
$$

}
\prf
Let $\set{\Th_s}{s\in\Si}$ be the smooth partition of unity  of the 
Fermi curve $F$ that was chosen just before (\eqnOSpartunit) and set
$$
\chi_{\La,s}(k) = \mu_\La(k)\,\Th_s\big(\pi_F(k)\big) 
$$
Then, by Lemma \lemOSsectorderiv\ and Proposition \propOSGenDecay.iii, 
as in Lemma \lemOSmorepartunity,
$$
\| \hat\chi_{\La,s}\|_{1,\infty}  \le \abcst\,\cb_{j(\La)}
$$
We treat $u*\hat\mu_\La$. The other case is similar. As
$$
u*\hat\mu_\La({\sst(\xi,s)},{\sst(\xi',s')})
=\sum_{s''\in\Si\atop s''\cap s'\ne\emptyset}
\int_\cB d\ze\ u({\sst(\xi,s)},{\sst(\ze,s')})\hat\chi_{\La,s''}(\ze,\xi')
$$
Lemma \lemOSelloneinfty\ implies that
$$
\| u*\hat\mu_\La({\sst(\,\cdot\,,s)},{\sst(\,\cdot\,,s')})\|_{1,\infty}
\le \sum_{s''\in\Si\atop s''\cap s'\ne\emptyset} \abcst\,\cb_{j(\La)}\ 
\| u({\sst(\,\cdot\,,s)},{\sst(\,\cdot\,,s')})\|_{1,\infty}
\le \abcst\,\cb_{j(\La)}\ 
\| u({\sst(\,\cdot\,,s)},{\sst(\,\cdot\,,s')})\|_{1,\infty}
$$
since, for each $s'\in\Si$, there are only three $s''\in\Si$ with
$s''\cap s'\ne\emptyset$. The Lemma follows.
\endproof

\remark{\STM\remOSumu}{ In the notation of Lemma \lemOSumu,
$$
(u*\hat\mu_\La){\check{\,}}(k)=
(\hat\mu_\La*u){\check{\,}}(k)=
\check u(k)\,\mu_\La(k)
$$

}

\vfill\eject

\chap{Ladders}\PG\pgOSXIV

In \S\CHestren, we will apply Theorem \theoremVa\ of [FKTr2]
to estimate the renormalization group map $\tilde \Om$ of Definition \defOSrengrpmap, with respect to the sectorized norms of Definition \defOSsectnorm. It will give ``improved power counting'' for two--legged
contributions and ``improved power counting'' for those four--legged 
contributions that are not ladders. A similar result using the norms of
\S\CHmomentum, will be derived in \S\CHmomestren. Depending on the geometry of the Fermi curve, ladders have different behaviour. We shall investigate ladders in \S\CHppladsect, [FKTf2, \S\CHphladders] and the paper [FKTl].
In this Section, we introduce notation for ladders that will be useful in all 
these investigations.

In this Section, the internal lines of ladders will be functions with arguments running over an arbitrary measure space $\fX$. We think of
$\fX$ as $\cB=\bbbr\times \bbbr^2\times \{\uparrow,\downarrow\}\times \{0,1\}$ 
or $\bbbr\times \bbbr^2\times \{\uparrow,\downarrow\}$ or
$\bbbr\times \bbbr^2$ or $\cB\times\Si$, where $\Si$ is a sectorization. 

\definition{\STM\defOSbubbleprop}{
\Item{i)} A complex valued function on $\fX\times \fX$ is called a propagator 
over $\fX$.
\Item{ii)} A four legged kernel over $\fX$ is a complex valued function 
on $\fX^2\times \fX^2$. We sometimes consider it as a bubble propagator over $\fX$,  graphically depicted by 

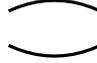
\centerline{\figplace{bubProp}{0 in}{0 in} }

\noindent
or as a rung over $\fX$, graphically depicted by 

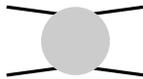
\centerline{\figplace{rung}{0 in}{0 in} }

\noindent
\Item{ iii)} If $A$ and $B$ are propagators over $\fX$ then the tensor product
$$
A\otimes B(x_1,x_2,x_3,x_4)=A(x_1,x_3)B(x_2,x_4)
$$
is a bubble propagator over $\fX$. We set
$$
\cC(A,B) = A\otimes A + A\otimes B + B\otimes A 
$$

\Item{ iv)} Let $F,\ F'$ be four legged kernels over $\fX$. We define the 
four legged kernel $F\fcirc F'$ as
$$
(F\fcirc F')(x_1,x_2;x_3,x_4) 
=  \int dx'_1 dx'_2 \ F(x_1,x_2;x'_1,x'_2)\ F'(x'_1,x'_2;x_3,x_4) 
$$
whenever the integral is well--defined. 

\Item {v)} Let $\ell \ge 1\,$. The ladder with rungs
$R_1,\cdots,R_{\ell+1}$ and  bubble propagators $P_1,\cdots,P_\ell$ is 
defined to be
$$
R_1\circ P_1 \circ R_2 \circ P_2\circ \cdots \circ P_{\ell-1}\circ
 R_\ell\circ P_\ell\circ  R_{\ell+1}
$$

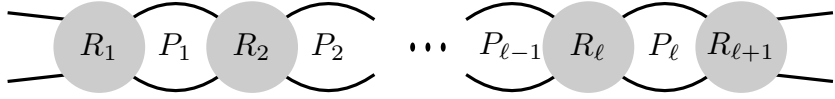
\centerline{\figplace{laddersNotnLadder1}{-.1 in}{0 in} }

\noindent
If $R$ is a rung and $A,B$ are propagators
we define $L_\ell(R;A,B)$ as the ladder with $\ell+1$ rungs $R$ and  $\ell$ bubble propagators $\cC(A,B)$.

}

The ladders contribute to the four point function, which is antisymmetric.
So the ladders must be antisymmetrized.
\definition{\STM\defOSantisymmFour}{
Let $F$ be a four legged kernel. 
The antisymmetrization of $F$ is the four legged kernel
$$\eqalign{
\big({\rm Ant\,} F\big) (x_1,x_2,x_3,x_4)
& = \sfrac{1}{4!} \sum_{\pi\in S_4} {\rm sign}(\pi)\,
F(x_{\pi(1)},x_{\pi(2)},x_{\pi(3)},x_{\pi(4)})
}$$
$F$ is called antisymmetric if $F={\rm Ant\,} F$.

}

In the direct application of Theorem \theoremVa\ of [FKTr2],
we will consider ladders with internal lines taking values in the measure space $\fX=\cB\times\Si$, where $\Si$ is a sectorization. However, the propagators are not naturally sectorized, and in [FKTf2, \S\CHphladders] we will combine bubble propagators of different scales. This motivates the following variant of the previous definitions. 

\definition{\STM\defOSsectbubbleprop}{
Let $S$ be a finite set\footnote{$^{(1)}$}{In practice, $S$ will be a set of sectors and $\fX$ will be $\cB$ 
or $\bbbr\times \bbbr^2\times \{\uparrow,\downarrow\}$ or
$\bbbr\times \bbbr^2$.}. It is endowed with the counting measure. Then $\fX\times S$ is also a measure space.

\Item{ i)} Let $P$ be a propagator over $\fX$, $f$ a four legged kernel over $\fX\times S$ and $F$ a function on $(\fX\times S)^2 \times \fX^2$. We define
$$\eqalign{
(f\bullet P)({\sst (x_1,s_1),(x_2,s_2);x_3,x_4}) 
&=  \smsum_{s'_1,s'_2\in S} \int {\sst dx'_1 dx'_2}\ 
f({\sst (x_1,s_1),(x_2,s_2),(x'_1,s'_1),(x'_2,s'_2)})\ 
P({\sst x'_1,x'_2;x_3,x_4}) \cr
(F\bullet f)({\sst(x_1,s_1),\cdots,(x_4,s_4)}) 
&=  \hskip-4pt\smsum_{s'_1,s'_2\in S}\int\hskip-4.2pt {\sst dx'_1 dx'_2 }\ 
F({\sst (x_1,s_1),(x_2,s_2);x'_1,x_2'})\ 
f({\sst (x_1',s_1'),(x_2',s_2'),(x_3,s_3),(x_4,s_4)}) \cr
}$$
whenever the integrals are well--defined. Observe that $(f\bullet P)$ is a function on $(\fX\times S)^2 \times \fX^2$ and $F\bullet f$ is a four legged kernel over $\fX\times S$.

\Item {ii)} Let
$\ell \ge 1\,$, $r_1,\cdots,r_{\ell+1}$ be rungs over $\fX\times S$  and 
$P_1,\cdots,P_\ell$  be bubble propagators over $\fX$.
The ladder with rungs $r_1,\cdots,r_{\ell+1}$ and 
bubble propagators $P_1,\cdots,P_\ell$ is defined to be
$$
r_1\bullet P_1 \bullet r_2 \bullet P_2\bullet \cdots \bullet
 r_{\ell}\bullet P_\ell\bullet r_{\ell+1}
$$
If $r$ is a  rung over $\fX\times S$
and $A,B$ are propagators over $\fX$, we define 
$L_\ell(r;A,B)$ as the ladder with   
$\ell+1$ rungs $r$ and  $\ell$ bubble propagators $\cC(A,B)$.

}

\lemma{\STM\lemOSladderfunctoriality}{
Let $c$ and $d$ be propagators over $\fX\times S$ and $r$ a rung over 
$\fX\times S$. Define the propagators $C$ and $D$ over $\fX$ by
$$
C(x_1,x_2)=\sum_{t_1,t_2\in S}c\big((x_1,t_1),(x_2,t_2)\big)\qquad\qquad
D(x_1,x_2)=\sum_{t_1,t_2\in S}d\big((x_1,t_1),(x_2,t_2)\big)
$$
and new propagators $\tilde c$ and $\tilde d$ over $\fX\times S$ by
$$
\tilde c\big((x_1,s_1),(x_2,s_2)\big)=C(x_1,x_2)\qquad\qquad
\tilde d\big((x_1,s_1),(x_2,s_2)\big)=D(x_1,x_2)
$$
Then 
$$
L_\ell(r;C,D)=L_\ell(r;\tilde c,\tilde d)
$$
for all  $\ell \ge 1$. Here, the ladder $L_\ell(r;\tilde c,\tilde d)$,
of the right hand side, is defined over the measure space $\fX\times S$ 
and uses the $\circ$ product
while the ladder on the left hand side is as in Definition 
\defOSsectbubbleprop.ii.
}

\prf For any rungs $r',r''$ over $\fX\times S$,
$$
r'\bullet\cC(C,D)\bullet r''=r'\circ\cC(\tilde c,\tilde d)\circ r''
$$
The Lemma now follows by induction on $\ell$.
\endproof

\lemma{\STM\lemOSsectorladderfunctoriality}{
Let $\Si$ be a sectorization at scale $j$ and $\varphi \in\cF_0(4,\Si)$.
Let $C(k)$ and $D(k)$ be functions on $\bbbr\times\bbbr^2$, that are supported in the $j^{\rm th}$ neighbourhood, and
$C(\xi,\xi'),\,D(\xi,\xi')$ their Fourier transforms as in Definition 
\defOSftcov. Furthermore, let $c\big((\cdot,s),(\cdot,s') \big)$ and $d\big((\cdot,s),(\cdot,s') \big)$ be the Fourier transforms of 
$\chi_s(k)\,C(k)\,\chi_{s'}(k)$  and $\chi_s(k)\,D(k)\,\chi_{s'}(k)$. 
Define propagators over $\cB\times\Si$ by
$$\eqalign{
c_\Si({\sst(\xi,s),(\xi',s')}) 
&= \sum_{t\cap s \ne \emptyset \atop t'\cap s' \ne \emptyset}
c({\sst (\xi,t),(\xi',t')}) \cr
d_\Si({\sst(\xi,s),(\xi',s')})  
&= \sum_{t\cap s \ne \emptyset \atop t'\cap s' \ne \emptyset}
d({\sst (\xi,t),(\xi',t')}) \cr
}$$
Then 
$$
L_\ell(\varphi;C,D)=L_\ell(\varphi;c_\Si,d_\Si)
$$
for all  $\ell \ge 1$. Here, the ladder on the right hand side is defined 
over the measure space $\cB\times \Si$ 
and uses the $\circ$ product
while the ladders on the left hand side are as in Definition 
\defOSsectbubbleprop.ii.
}

\prf 
Since $\sum_{s\in\Si}\chi_s(k)$ is identically one on the support of $C(k)$ and 
$D(k)$,
$$
C(\xi,\xi_2)=\sum_{t, t'\in\Si}c({\sst (\xi,t),(\xi',t')})
\qquad\qquad
D(\xi_1,\xi_2)=\sum_{t, t'\in\Si}d({\sst (\xi,t),(\xi',t')})
$$
As in Lemma \lemOSladderfunctoriality, set
$$
\tilde c\big((\xi_1,s_1),(\xi_2,s_2)\big)=C(\xi_1,\xi_2)\qquad\qquad
\tilde d\big((\xi_1,s_1),(\xi_2,s_2)\big)=D(\xi_1,\xi_2)
$$
Denote
$$
p=\cC(c,d)\qquad\qquad
p_\Si=\cC(c_\Si,d_\Si)\qquad\qquad
\tilde p=\cC(\tilde  c, \tilde  d)
$$
Then, $p$ is a $\Si$--sectorized bubble propagator and 
$$\eqalign{
p_\Si({\sst(\xi_1,s_1),(\xi_2,s_2),(\xi_3,s_3),(\xi_4,s_4)}) 
&= \sum_{t_i\cap s_i \ne \emptyset \atop 1\le i\le 4}
p({\sst(\xi_1,t_1),(\xi_2,t_2),(\xi_3,t_3),(\xi_4,t_4)})\cr
\tilde p({\sst(\xi_1,s_1),(\xi_2,s_2),(\xi_3,s_3),(\xi_4,s_4)}) 
&= \sum_{t_i\in\Si \atop 1\le i\le 4}
p({\sst(\xi_1,t_1),(\xi_2,t_2),(\xi_3,t_3),(\xi_4,t_4)})\cr
}$$
For any $ w\in\cF_0(4,\Si)$,
$$\eqalignno{
w\circ p_\Si\circ \varphi
&=\sum_{s'_i\in \Si\atop 1\le i\le 4}\int {\sst d\xi'_1 \cdots d\xi'_4}\ 
 w({\sst \,\cdot\,,\,\cdot\,;(\xi'_1,s'_1),(\xi'_2,s'_2)})
    \,p_\Si({\sst(\xi'_1,s'_1),\cdots,(\xi'_4,s'_4)}) 
    \,\varphi({\sst(\xi'_3,s'_3),(\xi'_4,s'_4),\,\cdot\,,\,\cdot\,}) \cr
&=\sum_{{s'_i,t_i\in \Si\atop s'_i\cap t_i\ne\emptyset }\atop 1\le i\le 4}
     \int {\sst d\xi'_1 \cdots d\xi'_4}\ 
 w({\sst \,\cdot\,,\,\cdot\,;(\xi'_1,s'_1),(\xi'_2,s'_2)})
    \,p({\sst(\xi'_1,t_1),\cdots,(\xi'_4,t_4)}) 
    \,\varphi({\sst(\xi'_3,s'_3),(\xi'_4,s'_4),\,\cdot\,,\,\cdot\,}) \cr
&=\sum_{s'_i,t_i\in \Si\atop 1\le i\le 4}
     \int {\sst d\xi'_1 \cdots d\xi'_4}\ 
 w({\sst \,\cdot\,,\,\cdot\,;(\xi'_1,s'_1),(\xi'_2,s'_2)})
    \,p({\sst(\xi'_1,t_1),\cdots,(\xi'_4,t_4)}) 
    \,\varphi({\sst(\xi'_3,s'_3),(\xi'_4,s'_4),\,\cdot\,,\,\cdot\,}) \cr
&= w\circ\tilde p\circ \varphi&\EQNO\eqnOSladderfunctC
}$$
because
$$
\int {\sst d\xi'_1 \cdots d\xi'_4}\ 
 w({\sst \,\cdot\,,\,\cdot\,;(\xi'_1,s'_1),(\xi'_2,s'_2)})
    \,p({\sst(\xi'_1,t_1),\cdots,(\xi'_4,t_4)}) 
    \,\varphi({\sst(\xi'_3,s'_3),(\xi'_4,s'_4),\,\cdot\,,\,\cdot\,})
$$
vanishes unless $s'_i\cap t_i\ne\emptyset $ for all $1\le i\le 4$.
Observe that $\tilde w\circ\tilde p\circ \varphi$ is again in $\cF_0(4,\Si)$.

It follows by induction from (\eqnOSladderfunctC) that
$$
L_\ell(\varphi;c_\Si,d_\Si)=L_\ell(\varphi;\tilde c,\tilde d)
$$
The lemma follows by Lemma \lemOSladderfunctoriality.
\endproof

\vfill\eject

\chap{ Norm Estimates on the Renormalization Group Map}\PG\pgOSXV

Again, let $j\ge 2$ and $\Si$ be a sectorization of scale $j$ and length
$\sfrac{1}{M^{j-3/2}}\le\fl\le\sfrac{1}{M^{(j-1)/2}}$. 
Fix a system 
$\vec \rho = (\rho_{m;n})$ of positive real numbers such that
$$\deqalign{
\rho_{m;n} & \le \rho_{m;n'} \qquad &{\rm if\ } n\le n' \cr
\rho_{m+m';n+n'-2} & \le \rho_{m;n}\,\rho_{m';n'}\cr
\rho_{m+1;n-1} & \le \rho_{m;n} \qquad &{\rm if\ } m\ge 1 \cr
\rho_{1;n-1} & \le \sqrt{\fl M^j}\ \rho_{0;n}  \cr
}\EQN\eqnOSrhomn$$

\definition{\STM\defOSscalednorms}{
\Item{i)}
For $\varphi\in \cF_m(n;\Si)$ set
$$
\v \varphi \v_\Si = \rho_{m;n}\cases{
\v \varphi \v_{1,\Si} + \sfrac{1}{\fl}\,\v \varphi \v_{3,\Si}
+ \sfrac{1}{\fl^2}\,\v \varphi \v_{5,\Si} 
    & if $m=0$ \cr
\sfrac{\fl}{M^{2j}}\,\v \varphi \v_{1,\Si} & if $m\ne0$ \cr
}$$
\Item{ii)}
We set, for 
$X = \smsum_{\de\in \bbbn_0\times\bbbn_0^2} X_\de\,t^\de \in \fN_{d+1}$
with $X_\0<\sfrac{1}{M^j}$,
$$
\fe_j(X) = \sfrac{\cb_j}{1-M^j X}
$$
\Item{iii)}
A sectorized Grassmann function $w$ can be uniquely written in the form
$$\eqalign{
w(\phi,\psi) = \sum_{m,n}\ \sum_{s_1,\cdots,s_n\in\Si}\ 
\int {\sst d\eta_1\cdots d\eta_m\,d\xi_1\cdots d\xi_n}\ & 
w_{m,n}({\sst \eta_1,\cdots, \eta_m\,(\xi_1,s_1),\cdots ,(\xi_n,s_n)})\cr
& \hskip 1cm \cdot\ \phi({\sst \eta_1})\cdots \phi({\sst \eta_m})\
\psi({\sst (\xi_1,s_1)})\cdots \psi({\sst (\xi_n,s_n)\,})\cr
}$$
with $w_{m,n}$ antisymmetric separately in the $\eta$ and in the $\xi$ variables.
Set, in analogy with Theorem \thmOSfirststep, for $\al >0$
and $X\in \fN_{d+1}$,
$$
N_j(w;\al;\,X,\Si,\vec\rho)
=\sfrac{M^{2j}}{\fl}\,\fe_j(X) 
\smsum_{m,n\ge 0}\,
\al^{n}\,\big(\sfrac{\fl\,\IB}{M^j}\big)^{n/2} \,\v w_{m,n}\v_\Si 
$$
The constant $\IB$ will be chosen in Definition \defOSmomscalednorms.iii.
It will obey $\IB>4\max\{8\IB_1,\IB_2\}$ with $\IB_1,\ \IB_2$ being 
the constants of Propositions \propOScontrintboundsectors\ and
 \propOSoverlapploops.

}

\goodbreak
\remark{\STM\remOSscalednorms}{
\Item{i)}
By definition, for even $w$
$$\eqalign{
N_j(w;\La,\al;\,X,\Si,\vec\rho)
&= \fe_j(X) \smsum_{n\ge 1}\,\IB^n\,\al^{2n}\,\sfrac{\fl^{n-1}}{M^{j(n-2)}} \
\rho_{0;2n}\,  \v w_{0,2n}\v_{1,\Si} \cr
&+\fe_j(X) \smsum_{n\ge 2}\,\IB^n\,\al^{2n}\,\sfrac{\fl^{n-2}}{M^{j(n-2)}} \
\rho_{0;2n} \,\v w_{0,2n}\v_{3,\Si} \cr
&+\fe_j(X) \smsum_{n\ge 3}\,\IB^n\,\al^{2n}\,\sfrac{\fl^{n-3}}{M^{j(n-2)}} \
\rho_{0;2n} \,\v w_{0,2n}\v_{5,\Si} \cr
&+ \fe_j(X)\,\smsum_{m\ge 1} \smsum_{n\ge 0} \al^n
\big(\sfrac{\fl\,\IB}{M^j}\big)^{n/2}\,\rho_{m;n}\,
  \v w_{m,n}\v_{1,\Si}
}$$
If, in a renormalization group analysis, $\rho_{0;2n}$ is independent of the 
scale number, j, then boundedness of the norms $N_j$ imply that 
$$
\v w_{0,2}\v_{1,\Si}=O\big(\sfrac{1}{M^j}\big)\qquad
\v w_{0,4}\v_{3,\Si}=O\big(1\big)
$$
modulo $t$.

\Item{ii)}
If $X \le \sfrac{1}{2M^{j}}\, \cb_j$ then
$\fe_j( X) \le \abcst\, \cb_j$.

\Item{iii)}
$\sfrac{\partial\hfill}{\partial t_0}\cb_j\le \abcst\, M^j\cb_j+\sum_{\de_0=r_0}\infty\,t^\de$.

\Item{iv)}
If $X$ is independent of $t_0$, then
$\sfrac{\partial\hfill}{\partial t_0}\fe_j( X) \le \abcst\, M^j\fe_j( X)
+\sum_{\de_0=r_0}\infty\,t^\de$.

\Item{v)} The $j$--dependent factors in the definition of 
$N_j$ were largely motivated by the discussion in [FKTf1, \S\CHintroOverview, subsection 8] and Remark \remNPnvsnj\ of [FKTf2]. 

}

The main result of this paper is, that the norms of Definition \defOSscalednorms\ 
are not changed very much by the renormalization group map $\tilde \Om_C$ of 
Definition \defOSrengrpmap, and that there is volume improvement for the 
two point function and all contributions to the four point function with 
the exception of ladders.

\theorem{\STM\thOSrengroupestimate}{
There are constants $\const,\ \cst{}{0},\ \al_0$ and $\tau_0$
that are independent of $j,\ \Si,\ \vec\rho$ such that for all $\al\ge \al_0$ 
 the following estimates hold:

{\parindent=.25in
\item{}
Let $u({\sst(\xi,s)},{\sst(\xi',s')}),v({\sst(\xi,s)},{\sst(\xi',s')})
\in\cF_0(2;\Si)$ be antisymmetric, spin independent, particle number 
conserving functions whose Fourier transforms  obey
 $|\check u(k)|,|\check v(k)| \le \half |\imath k_0-e(k)|$.
Furthermore, let
$X\in\fN_{d+1}$, $\mu,\La>0$ and assume that
$\ \v  u \v_{1,\Si} \le \mu(\La+X)\fe_j(X)\ $ and
$(1+\mu)(\La+X_\0)\le\sfrac{\tau_0}{M^j}$. Set
$$
C(k) = \sfrac{\nu^{(j)}(k)}{\imath k_0 -e(\k) -\check u(k)}
\qquad,\qquad 
D(k) = \sfrac{\nu^{(\ge j+1)}(k)}{\imath k_0 -e(\k) -\check v(k)}
$$
and let $C(\xi,\xi'),\,D(\xi,\xi')$ be the Fourier transforms of $C(k)$, $D(k)$ as
in Definition \defOSftcov. Let $\cW(\phi,\psi)$ be a Grassmann function and 
set\footnote{$^{(1)}$}{The definition of $\cW'$ as an analytic function, rather 
than merely a formal Taylor series will be explained in 
Remark \:\remOSchoiceofrep.}
$$
\lw \cW'(\phi,\psi) \rw_{\psi,D} 
= \tilde \Om_C \big( \lw \cW(\phi,\psi) \rw_{\psi,C+D} \big) 
$$
Assume that $\cW$ has a sectorized representative $w$
with $w_{0,2}=0$ and
$$
N_j(w;64\al;\,X,\Si,\vec\rho)
\le \cst{}{0}\,\al + \smsum_{\de \ne 0} \infty\,t^\de
$$
Then $\cW'$ has a sectorized representative $w'$ such that
$$
N_j(w'-\sfrac{1}{2}\phi JCJ\phi-w;\al;\,X,\Si,\vec\rho) 
\le \sfrac{\const}{\al}\,
\sfrac{N_j(w;64\al;\,X,\Si,\vec\rho)}{1-{\const \over\al}N_j(w;64\al;\,X,\Si,\vec\rho)}
$$
Furthermore
$$
\v w_{0,2}' \v_{1,\Si}  
\le  \sfrac{\const}{\al^8\,\rho_{0;2}}\sfrac{ \fl}{M^j}\,
\sfrac{N_j(w;64\al;\,X,\Si,\vec\rho)^2}
          {1-{\const \over\al}N_j(w;64\al;\,X,\Si,\vec\rho)}
$$
and
$$\eqalign{
&\V w_{0,4}'-w_{0,4} -
\sfrac{1}{4} \smsum_{\ell=1}^\infty (-1)^\ell(12)^{\ell+1} 
{\rm Ant}\,L_\ell(w_{0,4};C,D) \V_{3,\Si} \cr
&\hskip3in \le\sfrac{\const}{\al^{10}\rho_{0;4}}\,\fl\,
\sfrac{N_j(w;64\al;\,X,\Si,\vec\rho)^2}
           {1-{\const \over\al}N_j(w;64\al;\,X,\Si,\vec\rho)} 
}$$

}
}

\remark{\STM\remOSrengroupestimate}{
\Item{i)} 
When we use Theorem \thOSrengroupestimate\ in a
renormalization group analysis, $u$ will depend on counterterms that will  ultimately be generated at scales $j'>j$. Then the derivatives of $\check u(k)$ can
have a scaling behaviour characteristic of scale $j'$. In this case 
$\v u \v_{1,\Si}$ will not be of order $\cb_j$. This is why we
introduce the factor $\fe_j(X)$ in the definition of $N_j$.
\Item{ii)} 
The hypothesis that $w_{0,2}=0$ is used, in conjunction with Wick ordering, 
to ensure that all non--ladder contributions to $w_{0,2}'$ and $w_{0,4}'$
contain overlapping loops. See [FKTf1, \S \CHintroOverview, 
subsections 4 and 9].
\Item{iii)} In Appendix \APappNaiveladder, we give naive power--counting bounds for ladders $L_\ell(w_{0,4};C,D)$. These 
estimates are not good enough for a renormalization group analysis. 
They would lead to logarithmic 
divergences. Stronger estimates on the ``particle--particle'' part of the ladders are derived in Theorem \:\theoremOSLadA.
The ``particle--hole'' parts of the ladders are treated in [FKTl].

}

Most of the rest of this Chapter is devoted to the proof of Theorem \thOSrengroupestimate.
To simplify notation we write $N_j(w;\al)$ for
$N_j(w;\al;\,X,\Si,\vec\rho) $. We define a family of seminorms on the
spaces $\cF_m(n;\Si)$ by
$$
\v \varphi \v_p = \rho_{m;n} \cases{ \v \varphi\v_{p,\Si} & if $m=0$ \cr
  \sfrac{\fl}{M^{2j}} \v \varphi\v_{p,\Si} & if $m \ne 0$ \cr
}
$$
with $p=1,3,5$. As in Definition \defOStens, these norms induce a family of
symmetric seminorms on the spaces $A_m\otimes V_\Si^{\otimes n}$.
This family of seminorms will only appear in the proof of 
Theorem \thOSrengroupestimate\ and in the preliminary Lemma  \thOSrengroupestimate.

\noindent
Let $c\big((\cdot,s),(\cdot,s') \big)$ and $d\big((\cdot,s),(\cdot,s') \big)$ be
the Fourier transform of 
$\chi_s(k)\,C(k)\,\chi_{s'}(k)$  and $\chi_s(k)\,D(k)\,\chi_{s'}(k)$
in the sense of Definition \defOSftcov. As in Lemma \lemOSsectorladderfunctoriality, let
$$\eqalign{
c_\Si({\sst(\xi,s),(\xi',s')}) 
&= \sum_{t\cap s \ne \emptyset \atop t'\cap s' \ne \emptyset}
c({\sst (\xi,t),(\xi',t')}) \cr
d_\Si({\sst(\xi,s),(\xi',s')})  
&= \sum_{t\cap s \ne \emptyset \atop t'\cap s' \ne \emptyset}
d({\sst (\xi,t),(\xi',t')}) \cr
}$$
As in
Proposition \propOSfunctorialitySect,
$$\eqalign{
C_\Si\big(\psi({\sst\xi,s}),\psi({\sst\xi',s'}) \big) 
&= c_\Si({\sst(\xi,s),(\xi',s')}) \cr
D_\Si\big(\psi({\sst\xi,s}),\psi({\sst\xi',s'}) \big) 
&= d_\Si({\sst(\xi,s),(\xi',s')}) \cr
}$$
are covariances on $V_\Si$.

\lemma{\STM\lemOSconcreteintconst}{
Under the hypotheses of Theorem \thOSrengroupestimate, there exists a constant
$\cst{}{1}$ that is independent of $j$ and $\Si$ such that
the covariances $C_\Si,\,D_\Si$ have integration constants%
\footnote{$^{(2)}$}{We shall, in the proof of 
Theorem \:\thOSrengroupdiffestimate\ below, apply Theorem \theoremIVb\ 
of [FKTr1], which requires integral bounds $\half\ib$. Of course, then 
 $\ib$ is also an integral bound, as is required in 
the proof of the current Theorem \thOSrengroupestimate.} 
$$
\cb = \cst{}{1}\,M^j\,\fe_j(X),\qquad \half\ib =\sqrt{\sfrac{\IB\,\fl}{4M^j}}\ 
$$  
(in the sense of Definition \defimprconf\  of [FKTr2]) for the configuration $\v
\cdot\v_p$ of seminorms.  }

\prf
Clearly, the functions $C(k)$ and $D(k)$ are supported on the $j^{\rm th}$
neighbourhood, and $\ |C(k)|,\,|D(k)| \le \sfrac{2}{|\imath k_0 - e(\k)|}\ $. By
part (ii) of  Proposition \propOScontrintboundsectors\ and the first condition
of (\eqnOSrhomn), $\half\ib$ is 
an integral bound both for $C_\Si$ and $D_\Si$. 

We now verify the contraction estimates of Definition \defimprconf\  of [FKTr2].
Contraction by $c$ for functions on $\cB^m\times \big(\cB\times\Si\big)^m$ as
in Definition \defOSsectcontnorm\ corresponds to contraction by $C_\Si$
in the Grassmann algebra over $V_\Si$ as in Definition \defContract\ of [FKTr1]. Set 
$$
\cb' = 9\max \Big\{ \v c\v_{1,\Si},\ 
\sfrac{M^{2j}}{\fl}\,\sup_{\xi,\xi',s,s'} | c({\sst(\xi,s),(\xi',s')}) |
\big)  \Big\}
$$
It follows from part (i) of Proposition \propOScontrintboundsectors,
combined with the second property of $\vec\rho$, and
Proposition \propOSoverlapploops, combined with the first two properties of 
$\vec\rho$, that $C_\Si,\,D_\Si$ have integration
constants $\cb',\half\ib$. By Proposition \propOSrealpropbound, if $\tau_0$ is 
small enough,
$$
\cb' \le \const \max \Big\{ 
\sfrac{M^j\cb_j}{1-M^j\sv u \sv_{1,\Si}},\ M^j \Big\}
\le \sfrac{\const\,M^j\,\cb_j}{1-M^j\sv u \sv_{1,\Si}}
$$
Therefore, by the hypotheses on $u$,
$$
\cb' 
\le \frac{\const\,M^j\,\cb_j}{1-
\mu M^j(\La+X){\cb_j\over 1-M^jX}}
\le \frac{\const\,M^j\,\cb_j}{1-\mu
{M^j\cb_j(\La+X)\over 1-M^j\cb_j(\La+X)}}
=\const\,M^j\,\cb_j\,f(Y)
$$
where $Y=M^j\cb_j(\La+X)$ and 
$$
f(z)= \frac{1}{1-\mu{z\over 1-z}}
= \frac{1-z}{1-(1+\mu)z}
$$
By Lemma \lemOSappMonoidV, $f(Y)\le\abcst\sfrac{1}{1-Y}$ so that
$$
\cb' \le\sfrac{\const\,M^j\,\cb_j}{1-M^j\cb_j\La-M^j\cb_jX}
\le\const\,M^j\,\sfrac{\cb_j}{1-M^j\cb_j\La}\ \sfrac{1}{1-M^j\cb_jX}
\le\const\,M^j\,\sfrac{\cb_j}{1-\cb_j/3}\ \sfrac{1}{1-M^j\cb_jX}
$$
In the second inequality we used Lemma \lemOSappMonoidIV.ii.
As, by Corollary \corOSappMonoidIV.i, $\sfrac{\cb_j}{1-\cb_j/3}\le\abcst\,\cb_j$ and
$\sfrac{\cb_j}{1-M^j\cb_jX}\le \abcst\,\sfrac{\cb_j}{1-M^j X}$, we have
$$
\cb' \le \const\,M^j\,\fe_j(X)
\EQN\eqnOScbprimebnd$$
\endproof

\lemma{\STM\lemOSconcretesourcetermintconst}{
Let $g(\phi,\psi)$ be a sectorized Grassmann function.
Let $C(k)$ be  a function obeying $|C(k)|\le \sfrac{2}{|\imath k_0 -e(\k)|}$ 
and  $C(\xi,\xi')$, resp. $c({\sst (\xi,s),(\xi',s')})$, be the
Fourier transforms of $C(k)$, resp. $\chi_s(k)\,C(k)\,\chi_{s'}(k)$, in 
the sense of Definition \defOSftcov. Set
$$
g'(\phi,\psi)=g(\phi,\psi+C J\phi)
$$
If $\v c \v_{1,\Si} \le \const M^j+ \smsum_{\de\ne 0}\infty t^\de$, then
$$
N_j(g'-g;\al)\le \sfrac{\const}{\al}N_j(g;2\al)
$$
In particular, this bound is true under the hypotheses of Theorem \thOSrengroupestimate.

  }
\prf 
Let $\varphi\in\cF_m(n;\Si)$, $1\le i\le n$ and set
$$\eqalign{
&\varphi'({\sst\et_1,\cdots,\et_{m+1}\,;
\,(\xi_1,s_1),\cdots,(\xi_{n-1},s_{n-1})})\cr
&\hskip1cm= {\rm Ant_{ext}}\smsum_{s,t,t'\in\Si}\int\!\! d{\sst \ze}d{\sst \ze'}\ 
\varphi({\sst\et_1,\cdots,\et_m\,;\,(\xi_1,s_1),\cdots,
(\xi_{i-1},s_{i-1}),(\ze',t),(\xi_{i},s_i),\cdots,(\xi_{n-1},s_{n-1})})\cr
\noalign{\vskip-.1in}
&\hskip10cm \cdot\ c({\sst(\ze',t',),(\ze,s)})\,J({\sst \ze,\et_{m+1}})\cr
}$$
Under the hypotheses of Theorem \thOSrengroupestimate, 
 $\v c \v_{1,\Si} \le \const M^j+ \smsum_{\de\ne 0}\infty t^\de$
by Proposition \propOSrealpropbound.ii. Hence, by Lemma \lemOSsectextimpr,
$$
\v\varphi'\v_{1,\Si}
\le \const \v \varphi\v_{1,\Si}\cases{\sfrac{M^j}{\fl}& if $m=0$ \cr
                                \noalign{\vskip.05in}
                               \sfrac{\fl}{M^j}& if $m\ne 0$ \cr}
$$
Here we have used that the coefficient of $t^\de$ in $\v\varphi'\v_{1,\Si}$
vanishes for $\de\ne 0$ so that in Lemma \lemOSsectextimpr\ we may replace 
$\v c\v_{1,\Si}$ by its value at $t=0$. Hence, for $m=0$,
$$\eqalign{
\v\varphi'\v_\Si
=\rho_{1;n-1}\sfrac{\fl}{M^{2j}}\v\varphi'\v_{1,\Si}
&\le\const\rho_{1;n-1}\sfrac{1}{M^{j}}\v\varphi\v_{1,\Si}
\le\const\sfrac{\rho_{1;n-1}}{\rho_{0;n}}\sfrac{1}{M^{j}}\v\varphi\v_{\Si}\cr
&\le\const\sqrt{\fl M^j}\sfrac{1}{M^{j}}\v\varphi\v_{\Si}
\le\const\ib \v\varphi\v_{\Si}
}$$
and, for $m\ne 0$,
$$\eqalign{
\v\varphi'\v_\Si
=\rho_{m+1;n-1}\sfrac{\fl}{M^{2j}}\v\varphi'\v_{1,\Si}
&\le\const\rho_{m+1;n-1}\sfrac{\fl^2}{M^{3j}}\v\varphi\v_{1,\Si}
\le\const\sfrac{\rho_{m+1;n-1}}{\rho_{m;n}}\sfrac{\fl}{M^{j}}\v\varphi\v_{\Si}
\cr
&\le\const\sfrac{\fl}{M^{j}}\v\varphi\v_{\Si}
\le\const\ib \v\varphi\v_{\Si}
}$$
The Lemma now follows from $\v\varphi'\v_\Si\le\const\ib \v\varphi\v_{\Si}$
as Proposition \propOSextimpr\ follows from the bound of Definition
\defOSextimpr.
\endproof

\proof{ of  Theorem \thOSrengroupestimate}
For $\varphi\in \cF_m(n;\Si)$ set
$$
\v \varphi \v_{{\rm impr},\Si} = \rho_{m;n}\cases{
\v \varphi \v_{1,\Si} + \sfrac{1}{\fl}\,\v \varphi \v_{3,\Si} & if $m=0$ \cr
0 & if $m\ne 0$
}
$$
This family of seminorms will only appear in this proof.
By Lemma \lemOSconcreteintconst\ and Lemma \lemimprconfig\  of [FKTr2], 
with $q=5$, $J=\fl$ and $\|\ \cdot\ \|_p=\v\ \cdot\ \v_p$, the
covariances $(C_\Si, D_\Si)$ have improved integration constants $\cb,\ib,\fl$
for the families $\v\cdot\v_\Si$ and $\v\cdot\v_{{\rm impr},\Si}$ 
of seminorms (in the sense of Definition \defimprnorm\  of [FKTr2]).
For a sectorized Grassmann function $v=\smsum_{m,n} v_{m,n}$ with
$v_{m,n} \in A_m\otimes \bigwedge^nV_\Si$ let
$$\eqalign{ 
N(v;\al) &= \sfrac{1}{\ib^2}\cb\,\smsum_{m,n} \al^n\, \ib^n\, \v v_{m,n} \v_\Si \cr
N_{\rm impr}(v;\al) &= \sfrac{1}{\ib^2}\cb\,
\smsum_{n} \al^n\, \ib^n\, \v v_{0,n}\v_{{\rm impr},\Si}
}$$
be the quantities introduced in Definition \deffunctnorm\ of [FKTr1]
and just after Lemma \lemGrassimprnorm\ of [FKTr2]. Then
$$
N(v;\al) = \sfrac{\cst{}{1}}{\IB}\, N_j(v;\al;\,X,\Si,\vec\rho)
$$
where $\cst{}{1}$ is the constant of Lemma \lemOSconcreteintconst.

Set $\ \lw w''\rw_{\psi,D_\Si} = \Om_{C_\Si}(\lw w \rw_{\psi,C_\Si+D_\Si})\ $
and 
$$
w'=\half\phi JCJ\phi+w''(\phi,\psi+ C J\phi)
$$
By parts (ii) and (iii) of Proposition \propOSfunctorialitySect,
$\lw w' \rw_{\psi,D_\Si}$   is a sectorized representative for 
$\lw \cW'(\phi,\psi) \rw_{\psi,D}$.
Hence, by Proposition \propOSfunctorialitySect.i and Proposition \propBII.ii 
of [FKTr1], $w'$ is a sectorized representative for $\cW'$.
We apply Theorem \theoremVa\  of [FKTr2] to get estimates on $w''$. With 
$\cst{}{0} =\sfrac{\IB}{8\,\cst{}{1}}$ the hypotheses of this Theorem are fulfilled. Consequently
$$
N(w''-w;\al) \le \sfrac{1}{2\al^2}\,
\sfrac{N(w;32\al)^2}{1-{1 \over\al^2}N(w;32\al)}
\EQN\eqnOSrengroupestimate$$
and
$$\eqalign{
\al^2\, \cb\,\v w_{0,2}'' \v_{{\rm impr},\Si} 
& \le \sfrac{2^{10} \fl}{\al^6}\,\sfrac{N(w;64\al)^2}{1-{8 \over\al}N(w;64\al)}\cr
\al^4\,\ib^2 \cb\,\V w_{0,4}''-w_{0,4} -
\sfrac{1}{4} \smsum_{\ell=1}^\infty (-1)^\ell(12)^{\ell+1} {\rm Ant}\,L_\ell(w_{0,4};c_\Si,d_\Si)
\V_{{\rm impr},\Si}  
& \le \sfrac{2^{10} \fl}{\al^6}\,\sfrac{N(w;64\al)^2}{1-{8\over\al}N(w;64\al)}\cr
}$$ 
For the last estimate, we also used the description of ladders in terms of
kernels of Proposition \propAIV\  of [FKTr2]. As $w'_{0,2} = w_{0,2}''$ and
$w'_{0,4} = w_{0,4}''$ this implies that
$$
\fe_j(X)\,\v w_{0,2}' \v_{1,\Si}  
\le \sfrac{\const}{\al^8\,\rho_{0;2}}\sfrac{ \fl}{M^j}\,
\sfrac{N_j(w;64\al)^2}{1-{\const \over\al}N_j(w;64\al)}
$$
and, using Lemma \lemOSsectorladderfunctoriality
$$\eqalign{
&\fe_j(X)\,\V w_{0,4}'-w_{0,4} -
\sfrac{1}{4} \smsum_{\ell=1}^\infty (-1)^\ell(12)^{\ell+1} {\rm Ant}\,L_\ell(w_{0,4};C,D)
\V_{3,\Si} \cr
&\hskip.5in\le \sfrac{1}{\rho_{0;4}}\fl\,\fe_j(X)\,\V w_{0,4}''-w_{0,4} -
\sfrac{1}{4} \smsum_{\ell=1}^\infty (-1)^\ell(12)^{\ell+1} {\rm Ant}\,L_\ell(w_{0,4};c_\Si,d_\Si)
\V_{{\rm impr},\Si} \cr
&\hskip.5in \le \sfrac{\const}{\al^{10}\rho_{0;4}}\,\fl\,
\sfrac{N_j(w;64\al)^{2}}{1-{\const \over\al}N_j(w;64\al)} \cr
}$$
By Lemma \lemOSconcretesourcetermintconst,
$$\eqalign{
&N_j(w'-\sfrac{1}{2}\phi JCJ\phi-w;\al)
=N_j\big(w''(\phi,\psi+ C J\phi)-w(\phi,\psi);\al\big)\cr
&\hskip.5in
\le N_j\big(w''(\phi,\psi+ C J\phi)-w''(\phi,\psi);\al\big)
+N_j\big(w''(\phi,\psi)-w(\phi,\psi);\al\big)\cr
&\hskip.5in
\le \sfrac{\const}{\al} N_j\big(w'';2\al\big)+N_j\big(w''-w;\al\big)\cr
&\hskip.5in
\le \sfrac{\const}{\al} N_j\big(w;2\al\big)
+\big(1+\sfrac{\const}{\al}\big)\sfrac{\IB}{\cst{}{1}} N\big(w''-w;2\al\big)\cr
&\hskip.5in
\le \sfrac{\const}{\al} N_j\big(w;2\al\big)
+\big(1+\sfrac{\const}{\al}\big)\sfrac{\IB}{\cst{}{1}} \sfrac{1}{8\al^2}\,
\sfrac{N(w;64\al)^2}{1-{1 \over4\al^2}N(w;64\al)}\cr
&\hskip.5in
\le \sfrac{\const}{\al} N_j\big(w;2\al\big)
+ \sfrac{\const}{\al^2}\,
\sfrac{N_j(w;64\al)^2}{1-{\const \over\al^2}N_j(w;64\al)}\cr
&\hskip.5in
\le \sfrac{\const}{\al} N_j\big(w;64\al\big)
+ \sfrac{\const}{\al^2}\,
\sfrac{N_j(w;64\al)^2}{1-{\const \over\al}N_j(w;64\al)}\cr
&\hskip.5in
\le  \sfrac{\const}{\al}\,
\sfrac{N_j(w;64\al)}{1-{\const \over\al}N_j(w;64\al)}\cr
}$$
\endproof

We also wish to allow the functions $u$ and $v$ of Theorem
\thOSrengroupdiffestimate\ to depend on a parameter $\ka$.

\theorem{\STM\thOSrengroupdiffestimate}{
There are constants $\const,\ \cst{}{0},\ \al_0,\ \tau_0$ 
that are independent of $j,\ \Si,\ \vec\rho$ such that for all $\veps>0$ and $\al\ge \al_0$
the following estimates hold:

{\parindent=.25in
\item{}
Let, for $\ka$ in a neighbourhood of zero, $u_\ka,v_\ka\in\cF_0(2;\Si)$ be antisymmetric, spin independent,
particle number conserving functions whose Fourier transforms satisfy 
$|\check u_0(k)|, |\check v_0(k)| \le \half |\imath k_0-e(k)|$
and $\big|\sfrac{d\hfill}{d\ka}\check v_\ka(k)\big|_{\ka=0}\big| \le 
\veps |\imath k_0-e(k)|$.
Furthermore, let $X,Y\in\fN_{d+1}$, $\mu,\La>0$ and assume that  
$$
\v  u_0 \v_{1,\Si} \le \mu(\La+X)\fe_j(X)\qquad\qquad
\V \sfrac{d\hfill}{d\ka} u_\ka\big|_{\ka=0}\V_{1,\Si} \le \fe_j(X)Y
$$ 
and $(1+\mu)(\La+X_\0)\le\sfrac{\tau_0}{M^j}$.
Set
$$
C_\ka(k) = \sfrac{\nu^{(j)}(k)}{\imath k_0 -e(\k) -\check u_\ka(k)}
\qquad,\qquad 
D_\ka(k) = \sfrac{\nu^{(\ge j+1)}(k)}{\imath k_0 -e(\k) -\check v_\ka(k)}
$$
and let $C_\ka(\xi,\xi'),\,D_\ka(\xi,\xi')$ be the Fourier transforms of $C_\ka(k)$, $D_\ka(k)$. Let, for $\ka$ in a neighbourhood of zero, 
$\cW_\ka(\phi,\psi)$  be an even Grassmann function and set
$$
\lw \cW'_\ka(\psi) \rw_{\psi,D_\ka} 
= \tilde \Om_{C_\ka} \big( \lw \cW_\ka \rw_{\psi,C_\ka+D_\ka} \big)
$$
Assume that $\cW_\ka$ has a sectorized representative $w_\ka$
with
$$
\fn\equiv N_j(w_0;64\al;\,X,\Si,\vec\rho)
\le \cst{}{0}\,\al + \smsum_{\de \ne 0} \infty\,t^\de
$$
Then $\cW'_\ka$  has a sectorized representative 
$w'_\ka$ such that
$$\eqalign{
&N_j\big(\,\sfrac{d\hfill}{d\ka}
  \big[w'_\ka-\half\phi JC_\ka J\phi-w_\ka\big]_{\ka=0}
\,;\,\al\,;\,{\sst X,\Si,\vec\rho}\,\big)\cr
&\hskip0.75in\le \const
\Big\{\sfrac{1}{\al}+\sfrac{1}{\al^2}\sfrac{\fn}{1-{\const\over\al^2}\fn}\Big\}
N_j\big(\,\sfrac{d\hfill}{d\ka}w_\ka\big|_{\ka=0}
\,;\,16\al\,;\,{\sst X,\Si,\vec\rho}\,\big)\cr
&\hskip1.5in+\const\sfrac{\fn}{1-{\const\over\al^2}\fn}  \Big\{
\big(\sfrac{1}{\al}+\sfrac{\fn}{\al^2}\big)M^{j}Y+\sfrac{\veps}{\al^2}\fn\Big\}\cr
}$$

}
}

\lemma{\STM\lemOSconcretediffintconst}{
Under the hypotheses of Theorem \thOSrengroupdiffestimate, 
there exists a constant $\cst{}{2}$ that is independent of $j$ and $\Si$ 
such that
$C_{0,\Si}$ has contraction bound $\cb$, 
$C_{0,\Si}$ and $D_{0,\Si}$ have integral bound $\half\ib$
and 
$$\eqalign{
&\sfrac{d\hfill}{d\ka}C_{\ka,\Si}\big|_{\ka=0}\hbox{ has contraction bound\ \  }
 \cb'= \cst{}{2} M^{2j}\fe_j(X)Y\cr 
&\sfrac{d\hfill}{d\ka}D_{\ka,\Si}\big|_{\ka=0}\hbox{ has integral bound\ \  }
\half \ib'=\sqrt{\veps}\ \ib\cr
}$$
 for the family $\v \cdot\v_{\Si}$ of symmetric seminorms.  
}

\prf
The contraction and integral bounds on $C_{0,\Si}$ and $D_{0,\Si}$
were proven in Lemma \lemOSconcreteintconst.
Clearly, the function
$$
\sfrac{d\hfill}{d\ka}D_\ka(k)
= \sfrac{d\hfill}{d\ka}
        \sfrac{\nu^{(\ge j+1)}(k)}{\imath k_0 -e(\k) -\check v_\ka(k)}
= \sfrac{\nu^{(\ge j+1)}(k)}{[\imath k_0 -e(\k) -\check v_\ka(k)]^2}
 \sfrac{d\hfill}{d\ka}\check v_\ka(k)
$$  
is supported on the $j^{\rm th}$ neighbourhood and obeys 
$\ \big|\sfrac{d\hfill}{d\ka}D_\ka(k)\big|_{\ka=0}\big|
           \le \sfrac{4\veps}{|\imath k_0 - e(\k)|}\ $. By
part (ii) of  Proposition \propOScontrintboundsectors, $2\sqrt{4\IB_1\veps\sfrac{\fl}{M^j}}\le\sqrt{\veps}\ \ib $ 
is an integral bound for $\sfrac{d\hfill}{d\ka}D_{\ka,\Si}\big|_{\ka=0}\ $. 

 Set 
$$
\cb'' = 9\max \Big\{ \V \sfrac{d\hfill}{d\ka}c_\ka\big|_{\ka=0}\V_{1,\Si},\ 
\sfrac{M^{2j}}{\fl}\,\sup_{\xi,\xi',s,s'} 
       \big| \sfrac{d\hfill}{d\ka}c_\ka({\sst(\xi,s),(\xi',s')})\big|_{\ka=0} \big|
\big)  \big\}
$$
By part (i) of Proposition \propOScontrintboundsectors\ and
the second property of $\vec\rho$, 
$\big(\sfrac{d\hfill}{d\ka}c_\ka\big|_{\ka=0}\big)_\Si$ has contraction 
bound $\cb''$. By Lemma \lemOSdiffpropbound
$$\eqalign{
\cb'' 
&\le \const \max \Big\{ M^j\cb_j
\sfrac{M^j\v {d\hfill\over d\ka}{u_\ka|}_{\ka=0} \v_{1,\Si}}
{1-M^j\sv u_0\sv_{1,\Si}},
\ M^{2j}\V \sfrac{d\hfill}{d\ka}u_\ka\big|_{\ka=0} \V_{1,\Si} \Big\}\cr
&\le \const\,M^{j}\cb_j\,\sfrac{M^j\,\fe_j(X)}
{1-M^j\mu(\La+X)\fe_j(X)}Y\cr
&\le \const\,M^{2j}\cb_j\,\frac{\sfrac{1}{1-M^jX}}
{1-M^j\mu(\La+X)\sfrac{\cb_j}{1-M^jX}}\ Y\cr
&\le \const\,M^{2j}\cb_jY\,\frac{\sfrac{1}{1-M^j\cb_j(\La+X)}}
{1-\mu\sfrac{M^j\cb_j(\La+X)}{1-M^j\cb_j(\La+X)}}\cr
&=\const\, M^{2j}\cb_j\,Y\,f(Z)
}\EQN\eqnOSsectextimpr$$
where $Z=M^j\cb_j(\La+X)$ and 
$$
f(z)= \frac{{1\over 1-z}}{1-\mu{z\over 1-z}}
= \frac{1}{1-(1+\mu)z}
$$
By Lemma \lemOSappMonoidV, $f(Z)\le\abcst\sfrac{1}{1-Z}$ so that
$$
\cb'' \le \const\,M^{2j}\,\fe_j(X)Y
$$
as in Lemma \lemOSconcreteintconst.
\endproof

\lemma{\STM\lemOSconcretesourcetermderivconst}{
Let $g(\phi,\psi)$ be a sectorized Grassmann function and set
$$
g'_\ka(\phi,\psi)=g(\phi,\psi+C_\ka J\phi)
$$
Under the hypotheses of Theorem \thOSrengroupdiffestimate, 
$$
N_j(\sfrac{d\hfill}{d\ka}g'_\ka\big|_{\ka=0};\al;\,X,\Si,\vec\rho)
\le \sfrac{\const}{\al}\,M^jY_\0\,N_j(g;2\al;\,X,\Si,\vec\rho)
$$

  }
\prf Define
$$
\tilde c_z=c_0+z\sfrac{d\hfill}{d\ka}c_\ka\big|_{\ka=0}
$$
and
$$
\tilde g_z(\phi,\psi)
=g\big(\phi,\psi+\big[C_0+z\sfrac{d\hfill}{d\ka}C_\ka\big]_{\ka=0}
J\phi\big)
$$
Let $\varphi\in\cF_m(n;\Si)$, $1\le i\le n$ and set
$$\eqalign{
&\varphi'_z({\sst\et_1,\cdots,\et_{m+1}\,;
\,(\xi_1,s_1),\cdots,(\xi_{n-1},s_{n-1})})\cr
&\hskip1cm= {\rm Ant_{ext}}\smsum_{s,t,t'\in\Si}\int\!\! d{\sst \ze}d{\sst \ze'}\ 
\varphi({\sst\et_1,\cdots,\et_m\,;\,(\xi_1,s_1),\cdots,
(\xi_{i-1},s_{i-1}),(\ze',t),(\xi_{i},s_i),\cdots,(\xi_{n-1},s_{n-1})})\cr
\noalign{\vskip-.1in}
&\hskip10cm \cdot\ 
\tilde c_z({\sst(\ze',t',),(\ze,s)})\,J({\sst \ze,\et_{m+1}})\cr
}$$
By Lemma \lemOSdiffpropbound.i and (\eqnOSsectextimpr)
$$
\V \sfrac{d\hfill}{d\ka}c_\ka\V_{1,\Si}\Big|_{\ka=0\atop t=0}
\le \const M^{2j}Y_\0
$$
so that, using the  bound on $\v c_0\v_{1,\Si}$ that was derived in Lemma
\lemOSconcreteintconst,
$$
\v \tilde c_z\v_{1,\Si}\big|_{t=0}\le \const M^{j}
$$
for all $|z|\le \sfrac{1}{M^jY_\0}$. Consequently,
for all $|z|\le \sfrac{1}{M^jY_\0}$, 
Lemma \lemOSsectextimpr\ yields
$$
\v\varphi'_z\v_{1,\Si}
\le \const \v \varphi\v_{1,\Si}\cases{\sfrac{M^j}{\fl}& if $m=0$ \cr
                                \noalign{\vskip.05in}
                               \sfrac{\fl}{M^j}& if $m\ne 0$ \cr}
$$
so that, for $m=0$,
$$\eqalign{
\v\varphi'_z\v_\Si
=\rho_{1;n-1}\sfrac{\fl}{M^{2j}}\v\varphi'\v_{1,\Si}
&\le\const\rho_{1;n-1}\sfrac{1}{M^{j}}\v\varphi\v_{1,\Si}
\le\const\sfrac{\rho_{1;n-1}}{\rho_{0;n}}\sfrac{1}{M^{j}}\v\varphi\v_{\Si}\cr
&\le\const\sqrt{\fl M^j}\sfrac{1}{M^{j}}\v\varphi\v_{\Si}
\le\const\ib \v\varphi\v_{\Si}
}$$
and, for $m\ne 0$,
$$\eqalign{
\v\varphi'_z\v_\Si
=\rho_{m+1;n-1}\sfrac{\fl}{M^{2j}}\v\varphi'\v_{1,\Si}
&\le\const\rho_{m+1;n-1}\sfrac{\fl^2}{M^{3j}}\v\varphi\v_{1,\Si}
\le\const\sfrac{\rho_{m+1;n-1}}{\rho_{m;n}}\sfrac{\fl}{M^{j}}\v\varphi\v_{\Si}
\cr
&\le\const\sfrac{\fl}{M^{j}}\v\varphi\v_{\Si}
\le\const\ib \v\varphi\v_{\Si}
}$$
Hence, as in Lemma \lemOSconcretesourcetermintconst,
$$
N_j(\tilde g_z-g;\al;\,X,\Si,\vec\rho)
      \le \sfrac{\const}{\al}N_j(g,2\al;\,X,\Si,\vec\rho)
$$
for all $|z|\le \sfrac{1}{M^jY_\0}$ and, by the Cauchy integral theorem,
$$
N_j(\sfrac{d\hfill}{d\ka}g'_\ka\big|_{\ka=0};\al;\,X,\Si,\vec\rho)
=N_j(\sfrac{d\hfill}{dz}[\tilde g_z-g]\big|_{z=0};\al;\,X,\Si,\vec\rho)
\le \sfrac{\const}{\al}\,M^jY_\0\,N_j(g,2\al;\,X,\Si,\vec\rho)
$$

\endproof

\proof{ of  Theorem \thOSrengroupdiffestimate}
As in the proof of Theorem \thOSrengroupestimate,
let, for a sectorized Grassmann function $v=\smsum_{m,n} v_{m,n}$ with
$v_{m,n} \in A_m\otimes \bigwedge^nV_\Si$,
$$
N(v;\al) = \sfrac{1}{\ib^2}\cb\,\smsum_{m,n} \al^n\, \ib^n\, \v v_{m,n} \v_\Si
=\sfrac{\cst{}{1}}{\IB}\, N_j(v;\al;\,X,\Si,\vec\rho)
$$
and
$$
\ \lw w''_\ka\rw_{\psi,D_{\ka,\Si}} =
\Om_{C_{\ka,\Si}} \big(\lw w_\ka\rw_{\psi,C_{\ka,\Si}+D_{\ka,\Si}}\big)
$$
By Proposition \propOSfunctorialitySect, parts (ii) and (iii), 
\ and Proposition \propBII.ii of [FKTr1],
$$
w'_\ka=\half\phi JC_\ka J\phi+w''_\ka\big(\phi,\psi+C_\ka J\phi\big)
$$
is a sectorized representative for $\cW'_\ka$.
By the chain rule and the triangle inequality
$$\eqalign{
N\big(\,\sfrac{d\hfill}{d\ka}
     \big[w'_\ka-\half\phi JC_\ka J\phi-w_\ka\big]_{\ka=0}\,\cl \al\big)
\ \le\  &N\big(\,\sfrac{d\hfill}{d\ka}
     w''_0\big(\phi,\psi+C_\ka J\phi\big)\big|_{\ka=0}\,\cl \al\big)\cr
&+N\big(\,\sfrac{d\hfill}{d\ka}
  [w''_\ka\big(\phi,\psi+C_0 J\phi\big)-w''_\ka(\phi,\psi)]_{\ka=0}
                  \,\cl \al\big)\cr
&+N\big(\,\sfrac{d\hfill}{d\ka}
     [w''_\ka(\phi,\psi)-w_\ka(\phi,\psi)]_{\ka=0}\,\cl \al\big)\cr
}\EQN\eqnOSimprderivI$$
By Lemma \lemOSconcretesourcetermderivconst,
$$\eqalign{
N\big(\,\sfrac{d\hfill}{d\ka}
     w''_0\big(\phi,\psi+C_\ka J\phi\big)\big|_{\ka=0}\,\cl \al\big)
&\le \sfrac{\const}{\al}M^jY_\0 N_j(w''_0;2\al;\,X,\Si,\vec\rho)\cr
}$$
By (\eqnOSrengroupestimate),
$$\eqalign{
&N_j(w''_0;2\al;\,X,\Si,\vec\rho)
\le N_j\big(w_0;2\al;\,X,\Si,\vec\rho\big)
+N_j\big(w''_0-w_0;2\al;\,X,\Si,\vec\rho\big)\cr
&\hskip.5in
\le  N_j\big(w_0;2\al;\,X,\Si,\vec\rho\big)
+\sfrac{\IB}{\cst{}{1}} \sfrac{1}{8\al^2}\,
\sfrac{N(w_0;64\al)^2}{1-{1 \over4\al^2}N(w_0;64\al)}\cr
&\hskip.5in
\le N_j\big(w_0;64\al;\,X,\Si,\vec\rho\big)
+ \sfrac{\const}{\al^2}\,
\sfrac{N_j(w_0;64\al;\,X,\Si,\vec\rho)^2}
{1-{\const \over\al^2}N_j(w_0;64\al;\,X,\Si,\vec\rho)}\cr
&\hskip.5in
\le  \const\,\sfrac{N_j(w_0;64\al;\,X,\Si,\vec\rho)}
{1-{\const \over\al^2}N_j(w_0;64\al;\,X,\Si,\vec\rho)}\cr
}$$
so that
$$\eqalign{
N\big(\,\sfrac{d\hfill}{d\ka}
     w''_0\big(\phi,\psi+C_\ka J\phi\big)\big|_{\ka=0}\,\cl \al\big)
&\le \sfrac{\const}{\al}\,\sfrac{\fn}{1-{\const \over\al^2}\fn}M^jY_\0\cr
}\EQN\eqnOSimprderivIa$$
By Lemma \lemOSconcretesourcetermintconst, with $g=
\sfrac{d\hfill}{d\ka}w''_\ka\big|_{\ka=0}$,
$$\eqalign{
&N\big(\,\sfrac{d\hfill}{d\ka}
  [w''_\ka\big(\phi,\psi+C_0 J\phi\big)-w''_\ka(\phi,\psi)]_{\ka=0}
                  \,\cl \al\big)
\le \sfrac{\const}{\al}
N_j\big(\,\sfrac{d\hfill}{d\ka}w''_\ka\big|_{\ka=0}\,;2\al;\,X,\Si,\vec\rho)\cr
&\hskip.5in\le \sfrac{\const}{\al} \ 
N_j\big(\,\sfrac{d\hfill}{d\ka}w_\ka\big|_{\ka=0}\,;2\al;\,X,\Si,\vec\rho\big)
+ \sfrac{\const}{\al} \ 
N_j\big(\,\sfrac{d\hfill}{d\ka}
     [w''_\ka-w_\ka]_{\ka=0}\,;2\al;\,X,\Si,\vec\rho\big)\cr
}\EQN\eqnOSimprderivIb$$
By Theorem \theoremIVb\ of [FKTr1], with $\mu=\sfrac{1}{M^j}$,
$$\eqalign{
N\big(\,\sfrac{d\hfill}{d\ka}[w''_\ka-w_\ka]_{\ka=0}\,\cl \al\big)
&\le\ \sfrac{1}{2\al^2}\,
     \sfrac{N( w_0\cl 32\al)}{1-{1\over\al^2}N( w_0\cl 32\al)}
N\big(\,\sfrac{d\hfill}{d\ka}w_\ka\big|_{\ka=0}\,\cl 8\al\big) \cr
&\hskip1cm+\sfrac{1}{2\al^2}\,
     \sfrac{N( w_0\cl 32\al)^2}{1-{1\over\al^2}N( w_0\cl 32\al)}
\Big\{\sfrac{1}{4M^j}\cst{}{2}M^{2j}\fe_j(X)Y+4\veps\Big\}\cr
&\le \sfrac{\const}{\al^2}\,
     \sfrac{\fn}{1-{\const\over\al^2}\fn}
\Big\{N\big(\,\sfrac{d\hfill}{d\ka}w_\ka\big|_{\ka=0}\,\cl 8\al\big)
+M^{j}Y\fn+4\veps\fn\Big\}
}\EQN\eqnOSimprderivIc$$
since $\fe_j(X)N( w_0\cl 32\al)\le\const N( w_0\cl 32\al)$. Also
$$
N_j\big(\,\sfrac{d\hfill}{d\ka}
     [w''_\ka-w_\ka]_{\ka=0}\,;2\al;\,X,\Si,\vec\rho\big)
\le \sfrac{\const}{\al^2}\,
     \sfrac{\fn}{1-{\const\over\al^2}\fn}
\Big\{N\big(\,\sfrac{d\hfill}{d\ka}w_\ka\big|_{\ka=0}\,\cl 16\al\big)
+M^{j}Y\fn+4\veps\fn\Big\}
$$
Subbing (\eqnOSimprderivIa--\eqnOSimprderivIc) into (\eqnOSimprderivI),
$$\eqalign{
N\big(\,\sfrac{d\hfill}{d\ka}
     \big[w'_\ka-\half\phi JC_\ka J\phi-w_\ka\big]_{\ka=0}\,\cl \al\big)
&\le\ \sfrac{\const}{\al}\,\sfrac{\fn}{1-{\const \over\al^2}\fn}M^jY_\0
+\sfrac{\const}{\al} \ 
N\big(\,\sfrac{d\hfill}{d\ka}w_\ka\big|_{\ka=0}\,;2\al\big)\cr
&\ +\sfrac{\const}{\al^2}\,
     \sfrac{\fn}{1-{\const\over\al^2}\fn}
\Big\{\!N\big(\,\sfrac{d\hfill}{d\ka}w_\ka\big|_{\ka=0}\,\cl 16\al\big)
+M^{j}Y\fn+4\veps\fn\Big\}\cr
&\le \const
\Big\{\sfrac{1}{\al}+\sfrac{1}{\al^2}\sfrac{\fn}{1-{\const\over\al^2}\fn}\Big\}
N_j\big(\,\sfrac{d\hfill}{d\ka}w_\ka\big|_{\ka=0}
\,;\,16\al\,;\,{\sst X,\Si,\vec\rho}\,\big)\cr
&\ +\const\sfrac{\fn}{1-{\const\over\al^2}\fn}  \Big\{
\big(\sfrac{1}{\al}+\sfrac{\fn}{\al^2}\big)M^{j}Y+\sfrac{\veps}{\al^2}\fn\Big\}
\cr
}$$
\endproof

We also must control the pure $\phi$ contributions in a situation similar to
that of Theorem \thOSrengroupestimate. 

\proposition{\STM\propOSresidualrengroupest}{
There are constants $\al_0$ and $\tau_0$
that are independent of $j,\ \Si,\ \vec\rho$ such that for all $\al\ge \al_0$ 
 the following estimates hold:

{\parindent=.25in
\item{}
Let $u({\sst(\xi,s)},{\sst(\xi',s')}),v({\sst(\xi,s)},{\sst(\xi',s')})
\in\cF_0(2;\Si)$ be antisymmetric, spin independent, particle number 
conserving functions whose Fourier transforms  obey
 $|\check u(k)|,|\check v(k)| \le \half |\imath k_0-e(k)|$.
Furthermore, let
$X\in\fN_{d+1}$ and assume that
$\ X, \v  u \v_{1,\Si} \le \sfrac{\tau_0}{M^j} +\sum_{\de\ne 0}\infty t^\de$. Let $\il$ be a real number in $(j+1,j+2]$ and set
$$
S(k) = \sfrac{\nu^{(\ge j+1)}(k)-\nu^{(\ge \il)}(k)}
          {\imath k_0 -e(\k) -\check u(k)}
\qquad,\qquad 
D(k) = \sfrac{\nu^{(\ge j+1)}(k)}{\imath k_0 -e(\k) -\check v(k)}
$$
and let $S(\xi,\xi'),\,D(\xi,\xi')$ be the Fourier transforms of $S(k)$, $D(k)$ as
in Definition \defOSftcov. Let $\cW(\phi,\psi)$ be a Grassmann function obeying $\cW(\phi,0)=0$ and set
$$
\cG(\phi)
= \tilde \Om_S \big( \lw \cW(\phi,\psi) \rw_{\psi,D} \big)(\phi,0)
$$
Assume that $\cW$ has a sectorized representative $w$
with 
$$
N_j(w;\al;\,X,\Si,\vec\rho)
\le 2 + \smsum_{\de \ne 0} \infty\,t^\de
$$
Write
$$
\cG(\phi)-\sfrac{1}{2}\phi JSJ\phi = \smsum_{m} 
\int {\sst d\eta_1\cdots d\eta_m}\ 
G_{m}({\sst \eta_1,\cdots, \eta_m})
\ \phi({\sst \eta_1})\cdots \phi({\sst \eta_m})
$$
with $G_{m}$ antisymmetric.
Then 
$$
\sum_{m>0}\rho_{m;0}\tn G_m\tn_\infty 
\le 10
$$
}
}
\prf We use the notation of the proof of Theorem \thOSrengroupestimate.
As in Lemma \lemOSconcreteintconst, $\cb = \cst{}{1}\,M^j+ \smsum_{\de \ne 0} \infty\,t^\de$ is a contraction bound for $S_\Si$ and 
$\ib =\sqrt{\sfrac{\IB\,\fl}{M^j}}$ is an integral bound for both $S_\Si$ and $D_\Si$  (in the sense of Definition \defcontractintbound\  of [FKTr1]).  
Write
$$\eqalign{
 \lw w(\phi,\psi) \rw_{\psi,D_\Si}&= \lw \tilde w(\phi,\psi) \rw_{\psi,S_\Si}\cr
w''(\phi,\psi)&=\Om_{S_\Si}\big(\lw w(\phi,\psi) \rw_{\psi,D_\Si}\big)\cr
}$$
By Proposition \propOSfunctorialitySect
$$
\cG(\phi)=\half\phi JSJ\phi+w''(\phi,S J\phi)
$$
By Corollary \corwicknorm.ii of [FKTr1]
$$
N_j(\tilde w;\sfrac{\al}{2};\,X,\Si,\vec\rho)\le 
2N_j(w;\al;\,X,\Si,\vec\rho)
$$
By Theorem \theorII\  of [FKTr1], with $\al$ replaced by $\sfrac{\al}{16}$,
$$\eqalign{
N_j\Big(w''(\phi,\psi);\sfrac{\al}{16};\,X,\Si,\vec\rho\Big)
&\le N_j(\tilde w;\sfrac{\al}{16};\,X,\Si,\vec\rho)
+\sfrac{2^9}{\al^2}\,
     \sfrac{N_j(\tilde w;{1\over 2} \al;\,X,\Si,\vec\rho)^2}
      {1-{2^{10}\over\al^2}N_j(\tilde w;{1\over 2} \al;\,X,\Si,\vec\rho)}\cr
&\le 5+\sum_{\de\ne 0}\infty t^\de
}$$
By Lemma \lemOSconcretesourcetermintconst
$$\eqalign{
N_j\Big(w''(\phi,\psi+S J\phi);\sfrac{\al}{32} ;\,X,\Si,\vec\rho\Big)
&\le 2N_j(w''(\phi,\psi);\sfrac{\al}{16};\,X,\Si,\vec\rho)\cr
&\le 10+\sum_{\de\ne 0}\infty t^\de
}$$
so that 
$$\eqalign{
\fe_j(X)\sum_{m>0}\rho_{m;0}\tn G_m\tn_\infty
&=N_j\Big(\cG(\phi)-\half\phi JSJ\phi;\sfrac{\al}{32} ;\,X,\Si,\vec\rho\Big)\cr
&\le N_j\Big(w''(\phi,\psi+S J\phi);\sfrac{\al}{32} ;\,X,\Si,\vec\rho\Big)\cr
&\le 10+\sum_{\de\ne 0}\infty t^\de
}$$
\endproof

\remark{\STM\remOSchoiceofrep}{
In Theorem \thOSrengroupestimate, 
the sectorized representative $w'$ of $\cW'$ may be obtained from 
the sectorized representative $w$ of $\cW$ by
$$
\lw w'\rw_{\psi,D_\Si} = \half\phi JCJ\phi
+\Om_{C_\Si}(\lw w \rw_{\psi,C_\Si+D_\Si})(\phi,\psi+ C J\phi)
$$
Again, $w'$ is initially defined as a formal Taylor series in $w$. By Remark 
\remjointanalyticitywick\ of [FKTr1] and the observation that, as in 
Proposition \propOSrealfirstpropbound.i, $C_\Si$ and $D_\Si$
are analytic function of $u$ and $v$, respectively, this formal Taylor series
 converges to a function that is jointly analytic in $w$, $u$ and $v$. 
By the functoriality Lemma \propOSfunctorialitySect, 
if $w_1$ and $w_2$ are two sectorized representatives of $\cW$, then the corresponding $w'_1$ and $w'_2$ represent the same unsectorized Grassmann function $\cW'$. In this way one sees that the formal Taylor series for $\cW'$ converges.

\noindent
The obvious analogs of these statements apply to Theorem 
\thOSrengroupdiffestimate\ and Proposition \propOSresidualrengroupest.

}

\vfill\eject

\chap{ Sectorized Momentum Space Norms}\PG\pgOSXVI

Again, let  $\Si$ be a  sectorization of length 
$\sfrac{1}{M^{j-3/2}}\le\fl\le\sfrac{1}{M^{(j-1)/2}}$ at scale $j\ge 2$.
In \S\CHestren\ we described the renormalization group map $\tilde \Om_C$ using the algebra $\bigwedge_A V_\Si$, where $V_\Si$ is the vector space generated by $\psi({\sst \xi,s})$, $\xi \in\cB,\,s\in\Si$ (see Definition \defOStens) and $A$ is the Grassmann algebra in the external fields 
$\phi(\eta)$, $\eta \in \cB$. To deal with amputated Green's functions 
in momentum space, we set for $\check \eta=(k,\si,a) \in \check \cB$
$$
\check \phi(\check \eta) 
=  \int d^{d+1}x\ e^{-(-1)^a\imath<k,x>_-}\phi(x,\si,a)
$$
and denote by $V_{\rm ext}$ the vector space generated by 
$\check \phi(\check \eta),\ \check \eta \in \check \cB$. Furthermore set
$$
\tilde V = V_{\rm ext} \oplus V_\Si
$$
Then $\bigwedge_A V_\Si$ is canonically isomorphic to the Grassmann algebra
$\bigwedge \tilde  V$ over $\tilde V$ with complex coefficients. In terms of Grassmann functions, this isomorphism amounts to the following: 
A translation invariant sectorized Grassmann function $w$ can be uniquely written in the form
$$\eqalign{
w(\phi,\psi) = \smsum_{m,n} \smsum_{s_1,\cdots,s_n\in \Si} 
\int {\sst d\eta_1\cdots d\eta_m\,d\xi_1\cdots d\xi_n}\ & 
w_{m,n}({\sst \eta_1,\cdots, \eta_m\,(\xi_1,s_1),\cdots ,(\xi_n,s_n)})\cr
& \hskip 1.5cm \cdot\ \phi({\sst \eta_1})\cdots \phi({\sst \eta_m})\
\psi({\sst (\xi_1,s_1)})\cdots \psi({\sst (\xi_n,s_n)\,})\cr
}$$
with $w_{m,n}$ antisymmetric separately in the $\eta$ and in the $\xi$ variables. As well,
$$\eqalign{
w(\phi,\psi) &=
 \smsum_m 
\int {\sst d\check \eta_1\cdots d\check \eta_m}\  
w_{m,0}^\sim({\sst\check \eta_1,\cdots,\check \eta_m})\
{\sst (2\pi)^{d+1}}
 \de({\sst\check \eta_1+\cdots+\check \eta_m})\ 
 \check\phi({\sst \check\eta_1})\cdots \check\phi({\sst \check \eta_m}) \cr
&\hskip.25in+ \smsum_{m,n \atop n \ge 1} \smsum_{s_1,\cdots,s_n\in \Si} 
\int {\sst d\check \eta_1\cdots d\check \eta_m\,d\xi_1\cdots d\xi_n}\  
w_{m,n}^\sim({\sst\check \eta_1,\cdots,\check \eta_m\,(\xi_1,s_1),\cdots ,(\xi_n,s_n)})\cr
& \hskip 6.8cm\cdot\ 
 \check\phi({\sst \check\eta_1})\cdots \check\phi({\sst \check \eta_m})\
\psi({\sst (\xi_1,s_1)})\cdots \psi({\sst (\xi_n,s_n)\,})\cr
}$$
Here $w_{m,n}^\sim$ is the partial Fourier transform of Definition \defOSfourtrans.

The basis elements of the vector space $\tilde V = V_{\rm ext} \oplus V_\Si$ are in one to one correspondence with the points of the disjoint union $\fX_\Si$ of  $\check \cB$ and $\cB \times \Si$. To simplify notation, we make the

\definition{\STM\defOSdisjointfield}{
For $x \in \fX_\Si= \check \cB\dunion (\cB \times \Si)$ set
$$
\Psi(x) = \cases{ 
      \check \phi(\check \eta) & if\ $x = \check \eta \in \check B $  \cr 
      \psi(\xi,s) & if $x = (\xi,s)  \in \cB \times \Si$ 
}$$
}

The purpose of this Section is to define and analyze norms on functions on 
$\fX_\Si^n$ to which the results of [FKTr1,2] can be applied. First, we look at the structure of $\fX_\Si^n$ more carefully.

\definition{\STM\defOSdisjointOrd}{
Set $\fX_0 = \check \cB$ and $\fX_1 = \cB \times \Si$. 
Let $\vec\imath = (i_1,\cdots,i_n) \in \{0,1\}^n$. 

\Item i) The inclusions of $\fX_{i_j},\ j=1,\cdots,n,\ $ in $\fX_\Si$ induce an inclusion of 
$\fX_{i_1} \times\cdots\times \fX_{i_n}$ in $\fX_\Si^n$. We identify 
$\fX_{i_1} \times\cdots\times \fX_{i_n}$ with its image in $\fX_\Si^n$.

\Item ii) Set $m(\vec\imath) = n-(i_1+\cdots +i_n)$. Clearly, $m(\vec\imath)$ is the number of copies of $ \check \cB$ in $\fX_{i_1} \times\cdots\times \fX_{i_n}$.

\Item iii)
If $f$ is a function on $ \fX_{i_1} \times\cdots\times \fX_{i_n}$, then 
$\ord f$ is the function on 
$\check \cB^{m(\vec\imath)}\times (\cB \times \Si)^{n-m(\vec\imath)}$ 
obtained from $f$ by shifting all of the $\check B$ arguments before all of the
$\cB \times \Si$ arguments, while preserving the relative order of the 
$\check B$ arguments and the 
relative order of the $\cB \times \Si$ arguments and multiplying by the sign of the permutation that implements the reordering of the arguments. That is, 
$\ord f(x_1,\cdots,x_n)= \sgn \pi f(x_{\pi(1)},\cdots,x_{\pi(n)})$
where the permutation $\pi\in S_n$ is determined by
$\ 
\pi(j)<\pi(j')\hbox{ if } i_{j}<i_{j'}\hbox{ or } i_{j}=i_{j'}\ j<j'.
\ $
}

\remark{\STM\remOSbigdisjointunion}{
Using the identification of Definition \defOSdisjointOrd.i, 
$$
\fX_\Si^n\ =\ 
\bigcup_{i_1,\cdots,i_n \in \{0,1\}}\kern-2.8em\cdot\kern2.8em \fX_{i_1} \times\cdots\times \fX_{i_n}
$$
where, on the right hand side we have a disjoint union. If $f$ is a function on
$\fX_\Si^n$ and $\vec\imath = (i_1,\cdots,i_n) \in \{0,1\}^n$, we denote by $f\big|_{\vec\imath}$ the restriction of $f$ to 
$\fX_{i_1} \times\cdots\times \fX_{i_n}$.
}

\vskip.25in
To define norms for functions on $\fX_\Si^n$ it thus suffices to define norms
for functions on each of the spaces $\fX_{i_1} \times\cdots\times \fX_{i_n}$.
As we want these norms to be invariant under permutations, it suffices, 
using the map $\ord$ of Definition \defOSdisjointOrd.iii, to define norms for functions on the spaces
$\check \cB^m \times (\cB\times \Si)^{n-m}$.

\definition{\STM\defOSsectdiffdecaynorm}{
Let $p$ be a natural number.

\Item{(i)} 
For a function $f$ on $\check\cB_m$ we define
$$
\v f\tv_{p,\Si} 
= \cases{ \| f\tnorm & if $ p=m-1$, $m=2,4$ \cr
\ \ 0  & otherwise \cr
}$$
with $\| \ \cdot\ \tnorm$ being the norm of Definition \defOSdiffdecaynorm.
\Item{(ii)}
For a translation invariant function $f$ on 
$\check\cB^m \times (\cB\times \Si)^n$ with $n\ge 1$, 
we set $\v f\tv_{p,\Si} \ =\ 0 \ $ when $p>m+n$ or $p<m$, and
$$
\v f\tv_{p,\Si} 
= \smsum_{\de\in \bbbn_0\times\bbbn_0^2} 
\sup_{{1\le i_1<\cdots<i_{p-m}\le n \atop s_{i_1},\cdots,s_{i_{p-m}}\in\Si}
\atop \check\eta_1,\cdots,\check\eta_m \in \check \cB} 
\smsum_{s_i \in \Si \ {\rm for}\atop i\ne i_1,\cdots i_{p-m}} \hskip-5pt
\sfrac{1}{\de!}
\max_{\rD\, {\rm dd-operator} \atop{\rm with\ } \de(\rD) =\de} \TN\rD f\,({\sst\check\eta_1,\cdots,\check\eta_m;(\xi_1,s_1),\cdots,(\xi_n,s_n)})
\TN_{1,\infty}\ t^\de
$$
when $m\le p \le m+n\,$. The norm
$\tn\,\cdot\,\tn_{1,\infty}$ of Example \exOSSymmNorm\ refers to the variables 
${\sst \xi_1,\cdots,\xi_n}$. 

}

\remark{\STM\remOSsecdiffdecaynorm}{
In the case $m=0$ and $p$ odd, the norm $\v \,\cdot\,\v_{p,\Si}$ of Definition
\defOSsectnorm\ and the norm $\v \,\cdot\,\tv_{p,\Si}$ of Definition 
\defOSsectdiffdecaynorm\ agree.
}

\lemma{\STM\lemOSelloneinftyampsectors}{
Let $f$ be a translation invariant function on $\check\cB^{m}\times
\big(\cB\times\Si\big)^{n}$,
$f'$ a translation invariant function on $\check\cB^{m'}\times
\big(\cB\times\Si\big)^{n'}$ and 
$1\le i \le n$,  $1\le i'\le n'$.

\noindent 
If $n\ge 2$ or $n'\ge 2$ define the function 
$g$ on $\check \cB^{m+m'}\times \big(\cB\times\Si\big)^{n+n'-2}$ by
$$\eqalign{
&g({\sst \check\eta_1,\cdots,\check\eta_{m+m'};
(\xi_1,s_1),\cdots,(\xi_{i-1},s_{i-1}),\,(\xi_{i+1},s_{i+1}),\cdots,
(\xi_{n+i'-1},s_{n+i'-1}),\,
(\xi_{n+i'+1},s_{n+i'+1}),\cdots,(\xi_{n+n'},s_{n+n'})}) \cr
&\hskip .1cm  = \smsum_{s,s'\in\Si\atop s\cap s'\ne\emptyset}
\int_\cB\!\! {\sst d\zeta}\,
f( {\sst \check\eta_1,\cdots,\check\eta_m;\, 
(\xi_1,s_1),\cdots,(\xi_{i-1},s_{i-1}),\,(\zeta,s),\,(\xi_{i+1},s_{i+1}),
\cdots,(\xi_n,s_n) })\cr
& \hskip 1.5cm \cdot\, f'( {\sst \check\eta_{m+1},\cdots,\check\eta_{m+m'};
\,(\xi_{n+1},s_{n+1})\cdots,(\xi_{n+i'-1},s_{n+i'-1}),\,(\zeta,s'),\,
(\xi_{n+i'+1},s_{n+i'+1}),\cdots,(\xi_{n+n'},s_{n+n'})})
\cr
}$$
If $n=n'=1$, define the function $g$ on $\check \cB_{m+m'}$ by
$$
g({\sst \check\eta_1,\cdots,\check\eta_{m+m'}})\ (2\pi)^{d+1}
 \de({\sst \check\eta_1+\cdots+\check\eta_{m+m'}}) 
= \smsum_{s,s'\in\Si\atop s\cap s'\ne\emptyset}\ 
\int_\cB {\sst d\zeta}\,
f( {\sst \check\eta_1,\cdots,\check\eta_m;\,(\zeta,s)})\
 f'( {\sst\check\eta_{m+1},\cdots,\check\eta_{m+m'};\,(\zeta,s')})
$$

\noindent
Then, for all natural numbers $p$,
$$
\v g\,\tv_{p,\Si} 
\le \cases{
3\ \max\limits_{p_1+p_2 =p\atop {m\le p_1< m+n\atop m'\le p_2< m'+n'}}
\min\{\v f\tv_{p_1,\Si} \ \v f'\tv_{p_2+1,\Si},
\v f\tv_{p_1+1,\Si} \ \v f'\tv_{p_2,\Si}\}& if $(n,n')\ne (1,1)$\cr
\noalign{\vskip.1in}
4\ \v f\tv_{m,\Si} \ \v f'\tv_{m',\Si}& if $(n,n')= (1,1)$\cr
}
$$
}

\prf
\Item {\it The case $n \ge 2$ or $n' \ge 2$.}
The proof is analogous to that of Lemma \lemOSelloneinftysectors. 
The $\cB\times\Si$ indices for $g$ lie in the set $I\cup I'$, where
$$\eqalign{
I & = \{1,\cdots,i-1,\,i+1,\cdots,n\}\cr
I' & = \{n+1,\cdots,n+i'-1,\,n+i'+1,\cdots,n+n'\}\cr 
}$$ 
Let $q$ obey $0\le q-m\le n-1$   and $0\le p-q-m'\le n'-1$ or equivalently
$m\le q< n+m$ and $m'\le p-q< m'+n'$. Fix 
$u_1,\cdots,u_{q-m} \in I,\ u_{q-m+1},\cdots u_{p-m-m'} \in I'$ and fix sectors 
$s_{u_1},\cdots,s_{u_{p-m-m'}} \in \Si$. 
Let
$$\eqalign{
F( {\sst \check\eta_1,\cdots,\check\eta_m;\, s_1,\cdots,s_n })
= \smsum\limits_{\de \in \bbbn_0\times\bbbn_0^2} \sfrac{1}{\de !}
\max\limits_{\rD\ {\rm dd-operator} 
\atop{\rm with\ } \de(\rD)=\de}
\TN \rD f( {\sst \check\eta_1,\cdots,\check\eta_m;\, 
(\,\cdot\,,s_1),\cdots,(\,\cdot\,,s_n) }) \TN_{1,\infty} t^\de
}$$
so that
$$
\v f\tv_{p,\Si} 
= \hskip-7pt\sup_{{1\le i_1<\cdots<i_{p-m}\le n \atop s_{i_1},\cdots,s_{i_{p-m}}\in\Si}
\atop \check\eta_1,\cdots,\check\eta_m \in \check \cB} 
\smsum_{s_i \in \Si \ {\rm for}\atop i\ne i_1,\cdots i_{p-m}} \
F( {\sst \check\eta_1,\cdots,\check\eta_m;\, s_1,\cdots,s_n })
$$
and define $G({\sst \check\eta_1,\cdots,\check\eta_{m+m'};
s_1,\cdots,\not s_i,\cdots,\not s_{n+i'},\cdots,s_{n+n'}})$ and 
$F'( {\sst \check\eta_{m+1},\cdots,\check\eta_{m+m'};\, s_{n+1},\cdots,s_{n+n'} })$ similarly.
By Remark \remOSelloneinftyamp, for each choice of sectors 
$s_\nu,\ \nu \in I\cup I'$, one has
$$\eqalign{
&G({\sst \check\eta_1,\cdots,\check\eta_{m+m'};
s_1,\cdots,\not s_i,\cdots,\not s_{n+i'},\cdots,s_{n+n'}})  \cr
& \hskip 1cm \le\smsum_{s,s'\in\Si\atop s\cap s'\ne\emptyset}
F( {\sst \check\eta_1,\cdots,\check\eta_m;\, 
s_1,\,\cdots\,,s_{i-1},s,s_{i+1},\,\cdots\,,s_n }) 
F'( {\sst \check\eta_{m+1},\cdots,\check\eta_{m+m'};
\,s_{n+1},\,\cdots\,,s_{n+i'-1},s',\,\cdots\,,s_{n+n'}})  \cr
}$$
Observe that for every $s\in\Si$ there are at most three sectors $ s'$ 
such that $s'\cap s \ne \emptyset$. Consequently
$$\eqalign{
\sum_{s_\nu \in \Si \atop \nu \in I\cup I' \setminus \{u_1,\cdots,u_{p-m-m'}\}}
 \hskip -1cm &
G({\sst \check\eta_1,\cdots,\check\eta_{m+m'};
s_1,\cdots,\not s_i,\cdots,\not s_{n+i'},\cdots,s_{n+n'}}) \cr
& \hskip .5cm \le\ 3 \hskip -.6cm
\sum_{s_\nu  \in  \Si
 \atop   \nu\in I\setminus \{u_1,\cdots,u_{q-m}\}}  \hskip -.1cm
\smsum_{s\in\Si}
F( {\sst \check\eta_1,\cdots,\check\eta_m;\, 
s_1,\,\cdots\,,s_{i-1},s,s_{i+1},\,\cdots\,,s_n })  \cr
& \hskip 1.3cm \max_{s'\in \Si}  \hskip -0.5cm
\sum_{ s'_\mu \in  \Si
 \atop \mu\in I'\setminus \{u_{q-m+1},\cdots,u_{p-m-m'}\} }  \hskip -1cm
F'( {\sst \check\eta_{m+1},\cdots,\check\eta_{m+m'};
\,s_{n+1},\,\cdots\,,s_{n+i'-1},s',\,\cdots\,,s_{n+n'}})  \cr
& \hskip .5cm \le 3\ \v f\tv_{q,\Si}\ \  \v f'\tv_{p-q+1,\Si}
}$$
Taking the supremum over the $\check\et$'s and the remaining $s_\nu$'s gives
$$
\v g\tv_{p,\Si}\ \le \ 3\ \v f\tv_{q,\Si}\ \  \v f'\tv_{p-q+1,\Si}
$$
By interchanging the roles of $(f,q)$ and $(f',p-q)$, we get the bound 
$3\ \v f\tv_{q+1,\Si}\ \  \v f'\tv_{p-q,\Si}$.

\Item {\it The case $n=n'=1$.} In this
case, the norm  $\v g\tv_{p,\Si}$ is defined in Definition 
\defOSsectdiffdecaynorm.i. We need only consider the case $p=m+m'-1$. 

By Remark \remOSdiffdecay.iii
$$\eqalign{
&\smsum_{s,s'\in\Si\atop s\cap s'\ne\emptyset}\ 
\int_\cB {\sst d\xi}\,
f( {\sst \check\eta_1,\cdots,\check\eta_m;\,(\xi,s)})\,
 f'( {\sst\check\eta_{m+1},\cdots,\check\eta_{m+m'};\,(\xi,s')}) \cr
&\ \ =\int \hskip -.1cm{\sst dx_0} \hskip -.1cm
   \int\hskip -.1cm {\sst d\x} \hskip-.2cm 
\smsum_{\si \in \{\uparrow,\downarrow\} \atop b \in \{0,1\}}\,  
\smsum_{s,s'\in\Si} 
 f ({\sst \check\eta_1,\cdots,\check\eta_m;\,(0,\si,b,s)})\ 
 e^{\imath \<\check\eta_1+\cdots+\check\eta_{m+m'},(x_0,\x)\>_-}
 f' ({\sst \check\eta_{m+1},\cdots,\check\eta_{m+m'};\,(0,\si,b,s')}) \cr 
&\ \ = \smsum_{\si \in \{\uparrow,\downarrow\} \atop b \in \{0,1\}} \,
\smsum_{s,s'\in\Si} 
 f ({\sst \check\eta_1,\cdots,\check\eta_m;(0,\si,b,s)})\
f'({\sst \check\eta_{m+1},\cdots,\check\eta_{m+m'};(0,\si,b,s')}) \  
(2\pi)^{d+1}\de( {\sst\check\eta_1+\cdots+\check\eta_{m+m'}})
}$$
Consequently
$$
g({\sst \check\eta_1,\cdots,\check\eta_{m+m'}}) = 
\smsum_{\si \in \{\uparrow,\downarrow\} \atop b \in \{0,1\}}\,
\smsum_{s,s'\in\Si}
f ({\sst \check\eta_1,\cdots,\check\eta_m;(0,\si,b,s)})\
f' ({\sst \check\eta_{m+1},\cdots,\check\eta_{m+m'};\,(0,\si,b,s')})
$$
The claim now follows, as in Lemma \lemOSelloneinfty,
 by iterated application of the product rule for derivatives
and Remark \remOSdiffdecay.iii.
\endproof

\definition{\STM\defOSsectcheckcF }{
Let $m,n \ge 0$.

\Item{i)}  
For $n\ge 1$, denote by  $\check\cF_m(n;\Si)$ the space
of all translation invariant, complex valued functions 
$\ 
f({\sst\check\eta_1,\cdots,\check\eta_m;\,(\xi_1,s_1),\cdots,(\xi_n,s_n)} )
\ $
on $\check\cB^m \times  \big( \cB \times\Si \big)^n$ whose Fourier transform
$\check f({\sst\check\eta_1,\cdots,\check\eta_m;
\,(\check\xi_1,s_1),\cdots,(\check\xi_n,s_n)} )$
vanishes unless 
$ k_i\in \tilde s_i$ for all $1\le j\le n$. 
Here, $\check\xi_i=(k_i,\si_i,a_i)$.
Also, let $\check\cF_m(0;\Si)$  be the space of all momentum conserving,
complex valued functions 
$\ 
f({\sst\check\eta_1,\cdots,\check\eta_m} )
\ $
on $\check\cB^m $.

\Item{ii)}  
Let $c\big((\xi,s),(\xi',s')\big)$ be any skew symmetric function on $\big(\cB\times\Si\big)^2$.
Let $f\in \check\cF_m(n;\Si)$ and $1\le i < j\le n$. We define ``contraction'',
for $n\ge 2$, by
$$\eqalign{
&\Cont{i}{j}{c} f\, ({\sst \check\eta_1,\cdots,\check\eta_m;
(\xi_1,s_1),\cdots,(\xi_{i-1},s_{i-1})\,,\,
(\xi_{i+1},s_{i+1}),\cdots,(\xi_{j-1},s_{j-1})\,,\,
(\xi_{j+1},s_{j+1}),\cdots,(\xi_n,s_n)})
\cr
& \hskip1cm= (-1)^{j-i+1} 
\sum_{s_i,s_j,t_i,t_j\in\Si\atop 
            {t_i\cap s_i\ne\emptyset\atop t_j\cap s_j\ne\emptyset}}
\int {\sst d\xi_i\,d\xi_j}\  c({\sst (\xi_i,t_i),(\xi_j,t_j)})\,
f({\sst \check\eta_1,\cdots,\check\eta_m;
(\xi_1,s_1),\cdots,(\xi_n,s_n)}) 
}$$  
and, for $n=2$, by
$$\eqalign{
&\Cont{1}{2}{c} f\, ({\sst \check\eta_1,\cdots,\check\eta_m})\ 
(2\pi)^{d+1} \de({\sst \check\eta_1+\cdots+\check\eta_m})
\cr
& \hskip2cm=  
\sum_{s_1,s_2,t_1,t_2\in\Si\atop 
            {t_1\cap s_1\ne\emptyset\atop t_2\cap s_2\ne\emptyset}}
\int {\sst d\xi_1\,d\xi_2}\  c({\sst (\xi_1,t_1),(\xi_2,t_2)})\,
f({\sst \check\eta_1,\cdots,\check\eta_m;(\xi_1,s_1),(\xi_2,s_2)}) 
}$$ 

\Item iii) 
We denote by $\check \cF_{n;\Si}$ the set of functions on $\fX_\Si^n$
with the property that for each $\vec\imath = (i_1,\cdots,i_n) \in \{0,1\}^n$
with $m(\vec\imath) <n$
$$
\ord \big(f\big|_{\vec\imath} \big) \in \check \cF_{m(\vec\imath)}(n-m(\vec\imath);\Si)
$$
and such there is a function $g$ on $\check \cB_n$ such that
$$
f\big|_{(0,\cdots, 0)}(\check\eta_1,\cdots,\check\eta_n) = 
(2\pi)^3\de(\check\eta_1+\cdots+\check\eta_n)
\,g(\check\eta_1,\cdots,\check\eta_n)
$$
The map $\ord$ was introduced in Definition \defOSdisjointOrd.iii
and the restriction $f\big|_{\vec\imath}$ was introduced in Remark 
\remOSbigdisjointunion.

}

The partial Fourier transforms $\varphi^\sim$ (as in Definition \defOSfourtrans.ii) of functions $\varphi\in\cF_m(n;\Si)$ as in Definition \defOSsectrepr.ii are the functions in $\check\cF_m(n;\Si)$ that are antisymmetric in their external variables. Also, 
$\ \Cont{i}{j}{c} \varphi^\sim = (\Cont{i}{j}{c} \varphi)^\sim\ $, where
$\Cont{i}{j}{c} \varphi$ is defined in Definition \defOSsectcontnorm.

\proposition{\STM\propOSmomcontrintboundsectors}{
Let $c\big((\xi,s),(\xi',s')\big) \in \cF_0(2;\Si)$ be an antisymmetric function. 
\Item{i)}
Let $p$ be a natural number, $m,m'\ge 0$, $n,n'\ge 1$
and $f\in\check\cF_m(n;\Si)$, $f'\in\check\cF_{m'}(n',\Si)$. If $(n,n')\ne
(1,1)$ then
$$
\V \Cont{1}{n+1}{c}\, {\rm Ant}_{\rm ext}(f\otimes f')\tV_{p,\Si}\
\le\ 9\,\v c\v_{1,\Si}\,
\max\limits_{p_1+p_2 =p\atop {m\le p_1< m+n\atop m'\le p_2< m'+n'}}\!\!\!  
\min\{\v f \tv_{p_1+1,\Si}\ \v f'\tv_{p_2,\Si},
\v f \tv_{p_1,\Si}\ \v f'\tv_{p_2+1,\Si} \}
$$ 
and if $(n,n')= (1,1)$ then
$$
\V \Cont{1}{n+1}{c}\, {\rm Ant}_{\rm ext}(f\otimes f')\tV_{m+m'-1,\Si}\
\le\ 12\,\v c\v_{1,\Si}\,\v f \tv_{m,\Si}\ \v f'\tv_{m',\Si}
$$

\Item{ii)} 
Assume that there is a function $C(k)$ 
that is supported in the  $j^{\rm th}$ neighbourhood, such that $c\big((\cdot,s),(\cdot,s') \big)$ is the Fourier transform of
$\chi_s(k)\,C(k)\,\chi_{s'}(k)$ in the sense of Definition \defOSftcov\ 
and that $|C(k)| \le \sfrac{\veps}{|\imath k_0 -e(\k)|}$ for
some $\veps \ge 0$.

\noindent
Let  $f\in \check\cF_m(n;\Si)$, $n'\le n$ and set, when $n'<n$, as in 
Definition \defOSintbnd,
$$\eqalign{
f'&
({\sst \check\eta_1,\cdots,\check\eta_m;\,(\xi_{n'+1},s_{n'+1}),\cdots,(\xi_n,s_n)})\cr
& = 
\smsum_{s_i\in\Si \atop i=1,\cdots,n'}
\int \hskip -3pt \int {\sst d\xi_1\cdots d\xi_{n'}}\ 
f({\sst\check\eta_1,\cdots,\check\eta_m;\,(\xi_1,s_1),\cdots,(\xi_{n'},s_{n'}),
\cdots,(\xi_n,s_n)})\ 
\psi({\sst\xi_1,s_1})\cdots\psi({\sst \xi_{n'},s_{n'}})\,d\mu_{C_\Si}(\psi)
}$$
where
$$
C_\Si\big(\psi({\sst\xi,s}),\psi({\sst\xi',s'}) \big) = 
\sum_{t\cap s \ne \emptyset \atop t'\cap s' \ne \emptyset}
c\big((\xi,t),(\xi',t')\big)
$$
For $n'=n$, set
$$\eqalign{
&f'({\sst \check\eta_1,\cdots,\check\eta_m})\ 
(2\pi)^{d+1} \de({\sst \check\eta_1+\cdots+\check\eta_m})\cr
 &\hskip2cm= 
\smsum_{s_i\in\Si \atop i=1,\cdots,n}
\int \hskip -3pt \int {\sst d\xi_1\cdots d\xi_{n}}\ 
f({\sst\check\eta_1,\cdots,\check\eta_m;\,(\xi_1,s_1),\cdots,(\xi_{n},s_{n})})\ 
\psi({\sst\xi_1,s_1})\cdots\psi({\sst \xi_{n},s_{n}})\,d\mu_{C_\Si}(\psi)
}$$
Then, for all natural numbers $p$,
$$
\v f'\tv_{p,\Si}\ 
\le\ \big(\veps\,\IB_3\, \sfrac{\fl}{M^j}\big)^{n'/2}\,
\cases{
\v f\tv_{p,\Si}& if $n\ne n'$\cr
\noalign{\vskip.1in}
\v f\tv_{p+1,\Si}& if $n= n'$\cr
}
$$
with a constant $\IB_3$ that is independent of $j$ and $\Si$.

\Item{iii)}
Let $D(k),\,D'(k)$ be functions obeying 
$|D(k)|,\,|D'(k)| \le \sfrac{2}{|\imath k_0 -e(\k)|}$ and let 
$d\big((\cdot,s),(\cdot,s') \big)$ resp. 
$d'\big((\cdot,s),(\cdot,s') \big)$ be the Fourier transform of
$\chi_s(k)\,D(k)\,\chi_{s'}(k)$ resp. $\chi_s(k)\,D'(k)\,\chi_{s'}(k)$
in the sense of Definition \defOSftcov. 

\noindent
Let $1\le i_1,i_2,i_3 \le n$, $1\le i'_1,i'_2,i'_3 \le n'$ with
$i_1\ne i_2 \ne i_3 \ne i_1$, $i'_1\ne i'_2 \ne i'_3 \ne i'_1$, 
and let $p\ge 1$. 
Then there is a constant $\IB_4$ that is independent of $j$ and $\Si$
such that for all $f\in\check\cF_m(n;\Si)$, $f'\in \check\cF_{m'}(n',\Si)$
$$\eqalign{
&\V
\Cont{i_1}{n+i_1'}{c}\,\Cont{i_2}{n+i_2'}{d}\,\Cont{i_3}{n+i_3'}{d'}\, 
(f\otimes f')\tV_{p,\Si}\cr
&\hskip1cm\le\ 
\big(\IB_4\,\sfrac{\fl}{M^{j}} \big)^2\v c \v_{1,\Si}\hskip-10pt
\max\limits_{p_1+p_2 =p\atop {m\le p_1< m+n\atop m'\le p_2< m'+n'}}\hskip-10pt  
\min\big\{\v f \tv_{p_1+3,\Si}\ \v f'\tv_{p_2,\Si},
\v f \tv_{p_1,\Si}\ \v f'\tv_{p_2+3,\Si} \big\}
\cr
}$$
if $(n,n')\ne(3,3)$ and 
$$\eqalign{
&\V
\Cont{i_1}{n+i_1'}{c}\,\Cont{i_2}{n+i_2'}{d}\,\Cont{i_3}{n+i_3'}{d'}\, 
(f\otimes f')\tV_{m+m'-1,\Si}\cr
&\hskip1cm\le\ 
\big(\IB_4\,\sfrac{\fl}{M^{j}} \big)^2\v c \v_{1,\Si}  
\min\big\{\v f \tv_{m+2,\Si}\ \v f'\tv_{m',\Si},
\v f \tv_{m,\Si}\ \v f'\tv_{m'+2,\Si} \big\}
\cr
}$$
if $(n,n')=(3,3)$.
}

\prf The proofs of part (i) and part (ii), except for the case $n=n'$,
 is similar to that of parts (i) and (ii) of Proposition \propOScontrintboundsectors. The proof of part (iii) is similar to that
of Proposition \propOSoverlapploops. So we only give the proof of part (ii)
for the case $n=n'$.

 By translation invariance,
$$\eqalign{
f'({\sst \check\eta_1,\cdots,\check\eta_m})\ 
= \hskip -5pt
\smsum_{s_i\in\Si \atop i=1,\cdots,n}\hskip-7pt
\int \hskip -8pt \int \hskip -4pt{\sst d\xi_1\cdots d\xi_{n}}
\de{\sst (x_{n,0})}\de{\sst (\x_n)}
f({\sst\check\eta_1,\cdots,\check\eta_m;\,(\xi_1,s_1),\cdots,(\xi_{n},s_{n})}) 
\psi{\sst(\xi_1,s_1)}\cdots\psi{\sst (\xi_{n},s_{n})}\,d\mu_{C_\Si}{\sst(\psi)}
}$$
where $\xi_{n}=(x_{n,0},\x_n,\si_n,a_n)$. 
By part (ii) of Proposition \propIntBndsII, with
$$
G\ =
\ \int \sfrac{d^{d+1}k}{(2\pi)^{d+1}}\,\tilde\chi_s(k)^2\,|C(k)|
\ \le\ \veps \int \sfrac{d^{d+1}k}{(2\pi)^{d+1}}\,\sfrac{\tilde\chi_s(k)^2}{|\imath k_0 -e(\k)|}
\ \le \ \const\ \veps \,\sfrac{\fl}{M^j}
$$
we have, for any dd--operator $\rD$,
$$\eqalign{
\big|\rD f'({\sst \check\eta_1,\cdots,\check\eta_m}) \big|
&\le 4G^{n/2}
\smsum_{s_i\in\Si \atop i=1,\cdots,n}
\tn\rD f({\sst\check\eta_1,\cdots,\check\eta_m;\,(\,\cdot\,,s_1),\cdots,(\,\cdot\,,s_{n})})\tn_{1,\infty}\
\cr
}$$
Hence
$$\eqalign{
\v f\tv_{m-1,\Si}&\le  4\, G^{n/2}\v f\tv_{m,\Si}
}$$
\endproof

In Theorem \thOSrengroupestimate, ladders played a special role. Due to 
the ``external improvement" of Lemma \lemOSsectextimpr, we needed to consider only ladders all of whose ``ends" correspond to $\psi$ fields and are 
integrated out at a later scale. This is not the case when we use the norms 
developed in this chapter. We consider ladders some of whose ``ends" correspond to $\psi$ fields and have sectorized position space variables $(\xi,s)\in\cB\times \Si$, and some of whose ends correspond to $\phi$ fields and have momentum space variables $\check \eta \in \check\cB$. To do this, we extend the definitions and estimates of ladders 
from \S \CHladdersNotn.

\definition{\STM\defOSsectbubblepropII}{

\Item i) Let $C$ be a propagator over $\cB$. We define its extension 
 $\tilde C$ over the disjoint union $\check\cB \dunion \cB$ by
$$
\tilde C(x,y) = \cases{ C(x,y) & if $x,y \in \cB$ \cr
                    0   & if $x \in \check\cB$ or $y \in \check\cB$
}$$

\Item ii) Let $C,D$ be propagators over $\cB$ and $R$ a rung over 
$\check\cB \dunion \cB$. We set
$$
L_\ell(R;C,D) = L_\ell(R;\tilde C, \tilde D)
$$

\Item iii) Let $P$ be a bubble propagator over $\cB$, $r$
a rung over $\fX_\Si=\check\cB\dunion(\cB\times\Si)$.
We set
$$
(r\bullet P)({\sst y_1,y_2;x_3,x_4}) 
=  \smsum_{s'_1,s'_2\in \Si} \int_{\cB\times \cB} {\sst dx'_1 dx'_2}\ 
r({\sst y_1,y_2,(x'_1,s'_1),(x'_2,s'_2)})\ 
P({\sst x'_1,x'_2;x_3,x_4}) 
$$
$(r\bullet P)$ is a function on $\fX_\Si^2 \times \cB^2$. For a general function $F$ on $\fX_\Si^2 \times \cB^2$, define the rung $(F\bullet r)$ over $\fX_\Si$ by 
$$
(F\bullet r)({\sst y_1,y_2,y_3,y_4}) 
=  \smsum_{s'_1,s'_2\in \Si}\int_{\cB\times \cB} {\sst dx'_1 dx'_2 }\ 
F({\sst y_1,y_2;x'_1,x_2'})\ 
r({\sst (x_1',s_1'),(x_2',s_2'),y_3,y_4}) 
$$
if at least one of the arguments $y_1,\cdots,y_4$ lies in 
$\cB\times \Si \subset \fX_\Si$, and for 
$ \check\eta_1,\check\eta_2,\check\eta_3,\check\eta_4 
\in \check\cB\subset \fX_\Si$
$$
(F\bullet r)({\sst \check\eta_1,\check\eta_2,\check\eta_3,\check\eta_4})\,
(2\pi)^{d+1}\de({\sst\check\eta_1+\check\eta_2+\check\eta_3+\check\eta_4}) 
=  \smsum_{s'_1,s'_2\in \Si}\int_{\cB\times \cB} \hskip-8pt{\sst dx'_1 dx'_2 }\ 
F({\sst \check\eta_1,\check\eta_2;x'_1,x_2'})\ 
r({\sst (x_1',s_1'),(x_2',s_2'),\check\eta_3,\check\eta_4}) 
$$
\Item iv) Let $\ell \ge 1\,$, $r_1,\cdots,r_{\ell+1}$ rungs over $\fX_\Si$  and 
$P_1,\cdots,P_\ell$  bubble propagators over $\cB$.
The ladder with rungs $r_1,\cdots,r_{\ell+1}$ and 
bubble propagators $P_1,\cdots,P_\ell$ is defined to be
$$
r_1\bullet P_1 \bullet r_2 \bullet P_2\bullet \cdots \bullet
 r_{\ell}\bullet P_\ell\bullet r_{\ell+1}
$$
If $r$ is a  rung over $\fX_\Si$ and $A,B$ are propagators over $\cB$, we define $L_\ell(r;A,B)$ as the ladder with   
$\ell+1$ rungs $r$ and  $\ell$ bubble propagators $\cC(A,B)$.
}

\remark{\STM\remOSsectbubblepropII}{
In the situation of Definition \defOSsectbubblepropII.ii, let $R'$ be 
the restriction of $R$ to $\cB^4$, $R_{\rm left}$ the restriction of $R$ to $({\check\cB\dunion\cB})^2\times \cB^2$ and $R_{\rm right}$ the restriction 
of $R$ to $\cB^2\times({\check\cB\dunion\cB})^2$. Then
$$
L_\ell(R;C,D) = R_{\rm left} \circ \cC(C,D) \circ R' \circ \cdots \circ R' \circ 
\cC(C,D) \circ R_{\rm right}
$$
Similarly, in the situation of Definition \defOSsectbubblepropII.iv, let $r'$ 
be the restriction of $r$ to $(\cB\times \Si)^4$, $r_{\rm left}$ the 
restriction of $r$ to $\fX_\Si^2 \times (\cB \times \Si)^2$ and $r_{\rm right}$ the restriction of $r$ to 
$ (\cB \times \Si)^2\times \fX_\Si^2$. Then
$$
L_\ell(r;C,D) = r_{\rm left} \bullet \cC(C,D) \bullet r' \bullet \cdots 
\bullet r' \bullet  \cC(C,D) \bullet r_{\rm right}
$$

}

In analogy to Lemma \lemOSladderfunctoriality\ we have

\lemma{\STM\lemOSladderfunctorialityII}{
Let $c$ and $d$ be propagators over $\cB\times \Si$ and $r$ a rung over 
$\fX_\Si$. Define the propagators $C$ and $D$ over $\cB$ by
$$
C(x_1,x_2)=\sum_{t_1,t_2\in \Si}c\big((x_1,t_1),(x_2,t_2)\big)\qquad\qquad
D(x_1,x_2)=\sum_{t_1,t_2\in \Si}d\big((x_1,t_1),(x_2,t_2)\big)
$$
and new propagators $\tilde c$ and $\tilde d$ over $\cB\times \Si$ by
$$
\tilde c\big((x_1,s_1),(x_2,s_2)\big)=C(x_1,x_2)\qquad\qquad
\tilde d\big((x_1,s_1),(x_2,s_2)\big)=D(x_1,x_2)
$$
Then, for all  $\ell \ge 1$ 
$$
L_\ell(r;C,D)=L_\ell(r;\tilde c,\tilde d)
$$

}

In analogy to Lemma \lemOSsectorladderfunctoriality, we have

\lemma{\STM\lemOSsectorladderfunctorialityII}{
Let  $f \in\check\cF_{4;\Si}$.
Let $C(k)$ and $D(k)$ be functions on $\bbbr\times\bbbr^2$, that are supported in the $j^{\rm th}$ neighbourhood, and
$C(\xi,\xi'),\,D(\xi,\xi')$ their Fourier transforms as in Definition 
\defOSftcov. Furthermore, let $c\big((\cdot,s),(\cdot,s') \big)$ and $d\big((\cdot,s),(\cdot,s') \big)$ be the Fourier transform of 
$\chi_s(k)\,C(k)\,\chi_{s'}(k)$  and $\chi_s(k)\,D(k)\,\chi_{s'}(k)$. 
Define propagators over $\cB\times\Si$ by
$$\eqalign{
c_\Si({\sst(\xi,s),(\xi',s')}) 
&= \sum_{t\cap s \ne \emptyset \atop t'\cap s' \ne \emptyset}
c({\sst (\xi,t),(\xi',t')}) \cr
d_\Si({\sst(\xi,s),(\xi',s')})  
&= \sum_{t\cap s \ne \emptyset \atop t'\cap s' \ne \emptyset}
d({\sst (\xi,t),(\xi',t')}) \cr
}$$
Then 
$$
L_\ell(f;C,D)=L_\ell(f;c_\Si,d_\Si)
$$
for all $ \ell \ge 1$. Here, the ladder on the right hand side is defined 
as in Definition \defOSsectbubblepropII.ii, but with $\cB$ replaced by
$\cB\times\Si$, and uses the $\circ$ product of Definition \defOSbubbleprop.iv,
while the ladder on the left hand side is as in Definition \defOSsectbubblepropII.iv and uses the $\bullet$ product.
}

Also observe that, by Remark \remOSsectbubblepropII, for 
$f\in \check \cF_{4;\Si}$
$$
L_\ell(f; C, D) = \hskip -.3cm
\sum_{i_1,\cdots,i_4\in\{0,1\}}
f\big|_{(i_1,i_2,1,1)} \bullet \cC(C,D) \bullet f\big|_{(1,1,1,1)} \bullet
\cdots \bullet \cC(C,D) \bullet f\big|_{(1,1,i_3,i_4)}
\EQN\eqnOSdisjointladder$$

\vfill\eject

\chap{The Renormalization Group Map and Norms  }\PG\pgOSXVII
\vskip-.4in
{\centerline{\tafontt in Momentum Space }\vskip.5truein}

This section provides the analogue of \S\CHestren\ for the $\v\ \ \tv$--norms.
Again, let $j\ge 2$ and let $\Si$ be a sectorization of scale $j$ and length
$\sfrac{1}{M^{j-3/2}}\le\fl\le\sfrac{1}{M^{(j-1)/2}}$. 
Fix a system 
$\vec \rho = (\rho_{m;n})$ of positive real numbers such that
$$\deqalign{
\rho_{m;n} & \le \rho_{m;n'} \qquad &{\rm if\ } n\le n' \cr
\rho_{m+m';n+n'-2} & \le \rho_{m;n}\,\rho_{m';n'}\cr
\rho_{m+1;n-1} & \le \rho_{m;n} \cr
}\EQN\eqnOStilderhomn$$

\definition{\STM\defOSmomscalednorms}{
\Item i)
For a function $f\in \check \cF_{n;\Si}$ and a natural number $p$ we set
$$
\v f\tv_{p,\Si,\vec\rho} =  \rho_{n;0}\,\v g \tv_{p,\Si} +
\sum_{\vec\imath \in \{0,1\}^n \atop m(\vec\imath) <n} 
\rho_{m(\vec\imath);n-m(\vec\imath)}\,
\V \ord \big(f\big|_{\vec\imath} \big) \tV_{p,\Si} 
$$
where $g$ is the function on $\check \cB_n$ such that
$$
f\big|_{(0,\cdots, 0)}(\check\eta_1,\cdots,\check\eta_n) = 
(2\pi)^{d+1}
\de(\check\eta_1+\cdots+\check\eta_n)\,g(\check\eta_1,\cdots,\check\eta_n)
$$

\Item{ii)}
For $f\in \check\cF_m(n;\Si)$ set
$$\eqalign{
\v f \tv_\Si &= 
\rho_{m;n}\cases{
\v f \tv  _{1,\Si} +\v f \tv  _{2,\Si} 
+ \sfrac{1}{\fl}\,\v f \tv_{3,\Si}
+ \sfrac{1}{\fl}\,\v f \tv_{4,\Si}
+ \sfrac{1}{\fl^2}\,\v f \tv_{5,\Si}
+ \sfrac{1}{\fl^2}\,\v f \tv_{6,\Si}& if $m\ne 0$\cr
\noalign{\vskip.1in}
\v f \tv  _{1,\Si} + \sfrac{1}{\fl}\,\v f \tv_{3,\Si}
+ \sfrac{1}{\fl^2}\,\v f \tv_{5,\Si}& if $m= 0$\cr
}\cr
}$$
and for $f\in \check\cF_{n;\Si}$ set
$$\eqalign{
\v f \tv_\Si &= \v g \tv_\Si +
\sum_{\vec\imath \in \{0,1\}^n \atop m(\vec\imath) <n} 
\V \ord \big(f\big|_{\vec\imath} \big) \tV_\Si\cr
}$$
where, as in part (i), $g$ is the function on $\check \cB_n$ such that
$$
f\big|_{(0,\cdots, 0)}(\check\eta_1,\cdots,\check\eta_n) = 
(2\pi)^{d+1}
\de(\check\eta_1+\cdots+\check\eta_n)\,g(\check\eta_1,\cdots,\check\eta_n)
$$

\Item{iii)}
With the notation introduced in Definitions \defOSdisjointfield\ 
and \defOSsectcheckcF.iii, every translation invariant sectorized Grassmann function $w$ can be uniquely written in the form
$$\eqalign{
w(\phi,\psi) = \sum_n 
\int_{\fX_\Si^n} {\sst dx_1\cdots dx_n}\ 
f_{n}({\sst x_1,\cdots, x_n})\ 
\Psi({\sst x_1})\cdots \Psi({\sst x_n})\cr
}$$
where $f_n \in \check \cF_{n;\Si}$ is an antisymmetric function.
Set, in analogy with Definition \defOSscalednorms.iii, for $\al >0$
and $X\in \fN_{d+1}$,
$$
N_j^\sim(w;\al;\,X,\Si,\vec\rho)
=\sfrac{M^{2j}}{\fl}\,\fe_j(X) 
\smsum_{n\ge 0}\,
\al^{n}\,\big(\sfrac{\fl\,\IB}{M^j}\big)^{n/2} \,
\v f_n\tv_\Si 
$$
where $\IB=4\max\{8\IB_1,\IB_2,32\IB_3,4\IB_4\}$ with 
with $\IB_1,\ \IB_2$ being the constants of Propositions \propOScontrintboundsectors\ and \propOSoverlapploops\ and $\IB_3,\ \IB_4$ 
being the constants of Proposition \propOSmomcontrintboundsectors. 

}

\remark{\STM\remOSmomscalednorms}{
A sectorized Grassmann function $w$ can also be uniquely written in the form
$$\eqalign{
w(\phi,\psi) = \smsum_{m,n} 
\int {\sst d\eta_1\cdots d\eta_m\,d\xi_1\cdots d\xi_n}\ & 
w_{m,n}({\sst \eta_1,\cdots, \eta_m\,(\xi_1,s_1),\cdots ,(\xi_n,s_n)})\cr
& \hskip 2cm \cdot\ \phi({\sst \eta_1})\cdots \phi({\sst \eta_m})\
\psi({\sst (\xi_1,s_1)})\cdots \psi({\sst (\xi_n,s_n)\,})\cr
}$$
with $w_{m,n}$ antisymmetric separately in the $\eta$ variables
and in the $\xi$ variables. Then,
$$\eqalign{
&N_j^\sim(w;\al;\,X,\Si,\vec\rho)
=\sfrac{M^{2j}}{\fl}\,\fe_j(X) 
\smsum_{m,n\ge 0}\,
\al^{m+n}\,\big(\sfrac{\fl\,\IB}{M^j}\big)^{(m+n)/2} \,
\v w^\sim_{m,n}\tv_\Si \cr
&\hskip.2in=\sfrac{M^{2j}}{\fl}\,\fe_j(X) 
\smsum_{n\ge 0}\,
\al^{n}\,\big(\sfrac{\fl\,\IB}{M^j}\big)^{n/2} \,\rho_{0;n}
\Big[ \v w^\sim_{0,n}\tv_{1,\Si}
 + \sfrac{1}{\fl}\v w^\sim_{0,n}\tv_{3,\Si}
+ \sfrac{1}{\fl^2}\v w^\sim_{0,n}\tv_{5,\Si}\Big]\cr
&\hskip.3in+\sfrac{M^{2j}}{\fl}\,\fe_j(X) 
\smsum_{m,n\ge 0\atop m\ne 0}\,
\al^{m+n}\,\big(\sfrac{\fl\,\IB}{M^j}\big)^{(m+n)/2} \,\rho_{m;n}
\Big[ \smsum_{p=1}^6\sfrac{1}{\fl^{[(p-1)/2]}}\v w^\sim_{m,n}\tv_{p,\Si}
\Big]\cr
}$$
Here, $w_{m,n}^\sim$ is the partial Fourier transform of $w_{m,n}$ of 
Definition \defOSfourtrans.ii and $[(p-1)/2]$ is the integer part of $\sfrac{p-1}{2}$.  In particular,
$$
N_j^\sim(w(\phi,0) ;\al;\,X,\Si,\vec\rho)
=\fe_j(X) \bigg[ \al^2\rho_{2;0} \IB\, M^j \v w^\sim_{2,0}\tv_{1,\Si}
 + \al^4 \rho_{4;0}  \IB^2 \v w^\sim_{4,0}\tv_{3,\Si}
\bigg]
$$
and
$$
N_j^\sim(w(0,\psi);\al;\,X,\Si,\vec\rho)
=N_j(w(0,\psi);\al;\,X,\Si,\vec\rho)
$$
}

\theorem{\STM\thOSmomrengroupestimate}{
Let $c_B>0$. 
There are constants $\const,\ \cst{}{0},\ \al_0,\ \ga_0$ and $\tau_0$
that are independent of $j,\ \Si,\ \vec\rho$ such that for all $\al\ge \al_0$ 
and $\ga\le\ga_0$ the following holds:

{\parindent=.25in
\item{}
Let $u({\sst(\xi,s)},{\sst(\xi',s')}),v({\sst(\xi,s)},{\sst(\xi',s')})
\in\cF_0(2;\Si)$ be antisymmetric, spin independent, particle number 
conserving functions.
Set
$$
C(k) = \sfrac{\nu^{(j)}(k)}{\imath k_0 -e(\k) -\check u(k)}
\qquad,\qquad 
D(k) = \sfrac{\nu^{(\ge j+1)}(k)}{\imath k_0 -e(\k) -\check v(k)}
$$
and let $C(\xi,\xi'),\,D(\xi,\xi')$ be the Fourier transforms of $C(k)$, $D(k)$ as
in Definition \defOSftcov. 
Let $B(k)$ be a function on $\bbbr\times\bbbr^2$ and set
$$
(\hat B\phi)(\xi)=\int d\xi'\  \hat B(\xi,\xi')\phi(\xi')
$$
where $\hat B$ was defined in Definition \defOSfourtransII.
Furthermore, let $\cW(\phi,\psi)$ be an even Grassmann function and 
set\footnote{$^{(1)}$}{The definition of $\cW'$ as an analytic function, rather than merely a formal Taylor series was explained in Remark \remOSchoiceofrep.}
$$
\lw \cW'(\phi,\psi) \rw_{\psi,D} 
=  \Om_C \big( \lw \cW(\phi,\psi) \rw_{\psi,C+D} \big) (\phi,\psi+\hat B\phi)
$$

\vskip.3cm
Assume that the following estimates are fulfilled:
{\parindent = .5in
\item{$\bullet$} $\rho_{m+1;n-1}\le\ga\,\rho_{m;n}\ $
for all $m\ge 0$ and $n\ge 1$.
\item{$\bullet$}
$|\check u(k)|,|\check v(k)| \le \half |\imath k_0-e(k)|$
\item{$\bullet$}
$\ \v  u \v_{1,\Si} \le \mu(\La+X)\fe_j(X)\ $ with
$X\in\fN_{d+1}$, $\mu,\La>0$ such that
$(1+\mu)(\La+X_\0)\le\sfrac{\tau_0}{M^j}$ 
\item{$\bullet$}
$\ 
\| B(k)\tnorm \le  c_B \fe_j(X)
\ $
\item{$\bullet$}$\cW$ has a sectorized representative 
$$
w(\phi,\psi) = \sum_n 
\int_{\fX_\Si^n} {\sst dx_1\cdots dx_n}\ f_{n}({\sst x_1,\cdots, x_n})\ 
\Psi({\sst x_1})\cdots \Psi({\sst x_n})
$$
with antisymmetric functions $f_n \in \check \cF_{n;\Si}$ such that
$f_2=0$ and
$$
N_j^\sim(w;64\al;\,X,\Si,\vec\rho)
\le \cst{}{0}\,\al + \smsum_{\de \ne 0} \infty\,t^\de
$$

}
\vskip .3cm

\item{}
Then $\cW'$ has a sectorized representative $w'$ such that
$$
N_j^\sim(w'-w;\al;\,X,\Si,\vec\rho) 
\le\const\big(\sfrac{1}{\al}+\ga\big)\,
\sfrac{N^\sim_j(w;64\al;X,\Si,\vec\rho)}
{1-{\const \over\al}N^\sim_j(w;64\al;X,\Si,\vec\rho)}
$$
Furthermore, if one writes 
$\ 
w'(\phi,\psi) = \sum_n 
\int_{\fX_\Si^n} {\sst dx_1\cdots dx_n}\ f'_{n}({\sst x_1,\cdots, x_n})\ 
\Psi({\sst x_1})\cdots \Psi({\sst x_n})
\ $,
with antisymmetric functions $f'_n \in \check \cF_{n;\Si}$, then
$$
\v    f'_2 \tv _{1,\Si,\vec\rho}  
\le  \sfrac{\const}{\al^8}\sfrac{ \fl}{M^j}\,
\sfrac{N_j^\sim(w;64\al;\,X,\Si,\vec\rho)^2}
           {1-{\const \over\al}N_j^\sim(w;64\al;\,X,\Si,\vec\rho)} 
$$
If one writes 
$\ 
w'(\phi,\psi)=w''(\phi,\psi+\hat B\phi)
\ $ 
where
$\ 
(\hat B\phi)(\xi,s)=\int d\xi'\ \hat B(\xi,\xi')\phi(\xi')
$,
with abuse of notation, and expands 
$$ 
w''(\phi,\psi) = \sum_n 
\int_{\fX_\Si^n} {\sst dx_1\cdots dx_n}\ f''_{n}({\sst x_1,\cdots, x_n})\ 
\Psi({\sst x_1})\cdots \Psi({\sst x_n})
$$
with antisymmetric functions $f''_n \in \check \cF_{n;\Si}$, then
$$\eqalign{
&\V   f''_4-f_4 -
\sfrac{1}{4} 
\smsum_{\ell=1}^\infty (-1)^\ell(12)^{\ell+1}
{\rm Ant\,}L_\ell(f_4;\,C,D)\tV_{3,\Si,\vec\rho}
\le\sfrac{\const}{\al^{10}}\,\fl\,
\sfrac{N_j^\sim(w;64\al;\,X,\Si,\vec\rho)^2}
           {1-{\const \over\al}N_j^\sim(w;64\al;\,X,\Si,\vec\rho)} 
}$$
Here $L_\ell(f_4;\,C,D)$ is a ladder in the sense of Definition \defOSsectbubblepropII.iv.

}
}

The proof of Theorem \thOSmomrengroupestimate\ is similar that of Theorems
\thOSrengroupestimate\ and \thmOSTfirststep. Recall that $w$ and $w'$ are elements of the Grassmann algebra over the vector space, $\tilde V$,
generated by  $\check\phi(\check\eta)$, $\check\eta\in\check \cB$,
$\psi(\xi,s)$, $(\xi,s)\in(\cB \times \Si)$.
Let $c\big((\cdot,s),(\cdot,s') \big)$ and $d\big((\cdot,s),(\cdot,s') \big)$ be
the Fourier transform of 
$\chi_s(k)\,C(k)\,\chi_{s'}(k)$  and $\chi_s(k)\,D(k)\,\chi_{s'}(k)$
in the sense of Definition \defOSftcov. Then $c$ and $d$ define covariances on
$\tilde V$ by
$$\meqalign{
\tilde C_\Si\big( \check\phi({\sst \check\eta}), \check\phi({\sst \check\eta'}) \big) &= 0
\qquad,\qquad
&\tilde D_\Si\big( \check\phi({\sst \check\eta}), \check\phi({\sst \check\eta'}) \big)  &= 0  \cr
\tilde C_\Si\big( \check\phi({\sst \check\eta}), \psi({\sst (\xi,s)}) \big) &= 0
\qquad,\qquad
&\tilde D_\Si\big( \check\phi({\sst \check\eta}), \psi({\sst (\xi,s)}) \big)  &= 0  \cr
}$$
and
$$\deqalign{
\tilde C_\Si\big(\psi({\sst\xi,s}),\psi({\sst\xi',s'}) \big) 
&= c_\Si\big((\xi,s),(\xi',s')\big)
& =\sum_{t\cap s \ne \emptyset \atop t'\cap s' \ne \emptyset}
c\big((\xi,t),(\xi',t')\big) \cr
\tilde D_\Si \big(\psi({\sst\xi,s}),\psi({\sst\xi',s'}) \big) 
&= d_\Si\big((\xi,s),(\xi',s')\big)
& = \sum_{t\cap s \ne \emptyset \atop t'\cap s' \ne \emptyset}
d\big((\xi,t),(\xi',t')\big) \cr
}$$
The restriction of $\tilde C_\Si$ resp. $\tilde D_\Si$ to 
the vector space, $V_\Si$, generated by
$\psi(\xi,s)$, $(\xi,s)\in(\cB \times \Si)$,  coincides with the $C_\Si$
resp. $D_\Si$ of Proposition \propOSfunctorialitySect, 
while the subspace $V_{\rm ext}$, generated by  $\check\phi(\check\eta)$, $\check\eta\in\check \cB$,
is isotropic and perpendicular to $V_\Si$ 
with respect to both $\tilde C_\Si$ and $\tilde D_\Si$.

For $f\in \check\cF_m(n;\Si)$ set
$$\eqalign{
\v f \tv_{{\rm impr},\Si} &= 
\rho_{m;n}\cases{
\v f \tv  _{1,\Si} +\v f \tv  _{2,\Si} 
+ \sfrac{1}{\fl}\,\v f \tv_{3,\Si}
+ \sfrac{1}{\fl}\,\v f \tv_{4,\Si}& if $m\ne 0$\cr
\noalign{\vskip.1in}
\v f \tv  _{1,\Si} + \sfrac{1}{\fl}\,\v f \tv_{3,\Si}
& if $m= 0$\cr
}\cr
}\EQN\eqnOSmomimprnorom$$
and for $f\in \check\cF_{n;\Si}$ set
$$\eqalign{
\v f \tv_{{\rm impr},\Si} &= \v g \tv_{{\rm impr},\Si} +
\sum_{\vec\imath \in \{0,1\}^n \atop m(\vec\imath) <n} 
\V \ord \big(f\big|_{\vec\imath} \big) \tV_{{\rm impr},\Si}\cr
}$$
where $g$ is the function on $\check \cB_n$ such that
$$
f\big|_{(0,\cdots, 0)}(\check\eta_1,\cdots,\check\eta_n) = 
(2\pi)^{d+1}
\de(\check\eta_1+\cdots+\check\eta_n)\,g(\check\eta_1,\cdots,\check\eta_n)
$$
The seminorms $\v \ \cdot\  \tv_{{\rm impr},\Si}$
(and $\v \ \cdot\  \v'_{{\rm impr}}\ $, $\ N^\sim_{{\rm impr}}( \ \cdot\  ;\al)$, to be introduced
shortly) are used only locally,
between this point and the end of the proof of Theorem \thOSmomrengroupestimate.

\lemma{\STM\lemOSmomconcreteintconst}{
Under the hypotheses of Theorem \thOSmomrengroupestimate, there exists a constant
$\cst{}{1}$ that is independent of $j$ and $\Si$ such that
the covariances $(\tilde C_\Si,\,\tilde D_\Si)$ have improved 
integration constants
$$
\cb = \cst{}{1}\,M^j\,\fe_j(X),\qquad \half\ib =\sqrt{\sfrac{\IB\,\fl}{4M^j}},
\qquad J=\fl 
$$ 
for the families $\v\cdot\tv_\Si$ and $\v\cdot\tv_{{\rm impr},\Si}$ 
of seminorms (in the sense of Definition \defimprnorm\  of [FKTr2]).  }

\prf
By Proposition \propOSmomcontrintboundsectors\ and (\eqnOStilderhomn), the covariances 
$(\tilde C_\Si,\,\tilde D_\Si)$ have integration constants
$$
\cb' = 12 \v c\v_{1,\Si}
\qquad \ib'=\sqrt{\max\{8\IB_3,\IB_4\}\sfrac{\fl}{M^j}}
$$
for the configuration $\v\ \cdot\  \tv_{1,\Si,\vec\rho}$, 
$\v\ \cdot\  \tv_{2,\Si,\vec\rho}$,
$\cdots$, $\v\ \cdot\  \tv_{6,\Si,\vec\rho}$ of seminorms, in the sense of 
Definition \deftildeimprconf\  of [FKTr2]. Hence, by Lemma \lemtildeimprconfig\
of [FKTr2],  $(\tilde C_\Si,\,\tilde D_\Si)$ have improved 
integration constants $\cb'$, $\ib'$ and $J=\fl$ for the families 
$$\eqalign{
\v f\v'&=
\v f \tv  _{1,\Si,\vec\rho} +\v f \tv_{2,\Si,\vec\rho} 
+ \sfrac{1}{\fl}\,\v f \tv_{3,\Si,\vec\rho}
+ \sfrac{1}{\fl}\,\v f \tv_{4,\Si,\vec\rho}
+ \sfrac{1}{\fl^2}\,\v f \tv_{5,\Si,\vec\rho}
+ \sfrac{1}{\fl^2}\,\v f \tv_{6,\Si,\vec\rho}\cr
\v f\v'_{{\rm impr}}&=
\v f \tv  _{1,\Si,\vec\rho} +\v f \tv_{2,\Si,\vec\rho} 
+ \sfrac{1}{\fl}\,\v f \tv_{3,\Si,\vec\rho}
+ \sfrac{1}{\fl}\,\v f \tv_{4,\Si,\vec\rho}\cr
}$$ 
When $f\in\check\cF_0(n;\Si)$, $\v f \tv_{p+1,\Si,\vec\rho}\le
\v f \tv_{p,\Si,\vec\rho}$ for all odd $p$ so that
$$
\v f \tv_{\Si}\le \v f\v' \le 2\v f \tv_{\Si}
\qquad
\v f \tv_{{\rm impr},\Si}\le \v f\v'_{{\rm impr}} \le 2\v f \tv_{{\rm impr},\Si}
$$
Hence  $(\tilde C_\Si,\,\tilde D_\Si)$ have improved 
integration constants $4\cb'$, $2\ib'$ and $J=\fl$ for the families 
$\v\ \cdot\ \tv_{\Si}$ and $\v\ \cdot\ \tv_{{\rm impr},\Si}$ of seminorms.
As in Lemma \lemOSconcreteintconst,
$\cb'\le\const M^j\,\fe_j(X) $ and the Lemma follows.
\endproof

\lemma{\STM\lemOStildesourceterm}{
Let $c_B>0$. Then there are constants $\abcst$ and $\ga_0$, independent
of $M$, $j$, $\Si$, $\vec\rho$ such that the following holds for 
all $\ga\le\ga_0$ and all $X,X_B\in\fN_{d+1}$.
Let $g(\phi,\psi)$ be a sectorized Grassmann function and set
$$
g'(\phi,\psi)=g(\phi,\psi+\hat B\phi)
$$
Assume that 
$\ 
\| B(k)\tnorm \le  c_B \fe_j(X)
\ $ and
$\ 
\| B(k)\tnorm \le  c_B X_B\fe_j(X)
$.
If $\rho_{m+1;n-1}\le\ga\,\rho_{m;n}\ $ for all $m\ge 0$ and $n\ge 1$, then 
$$
N^\sim_j(g'-g;\al;\,X,\Si,\vec\rho)
\le \abcst\,\ga X_B\, N^\sim_j(g;2\al;\,X,\Si,\vec\rho)
$$
Let $G_{m,n}$, resp. $G'_{m,n}$, be the kernel of the part of $g$, resp. $g'$,
that is of degree $m$ in $\phi$ and degree $n$ in $\psi$.
Then, for $p\in\{1,3\}$,
$$
\sum_{m,n\atop m+n=p+1}\v G^{\prime\sim}_{m,n}-G^{\sim}_{m,n}\tv_{p,\Si,\vec\rho}
\le\abcst\,\ga X_B\, \fe_j(X)\ 
\sum_{m,n\atop m+n=p+1}\v G^{\sim}_{m,n}\tv_{p,\Si,\vec\rho}
$$
  }
\prf 
Let $\varphi\in\check\cF_m(n;\Si)$, $1\le i\le n$ and set,
for $\check\et_{m+1}=\big(k_{m+1},\si_{m+1},a_{m+1}\big)$,
$$\eqalign{
&\varphi'({\sst\check\et_1,\cdots,\check\et_{m+1}\,;
\,(\xi_1,s_1)\cdots,(\xi_{n-1},s_{n-1})})\cr
&\hskip.3cm= {\rm Ant_{ext}}\smsum_{s\in\Si}\int\!\! d{\sst \ze}\ 
B({\sst k_{m+1}})\,E_+({\sst \check\et_{m+1},\ze})
\varphi({\sst\check\et_1,\cdots,\check\et_m\,;\,(\xi_1,s_1)\cdots,
(\xi_{i-1},s_{i-1}),(\ze,s),(\xi_{i},s_i),\cdots,(\xi_{n-1},s_{n-1})}) \cr
}$$
if $n\ge 2$ and 
$$\eqalign{
&\varphi'({\sst\check\et_1,\cdots,\check\et_{m+1}})
{\sst (2\pi)^{d+1}\de(k_1+\cdots+k_{m+1})}
= {\rm Ant_{ext}}\smsum_{s\in\Si}\int\!\! d{\sst \ze}\ 
B({\sst k_{m+1}})\,E_+({\sst \check\et_{m+1},\ze})
\varphi({\sst\check\et_1,\cdots,\check\et_m\,;\,(\ze,s)}) \cr
&\hskip1in= {\rm Ant_{ext}}\smsum_{s\in\Si}
B({\sst k_{m+1}})
\varphi({\sst\check\et_1,\cdots,\check\et_m\,;\,(0,\si_{m+1}, a_{m+1},s)})\ 
{\sst (2\pi)^{d+1}\de(k_1+\cdots+k_{m+1})} \cr
}$$
if $n=1$.
For any fixed $\check\et_1,\cdots,\check\et_{m+1}$
$$\eqalign{
&\TN \varphi'({\sst\check\et_1,\cdots,\check\et_{m+1}\,;
\,(\xi_1,s_1)\cdots,(\xi_{n-1},s_{n-1})})\TN_{1,\infty}\cr
&\hskip1in\le 2\sup_{k,s} \big|B(k)\big|\,\TN \varphi({\sst\check\et_1,\cdots,\check\et_m\,;\,(\xi_1,s_1)\cdots
,(\ze,s),\cdots,(\xi_{n-1},s_{n-1})})
\TN_{1,\infty}\cr
}$$
when $n\ge 2$, since $\big|E_+({\sst \check\et_{m+1},\ze})\big|\le 1$
and the requirement that $k_{m+1}$ be in the sector $s$ restricts the choice
of $s$ to at most two different sectors. For $n=1$,
$$\eqalign{
&\big| \varphi'({\sst\check\et_1,\cdots,\check\et_{m+1}})\big|
\le 2\sup_{k,s} \big|B(k)\big|\,\big| \varphi({\sst\check\et_1,\cdots,\check\et_m\,;\,(0,\si_{m+1}, a_{m+1},s)})
\big|\cr
}$$
Since $\rD^\de_{m+1}E_+({\sst \check\et_{m+1},\ze})
=\ze^\de E_+({\sst \check\et_{m+1},\ze})$, Leibniz and Corollary 
\corOSappMonoidIV.ii of [FKTo1] implies that, for both $X'=X_B$ and $X'=1$,
$$
\fe_j(X)\v\varphi'\tv_{p,\Si}
\le \abcst\, \fe_j(X)\| B(k)\tnorm\ \ \v \varphi\tv_{p,\Si}
\le\abcst\, c_BX' \fe_j(X)\ \v \varphi\tv_{p,\Si}
\EQN\eqnNPderivampprelim$$
so that $\fe_j(X)\v\varphi'\tv_{p,\Si,\vec\rho}
\le\abcst\, c_B\ga X'\, \fe_j(X)\ \v \varphi\tv_{p,\Si,\vec\rho}$ and 
$$
\fe_j(X)\v\varphi'\tv_\Si\le\abcst\, c_B\ga X'\, \fe_j(X)\ \v \varphi\tv_\Si
\EQN\eqnNPderivamp$$

Write $g(\phi,\psi) = \sum_{m,n}g_{m,n}(\phi,\psi)$, 
with $g_{m,n}$ of degree $m$ in $\phi$ and degree $n$ in $\psi$, and 
$$
g(\phi,\psi+\ze) =  \sum_{m,n}g_{m,n}(\phi,\psi+\ze)
=\sum_{m,n}\smsum_{\ell=0}^ng_{m,n-\ell,\ell}(\phi,\psi,\ze)
$$ 
with $g_{m,n-\ell,\ell}$ of degrees $m$ in $\phi$, $n-\ell$ in $\psi$  and $\ell$
in $\ze$. Let $G_{m,n}$ and $G_{m,n-\ell,\ell}$ be the kernels of 
$g_{m,n}(\phi,\psi)$ and $g_{m,n-\ell,\ell}(\phi,\psi,\hat B\phi)$ respectively. 
By the binomial theorem and repeated application of (\eqnNPderivamp), 
$\ell-1$ times with $X'=1$ and once with $X'=X_B$,
$$
\fe_j(X)\v G^\sim_{m,n-\ell,\ell}\tv_\Si\le 
(\abcst\,c_B\ga)^\ell {\tst{n\choose \ell}}\,X_B \fe_j(X)\v G^\sim_{m,n}\tv_\Si
$$
if $\ell\ge 1$.
 Then, 
$$\meqalign{
g'(\phi,\psi)-g(\phi,\psi)
 = g(\phi,\psi+\hat B\phi)-g(\phi,\psi)
=\sum_{m,n\ge 0}\smsum_{\ell=1}^ng_{m,n-\ell,\ell}(\phi,\psi,\hat B\phi)
}$$
and
$$\eqalign{
N^\sim_j\big(g'-g;\al;X,\Si,\vec\rho\big)
&\le \sfrac{M^{2j}}{\fl}\fe_j(X)\sum_{m,n\ge 0}\smsum_{\ell=1}^n
\al^{m+n}\,\big(\sfrac{\fl\,\IB}{M^j}\big)^{(m+n)/2} \,
\v G^\sim_{m,n-\ell,\ell}\tv_\Si
\cr
&\hskip-0.5in\le \sfrac{M^{2j}}{\fl}X_B\fe_j(X)\sum_{m,n\ge 0}\smsum_{\ell=1}^n
{\tst{n\choose \ell}}(\abcst\,c_B\ga)^\ell
\al^{m+n}\,\big(\sfrac{\fl\,\IB}{M^j}\big)^{(m+n)/2} \,
\v G^\sim_{m,n}\tv_\Si
\cr
&\hskip-0.5in= \sfrac{M^{2j}}{\fl}X_B\fe_j(X)\sum_{m,n\ge 0}\big[(1+\abcst\,c_B\ga)^n-1\big]
\al^{m+n}\,\big(\sfrac{\fl\,\IB}{M^j}\big)^{(m+n)/2} \,
\v G^\sim_{m,n}\tv_\Si
\cr
}$$
If $\abcst\,c_B\ga\le\sfrac{1}{3} $
$$\deqalign{
\big(1+\abcst\,c_B\ga\big)^{n}-1
&\le \abcst\,c_B\ga\,n\big(1+\abcst\,c_B\ga\big)^{n-1}
&\le \abcst\,c_B\ga\,\big(\sfrac{3}{2}\big)^n\big(1+\abcst\,c_B\ga\big)^{n-1}\cr
&\le\abcst\,c_B\ga\,2^{n}
&\le\abcst\,\ga\,2^{n}
}$$
and
$$
N^\sim_j(g'-g;\al;\,X,\Si,\vec\rho)
\le \abcst\,\ga\,X_B\, N^\sim_j(g;2\al;\,X,\Si,\vec\rho)
$$
The proof of the second claim is similar but uses 
$$ 
\v G^\sim_{m,n-\ell,\ell}\tv_{p,\Si,\vec\rho}\le 
(\abcst\,c_B\ga)^\ell {\tst{n\choose \ell}}\,X_B \fe_j(X)\v G^\sim_{m,n}
\tv_{p,\Si,\vec\rho}
$$
and $(\abcst\,c_B\ga)^\ell {\tst{n\choose \ell}}\le\abcst\,\ga$ for 
$\ell\ge 1$, $n\le 4$.
\endproof

\proof{ of  Theorem \thOSmomrengroupestimate}
For a sectorized Grassmann function $v=\smsum_{n} v_{n}$ with
$v_{n} \in \bigwedge^n\tilde V$ let
$$\eqalign{ 
N^\sim(v;\al) &= \sfrac{1}{\ib^2}\cb\,\smsum_{n} \al^{n}\, \ib^{n}\, 
\v v_{n} \tv  _\Si
\cr N^\sim_{\rm impr}(v;\al) &= \sfrac{1}{\ib^2}\cb\,
\smsum_{n} \al^{n}\, \ib^{n}\,  \v v_{n}\tv  _{{\rm impr},\Si}
}$$
be the quantities introduced in Definition \deffunctnorm\ of [FKTr1]
and just after Lemma \lemGrassimprnorm\ of [FKTr2]. Then
$$
N^\sim(v;\al) = \sfrac{\cst{}{1}}{\IB}\, N^\sim_j(v;\al;\,X,\Si,\vec\rho)
$$
where $\cst{}{1}$ is the constant of Lemma \lemOSmomconcreteintconst. 

If $\ \lw w''\rw_{\psi,\tilde D_\Si} = \Om_{\tilde C_\Si}
           (\lw w \rw_{\psi,\tilde C_\Si+\tilde D_\Si})\ $, then,
by Proposition \propOSfunctorialitySect, parts (ii) and (iii), 
\ and Proposition \propBII.ii of [FKTr1]
$$
w'=w''(\phi,\psi+\hat B\phi)
$$
is a sectorized representative for $\cW'$.
We apply Theorem \theoremVa\  of [FKTr2] to get estimates on $w''$. 
Choosing $\cst{}{0} =\sfrac{\IB}{8\,\cst{}{1}}$, the hypotheses of this Theorem 
are fulfilled by Lemma \lemOSmomconcreteintconst. Consequently,
$$\eqalignno{
N^\sim(w''-w;\al) &\le \sfrac{1}{2\al^2}\,
\sfrac{N^\sim(w;32\al)^2}{1-{1 \over\al^2}N^\sim(w;32\al)}
&\EQNO\eqnOSmomsectgeneral\cr
\al^2\, \cb\,\V f''_2 \tV_{{\rm impr},\Si} 
& \le \sfrac{2^{10} \fl}{\al^6}\,\sfrac{N^\sim(w;64\al)^2}{1-{8 \over\al}N^\sim(w;64\al)}&\EQNO\eqnOSmomsecttwopoint\cr
\al^4\,\ib^2 \cb\,
\V f''_4-f_4
 -   \sfrac{1}{4}\smsum_{\ell\ge 1} {\sst (-1)^\ell(12)^{\ell+1}}
{\rm Ant\,}L_\ell(f_4;\,c_\Si,d_\Si)
\tV_{{\rm impr},\Si}
&\le\sfrac{2^{10}\fl}{\al^6}\,\sfrac{N^\sim(w;64\al)^2}
{1-{8\over\al}N^\sim(w;64\al)}\hskip.8in&\EQNO\eqnOSmomsectfourpoint\cr
}$$
In (\eqnOSmomsectfourpoint), we used  the description of ladders in terms of
kernels given in Proposition \propAIV\  of [FKTr2].

\vskip.3cm
By Lemma \lemOStildesourceterm, with $X_B=1$,
$$\eqalign{
&N^\sim(w'-w;\al)
=N^\sim\big(w''(\phi,\psi+\hat B\phi)-w(\phi,\psi);\al\big)\cr
&\hskip.5in
\le N^\sim\big(w''(\phi,\psi+ \hat B\phi)-w''(\phi,\psi);\al\big)
+N^\sim\big(w''(\phi,\psi)-w(\phi,\psi);\al\big)\cr
&\hskip.5in
\le \const\ga\,N^\sim\big(w'';2\al\big)+N^\sim\big(w''-w;\al\big)\cr
&\hskip.5in
\le \const\ga\,N^\sim\big(w;2\al\big)
+\big(1+\const\ga \big) N^\sim\big(w''-w;2\al\big)\cr
&\hskip.5in
\le \const\ga\,N^\sim\big(w;2\al\big)
+\big(1+\const\ga\big) \sfrac{1}{8\al^2}\,
\sfrac{N^\sim(w;64\al)^2}{1-{1 \over\al^2}N^\sim(w;64\al)}\cr
&\hskip.5in
\le  \const\big(\sfrac{1}{\al}+\ga\big)\,
\sfrac{N^\sim_j(w;64\al;X,\Si,\vec\rho)}{1-{\const \over\al}N^\sim_j(w;64\al;X,\Si,\vec\rho)}\cr
}$$
By (\eqnOSmomsecttwopoint), 
$$
M^j\al^2\fe_j(X)\v  f''_2 \tv  _{1,\Si,\vec\rho} 
\le M^j\al^2\fe_j(X)\v  f''_2 \tv_{{\rm impr},\Si}  
\le  \const\sfrac{\fl}{\al^6}\,
\sfrac{N^\sim(w;64\al)^2}{1-{8\over\al}N^\sim(w;64\al)}
$$
Applying Lemma \lemOStildesourceterm\ to the part of $w''$ that
is homogeneous of degree two in $\psi$ and $\phi$ combined yields
$$
\v  f'_2 \tv  _{1,\Si,\vec\rho}  
\le  4\,\fe_j(X)\v  f''_2 \tv  _{1,\Si,\vec\rho} 
$$
and hence
$$
\v  f'_2 \tv  _{1,\Si,\vec\rho}  
\le  \sfrac{\const}{\al^8}\,\sfrac{\fl}{M^j}\,
\sfrac{{N^\sim_j(w;64\al;X,\Si,\vec\rho)}^2}{1-{\const\over\al}N^\sim_j(w;64\al;X,\Si,\vec\rho)}
$$
By Lemma \lemOSsectorladderfunctorialityII\ and (\eqnOSmomsectfourpoint)
$$\eqalign{
&\V f''_4-f_4 
- \sfrac{1}{4}\smsum_{\ell=1}^\infty {\sst (-1)^\ell(12)^{\ell+1}}
{\rm Ant\,}L_\ell(f_4;\,C,D)\tV_{3,\Si,\vec\rho} 
\le\sfrac{\const}{\al^{10}}\,\fl\,
\sfrac{N_j^\sim(w;64\al)^{2}}{1-{\const \over\al}N^\sim_j(w;64\al)} \cr
}$$

\endproof

\theorem{\STM\thOSmomrengroupdiffestimate}{
Let $c_B>0$. 
There are constants $\const,\ \cst{}{0},\ \al_0,\ \ga_0$ and $\tau_0$
that are independent of $j,\ \Si,\ \vec\rho$ such that for all $\al\ge \al_0$,
$\veps>0$ 
and $\ga\le\ga_0$ the following holds:

{\parindent=.25in
\item{}
Let, for $\ka$ in a neighbourhood of zero, $u_\ka,v_\ka\in\cF_0(2;\Si)$ 
be antisymmetric, spin independent, particle number conserving functions.
Set
$$
C_\ka(k) = \sfrac{\nu^{(j)}(k)}{\imath k_0 -e(\k) -\check u_\ka(k)}
\qquad,\qquad 
D_\ka(k) = \sfrac{\nu^{(\ge j+1)}(k)}{\imath k_0 -e(\k) -\check v_\ka(k)}
$$
and let $C_\ka(\xi,\xi'),\,D_\ka(\xi,\xi')$ be the Fourier transforms 
of $C_\ka(k)$, $D_\ka(k)$. 
Let $B_\ka(k)$ be a function on $\bbbr\times\bbbr^2$ and set
$$
(\hat B_\ka\phi)(\xi)=\int d\xi'\  \hat B_\ka(\xi,\xi')\phi(\xi')
$$
Furthermore, let, for $\ka$ in a neighbourhood of zero, $\cW_\ka(\phi,\psi)$    
be an even Grassmann function and set
$$
\lw \cW'_\ka(\phi,\psi) \rw_{\psi,D_\ka} 
=  \Om_{C_\ka} \big( \lw \cW_\ka(\phi,\psi) \rw_{\psi,C_\ka+D_\ka} \big) (\phi,\psi+\hat B_\ka\phi)
$$

\vskip.3cm
Assume that the following estimates are fulfilled:
{\parindent = .5in
\item{$\bullet$} $\rho_{m+1;n-1}\le\ga\,\rho_{m;n}\ $
for all $m\ge 0$ and $n\ge 1$.
\item{$\bullet$}
$|\check u_0(k)|, |\check v_0(k)| \le \half |\imath k_0-e(k)|$
and $\big|\sfrac{d\hfill}{d\ka}\check v_\ka(k)\big|_{\ka=0}\big| 
      \le \veps |\imath k_0-e(k)|$
\item{$\bullet$}
$\v  u_0 \v_{1,\Si} \le \mu(\La+X)\fe_j(X)$ and
$\V \sfrac{d\hfill}{d\ka} u_\ka\big|_{\ka=0}\V_{1,\Si} \le \fe_j(X)Y$ with
$X,Y\in\fN_{d+1}$, $\mu,\La>0$ such that
$(1+\mu)(\La+X_\0)\le\sfrac{\tau_0}{M^j}$ 
\item{$\bullet$}
$\ \| B_0(k)\tnorm \le c_B \fe_j(X)\ $ and
$\ \big\|\sfrac{d\hfill}{d\ka} B_\ka(k)\Tnorm \le c_B \fe_j(X)Z\ $
with $Z\in\fN_{d+1}$
\item{$\bullet$}$\cW_\ka$ has a sectorized representative $w_\ka$ with
$$
\fn\equiv N_j^\sim(w_0;64\al;\,X,\Si,\vec\rho)
\le \cst{}{0}\,\al + \smsum_{\de \ne 0} \infty\,t^\de
$$

}
\vskip .3cm

\item{}
Then $\cW'_\ka$ has a sectorized representative $w'_\ka$ such that
$$\eqalign{
&N_j^\sim\big(\,\sfrac{d\hfill}{d\ka}
  \big[w'_\ka-w_\ka\big]_{\ka=0}
\,;\,\al\,;\, X,\Si,\vec\rho\,\big)\cr
&\hskip0.5in\le \const\,
\Big\{\ga+\sfrac{1}{\al^2}\sfrac{\fn}{1-{\const\over\al^2}\fn}\Big\}
N^\sim_j\big(\,\sfrac{d\hfill}{d\ka}w_\ka\big|_{\ka=0}
\,;\,16\al\,;\,{\sst X,\Si,\vec\rho}\,\big)\cr
&\hskip0.7in+ \const\,
     \sfrac{\fn}{1-{\const\over\al^2}\fn}
\Big\{\sfrac{1}{\al^2}M^{j}Y\fn+\sfrac{\veps}{\al^2}\fn+\ga Z\Big\}\cr
}$$

}
}

\lemma{\STM\lemOSmomconcretediffintconst}{
Under the hypotheses of Theorem \thOSmomrengroupdiffestimate, 
there exists a constant $\cst{}{2}$ that is independent of $j$ and $\Si$ 
such that
$\tilde C_{0,\Si}$ has contraction bound $\cb$, 
$\tilde C_{0,\Si}$ and $D_{0,\Si}$ have integral bound $\half\ib$
and 
$$\eqalign{
&\sfrac{d\hfill}{d\ka}\tilde C_{\ka,\Si}\big|_{\ka=0}\hbox{ has contraction bound\ \  }
 \cb'= \cst{}{2} M^{2j}\fe_j(X)Y\cr 
&\sfrac{d\hfill}{d\ka}\tilde D_{\ka,\Si}\big|_{\ka=0}\hbox{ has integral bound\ \  }
\half \ib'=\sqrt{\veps}\ \ib\cr
}$$
 for the family $\v \cdot\tv_{\Si}$ of symmetric seminorms.  
}

\prf
The contraction and integral bounds on $\tilde C_{0,\Si}$ and $\tilde D_{0,\Si}$
were proven in Lemma \lemOSmomconcreteintconst.
Clearly, the function
$$
\sfrac{d\hfill}{d\ka}D_\ka(k)
= \sfrac{d\hfill}{d\ka}
        \sfrac{\nu^{(\ge j+1)}(k)}{\imath k_0 -e(\k) -\check v_\ka(k)}
= \sfrac{\nu^{(\ge j+1)}(k)}{[\imath k_0 -e(\k) -\check v_\ka(k)]^2}
 \sfrac{d\hfill}{d\ka}\check v_\ka(k)
$$  
is supported on the $j^{\rm th}$ neighbourhood and obeys 
$\ \big|\sfrac{d\hfill}{d\ka}D_\ka(k)\big|_{\ka=0}\big|
           \le \sfrac{4\veps}{|\imath k_0 - e(\k)|}\ $. By
part (ii) of  Proposition \propOSmomcontrintboundsectors\ and the first
property of (\eqnOStilderhomn), $2\sqrt{4\IB_3\veps\sfrac{\fl}{M^j}}\le\sqrt{\veps}\ \ib $ 
is an integral bound for 
$\sfrac{d\hfill}{d\ka}\tilde D_{\ka,\Si}\big|_{\ka=0}\ $. 

 Set 
$\ 
\cb'' = 12\,\V \sfrac{d\hfill}{d\ka}c_\ka\big|_{\ka=0}\V_{1,\Si}
.\ $
By part (i) of Proposition \propOSmomcontrintboundsectors\ (see also Lemma
\lemimprconfig\ of [FKTr2]) and
the second property of $\vec\rho$ in (\eqnOStilderhomn), 
$\big(\sfrac{d\hfill}{d\ka}c_\ka\big|_{\ka=0}\big)_\Si$ has contraction 
bound $\cb''$. We showed in Lemma \lemOSconcretediffintconst\ that
$$
\cb'' \le \const\,M^{2j}\,\fe_j(X)Y
$$
\endproof

\lemma{\STM\lemOStildesourcetermderiv}{
Let $g(\phi,\psi)$ be a sectorized Grassmann function and set
$$
g'_\ka(\phi,\psi)=g(\phi,\psi+\hat B_\ka \phi)
$$
Under the hypotheses of Theorem \thOSmomrengroupdiffestimate, 
$$
N^\sim_j(\sfrac{d\hfill}{d\ka}g'_\ka\big|_{\ka=0};\al;\,X,\Si,\vec\rho)
\le \const\ga\,Z\,N^\sim_j(g;2\al;\,X,\Si,\vec\rho)
$$

  }
\prf Define, as in Lemma \lemOStildesourceterm, 
for $\check\et_{m+1}=\big(k_{m+1},\si_{m+1},a_{m+1}\big)$,
$$\eqalign{
&\varphi'_\ka({\sst\check\et_1,\cdots,\check\et_{m+1}\,;
\,(\xi_1,s_1)\cdots,(\xi_{n-1},s_{n-1})})\cr
&\hskip.2cm= {\rm Ant_{ext}}\smsum_{s\in\Si}\int\!\! d{\sst \ze}\ 
B_\ka({\sst k_{m+1}})\,E_+({\sst \check\et_{m+1},\ze})
\varphi({\sst\check\et_1,\cdots,\check\et_m\,;\,(\xi_1,s_1)\cdots,
(\xi_{i-1},s_{i-1}),(\ze,s),(\xi_{i},s_i),\cdots,(\xi_{n-1},s_{n-1})}) \cr
}$$
if $n\ge 2$ and 
$$\eqalign{
&\varphi'_\ka({\sst\check\et_1,\cdots,\check\et_{m+1}})
{\sst (2\pi)^{d+1}\de(k_1+\cdots+k_{m+1})}
= {\rm Ant_{ext}}\smsum_{s\in\Si}\int\!\! d{\sst \ze}\ 
B_\ka({\sst k_{m+1}})\,E_+({\sst \check\et_{m+1},\ze})
\varphi({\sst\check\et_1,\cdots,\check\et_m\,;\,(\ze,s)}) \cr
&\hskip1in= {\rm Ant_{ext}}\smsum_{s\in\Si}
B_\ka({\sst k_{m+1}})
\varphi({\sst\check\et_1,\cdots,\check\et_m\,;\,(0,\si_{m+1}, a_{m+1},s)})\ 
{\sst (2\pi)^{d+1}\de(k_1+\cdots+k_{m+1})} \cr
}$$
if $n=1$.
By (\eqnNPderivamp), with $X'=X_B=1$,
$$
\fe_j(X)\,\v\varphi'_0\tv_\Si\le\const\ga\, \fe_j(X)\ \v \varphi\tv_\Si
\EQN\eqnNPderivampB$$
 and by the same derivation as led to (\eqnNPderivamp), but with $X'=X_B=Z$,
$$
\fe_j(X)\,\V \sfrac{d\hfill}{d\ka}\varphi'_\ka\big|_{\ka=0}\tV_\Si\le\const\ga\, \fe_j(X)Z\ \v \varphi\tv_\Si
\EQN\eqnNPderivampC$$

As in Lemma \lemOStildesourceterm,
write $g(\phi,\psi) = \sum_{m,n}g_{m,n}(\phi,\psi)$, 
with $g_{m,n}$ of degree $m$ in $\phi$ and degree $n$ in $\psi$, and 
$$
g(\phi,\psi+\ze) =  \sum_{m,n}g_{m,n}(\phi,\psi+\ze)
=\sum_{m,n}\smsum_{\ell=0}^ng_{m,n-\ell,\ell}(\phi,\psi,\ze)
$$ 
with $g_{m,n-\ell,\ell}$ of degrees $m$ in $\phi$, $n-\ell$ in $\psi$  and $\ell$
in $\ze$. Let $G_{m,n}$ and $G_{\ka;m,n-\ell,\ell}$ be the kernels of 
$g_{m,n}(\phi,\psi)$ and $g_{m,n-\ell,\ell}(\phi,\psi,\hat B_\ka\phi)$ respectively. 
By the binomial theorem, Leibniz, one application of (\eqnNPderivampC) and 
$\ell-1$ applications of (\eqnNPderivampB), 
$$
\fe_j(X)\V\sfrac{d\hfill}{d\ka} G^\sim_{\ka;m,n-\ell,\ell}\big|_{\ka=0}\tV_\Si
\le 
(\const\ga)^\ell\,\ell\, {\tst{n\choose \ell}} \fe_j(X)Z\,\v G^\sim_{m,n}\tv_\Si
$$
Since $G^\sim_{\ka;m,n,0}$ is independent of $\ka$,
$$\eqalign{
N^\sim_j(\sfrac{d\hfill}{d\ka}g'_\ka\big|_{\ka=0};\al;&\,X,\Si,\vec\rho)\le \sfrac{M^{2j}}{\fl}\fe_j(X)\sum_{m,n\ge 0}\smsum_{\ell=1}^n
\al^{m+n}\,\big(\sfrac{\fl\,\IB}{M^j}\big)^{(m+n)/2} \,
\V\sfrac{d\hfill}{d\ka} G^\sim_{\ka;m,n-\ell,\ell}\big|_{\ka=0}\tV_\Si
\cr
&\le \sfrac{M^{2j}}{\fl}\fe_j(X)Z\sum_{m,n\ge 0}\smsum_{\ell=1}^n
\ell{\tst{n\choose \ell}}(\const\ga)^\ell
\al^{m+n}\,\big(\sfrac{\fl\,\IB}{M^j}\big)^{(m+n)/2} \,
\v G^\sim_{m,n}\tv_\Si
\cr
&= \sfrac{M^{2j}}{\fl}\fe_j(X)Z\sum_{m,n\ge 0}\smsum_{\ell=1}^n
n{\tst{n-1\choose \ell-1}}(\const\ga)^\ell
\al^{m+n}\,\big(\sfrac{\fl\,\IB}{M^j}\big)^{(m+n)/2} \,
\v G^\sim_{m,n}\tv_\Si
\cr
&= \sfrac{M^{2j}}{\fl}\fe_j(X)Z\sum_{m,n\ge 0}\const\ga\,
n(1+\const\ga)^{n-1}
\al^{m+n}\,\big(\sfrac{\fl\,\IB}{M^j}\big)^{(m+n)/2} \,
\v G^\sim_{m,n}\tv_\Si
\cr
&\le \sfrac{M^{2j}}{\fl}\fe_j(X)Z \sum_{m,n\ge 0} 
\const\ga\,2^{n}
\al^{m+n}\,\big(\sfrac{\fl\,\IB}{M^j}\big)^{(m+n)/2} \,
\v G^\sim_{m,n}\tv_\Si
\cr
&\le \const\ga\,Z\, N^\sim_j(g;2\al;\,X,\Si,\vec\rho)\cr
}$$
\endproof

\proof{ of  Theorem \thOSmomrengroupdiffestimate}
As in the proof of Theorem \thOSmomrengroupestimate,
let, for a sectorized Grassmann function $v=\smsum_{n} v_{n}$ with
$v_{n} \in \bigwedge^n\tilde V$,
$$
N^\sim(v;\al) = \sfrac{1}{\ib^2}\cb\,\smsum_{n} \al^n\, \ib^n\, \v v_{n} \tv_\Si
=\sfrac{\cst{}{1}}{\IB}\, N^\sim_j(v;\al;\,X,\Si,\vec\rho)
$$
and
$$
\ \lw w''_\ka\rw_{\psi,\tilde D_{\ka,\Si}} =
\Om_{\tilde C_{\ka,\Si}} \big(\lw w_\ka
\rw_{\psi,\tilde C_{\ka,\Si}+\tilde D_{\ka,\Si}}\big)
$$
By Proposition \propOSfunctorialitySect, parts (ii) and (iii), 
\ and Proposition \propBII.ii of [FKTr1],
$$ 
w'_\ka=w''_\ka\big(\phi,\psi+\hat B_\ka\phi\big)
$$
is a sectorized representative for $\cW'_\ka$.
By the chain rule and the triangle inequality
$$\eqalign{
N^\sim\big(\,\sfrac{d\hfill}{d\ka}
     \big[w'_\ka-w_\ka\big]_{\ka=0}\,\cl \al\big)
\ \le\  &N^\sim\big(\,\sfrac{d\hfill}{d\ka}
     w''_0\big(\phi,\psi+\hat B_\ka\phi\big)\big|_{\ka=0}\,\cl \al\big)\cr
&+N^\sim\big(\,\sfrac{d\hfill}{d\ka}
  [w''_\ka\big(\phi,\psi+\hat B_0\phi\big)-w''_\ka(\phi,\psi)]_{\ka=0}
                  \,\cl \al\big)\cr
&+N^\sim\big(\,\sfrac{d\hfill}{d\ka}
     [w''_\ka(\phi,\psi)-w_\ka(\phi,\psi)]_{\ka=0}\,\cl \al\big)\cr
}\EQN\eqnOSmomimprderivI$$
By Lemma \lemOStildesourcetermderiv,
$$\eqalign{
N^\sim\big(\,\sfrac{d\hfill}{d\ka}
     w''_0\big(\phi,\psi+\hat B_\ka\phi\big)\big|_{\ka=0}\,\cl \al\big)
&\le \const\ga\,Z\,N^\sim_j(w''_0;2\al;\,X,\Si,\vec\rho)\cr
}$$
By (\eqnOSmomsectgeneral),
$$\eqalign{
&N^\sim_j(w''_0;2\al;\,X,\Si,\vec\rho)
\le N^\sim_j\big(w_0;2\al;\,X,\Si,\vec\rho\big)
+N^\sim_j\big(w''_0-w_0;2\al;\,X,\Si,\vec\rho\big)\cr
&\hskip.5in
\le  N^\sim_j\big(w_0;2\al;\,X,\Si,\vec\rho\big)
+\sfrac{\IB}{\cst{}{1}} \sfrac{1}{8\al^2}\,
\sfrac{N^\sim(w_0;64\al)^2}{1-{1 \over4\al^2}N^\sim(w_0;64\al)}\cr
&\hskip.5in
\le N^\sim_j\big(w_0;2\al;\,X,\Si,\vec\rho\big)
+ \sfrac{\const}{\al^2}\,
\sfrac{N^\sim_j(w_0;64\al;\,X,\Si,\vec\rho)^2}
{1-{\const \over\al^2}N^\sim_j(w_0;64\al;\,X,\Si,\vec\rho)}\cr
&\hskip.5in
\le  \const\,\sfrac{N^\sim_j(w_0;64\al;\,X,\Si,\vec\rho)}
{1-{\const \over\al^2}N^\sim_j(w_0;64\al;\,X,\Si,\vec\rho)}\cr
}$$
so that
$$\eqalign{
N^\sim_j\big(\,\sfrac{d\hfill}{d\ka}
     w''_0\big(\phi,\psi+\hat B_\ka\phi\big)\big|_{\ka=0}\,\cl \al\,;\,
     X,\Si,\vec\rho\big)
&\le \const\ga\ \sfrac{\fn}{1-{\const \over\al^2}\fn}\,Z\cr
}\EQN\eqnOSmomimprderivIa$$
By Lemma \lemOStildesourceterm, with 
$g=\sfrac{d\hfill}{d\ka}w''_\ka\big|_{\ka=0}$, $B=B_0$ and $X_B=1$,
$$\eqalign{
&N^\sim_j\big(\,\sfrac{d\hfill}{d\ka}
  [w''_\ka\big(\phi,\psi+\hat B_0\phi\big)-w''_\ka(\phi,\psi)]_{\ka=0}
                  ; \al;X,\Si,\vec\rho\big)
\le \const\ga\,
N^\sim_j\big(\,\sfrac{d\hfill}{d\ka}w''_\ka\big|_{\ka=0}\,
              ;2\al;X,\Si,\vec\rho)\cr
&\hskip.5in\le \const\ga\,
N^\sim_j\big(\,\sfrac{d\hfill}{d\ka}w_\ka\big|_{\ka=0}\,;2\al
               ;\,X,\Si,\vec\rho\big)
+ \const\ga\,
N^\sim_j\big(\,\sfrac{d\hfill}{d\ka}
     [w''_\ka-w_\ka]_{\ka=0}\,;2\al;\,X,\Si,\vec\rho\big)\cr
}\EQN\eqnOSmomimprderivIb$$
By Theorem \theoremIVb\ of [FKTr1], with $\mu=M^j$ (and assuming that we
have chosen $\cst{}{1}\ge 1$),
$$\eqalign{
N^\sim\big(\,\sfrac{d\hfill}{d\ka}[w''_\ka-w_\ka]_{\ka=0}\,\cl \al\big)
&\le\ \sfrac{1}{2\al^2}\,
     \sfrac{N^\sim( w_0\cl 32\al)}{1-{1\over\al^2}N^\sim( w_0\cl 32\al)}
N^\sim\big(\,\sfrac{d\hfill}{d\ka}w_\ka\big|_{\ka=0}\,\cl 8\al\big) \cr
&\hskip1cm+\sfrac{1}{2\al^2}\,
     \sfrac{N^\sim( w_0\cl 32\al)^2}{1-{1\over\al^2}N^\sim( w_0\cl 32\al)}
\Big\{\sfrac{1}{4M^j}\cst{}{2}M^{2j}\fe_j(X)Y+4\veps\Big\}\cr
&\le \sfrac{\const}{\al^2}\,
     \sfrac{\fn}{1-{\const\over\al^2}\fn}
\Big\{N^\sim\big(\,\sfrac{d\hfill}{d\ka}w_\ka\big|_{\ka=0}\,\cl 8\al\big)
+M^{j}Y\fn+4\veps\fn\Big\}
}\EQN\eqnOSmomimprderivIc$$
since $\fe_j(X)N^\sim( w_0\cl 32\al)\le\const N^\sim( w_0\cl 32\al)$. Also
$$
N^\sim_j\big(\,\sfrac{d\hfill}{d\ka}
     [w''_\ka-w_\ka]_{\ka=0}\,;2\al;\,X,\Si,\vec\rho\big)
\le \sfrac{\const}{\al^2}\,
     \sfrac{\fn}{1-{\const\over\al^2}\fn}
\Big\{N^\sim\big(\,\sfrac{d\hfill}{d\ka}w_\ka\big|_{\ka=0}\,\cl 16\al\big)
+M^{j}Y\fn+4\veps\fn\Big\}
$$
Subbing (\eqnOSmomimprderivIa--\eqnOSmomimprderivIc) into (\eqnOSmomimprderivI),
$$\eqalign{
N^\sim\big(\,\sfrac{d\hfill}{d\ka}
     \big[w'_\ka-w_\ka\big]_{\ka=0}\,\cl \al\big)
&\le\ \const\ga\,\,\sfrac{\fn}{1-{\const \over\al^2}\fn}Z
+\const\ga\, \ 
N^\sim\big(\,\sfrac{d\hfill}{d\ka}w_\ka\big|_{\ka=0}\,;2\al\big)\cr
&\ \ \ +(1+\ga\,)\sfrac{\const}{\al^2}\,
     \sfrac{\fn}{1-{\const\over\al^2}\fn}
\Big\{N^\sim\big(\,\sfrac{d\hfill}{d\ka}w_\ka\big|_{\ka=0}\,\cl 16\al\big)
+M^{j}Y\fn+4\veps\fn\Big\}\cr
&\hskip-1in\le \const\,
\Big\{\ga+\sfrac{1}{\al^2}\sfrac{\fn}{1-{\const\over\al^2}\fn}\Big\}
N^\sim_j\big(\,\sfrac{d\hfill}{d\ka}w_\ka\big|_{\ka=0}
\,;\,16\al\,;\,{\sst X,\Si,\vec\rho}\,\big)\cr
&\hskip0.7in+ \const\,
     \sfrac{\fn}{1-{\const\over\al^2}\fn}
\Big\{\sfrac{1}{\al^2}M^{j}Y\fn+\sfrac{\veps}{\al^2}\fn+\ga Z\Big\}\cr\cr
}$$

\endproof

\remark{\STM\remOSmomchoiceofrep}{
In Theorem \thOSmomrengroupestimate, 
the sectorized representative $w'$ of $\cW'$ may be obtained from 
the sectorized representative $w$ of $\cW$ by
$$
\lw w'\rw_{\psi,\tilde D_\Si} = 
\Om_{\tilde C_\Si}(\lw w \rw_{\psi,\tilde C_\Si+\tilde D_\Si})
(\phi,\psi+ \hat B\phi)
$$
The obvious analog of this statement applies to Theorem 
\thOSmomrengroupdiffestimate.

}

\vfill\eject

\appendix{\APappNaiveladder}{Naive Ladder Estimates}\PG\pgOSD

Let $j\ge 2$ and let $\Si$ be a sectorization of scale $j$ and length
$\sfrac{1}{M^{j-3/2}}\le\fl\le\sfrac{1}{M^{(j-1)/2}}$. 
To systematically treat ladders, we introduce an auxiliary channel norm,
similar to the $\v\ \cdot\ \tv_{2,\Si}$ norm, but with only the leftmost 
momenta held fixed.

\definition{\STM\defOSchannelnorm}{
\Item i)
Let $0\le r\le 2$ and $f\in\check\cF_r(4-r,\Si)$. 
We set
$$
\v f\tv_{ch,\Si} 
= \hskip-7pt\sup_{\check\eta_1,\cdots,\check\eta_r \in \check \cB\atop
 s_{1},\cdots,s_{2-r}\in\Si} 
\smsum_{s_{3-r},s_{4-r} \in \Si} \
\smsum_{\de\in \bbbn_0\times\bbbn_0^2} \sfrac{1}{\de!}
\max_{\rD\, {\rm dd-operator} \atop{\rm with\ } \de(\rD) =\de} \TN\rD f\,({\sst\check\eta_1,\cdots,\check\eta_r;(\xi_1,s_1),\cdots,(\xi_{4-r},s_{4-r})})
\TN_{1,\infty}\ t^\de
$$
The norm
$\tn\,\cdot\,\tn_{1,\infty}$ of Example \exOSSymmNorm\ refers to the variables 
${\sst \xi_1,\cdots,\xi_{4-r}}$. 
If $r=0$, we also write $\v f\v_{ch,\Si}$ instead of $\v f\tv_{ch,\Si}$.
\Item ii)
If $f\in\check\cF_{4,,\Si}$, we set
$$
\v f\tv_{ch,\Si} 
=\sum_{i_1,i_2\in\{0,1\}}\V\ord f\big|_{(i_1,i_2,1,1)}\tV_{ch,\Si}
$$

}
\lemma{\STM\lemchannelnorm}{ There is a constant $\abcst$, independent of $j$ and 
$M$ such that the following hold. 
Let $0\le r\le 2$ and $f\in\check\cF_r(4-r,\Si)$.
\Item i)
$$\eqalign{
\v f\tv_{ch,\Si}&\le \v f\tv_{1,\Si}\qquad\qquad\hbox{if $r\le 1$}\cr
\v f\tv_{ch,\Si}&\le \sfrac{\abcst}{\fl}\v f\tv_{3,\Si}\cr
}$$

\Item ii)
$$\eqalign{
\v f\tv_{4,\Si} &\le \v f\tv_{3,\Si}\cr
\v f\tv_{3,\Si} &\le \abcst\,\v f\tv_{4,\Si}\cr
}$$

\Item iii) If $r=1$ or if $r=0$ and $f$ is antisymmetric, then
$$
\v f\tv_{1,\Si}\le \sfrac{\abcst}{\fl}\v f\tv_{ch,\Si}
$$
}
\prf Set
$$
F({\sst\check\eta_1,\cdots,\check\eta_r;s_1,\cdots,s_{4-r}}) 
= 
\smsum_{\de\in \bbbn_0\times\bbbn_0^2} \sfrac{1}{\de!}
\max_{\rD\, {\rm dd-operator} \atop{\rm with\ } \de(\rD) =\de} \TN\rD f\,({\sst\check\eta_1,\cdots,\check\eta_r;(\xi_1,s_1),\cdots,(\xi_{4-r},s_{4-r})})
\TN_{1,\infty}\ t^\de
$$
\Item i)
Then
$$\eqalign{
\v f\tv_{ch,\Si} 
&= \sup_{\check\eta_1,\cdots,\check\eta_r \in \check \cB
                              \atop s_{1},\cdots,s_{2-r}\in\Si} 
   \smsum_{s_{3-r},s_{4-r} \in \Si} \
   F({\sst\check\eta_1,\cdots,\check\eta_r;s_1,\cdots,s_{4-r}})\cr 
&\le  \sup_{{1\le i_1<\cdots<i_{1-r}\le 4-r \atop s_{i_1},\cdots,s_{i_{1-r}}\in\Si}
\atop \check\eta_1,\cdots,\check\eta_r \in \check \cB} 
\smsum_{s_i \in \Si \ {\rm for}\atop i\ne i_1,\cdots i_{1-r}} \
   F({\sst\check\eta_1,\cdots,\check\eta_r;s_1,\cdots,s_{4-r}})
=\v f\tv_{1,\Si}\cr 
}$$
if $r\le 1$ and, since $\Si$ contains at most $\sfrac{\abcst}{\fl}$ elements,
$$\eqalign{
\v f\tv_{ch,\Si} 
&= \sup_{\check\eta_1,\cdots,\check\eta_r \in \check \cB
                              \atop s_{1},\cdots,s_{2-r}\in\Si} 
   \smsum_{s_{3-r},s_{4-r} \in \Si} \
   F({\sst\check\eta_1,\cdots,\check\eta_r;s_1,\cdots,s_{4-r}})\cr 
&\le \sfrac{\abcst}{\fl}\sup_{\check\eta_1,\cdots,\check\eta_r \in \check \cB
                              \atop s_{1},\cdots,s_{3-r}\in\Si} 
   \smsum_{s_{4-r} \in \Si} \
   F({\sst\check\eta_1,\cdots,\check\eta_r;s_1,\cdots,s_{4-r}})\cr 
&\le  \sfrac{\abcst}{\fl}\sup_{{1\le i_1<\cdots<i_{3-r}\le 4-r \atop s_{i_1},\cdots,s_{i_{3-r}}\in\Si}
\atop \check\eta_1,\cdots,\check\eta_r \in \check \cB} 
\smsum_{s_i \in \Si \ {\rm for}\atop i\ne i_1,\cdots i_{3-r}} \
   F({\sst\check\eta_1,\cdots,\check\eta_r;s_1,\cdots,s_{4-r}})
=\sfrac{\abcst}{\fl}\v f\tv_{3,\Si}\cr 
}$$
\Item ii)
$$\eqalign{
\v f\tv_{4,\Si} 
&= \sup_{ s_{1},\cdots,s_{4-r}\in\Si
\atop \check\eta_1,\cdots,\check\eta_r \in \check \cB} 
F({\sst\check\eta_1,\cdots,\check\eta_r;s_1,\cdots,s_{4-r}})\cr
&\le \sup_{{1\le i_1<\cdots<i_{3-r}\le 4-r \atop s_{i_1},\cdots,s_{i_{3-r}}\in\Si}
\atop \check\eta_1,\cdots,\check\eta_r \in \check \cB} 
\smsum_{s_i \in \Si \ {\rm for}\atop i\ne i_1,\cdots i_{3-r}} \
F({\sst\check\eta_1,\cdots,\check\eta_r;s_1,\cdots,s_{4-r}})
=\v f\tv_{3,\Si}\cr
&\le\abcst\, \sup_{ s_{1},\cdots,s_{4-r}\in\Si
\atop \check\eta_1,\cdots,\check\eta_r \in \check \cB} 
F({\sst\check\eta_1,\cdots,\check\eta_r;s_1,\cdots,s_{4-r}})
=\abcst\,\v f\tv_{4,\Si} \cr
}$$
since, by conservation of momentum, for any fixed 
$s_{i_1},\cdots,s_{i_{3-r}}$ and $\check\eta_1,\cdots,\check\eta_r$, there
at most $\abcst$ choices of $s_i,\ i\ne i_1,\cdots i_{3-r}$ for which 
$F({\sst\check\eta_1,\cdots,\check\eta_r;s_1,\cdots,s_{4-r}})$ does not vanish.
\Item iii)  If $r=1$ or if $r=0$ and $f$ is antisymmetric, then
$$\eqalign{
\v f\tv_{1,\Si} 
&= \sup_{{1\le i_1<\cdots<i_{1-r}\le 4-r \atop s_{i_1},\cdots,s_{i_{1-r}}\in\Si}
\atop \check\eta_1,\cdots,\check\eta_r \in \check \cB} 
\smsum_{s_i \in \Si \ {\rm for}\atop i\ne i_1,\cdots i_{1-r}} \
F({\sst\check\eta_1,\cdots,\check\eta_r;s_1,\cdots,s_{4-r}})\cr
&= \sup_{ s_1,\cdots,s_{1-r}\in\Si
\atop \check\eta_1,\cdots,\check\eta_r \in \check \cB} 
\smsum_{s_{2-r},\cdots,s_{4-r} \in \Si} \
F({\sst\check\eta_1,\cdots,\check\eta_r;s_1,\cdots,s_{4-r}})\cr
&\le\sfrac{\abcst}{\fl} \sup_{ s_1,\cdots,s_{2-r}\in\Si
\atop \check\eta_1,\cdots,\check\eta_r \in \check \cB} 
\smsum_{s_{3-r},s_{4-r} \in \Si} \
F({\sst\check\eta_1,\cdots,\check\eta_r;s_1,\cdots,s_{4-r}})
=\sfrac{\abcst}{\fl}\v f\tv_{ch,\Si} 
}$$
\endproof
\corollary{\STM\corchannelnorm}{ There is a constant $\abcst$, independent of 
$j$ and $M$ such that for all $f\in\check\cF_{4;\Si}$
$$\eqalign{
\v f\tv_{ch,\Si}&\le \sfrac{\abcst}{\fl}\v f\tv_{3,\Si}\cr
}$$

}

\lemma {\STM\lemOSNaivetensor}{
Let $f_1,f_2\in\check\cF_{4;\Si}$ and $c,d \in \cF_0(2;\Si)$. 
Define propagators
$$\eqalign{
 c_\Si\big((\xi,s),(\xi',s')\big)
& =\sum_{t\cap s \ne \emptyset \atop t'\cap s' \ne \emptyset}
c\big((\xi,t),(\xi',t')\big) \cr
d_\Si\big((\xi,s),(\xi',s')\big)
& = \sum_{t\cap s \ne \emptyset \atop t'\cap s' \ne \emptyset}
d\big((\xi,t),(\xi',t')\big) \cr
}$$
over $\cB\times\Si$. Then
$$\eqalign{
\V f_1 \circ (c_\Si\otimes d_\Si) \circ f_2 \tV_{1,\Si} 
&\le \abcst\,  
 \tn d\tn_\infty\,
\v c \v_{1,\Si}\ \v f_1 \tv_{1,\Si} \v f_2 \tv_{1,\Si} \cr
\V f_1 \circ (c_\Si\otimes d_\Si) \circ f_2 \tV_{ch,\Si} 
&\le \abcst\,  
 \tn d\tn_\infty\,
\v c \v_{1,\Si}\ \v f_1 \tv_{ch,\Si} \v f_2 \tv_{ch,\Si} \cr
\V f_1 \circ (c_\Si\otimes d_\Si) \circ  f_2 \tV_{3,\Si} 
&\le \abcst\,  
 \tn d\tn_\infty\,
\v c \v_{1,\Si}\ \v f_1 \tv_{ch,\Si} \v f_2 \tv_{3,\Si}  \cr
}$$
where $\tn d\tn_\infty
 = \max_{s,s'\in\Si} \sup_{\xi,\xi'} |d({\sst (\xi,s),(\xi',s')})|$.

}

\prf Set
$$\eqalign{
f({\sst \,\cdot\,,\,\cdot\,,\,\cdot\,,\,\cdot\,;(\xi,s),(\xi',s')}) 
&= \smsum_{s'_1,s_3' \in \Si}
 \int {\sst d\ze_1\,d\ze_3}\ 
f_1({\sst  \,\cdot\,, \,\cdot\,,(\ze_3,s_3'),(\xi,s)})\
c_\Si( {\sst (\ze_3,s_3'),(\ze_1,s_1')})\
f_2({\sst (\ze_1,s_1'), (\xi',s'), \,\cdot\,, \,\cdot\,})\cr
&= \smsum_{s'_3,s_3'' \in \Si \atop s'_1,s_1'' \in \Si }
 \int {\sst d\ze_1\,d\ze_3}\ 
f_1({\sst \,\cdot\,, \,\cdot\,,(\ze_3,s_3'),(\xi,s)})\
c( {\sst (\ze_3,s_3''),(\ze_1,s_1'')})\
f_2({\sst (\ze_1,s_1'), (\xi',s'), \,\cdot\,, \,\cdot\,})\cr
}
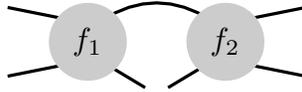
$$

\centerline{\figput{brokenbubble}}

\noindent
By iterated application of Lemma \lemOSelloneinftyampsectors,
$$
\v f\tv_{1,\Si}
\le \abcst\, \v c \v_{1,\Si} \v f_1 \tv_{1,\Si} \v f_2 \tv_{1,\Si} 
$$
Since
$$\eqalign{
f_1 \circ (c_\Si\otimes d_\Si) \circ f_2 
&=\smsum_{s,s'\in\Si}
 \int {\sst d\xi\,d\xi'}\
f({\sst \,\cdot\,,\,\cdot\,,\,\cdot\,,\,\cdot\,;(\xi,s),(\xi',s')}) \
d_\Si({\sst (\xi,s),(\xi',s')})\cr
&=\smsum_{s,s',t,t'\in\Si \atop \tilde s\cap \tilde t \ne \emptyset,\ 
\tilde s'\cap \tilde t' \ne \emptyset}
 \int {\sst d\xi\,d\xi'}\
f({\sst \,\cdot\,,\,\cdot\,,\,\cdot\,,\,\cdot\,;(\xi,s),(\xi',s')}) \
d({\sst (\xi,t),(\xi',t')})\cr
}$$
we have
$$
\V f_1 \circ (c_\Si\otimes d_\Si) \circ f_2 \tV_{1,\Si} 
\le \abcst\,\tn d\tn_\infty\,\v f\tv_{1,\Si}
$$
This proves the first inequality of the Lemma. To prove the third inequality,
set 
$$
g({\sst \,\cdot\,,\,\cdot\,,\,\cdot\,,\,\cdot\,;(\xi,s),\xi'})
= \smsum_{\tilde s'\cap \tilde s \ne \emptyset}
f({\sst \,\cdot\,,\,\cdot\,,\,\cdot\,,\,\cdot\,;(\xi,s),(\xi',s')})
$$
By conservation of momentum 
$$
f_1 \circ (c_\Si\otimes d_\Si) \circ f_2 
=\smsum_{s,t,t'\in\Si \atop \tilde s\cap \tilde t \ne \emptyset,\ 
\tilde t\cap \tilde t' \ne \emptyset}
 \int {\sst d\xi\,d\xi'}\
g({\sst \,\cdot\,,\,\cdot\,,\,\cdot\,,\,\cdot\,;(\xi,s),\xi'}) \
d({\sst (\xi,t),(\xi',t')})
\EQN\eqOSNaivetensor $$
Fix any $\vec\imath = (i_1,\cdots,i_4) \in \{0,1\}^4$ and let
$$\eqalign{
f_{\vec\imath}({\sst \,\cdot\,,\,\cdot\,,\,\cdot\,,\,\cdot\,;(\xi,s),(\xi',s')}) 
&= \smsum_{s'_3,s_3'' \in \Si \atop s'_1,s_1'' \in \Si }
 \int {\sst d\ze_1\,d\ze_3}\ 
f_1\big|_{(i_1,i_2,1,1)}({\sst \,\cdot\,, \,\cdot\,,(\ze_3,s_3'),(\xi,s)})\
c( {\sst (\ze_3,s_3''),(\ze_1,s_1'')})\cr
\noalign{\vskip-.2in}
&\hskip2.4in f_2\big|_{(1,1,i_3,i_4)}({\sst (\ze_1,s_1'), (\xi',s'), \,\cdot\,, \,\cdot\,})\cr
g_{\vec\imath}({\sst \,\cdot\,,\,\cdot\,,\,\cdot\,,\,\cdot\,;(\xi,s),\xi'})
&= \smsum_{\tilde s'\cap \tilde s \ne \emptyset}
f_{\vec\imath}({\sst \,\cdot\,,\,\cdot\,,\,\cdot\,,\,\cdot\,;(\xi,s),(\xi',s')})\cr
}$$
By (\eqOSNaivetensor)
$$
f_1 \circ (c_\Si\otimes d_\Si) \circ f_2 \big|_{\vec\imath}
=\smsum_{s,t,t'\in\Si \atop \tilde s\cap \tilde t \ne \emptyset,\ 
\tilde t\cap \tilde t' \ne \emptyset}
 \int {\sst d\xi\,d\xi'}\
g_{\vec\imath}({\sst \,\cdot\,,\,\cdot\,,\,\cdot\,,\,\cdot\,;(\xi,s),\xi'}) \
d({\sst (\xi,t),(\xi',t')})
\EQN\eqOSNaivetensorII$$
For each $\nu=1,\cdots,4$,
fix $ \check\eta_\nu\in \check\cB$ when $i_\nu=0$ and
$s_\nu \in \Si$  when $i_\nu=1$. Let
$$
z_\nu=\cases{\check\eta_\nu & if $i_\nu=0$\cr
             (\xi_\nu,s_\nu) & if $i_\nu=1$\cr}
$$
and
$$
G_{\vec\imath}= \smsum_{s\in \Si} \
\smsum_{\de\in \bbbn_0\times\bbbn_0^2} \sfrac{1}{\de!}
\max_{\rD\, {\rm dd-operator} \atop{\rm with\ } \de(\rD) =\de} 
\TN\rD\,
g_{\vec\imath}({\sst z_1,\cdots,z_4;\,(\xi,s),\xi'})
\TN_{1,\infty}\ t^\de
$$
By iterated application of Leibniz's rule and Lemma \lemchannelnorm.ii,
$$\eqalign{
G_{\vec\imath} &\le \abcst\, \v c \v_{1,\Si}\  
\V f_1\big|_{(i_1,i_2,1,1)} \tV_{ch,\Si}\ 
\V f_2\big|_{(1,1,i_3,i_4)} \tV_{4,\Si} \cr
&\le \abcst\, \v c \v_{1,\Si}\  
\V f_1\big|_{(i_1,i_2,1,1)} \tV_{ch,\Si}\ 
\V f_2\big|_{(1,1,i_3,i_4)} \tV_{3,\Si} \cr
&\le \abcst\, \v c \v_{1,\Si}\  \v f_1\tv_{ch,\Si} \ \v f_2\tv_{3,\Si} 
}$$
Furthermore, by (\eqOSNaivetensorII),
$$
\smsum_{\de\in \bbbn_0\times\bbbn_0^2}\sfrac{1}{\de!}
\max_{\rD\, {\rm dd-operator} \atop{\rm with\ } \de(\rD) =\de} 
\TN\rD\,
f_1 \circ (c_\Si\otimes d_\Si) \circ f_2\big|_{\vec\imath}
({\sst z_1,\cdots,z_4})
\TN_{1,\infty}\ t^\de 
 \le 9\,\tn d\tn_\infty G_{\vec\imath}
$$
This, together with part (ii) of Lemma \lemchannelnorm, shows that
$$
\V f_1 \circ (c_\Si\otimes d_\Si) \circ  f_2 \big|_{\vec\imath}\tV_{3,\Si} 
\ \le\ \abcst\,\V f_1\circ(c_\Si\otimes d_\Si)\circ f_2\big|_{\vec\imath}  
\tV_{4,\Si}   
\ \le\ \abcst\,\tn d\tn_\infty\,
\v c \v_{1,\Si}\ \v f_1 \tv_{ch,\Si} \v f_2 \tv_{3,\Si}
$$

The proof of the second inequality is similar to that of the third.
Choose $\vec\imath = (i_1,i_2,1,1)$ with $i_1,i_2 \in \{0,1\}$. For each $\nu=1,2$,
fix $ \check\eta_\nu\in \check\cB$ when $i_\nu=0$ and
$s_\nu \in \Si$  when $i_\nu=1$. Let
$$
z_\nu=\cases{\check\eta_\nu & if $i_\nu=0$\cr
             (\xi_\nu,s_\nu) & if $i_\nu=1$\cr}
$$
and
$$
G'_{\vec\imath}= \smsum_{s_3,s_4,s\in \Si} \
\smsum_{\de\in \bbbn_0\times\bbbn_0^2} \sfrac{1}{\de!}
\max_{\rD\, {\rm dd-operator} \atop{\rm with\ } \de(\rD) =\de} 
\TN\rD\,
g_{\vec\imath}({\sst z_1,z_2,(\xi_3,s_3),(\xi_4,s_4);\,(\xi,s),\xi'})
\TN_{1,\infty}\ t^\de
$$
By iterated application of Leibniz's rule,
$$\eqalign{
G'_{\vec\imath} &\le \abcst\, \v c \v_{1,\Si}\  
\V f_1\big|_{(i_1,i_2,1,1)} \tV_{ch,\Si}\ 
\V f_2\big|_{(1,1,1,1)} \tV_{ch,\Si} \cr
&\le \abcst\, \v c \v_{1,\Si}\  \v f_1\tv_{ch,\Si} \ \v f_2\tv_{ch,\Si} 
}$$
Furthermore, by (\eqOSNaivetensorII),
$$
\smsum_{s_3,s_4\in \Si}\smsum_{\de\in \bbbn_0\times\bbbn_0^2}\sfrac{1}{\de!}
\max_{\rD\, {\rm dd-operator} \atop{\rm with\ } \de(\rD) =\de} 
\TN\rD\,
f_1 \circ (c_\Si\otimes d_\Si) \circ f_2\big|_{\vec\imath}
({\sst z_1,z_2,(\xi_3,s_3),(\xi_4,s_4)})
\TN_{1,\infty}\ t^\de 
 \le 9\,\tn d\tn_\infty G'_{\vec\imath}
$$
This shows that
$$
\V f_1 \circ (c_\Si\otimes d_\Si) \circ  f_2 \big|_{\vec\imath}\tV_{ch,\Si} 
\ \le\ \abcst\,\tn d\tn_\infty\,
\v c \v_{1,\Si}\ \v f_1 \tv_{ch,\Si} \v f_2 \tv_{ch,\Si}
$$
\endproof

\lemma {\STM\lemOSNaiveBubble}{
Let $f_1,f_2\in\check\cF_{4;\Si}$ be momentum conserving functions. Also 
let $u,v,u',v' \in \cF_0(2;\Si)$ be antisymmetric, spin independent, particle number conserving functions whose Fourier transforms  obey
$|\check u(k)|,|\check v(k)|,|\check u'(k)|,|\check v'(k)|
\le \half |\imath k_0-e(k)|$. Let $X\in\fN_{3}$ and $0<\veps\le 1$ such that
$$
\v u\v_{1,\Si},\,\v u'\v_{1,\Si} \le \sfrac{1}{2M^j}\,X \qquad\qquad 
\v u-u'\v_{1,\Si}\le \sfrac{\veps}{M^j}\, X
$$
and 
$$
|\check u(k) -\check u'(k)|,\, |\check v(k) -\check v'(k)|
\le \veps |\imath k_0-e(k)|
$$
Set
$$\meqalign{
C(k) &= \sfrac{\nu^{(j)}(k)}{\imath k_0 -e(\k) -\check u(k)}
\qquad,\qquad 
D(k) &= \sfrac{\nu^{(\ge j+1)}(k)}{\imath k_0 -e(\k) -\check v(k)} \cr
C'(k) &= \sfrac{\nu^{(j)}(k)}{\imath k_0 -e(\k) -\check u'(k)}
\qquad,\qquad 
D'(k) &= \sfrac{\nu^{(\ge j+1)}(k)}{\imath k_0 -e(\k) -\check v'(k)} \cr
}$$
and let $C(\xi,\xi'),\,D(\xi,\xi'),C'(\xi,\xi'),\,D'(\xi,\xi')$ be their Fourier transforms as in Definition \defOSftcov. Assume that
$X_\0 \le \min\{\tau_1,\tau_2\}$, where $\tau_1$ and $\tau_2$ are
the constants of Proposition \propOSrealpropbound\ and
Lemma \lemOSdiffpropbound, respectively.
\Item i)
$$\eqalign{
\V f_1 \bullet \cC(C,D)\bullet f_2  \tV_{1,\Si}
& \le \const \sfrac{\fl\,\cb_j}{1-X}\
\v f_1 \tv_{1,\Si}  \v f_2\tv_{1,\Si}   \cr 
\V f_1 \bullet \cC(C,D)\bullet  f_2  \tV_{ch,\Si}
& \le \const \sfrac{\fl\,\cb_j}{1-X}\
\v f_1 \tv_{ch,\Si}  \v f_2 \v_{ch,\Si}   \cr 
\V f_1 \bullet \cC(C,D)\bullet f_2 \tV_{3,\Si}
&\le  \const \sfrac{\fl\,\cb_j}{1-X}\
\v f_1 \tv_{ch,\Si} \v f_2 \tv_{3,\Si} \cr 
}$$
\Item ii)
$$\eqalign{
\V f_1 \!\bullet\! \cC({\sst C,D})\!\bullet\! f_2 
-  f_1 \!\bullet\! \cC({\sst C',D'})\!\bullet\! f_2 \tV_{1,\Si}
&\le \const \,\veps\,\fl\,\sfrac{\cb_j(1+X)}{1-X}\,
\v f_1 \tv_{1,\Si} \v f_2\tv_{1,\Si}   \cr 
\V f_1 \!\bullet\! \cC({\sst C,D})\!\bullet\! f_2
-  f_1 \!\bullet\! \cC({\sst C',D'})\!\bullet\! f_2 \tV_{ch,\Si}
&\le \const \,\veps\,\fl\,\sfrac{\cb_j(1+X)}{1-X}\,
\v f_1 \tv_{ch,\Si} \v f_2 \v_{ch,\Si}   \cr
\V f_1\! \bullet\! \cC({\sst C,D})\!\bullet \!f_2 
-  f_1 \!\bullet\!  \cC({\sst C',D'}) \!\bullet\! f_2 \tV_{3,\Si}
&\le  \const \,\veps\,\fl\,\sfrac{\cb_j(1+X)}{1-X}
\v f_1 \tv_{ch,\Si} \v f_2 \tv_{3,\Si}  \cr 
}$$
}

\prf
Let $c({\sst (\cdot,s),(\cdot,s')})$, $d({\sst (\cdot,s),(\cdot,s')})$,
$c'({\sst (\cdot,s),(\cdot,s')})$, $d'({\sst (\cdot,s),(\cdot,s')})$
be the Fourier transforms of 
$\chi_s(k)\,C(k)\,\chi_{s'}(k)$, $\chi_s(k)\,D(k)\,\chi_{s'}(k)$, 
$\chi_s(k)\,C'(k)\,\chi_{s'}(k)$, $\chi_s(k)\,D'(k)\,\chi_{s'}(k)$
in the sense of Definition \defOSftcov. By Proposition \propOSrealpropbound.ii
and Lemma \lemOSdiffpropbound.i
$$\eqalign{
\v c \v_{1,\Si} & \le \const \sfrac{M^j\cb_j}{1-X} \cr
\v c-c' \v_{1,\Si} &\le \const \veps  \sfrac{M^j\cb_j X}{1- X} \cr
}\EQN\eqnOSNaiveII$$
For all $s,s'\in\Si$, the $L^1$--norm of $\chi_s(k)\,D(k)\,\chi_{s'}(k)$ is
bounded by $ \const \sfrac{\fl}{M^{2j}}\,M^j = \const \sfrac{\fl}{M^j} $. The
same holds for $\chi_s(k)\,D'(k)\,\chi_{s'}(k)$. Also, the $L^1$--norm of 
$$ 
\chi_s(k)\,D(k)\,\chi_{s'}(k)-\chi_s(k)\,D'(k)\,\chi_{s'}(k)\ 
= \chi_s(k)\,\big(v(k)-v'(k)\big)D(k)D'(k)\,\chi_{s'}(k)
$$
is bounded by $\veps\, \const \sfrac{\fl}{M^j} $. The same bounds
apply when $D$ is replaced by $C$. Consequently
$$\meqalign{
&\tn d \tn_{\infty} \le \const \sfrac{\fl}{M^j} \qquad,\quad&&
&\tn d' \tn_{\infty} \le \const \sfrac{\fl}{M^j} \qquad,\quad&&
&\tn d-d' \tn_{\infty} \le \veps\,\const \sfrac{\fl}{M^j}\cr 
&\tn c \tn_{\infty} \le \const \sfrac{\fl}{M^j} \qquad,\quad&&
&\tn c' \tn_{\infty} \le \const \sfrac{\fl}{M^j} \qquad,\quad&&
&\tn c-c' \tn_{\infty} \le \veps\,\const \sfrac{\fl}{M^j}\cr 
}\EQN\eqnOSNaiveIII$$
Also recall from Lemma \lemOSsectorladderfunctorialityII\ that
if
$$\meqalign{
 c_\Si\big((\xi,s),(\xi',s')\big)
& =\sum_{t\cap s \ne \emptyset \atop t'\cap s' \ne \emptyset}
c\big((\xi,t),(\xi',t')\big)  &&
 c'_\Si\big((\xi,s),(\xi',s')\big)
& =\sum_{t\cap s \ne \emptyset \atop t'\cap s' \ne \emptyset}
c'\big((\xi,t),(\xi',t')\big)  &&
\cr
d_\Si\big((\xi,s),(\xi',s')\big)
& = \sum_{t\cap s \ne \emptyset \atop t'\cap s' \ne \emptyset}
d\big((\xi,t),(\xi',t')\big) &&
d'_\Si\big((\xi,s),(\xi',s')\big)
& = \sum_{t\cap s \ne \emptyset \atop t'\cap s' \ne \emptyset}
d'\big((\xi,t),(\xi',t')\big)
\cr
}$$
then
$$\eqalign{
&f_1 \bullet ( C\otimes D ) \bullet f_2 
 = f_1 \circ ( c_\Si\otimes d_\Si ) \circ f_2  \qquad,\qquad
f_1 \bullet ( C'\otimes D' )\bullet f_2 
 = f_1 \circ ( c'_\Si\otimes d'_\Si )\circ f_2\cr
}$$

To prove the first inequality in part (i), observe that by Lemma \lemOSNaivetensor,
(\eqnOSNaiveII) and (\eqnOSNaiveIII)
$$
\V f_1 \circ ( c_\Si\otimes d_\Si )\circ f_2 \tV_{1,\Si} 
\le  \const \sfrac{M^j\cb_j}{1-X}\,\sfrac{\fl}{M^j}\
\v f_1 \tv_{1,\Si}  \v f_2\tv_{1,\Si} 
=   \const \sfrac{\fl\,\cb_j}{1-X}\ 
\v f_1 \tv_{1,\Si}  \v f_2\tv_{1,\Si}     
$$
Similarly
$$\eqalign{
\V f_1 \circ ( d_\Si\otimes c_\Si ) \circ f_2 \tV_{1,\Si} 
& \le  \const \sfrac{\fl\,\cb_j}{1-X}\
\v f_1 \tv_{1,\Si}  \v f_2\tv_{1,\Si}            \cr
\V f_1 \circ ( c_\Si\otimes c_\Si ) \circ f_2 \tV_{1,\Si} 
& \le \const \sfrac{\fl\,\cb_j}{1-X}\
\v f_1 \tv_{1,\Si}  \v f_2\tv_{1,\Si}             \cr
}$$    
By Definition \defOSbubbleprop.iii, 
$\cC(C,D)= C\otimes D + D\otimes C +C\otimes C$. Therefore, the first inequality of
part (i) follows. The proof of the other inequalities in part (i) is similar.

To prove the first inequality of part (ii), it suffices by Definition
\defOSbubbleprop.iii to bound each of the quantities 
$$\eqalign{
&\V f_1 \bullet ( C\otimes D )\bullet f_2 
-  f_1 \bullet ( C'\otimes D' )\bullet f_2 \tV_{1,\Si} \cr
&\V f_1 \bullet ( D\otimes C ) \bullet f_2 
-  f_1 \bullet ( D'\otimes C' )\bullet f_2 \tV_{1,\Si} \cr
&\V f_1 \bullet ( C\otimes C ) \bullet f_2 
-  f_1 \bullet ( C'\otimes C' )\bullet f_2 \tV_{1,\Si}  \cr
}$$    
by $\const \,\veps\,\fl\,\sfrac{\cb_j(1+X)}{1-X}\,
\v f_1 \tv_{1,\Si} \v f_2\tv_{1,\Si} $. Again we only bound the first
quantity; the other two are similar. As above,
by Lemma \lemOSNaivetensor, (\eqnOSNaiveII) and (\eqnOSNaiveIII)
$$\eqalign{
&\V f_1 \bullet ( C\otimes D ) \bullet f_2 - 
f_1 \bullet ( C'\otimes D')\bullet f_2 \tV_{1,\Si} \cr
&\hskip 2cm \le 
\V f_1 \circ  \big( c_\Si\otimes (d_\Si -d'_\Si)\big) \circ f_2 \tV_{1,\Si}
+ \V f_1 \circ \big( (c_\Si-c'_\Si)\otimes d'_\Si \big) \circ f_2 \tV_{1,\Si}\cr
&\hskip 2cm \le \const \Big( \sfrac{M^j\cb_j}{1-X}\,\veps\sfrac{\fl}{M^j} +  
\veps  \sfrac{M^j\cb_j X}{1- X}\,\sfrac{\fl}{M^j} \Big)\,
\v f_1 \tv_{1,\Si} \v f_2\tv_{1,\Si}  \cr
&\hskip 2cm \le \const \,\veps\,\fl\,\sfrac{\cb_j(1+X)}{1-X}\,
\v f_1 \tv_{1,\Si} \v f_2\tv_{1,\Si}  \cr
}$$
The proof of the other inequalities in part (ii) of the Lemma is similar.
\endproof

\corollary{\STM\corOSchoppedladder}{
Let $f\in\check\cF_{4;\Si}$ and let $C,\ D,\ C',\ D'$ be as in Lemma \lemOSNaiveBubble. Then
\Item i)
$$\eqalign{
\V L_\ell(f;C,D)\tV_{1,\Si} 
&\le  \Big( \const \sfrac{\fl\,\cb_j}{1-X}\Big)^\ell\ 
\v f\tv_{1,\Si}^{\kern5pt\ell+1}\cr
\V L_\ell(f;C,D)\tV_{ch,\Si} 
&\le  \Big( \const \sfrac{\fl\,\cb_j}{1-X}\Big)^\ell\ 
\v f\tv_{ch,\Si}^{\kern5pt\ell+1}\cr
}$$

\Item ii)
$$\eqalign{
\V L_\ell(f;C,D)-L_\ell(f;C',D')\tV_{1,\Si} 
&\le  \veps\,(1+X)\,
\Big( \const \sfrac{\fl\,\cb_j}{1-X}\Big)^\ell\
\v f\tv_{1,\Si}^{\kern5pt\ell+1}\cr 
\V L_\ell(f;C,D)-L_\ell(f;C',D')\tV_{ch,\Si} 
&\le  \veps\,(1+X)\,
\Big( \const \sfrac{\fl\,\cb_j}{1-X}\Big)^\ell\
\v f\tv_{ch,\Si}^{\kern5pt\ell+1}\cr 
}$$
}

\prf Part (i) follows by induction on $\ell$ from the first two inequalities of
Lemma \lemOSNaiveBubble.i using
$$
L_\ell(f;C,D)
   =L_{\ell-1}(f;C,D)\bullet\cC(C,D)\bullet f 
\EQN\eqnOSchopladder$$
To prove part (ii), observe that
$$\eqalign{
L_\ell(f;C,D)-L_\ell(f;C',D')
&= \Big[L_{\ell-1}(f;C,D)-L_{\ell-1}(f;C',D')\Big]
\bullet\cC(C,D)\bullet f\cr
&\hskip.5in +L_{\ell-1}(f;C',D')
\bullet\Big[\cC(C,D)-\cC(C',D')\Big]\bullet f
}\EQN\eqnOSladderdiff$$
and again apply induction on $\ell$, using part (i) and the first two inequalities
of Lemma \lemOSNaiveBubble.ii.

\endproof

\proposition{\STM\propOSNaiveLadder}{
Let $f\in\check\cF_{4;\Si}$.
Also let $u,v,u',v' \in \cF_0(2;\Si)$ be antisymmetric, spin independent, particle 
number conserving functions whose Fourier transforms  obey
$|\check u(k)|,|\check v(k)|,|\check u'(k)|,|\check v'(k)|
\le \half |\imath k_0-e(k)|$. Let $X\in\fN_{3}$ and $0<\veps\le 1$ such that
$$
\v u\v_{1,\Si},\,\v u'\v_{1,\Si} \le \sfrac{1}{2M^j}\,X \qquad\qquad 
\v u-u'\v_{1,\Si}\le \sfrac{\veps}{M^j}\, X
$$
and 
$$
|\check u(k) -\check u'(k)|,\, |\check v(k) -\check v'(k)|
\le \veps |\imath k_0-e(k)|
$$
Set
$$\meqalign{
C(k) &= \sfrac{\nu^{(j)}(k)}{\imath k_0 -e(\k) -\check u(k)}
\qquad,\qquad 
D(k) &= \sfrac{\nu^{(\ge j+1)}(k)}{\imath k_0 -e(\k) -\check v(k)} \cr
C'(k) &= \sfrac{\nu^{(j)}(k)}{\imath k_0 -e(\k) -\check u'(k)}
\qquad,\qquad 
D'(k) &= \sfrac{\nu^{(\ge j+1)}(k)}{\imath k_0 -e(\k) -\check v'(k)} \cr
}$$
and let $C(\xi,\xi'),\,D(\xi,\xi'),C'(\xi,\xi'),\,D'(\xi,\xi')$ be their Fourier 
transforms as in Definition \defOSftcov. Assume that
$X_\0 \le \min\{\tau_1,\tau_2\}$, where $\tau_1$ and $\tau_2$ are
the constants of Proposition \propOSrealpropbound\ and
Lemma \lemOSdiffpropbound, respectively. Then for all $\ell \ge 1$
\Item i)
$$\eqalign{
\V L_\ell(f;C,D)\tV_{1,\Si} 
&\le  \Big( \const \sfrac{\fl\,\cb_j}{1-X}\Big)^\ell\ 
\v f\tv_{1,\Si}^{\kern5pt\ell+1} 
\cr
\V L_\ell(f;C,D)\tV_{3,\Si} 
&\le  \Big( \const \sfrac{\fl\,\cb_j}{1-X}\Big)^\ell\
\v f\tv_{ch,\Si}^{\kern5pt \ell}\,\v f\tv_{3,\Si}\cr
}$$

\Item ii)
$$
\V L_\ell(f;C,D)
    -L_\ell(f;C',D') \tV_{3,\Si} 
\le \veps\,(1+X)\,
\Big( \const \sfrac{\cb_j}{1-X}\Big)^\ell\
\v f\tv_{3,\Si}^{\kern5pt \ell+1}
$$
}

\prf
The first inequality of part (i) was already stated in Corollary 
\corOSchoppedladder.i. By (\eqnOSchopladder), the second inequality of part (i)
follows from Corollary  \corOSchoppedladder.i and the third inequality of 
Lemma \lemOSNaiveBubble.i.

With an argument as above, using Lemma \lemOSNaiveBubble\ and Corollary
\corOSchoppedladder\ one deduces from (\eqnOSladderdiff) that
$$
\V L_\ell(f;C,D)
    -L_\ell(f;C',D') \tV_{3,\Si} 
\le \veps\,(1+X)\,
\Big( \const \sfrac{\fl\,\cb_j}{1-X}\Big)^\ell\
\v f\tv_{ch,\Si}^{\kern5pt \ell}\ \v f\tv_{3,\Si}
$$
the claim now follows Corollary \corchannelnorm.
\endproof

\remark{\STM\remOSnaiveladderest}{
Using Corollary \corchannelnorm, one also sees that in the
situation of  Proposition \propOSNaiveLadder,
$$\eqalign{
\V L_\ell(f;C,D)\tV_{3,\Si} 
&\le  \Big( \const \sfrac{\cb_j}{1-X}\Big)^\ell\
\v f\tv_{3,\Si}^{\kern5pt \ell+1}\cr
}$$

}

\vfill\eject

\titlea{ References}\PG\pgOSIIIref

\item{[FKTf1]} J. Feldman, H. Kn\"orrer, E. Trubowitz, 
{\bf A Two Dimensional Fermi Liquid, Part 1: Overview}, preprint.
\smallskip%
\item{[FKTf2]} J. Feldman, H. Kn\"orrer, E. Trubowitz, 
{\bf A Two Dimensional Fermi Liquid, Part 2: Convergence}, preprint.
\smallskip%
\item{[FKTl]} J. Feldman, H. Kn\"orrer, E. Trubowitz, 
{\bf  Particle--Hole Ladders}, preprint.
\smallskip%
\item{[FKTr1]} J. Feldman, H. Kn\"orrer, E. Trubowitz, 
{\bf Convergence of Perturbation Expansions in Fermionic Models, Part 1: Nonperturbative Bounds}, preprint.
\smallskip%
\item{[FKTr2]} J. Feldman, H. Kn\"orrer, E. Trubowitz, 
{\bf Convergence of Perturbation Expansions in Fermionic Models, Part 2: Overlapping Loops}, preprint.

\vfill\eject

\vsize = 9.3truein
\hoffset=-0.1in
\titlea{Notation }\PG\pgOSIIInot
\null\vskip-0.5in
\titleb{Norms}
\centerline{
\vbox{\offinterlineskip
\hrule
\halign{\vrule#&
         \strut\hskip0.05in\hfil#\hfil&
         \hskip0.05in\vrule#\hskip0.05in&
          #\hfil\hfil&
         \hskip0.05in\vrule#\hskip0.05in&
          #\hfil\hfil&
           \hskip0.05in\vrule#\cr
height2pt&\omit&&\omit&&\omit&\cr
&Norm&&Characteristics&&Reference&\cr
height2pt&\omit&&\omit&&\omit&\cr
\noalign{\hrule}
height2pt&\omit&&\omit&&\omit&\cr
&$\tn\ \cdot\ \tn_{1,\infty}$&&no derivatives, external positions, acts on functions&&Example \exOSSymmNorm&\cr
height4pt&\omit&&\omit&&\omit&\cr
&$\|\ \cdot\ \|_{1,\infty}$&&derivatives, external positions, acts on functions&&Example \exOSSymmNorm&\cr
height4pt&\omit&&\omit&&\omit&\cr
&$\|\ \cdot\ \cnorm_\infty$&&derivatives, external momenta, acts on functions
&&Definition \defOSderivmom&\cr
height4pt&\omit&&\omit&&\omit&\cr
&$\tn\ \cdot\ \tn_{\infty}$&&no derivatives, external positions, acts on functions&&Example \exOSelloneinftycontr&\cr
height4pt&\omit&&\omit&&\omit&\cr
&$\|\ \cdot\ \cnorm_1$&&derivatives, external momenta, acts on functions
&&Definition \defOSderivmom&\cr
height4pt&\omit&&\omit&&\omit&\cr
&$\|\ \cdot\ \cnorm_{\infty,B}$&&derivatives, external momenta, $B\subset\bbbr\times\bbbr^d$
&&Definition \defOSderivmom&\cr
height4pt&\omit&&\omit&&\omit&\cr
&$\|\ \cdot\ \cnorm_{1,B}$&&derivatives, external momenta, $B\subset\bbbr\times\bbbr^d$
&&Definition \defOSderivmom&\cr
height4pt&\omit&&\omit&&\omit&\cr
&$\|\ \cdot\ \|$&&$\rho_{m;n}\|\ \cdot\ \|_{1,\infty}$&&Lemma \lemOSscalednorm&\cr
height4pt&\omit&&\omit&&\omit&\cr
&$N(\cW;\cb,\ib,\al)$&&$\sfrac{1}{\ib^2}\,\cb\!\sum_{m,n\ge 0}\,
\al^{n}\,\ib^{n} \,\|\cW_{m,n}\|$&&Definition \defOSgrnorm&\cr
height4pt&\omit&&\omit&&\omit&\cr
& && &&Theorem  \thmOSinsulators&\cr
height4pt&\omit&&\omit&&\omit&\cr
&$N_0(\cW;\be;X,\vec\rho)$&&$\fe_0(X)\ \sum_{m+n\in 2\bbbn}\,
\be^{n}\rho_{m;n} \,\|\cW_{m,n}\|_{1,\infty}$
&&Theorem  \thmOSfirststep&\cr
height4pt&\omit&&\omit&&\omit&\cr
&$\|\ \cdot\ \|_{L^1}$&&derivatives, acts on functions on $\bbbr\times\bbbr^d$
&&before Lemma \lemOSprepintup&\cr
height4pt&\omit&&\omit&&\omit&\cr
&$\|\ \cdot\ \tnorm$&&derivatives, external momenta, acts on functions
&&Definition \defOSdiffdecaynorm&\cr
height4pt&\omit&&\omit&&\omit&\cr
&$N^\sim_0(\cW ;\be;X,\vec\rho)$&&$\fe_0(X)
\!\sum_{m+n\in 2\bbbn}\,\be^{m+n}\rho_{m;n} \,\| W^\sim_{m,n}\tnorm$
&&before Lemma \lemOSTZsourceterm&\cr
height4pt&\omit&&\omit&&\omit&\cr
&$\v \ \cdot\  \tv$&&like $\rho_{m;n}\|\ \cdot\ \tnorm$ but acts on $\tilde V^{\otimes n}$
&&Theorem  \thmOSTfirststep&\cr
height4pt&\omit&&\omit&&\omit&\cr
&$N^\sim(\cW ;\cb,\ib, \al)$&&$\sfrac{1}{\ib^2}\cb\,\smsum_{m,n} 
\al^{m+n}\, \ib^{m+n}\, \v W^\sim_{m,n} \tv$
&&Theorem  \thmOSTfirststep&\cr
height4pt&\omit&&\omit&&\omit&\cr
&$\v \ \cdot\ \v_{p,\Si}$&&derivatives, external positions, 
all but $p$ sectors summed
&&Definition \defOSsectnorm&\cr
height4pt&\omit&&\omit&&\omit&\cr
&$\v \varphi \v_{\Si}$&&$\rho_{m;n}\cases{
\v \varphi \v_{1,\Si} + \sfrac{1}{\fl}\,\v \varphi \v_{3,\Si}
+ \sfrac{1}{\fl^2}\,\v \varphi \v_{5,\Si} 
    & if $m=0$ \cr
\sfrac{\fl}{M^{2j}}\,\v \varphi \v_{1,\Si} & if $m\ne0$} $
&&Definition \defOSscalednorms&\cr
height4pt&\omit&&\omit&&\omit&\cr
&$N_j(w;\al;\,X,\Si,\vec\rho)$&&$\sfrac{M^{2j}}{\fl}\,\fe_j(X) 
\smsum_{m,n\ge 0}\,
\al^{n}\,\big(\sfrac{\fl\,\IB}{M^j}\big)^{n/2} \,\v w_{m,n}\v_\Si$
&&Definition \defOSscalednorms&\cr
height4pt&\omit&&\omit&&\omit&\cr
&$\v \ \cdot\ \tv_{p,\Si}$&&derivatives, external momenta, all but $p$ sectors
summed
&&Definition \defOSsectdiffdecaynorm&\cr
height4pt&\omit&&\omit&&\omit&\cr
&$\v \ \cdot\ \tv_{p,\Si,\vec\rho}$&&weighted variant of $\v \ \cdot\ \tv_{p,\Si}$
&&Definition \defOSmomscalednorms.i&\cr
height4pt&\omit&&\omit&&\omit&\cr
&$\v f \tv_\Si$&&$\rho_{m;n}\cases{
\v f\tv_{1,\Si}+\sfrac{1}{\fl}\,\v f\tv_{3,\Si}
              +\sfrac{1}{\fl^2}\,\v f \tv_{5,\Si} 
    & if $m=0$ \cr
\smsum_{p=1}^6\sfrac{1}{\fl^{[(p-1)/2]}}\v f\tv_{p,\Si}
  & if $m\ne0$} $
&&Definition \defOSmomscalednorms.ii&\cr
height4pt&\omit&&\omit&&\omit&\cr
&$N_j^\sim(w;\al;\,X,\Si,\vec\rho)$&&$\sfrac{M^{2j}}{\fl}\,\fe_j(X) 
\smsum_{n\ge 0}\,
\al^{n}\,\big(\sfrac{\fl\,\IB}{M^j}\big)^{n/2} \,
\v f_n\tv_\Si$
&&Definition \defOSmomscalednorms.iii&\cr
height4pt&\omit&&\omit&&\omit&\cr
&$\v\ \cdot\ \tv_{ch,\Si}$&&channel variant of $\v\ \cdot\ \tv_{2,\Si}$
for ladders
&&Definition \defOSchannelnorm&\cr
height4pt&\omit&&\omit&&\omit&\cr
&$\v\ \cdot\ \v_{ch,\Si}$&&channel variant of $\v\ \cdot\ \v_{2,\Si}$
for ladders
&&Definition \defOSchannelnorm&\cr
height4pt&\omit&&\omit&&\omit&\cr
}\hrule}}

\vfil
\goodbreak
\titleb{Other Notation}
\centerline{
\vbox{\offinterlineskip
\hrule
\halign{\vrule#&
         \strut\hskip0.05in\hfil#\hfil&
         \hskip0.05in\vrule#\hskip0.05in&
          #\hfil\hfil&
         \hskip0.05in\vrule#\hskip0.05in&
          #\hfil\hfil&
           \hskip0.05in\vrule#\cr
height2pt&\omit&&\omit&&\omit&\cr
&Not'n&&Description&&Reference&\cr
height2pt&\omit&&\omit&&\omit&\cr
\noalign{\hrule}
height2pt&\omit&&\omit&&\omit&\cr
&$\Om_S(\cW)(\phi,\psi)$
&&$\log\sfrac{1}{Z} \int  e^{\cW(\phi,\psi+\ze)}\,d\mu_{S}(\ze)$
&&before (\eqnOSintroI)&\cr
height2pt&\omit&&\omit&&\omit&\cr
&$J$&&particle/hole swap operator&&(\eqnOSjdef)&\cr
height2pt&\omit&&\omit&&\omit&\cr
&$\tilde \Om_C(\cW)(\phi,\psi)$
&&$\log \sfrac{1}{Z}\int e^{\phi J\ze}\,e^{\cW(\phi,\psi +\ze)} d\mu_C(\ze)$
&&Definition \defOSrengrpmap&\cr
height2pt&\omit&&\omit&&\omit&\cr
&$r_0$&&number of $k_0$ derivatives tracked&&\S\CHintroII&\cr
height2pt&\omit&&\omit&&\omit&\cr
&$r$&&number of $\k$ derivatives tracked&&\S\CHintroII&\cr
height2pt&\omit&&\omit&&\omit&\cr
&$M$&&scale parameter, $M>1$&&before Definition \defOSscales&\cr
height2pt&\omit&&\omit&&\omit&\cr
&$\const$&&generic constant, independent of scale&& &\cr
height2pt&\omit&&\omit&&\omit&\cr
&$\abcst$&&generic constant, independent of scale and $M$&& &\cr
height2pt&\omit&&\omit&&\omit&\cr
&$\nu^{(j)}(k)$&&$j^{\rm th}$ scale function&&Definition \defOSscales&\cr
height2pt&\omit&&\omit&&\omit&\cr
&$\tilde\nu^{(j)}(k)$&&$j^{\rm th}$ extended scale function
&&Definition \defOSextendedshell.i&\cr
height2pt&\omit&&\omit&&\omit&\cr
&$\nu^{(\ge j)}(k)$&&$\varphi\big(M^{2j-1}(k_0^2+e(\k)^2)\big)$&&Definition \defOSscales&\cr
height2pt&\omit&&\omit&&\omit&\cr
&$\tilde\nu^{(\ge j)}(k)$
&&$\varphi\big(M^{2j-2}(k_0^2+e(\k)^2)\big)$
&&Definition \defOSextendedshell.ii&\cr
height2pt&\omit&&\omit&&\omit&\cr
&$\bar\nu^{(\ge j)}(k)$
&&$\varphi\big(M^{2j-3}(k_0^2+e(\k)^2)\big)$
&&Definition \defOSextendedshell.iii&\cr
height2pt&\omit&&\omit&&\omit&\cr
&$\fl$&&length of sectors&&Definition \defOSsectors&\cr
height2pt&\omit&&\omit&&\omit&\cr
&$\Si$&&sectorization &&Definition \defOSsectors&\cr
height2pt&\omit&&\omit&&\omit&\cr
&$S(C)$&&$\sup_m\sup_{\xi_1,\cdots,\xi_m \in \cB}\
\Big(\ \Big| \int \psi(\xi_1)\cdots\psi(\xi_m)\,d\mu_C(\psi) \Big|\ \Big)^{1/m}$&&Definition \defIntBndsS&\cr
height2pt&\omit&&\omit&&\omit&\cr
&$\IB$&&$j$--independent constant&&Definitions \defOSscalednorms,\defOSmomscalednorms&\cr
height2pt&\omit&&\omit&&\omit&\cr
&$\cb_j$&& $
=\sum_{|\bde|\le r\atop |\de_0|\le r_0}  M^{j|\de|}\,t^\de
+\sum_{|\bde|> r\atop {\rm or\ }|\de_0|> r_0}\infty\, t^\de
\in\fN_{d+1}
$&&Definition \defOScbj&\cr
height2pt&\omit&&\omit&&\omit&\cr
&$\fe_j(X)$&& $= \sfrac{\cb_j}{1-M^j X}$&&Definition \defOSscalednorms.ii&\cr
height2pt&\omit&&\omit&&\omit&\cr
&$*$&& convolution&&before (\eqnOSexpandc)&\cr
height2pt&\omit&&\omit&&\omit&\cr
&$\circ$&& ladder convolution&&Definition \defOSbubbleprop.iv&\cr
height2pt&\omit&&\omit&&\omit&\cr
&$\bullet$&& ladder convolution&&Definitions \defOSsectbubbleprop,\defOSsectbubblepropII&\cr
height2pt&\omit&&\omit&&\omit&\cr
&$\check f$&&Fourier transform&&Definition \defOSfourtrans.i&\cr
height2pt&\omit&&\omit&&\omit&\cr
&$\check u$&&Fourier transform for sectorized $u$&&Definition \defOSsectrepr.iv&\cr
height2pt&\omit&&\omit&&\omit&\cr
&$f^\sim$&&partial Fourier transform&&Definition \defOSfourtrans.ii&\cr
height2pt&\omit&&\omit&&\omit&\cr
&$\hat\chi$&&Fourier transform&&Definition \defOSfourtransII&\cr
height2pt&\omit&&\omit&&\omit&\cr
&$\cB$&&$\bbbr \times \bbbr^d \times \{\uparrow, \downarrow\}\times\{0,1\}$ 
viewed as position space&&beginning of \S\CHnorms&\cr
height2pt&\omit&&\omit&&\omit&\cr
&$\check \cB$&&$\bbbr\times\bbbr^d\times\{\uparrow, \downarrow\}\times\{0,1\}$ 
viewed as momentum space&&beginning of \S \CHfourier&\cr
height2pt&\omit&&\omit&&\omit&\cr
&$\check \cB_m$&&$\set{(\check \eta_1,\cdots,\check \eta_m)\in \check \cB^m}
{\check \eta_1+\cdots+\check \eta_m=0}$&&before Definition \defOSamptransinv&\cr
height2pt&\omit&&\omit&&\omit&\cr
&$\fX_\Si$&&$\check \cB\dunion(\cB\times\Si)$&&Definition \defOSdisjointfield&\cr
height2pt&\omit&&\omit&&\omit&\cr
&$\cF_m(n)$&&functions on $\cB^m \times \cB^n$, antisymmetric in $\cB^m$
arguments&&Definition \defOSFmn&\cr
height2pt&\omit&&\omit&&\omit&\cr
&$\check\cF_m(n)$&&functions on $\check\cB^m \times \cB^n$, antisymmetric in $\check\cB^m$
arguments&&Definition \defOScheckcF&\cr
height2pt&\omit&&\omit&&\omit&\cr
&$\cF_m(n;\Si)$&&functions on $\cB^m \times  \big( \cB \times\Si \big)^n$,
internal momenta in sectors&&Definition \defOSsectrepr.ii&\cr
height2pt&\omit&&\omit&&\omit&\cr
&$\check\cF_m(n;\Si)$&&functions on $\check\cB^m \times  \big( \cB \times\Si \big)^n$,
internal momenta in sectors&&Definition \defOSsectcheckcF.i&\cr
height2pt&\omit&&\omit&&\omit&\cr
&$\check \cF_{n;\Si}$&&functions on $\fX_\Si^n$ that reorder to 
$\check \cF_{m}(n-m;\Si)$'s&&Definition \defOSsectcheckcF.iii&\cr
height4pt&\omit&&\omit&&\omit&\cr
}\hrule}}

\end